\documentclass[rmp,showpacs,preprintnumbers,amsmath,amssymb]{revtex4}

\usepackage[english]{babel} 
\usepackage[latin1]{inputenc} 
\usepackage[T1]{fontenc}
\usepackage{color}
\usepackage{latexsym} 
\usepackage{epsfig} 
\usepackage{graphicx}
\usepackage{lscape}
\usepackage{color}
\usepackage{graphicx}
\usepackage{bm}
\usepackage{upgreek}
\usepackage{epsfig} 
\usepackage{ulem}

 \definecolor{violet}{rgb}{0.5,0,0.5}
\definecolor{marron}{rgb}{0.8,0.6,0.4}
\definecolor{jaune}{rgb}{0.98,0.98,0.4}
\definecolor{gris}{rgb}{0.6,0.6,0.6}

        \makeatletter
        \let\@vieuxmakecaption=\@makecaption
        \newcommand{\@makefigcaption}[2]{{\small\@vieuxmakecaption{#1}{#2}}}

        \newenvironment{figuresmall}
        {\let\@makecaption\@makefigcaption\begin{figure}}
        {\end{figure}}
        \makeatother

\newcommand\po[1]
{\paragraph{#1}

 ~

}

\newcommand\popetit[1]
{
~

\textit{#1}

}

\newcommand\doublimage[6]
{
\begin{figuresmall}[h!]
 \small

\begin{center}

\begin{tabular}{c c}
\includegraphics[width=6cm]{#1} & \includegraphics[width=6cm]{#3} \\
#2 & #4 \\
\end{tabular}

\caption{#5}
\label{#6}
\end{center}
\normalsize
\end{figuresmall}
}

\newcommand\disp{\displaystyle}

\newcommand\refs[1]
{(see section~\ref{#1})}

\newcommand\reft[1]
{Table \ref{#1}}

\newcommand\refm[1]
{(\ref{#1})}

\newcommand\finsection{
\cleardoublepage 
}

\newcommand\imagea[4]
{\begin{figuresmall}[h!]
\begin{center}
\includegraphics[height=#4 cm]{#1}
\small
\caption{#2}
\normalsize
\label{#3}
\end{center}
\end{figuresmall}}

\newcommand\imageal[4]
{\begin{figuresmall}[h!]
\begin{center}
\includegraphics[height=#4]{#1}
\small
\caption{#2}
\normalsize
\label{#3}
\end{center}
\end{figuresmall}}

\newcommand\refi[1]
{(see figure~\ref{#1})}

\newcommand\doublimagem[6]
{

\begin{figuresmall}[h!]\small
   \begin{minipage}[c]{.46\linewidth}
\begin{center}
      \includegraphics[width=.95\linewidth]{#1}

#2
\end{center}
   \end{minipage} \hfill
   \begin{minipage}[c]{.46\linewidth}
\begin{center}
      \includegraphics[width=.95\linewidth]{#3}

 #4
\end{center}
   \end{minipage}\hfill

\caption{ #5}\label{#6}\normalsize
\end{figuresmall} 

}

\newcommand\old[1]
{{\textcolor{gris}{}}}

\newcommand\new[1]
{#1}

\newcommand\bla[1]
{\textcolor{red}{}}

\begin{document}

\title{Intermittent search strategies}

\author{O. B\'enichou}
\affiliation{UPMC Univ Paris 06, UMR 7600 Laboratoire de Physique Th\'eorique de la Mati\`ere Condens\'ee,
 4 Place Jussieu, F-75005 Paris, France.}

\author{C. Loverdo}
\affiliation{UPMC Univ Paris 06, UMR 7600 Laboratoire de Physique Th\'eorique de la Mati\`ere Condens\'ee,
 4 Place Jussieu, F-75005 Paris, France.}

\author{M. Moreau}
\affiliation{UPMC Univ Paris 06, UMR 7600 Laboratoire de Physique Th\'eorique de la Mati\`ere Condens\'ee,
 4 Place Jussieu, F-75005 Paris, France.}

\author{R. Voituriez}
\affiliation{UPMC Univ Paris 06, UMR 7600 Laboratoire de Physique Th\'eorique de la Mati\`ere Condens\'ee,
 4 Place Jussieu, F-75005 Paris, France.}

\date{\today}

\begin{abstract}

This review examines intermittent target search strategies, which  combine  phases of slow motion, allowing the searcher to detect the target, and phases of fast motion during which targets cannot be detected.
We first show that intermittent search strategies are actually widely observed at various scales. At the macroscopic scale, this is for example the case of  animals looking for food ; at the microscopic scale, intermittent transport patterns are involved in reaction pathway of DNA binding proteins as well as in intracellular transport. Second, we introduce generic stochastic models, which show that intermittent strategies are efficient strategies, which enable to minimize the search time.  This suggests that the intrinsic efficiency of intermittent search strategies could justify their frequent observation in nature. Last, beyond these modeling aspects, we propose that intermittent strategies could be used also in a broader context to design and accelerate search processes.

\end{abstract}


\maketitle

\tableofcontents

\section{Introduction}

\subsection{General scope and outline}

What is the best strategy to find a missing object ? Anyone who has ever lost his keys has already been facing this problem. This  every day life situation is a prototypical example of search problem, which under its simplest form  involves a searcher -- either a person, an animal, or any kind of organism or particle -- in general able to move across the search domain, and one or several  targets.  Even if very schematic, the search problem as stated   turns out to be a quite universal question, which pops up at different scales and in  various fields,  and has generated an increasing number of works in recent years, notably in the  physics community.

Theoretical studies of search strategies can be traced back to World War II, during which the US navy tried to most efficiently hunt for submarines and developed rationalized search procedures \cite{Uboats_patterns, Shlesinger:2009gf}. Similar search algorithms have then been developed and utilized in the context of castaway rescue operations \cite{gardecote}, or even for  the recovery of  an atomic bomb lost in the Mediterranean sea near Palomares in 1966. One can cite for instance the  rescue of the Scorpion, a nuclear submarine lost near the Azores in 1968  \cite{scorpion}. At the macroscopic scale, other important and  widely studied examples of search processes concern animals searching for mate, food or shelter   \cite{charnov,animauxObrien,Bell,viswaNat,PRE2006,NewsViewsNature,revisitingViswaNat} which will be discussed in more details in this review. One can even mention pre-historic migrations, which, apart from classical archaeological literature, 
have also been studied  as a search problem, in which  human groups search for new profitable territories \cite{Flores:2007tv}. At the microscopic scale, search processes naturally pop up in the context of chemical reactions, for which the encounter of reactive molecules -- or in other words the fact that one searcher molecule finds a reactive target site --  is a required first  step. One should obviously mention the theory of diffusion-controlled reactions, initiated many years ago by the celebrated theory of \textcite{smolu} and developed by innumerable researchers (see for instance the review by  \textcite{revueHanggiReacDiff}).  More recently, this field has regained interest in the context of biochemical reactions in cells, where the sometimes very small number of reactive molecules makes this first step of search for a reaction partner crucial for the kinetics. One can think for instance of reactions involved in genomic transcription,  a representative example of which being  the search for   specific DNA sequences  by transcription factors \cite{winter1,revue_hippel,revue_dna_tirfm,NAR08,natphys08comment}.

\begin{figure*}
\begin{center}
\includegraphics[scale = 0.4]{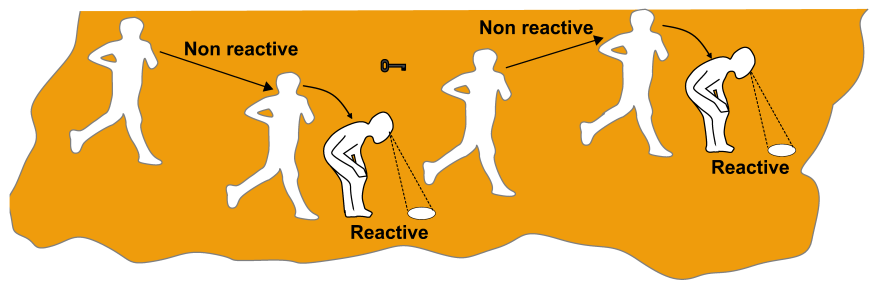}
\caption{\small Illustration of intermittent reaction paths by an every-day life example of search problem. The searcher looks for a target. The searcher  alternates fast relocation phases, which are not reactive as they do not allow for target detection, and slow reactive phases which permit target detection. }\label{key}
\end{center}
\end{figure*}

In all these examples, the time needed to discover a target  is a limiting quantity, and consequently minimizing this search time often appears as essential.  In order to get an intuition of what could be an efficient search strategy on general grounds, let us go back to
the everyday-life example mentioned above  (see figure \ref{key}). We consider a searcher  who lost a tiny object -- let us say a key  --  in a  large sandy beach, where the key  is so small that it cannot be detected  if the searcher  passes by too fast.
In addition, we assume that the searcher  has no prior information on the position of the key, 
except that the key is in a bounded domain (the beach). What 
is then the best strategy for the searcher to find the key  as fast as possible?
A first strategy  consists in a slow and careful exploration (to make sure that the key will be detected upon encounter) of the sand all along the beach. In the case of a very large beach, the search time can then be very long. An alternative strategy one
can think of consists in interrupting the slow and careful exploration of the sand by mere displacement
phases, during which the searcher  relocates on the beach very fast, but without even trying to detect  the key (typically the searcher "runs"). 
We will call hereafter intermittent search strategies  such processes that combine two distinct phases : a phase of slow displacement which enables target detection, and a phase of faster motion during which the target cannot be detected (note that the word intermittent has also been used more recently by \textcite{BartumeusDefiniSonIntermittent}  with another definition). 

The efficiency of such intermittent strategies results from a trade--off between speed and detection and  can be qualitatively discussed. Intuitively, the advantage of the fast relocation phases for the searcher is to reach unvisited regions.  The drawback is however that during these phases  time is consumed without any chance of detecting the target. The net efficiency of this strategy is therefore not trivial, and 
these last years, many works have focused on the following questions: (i) Can phases of fast motion which disable detection make the global search more efficient? (ii) If so, is there an optimal way for the searcher to share the time between the two phases ? (iii) Are these intermittent search patterns relevant to
the description of real situations ?

The goal of this article is to  review these works while bringing explicit answers to these questions. More precisely, we will first show that   intermittent transport  patterns  are actually widely {\it observed}
at various scales. At the macroscopic scale, this is for example the case of foraging animals (see section \ref{section_animaux}) ; at the microscopic scale, we will show that intermittent transport patterns are involved in reaction pathway of DNA binding proteins as well as 
in intracellular transport (see section \ref{section_ADN}). Second, we will show on generic stochastic models that intermittent strategies are {\it efficient} strategies which enable to minimize the search time (see sections \ref{section_animaux},\ref{section_ADN},\ref{section_generic}), and therefore suggest that this efficiency could justify their frequent observation in nature. Last,  beyond these modeling aspects, we will   propose that intermittent strategies  could be used also in a broader context to {\it design and  accelerate} search processes.

\subsection{General framework and first definitions}

The search problem can take multiple forms  \cite{revuejpA}; in this section we define more precisely the framework of this review -- namely the random intermittent search strategies -- and introduce the main hypothesis which will be made.

\subsubsection{Searching with or without cues}

Although in essence in a search problem  the target location is unknown and cannot be found from a rapid inspection of the search domain,  in practical cases there are often cues which restrict the territory to explore, or give indications on how to explore it.  We can quote the very classical example  of chemotaxis  \cite{ecoliinmotion}, 
which keeps arising  interest in the biological and physical communities (see for example \textcite{ecoli_topo,yariv_chemotaxis, Tailleur:2008rr}).
Bacteria like E.coli swim with a succession of ``runs'' (approximately straight moves) 
and ``tumbles'' (random changes of direction). 
When they sense a gradient of chemical concentration, 
they swim up or down the gradient by adjusting their tumbling rate~: 
when the environment is becoming more favorable, they tumble less, 
whereas they tumble more when their environment is degrading. This behavior results in a bias
towards the most favorable locations of high concentration of chemoattractant which can be  as varied as salts, glucose, amino-acids, oxygen, etc...
More recently it has been shown that a similar  behavior can also be triggered by other kinds of external signals such as 
temperature gradients \cite{thermotaxie,PRLlibchaber,Hanna_natcell} or light intensity \cite{lighttaxie}.

Chemotactic search requires a well defined gradient of chemoattractant, and is therefore applicable only when the concentration of cues is sufficient. On the contrary, at low concentrations cues can be sparse, or even discrete signals which do not allow for a gradient based strategy. It is for example the case of animals sensing odors in air or water where  the mixing in the potentially turbulent flow breaks up the chemical signal into  random and disconnected patches of high concentration. 
\textcite{infotaxis} proposed a search algorithm, which they called 'infotaxis', designed to work in this case of sparse and fluctuating cues. This algorithm, based on a maximization of  the expected rate of information gain produces trajectories such as 
 'zigzagging' and 'casting' paths which are  similar to those observed in the flight of moths \cite{Balkovsky:2002yj}. 
 
 In this review we focus on the extreme case where no cue is present that could lead the searcher to the target. This assumption applies to targets which  can be detected only if the searcher is within a given detection radius $a$ which is much smaller than the typical extension of the search domain. In particular this assumption clearly covers the case of search problems at the scale of chemical reactions, and more generally the case of searchers whose motion is independent of any exterior cue that could be emitted by the target.

\subsubsection{Systematic vs random strategies}

Whatever the scale, the behavior of a searcher relies strongly on his ability, or incapability, to keep memories of his past explorations. Depending on the searcher and on the space to explore, such kind of spatial memory can play a more or less important role \cite{Moreau:2009}. In an extreme case the searcher, for instance  human or animal, can have a  mental map of the exploration space and can thus perform a systematic search. Figure 2 presents several systematic patterns : lawn-mower, expanding square, spiral (for more patterns, see for example \textcite{Uboats_patterns}). These type of search have been extensively studied, in particular for designing efficient search operated by humans (Dobbie, 1968; Stone, 1989). 

\begin{figure*}
\begin{center}
\includegraphics[height=4 cm]{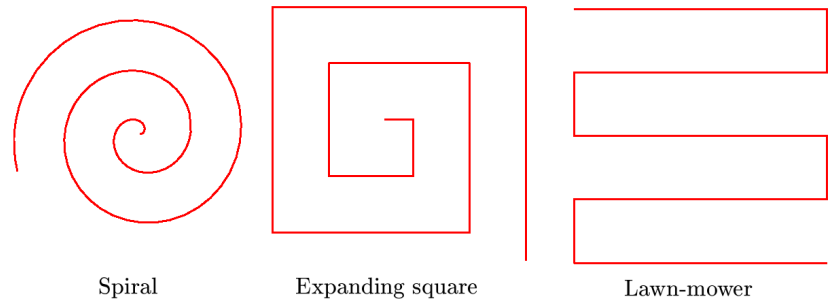}
\caption{Examples of patterns for systematic exploration of space}
\label{patterns_systematiques}
\end{center}
\end{figure*}

In the opposite case where the searcher has low  -- or no -- spatial memory abilities the search trajectories can  be qualified as random, and the theory of stochastic processes 
provides powerful tools for their quantitative analysis (for a short reminder on random walks, see Appendix \ref{resume_random_walks}). 
 This is obviously the case for ``molecular'' searchers at the microscopic scale which are subject to thermal Brownian motion, 
but also at  larger scales of animals with low cognition skills. This review is mainly focused on random search problems, and effects of spatial memory will be briefly discussed in the last section.

Note that we will keep using  the word "strategy"   for  animals with low cognitive abilities and even for  molecules, although such searchers are not able to design strategies themselves since of course their dynamics are simply governed by the laws of physics. 
In the context of proteins searching for targets on DNA, we mean  that the search time depends on parameters such as the ionic strength or the protein/DNA affinity which, if varied, can lead to a minimization of the search time. In the case where the search kinetics is a limiting constraint,  such good or even optimal values of these parameters  might have been selected  in the course of evolution. We call strategy this very fact that physical parameters can be tuned (implicitly by evolution) to optimize a biological function.
Note however that the real optimization problem depends on many parameters and constraints. The  models  studied in this review 
are restricted to kinetic constraints, which can be dominant both at the microscopic and macroscopic scale, as discussed in sections II and III. This key assumption will be used throughout this review.

\subsubsection{Framework of this review}

To summarize, in this review we shall focus on intermittent search strategies for targets which emit no cue. 
The searchers will be assumed to have  no (or low) memory skills,
resulting in their trajectories being intermittent random walks.
Depending on the example to be treated, different quantities can be used to assess the efficiency of search strategies,  such as the energy necessary for reaching the first prey, the number of preys collected in a given time, or the time for encountering the first prey. In this review we will discuss the efficiency of search strategies uniquely from a kinetic point of view. We shall mainly consider the mean first passage time to a target as a quantitative measure of the search efficiency,  and study the minimization of  this quantity. Note that the full distribution of the first-passage time is a priori needed to quantify the search kinetics on all time scales. However, in most of the situations that we will consider in this review, in can be checked numerically that the distributions of the search time can be well approximated by an exponential, which means that the kinetics is fully characterized by the mean first-passage time.

\section{Intermittent search strategies at the macroscopic scale}
 \label{animaux1Dcorrel}

\label{section_animaux}

 Searching for a randomly located object is one of the most frequent  tasks of living organisms, be it    for obtaining food, a sexual partner or a shelter  \cite{Bell}. In these examples, the search time is generally a limiting factor which has to be optimized for the  survival of the species. The question of determining the efficiency of  a search behavior    is thus  a crucial problem of behavioral ecology, which has inspired numerous experimental \cite{Bell, animauxObrien,OBrien:1989,kramer} and theoretical  \cite{Viswanathan:1996, viswaNat,Animaux1D, PRE2006,spidermonkeys, intermittentlevy} works . 
 In this context, L\'evy walk strategies have been proved to play a crucial role in such optimization problems. In this section, we first remind why these L\'evy walks are advantageous with respect to simple random walks when searching randomly, as
first  put forward by \textcite{levyKlafterTot}. We also remind the pioneering model of Viswanathan et al., which has played a major role in the development of ideas on random search strategies. 
On the other hand,  we  show how intermittent strategies  are naturally  involved as soon as hidden targets are considered and define a basic model relying on  intermittent strategies,  introduced to account for the search behavior of "saltatory" animals. This one-dimensional model is then extended to a bi-dimensional model, which is shown to be a minimal model optimizing the search time. Last, we  discuss the relationships between these two main classes of search strategies -- L\'evy and intermittent -- and come back on  the famous "albatross story".

\subsection{The L\'evy strategies}

\label{subsection_Levy}


\subsubsection{The advantage of L\'evy walks with respect to simple random walks}

\begin{figure}[h!]\label{illustration_Levy}
\begin{center}
\includegraphics[height=10 cm]{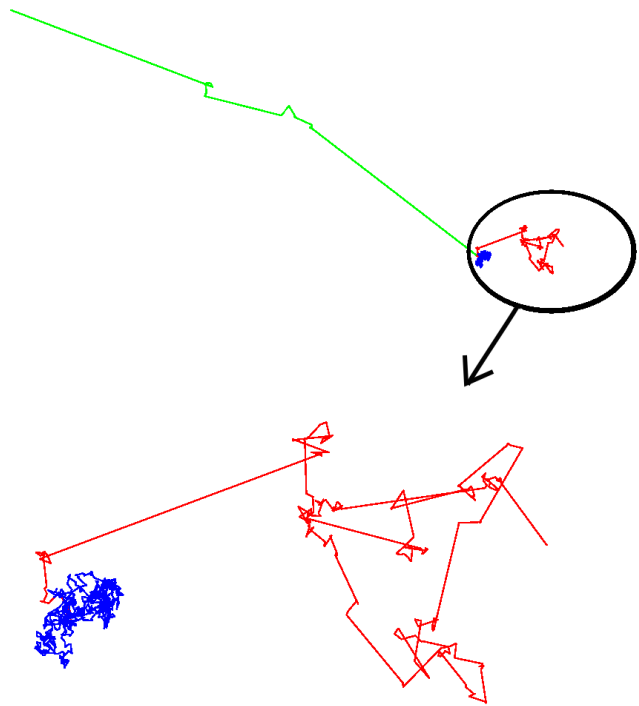}

\caption{Example of L\'evy walks, 
with $\mu=1.5$ (green, not present on the zoom), $\mu=2$ (red), $\mu=3$ (blue). 
The total path length is the same for the 3 examples.}
\end{center}
\end{figure}

The ballistic phases interspersed with turns of animals trajectories 
have often been  interpreted as L\'evy walks \cite{viswaNat,revueViswa}. 
 Actually, \textcite{levyKlafterTot} were the first to report that, 
due to their weak oversampling properties \refi{illustration_Levy},  L\'evy  \new{walks} could be an efficient way to explore 
space and could be used to model in particular trajectories of foraging  animals. 
As a matter of fact, the mean number of distinct sites visited in $n$ steps -- which is  a measure of the territory explored  -- is known to behave for a standard random walk like
$n$ in dimension $d>2$, and like $n^{d/2}$ if $d\le2$. This is  less efficient in low dimension than for a L\'evy  \new{walk}, of jump probability in dimension $d$ of the form $p(r)\propto r^{-\beta-1}$ (where $\beta$ is the index of the walk), for which the mean number of distinct sites visited in $n$ steps
behaves like $n$, as soon as $\beta<d$. From the point of view of the extension of the territory explored after a given number of steps, the  advantage of L\'evy  \new{walk} patterns over standard random walks is thus clear, and 
this effect  turns out to be as much marked as the number of searchers involved in the process is high \cite{Viswanathan:1996}.

\subsubsection{Optimizing the encounter rate  with L\'evy walks~: how and when?}\label{optimizing}

These  observations led   \textcite{viswaNat}  to propose
the following L\'evy search model, in the presence of fixed  targets randomly and sparsely distributed~: they consider
a searcher performing ballistic
 \new{step} at constant speed, and detecting targets closer than $r_v$.
A target is found when the searcher encounters it for the first time.  
The  \new{step} lengths are drawn from a L\'evy distribution $p(l)\propto l^{-\mu}$, with 
$1<\mu<3$. For $\mu \le 1$, the probability distribution is not defined. 
For $1<\mu\leq 2 $, the distribution has no mean and no variance. 
For $2 <\mu < 3 $, the distribution has a mean but no variance. 
For $\mu \geq 3$, the distribution has both a mean and a variance, 
thus it obeys the central limit theorem~: 
after enough  \new{steps}, the probability distribution of the difference between the starting point and the last position  
 is a Gaussian, as if it were diffusion, 
with the mean square distance scaling linearly with time. 

\textcite{viswaNat} are interested in the mean number of targets detected after a large observation time $t$. More precisely, they ask the following question  : is it possible
to optimize this number with respect to  the exponent $\mu$ characterizing the motion of the searcher ?
To answer, they 
actually consider   two  different types of targets, 
which lead to two very different optimal strategies.
\begin{itemize}
\item  In the first case of what they call ``revisitable targets'' - meaning that, as soon as detected, a target reappears \emph{at the same location} -
 they rely on a mean-field approximation of the problem and find that the encounter rate is optimized for a L\'evy exponent $\mu\simeq2$. 
\item In the second 
 case of "non revisitable targets"  (or  destructive search) where each target can
be found only once, or in the case of a single available target, the optimal
strategy proposed in \textcite{viswaNat} is  not anymore of L\'evy type, but reduces to a simple linear ballistic
motion. 
\end{itemize}

Several extensions of this pioneering model have been proposed.

 \textcite{BartViswa_critique_par_james} studied the  case of non-revisitable  moving targets. 
They showed that a L\'evy strategy with $\mu =2$ 
is often better than a ``Brownian'' one ($\mu \geq 3$).
However, \textcite{james} extended the study to ballistic motion, that outperformed these L\'evy strategies. 

An intermediate situation between revisitable and non revisitable targets has been studied by \textcite{ViswaPRL2003,ViswaRe}. In these works,   
the immobile target is  \new{destroyed} upon encounter, but regenerates after a time $\tau$ at the same place 
\new{(for example a plant bearing new fruits after previous fruits have been eaten)}.
Two regimes are found. 
When $\tau$ is large ($> \tau_c$, a critical time evaluated in \textcite{ViswaPRL2003,ViswaRe}), the simple ballistic motion remains the best strategy.
When  $\tau<\tau_c$, 
the best $\mu$ is between $1$ and $2$.
However, one could argue that in this regime the simple strategy where the searcher does not move and waits  for the renewal of the target 
outperforms searching for an hypothetical other target.  

In \textcite{BartumeusPNASspecialIssue08}, the targets are in patches \new{(such as fish schools)} or L\'evy distributed. 
Even if the targets are destroyed upon encounter, finding a target means that 
the presence of other targets in the vicinity is likely, 
which is close to the case of revisitable targets. 
 \new{Hence, like for revisitable targets,} the optimum is achieved for a L\'evy distribution, with $\mu \simeq 2$. 

In \textcite{BartumeusReynolds}, the optimum for destructive targets  is $\mu \to 1$ except in two cases  
(where $1< \mu^{opt} \leq 2$). 
On the one hand the optimum is not ballistic when the 
searcher can fail in capturing a detected target. 
On the other hand, for targets destroyed upon encounter, 
and for the very specific one-dimensional case, 
as the measure of  efficiency is  the number of targets captured during a long time, 
the searcher is after some time in a situation with a target close on one side, 
but the next target on the other side very far away~:
a pure ballistic motion is not favored because it can take the wrong direction.  

Finally, in the case of revisitable targets and the related cases 
(regenerating targets, patches, failed capture), 
the L\'evy strategy $\mu=2$ emerges   as a compromise between 
trajectories returning to one and the same ever target zone, and 
straight ballistic motion which is indeed the best way to explore space. Note however that, as stated above, in this case the strategy which consists simply  in waiting for target renewal performs even better.
In the case of non revisitable targets -- the generic situation considered hereafter -- the best strategy for the searcher is a mere ballistic motion 
without reorientations.

In all these L\'evy walks models, 
the searcher is assumed to be able to detect targets all along its trajectory. Qualitatively,  it corresponds to the case of   targets  "not too difficult" to detect. However, as it was the case with the example of the lost small key given in introduction, 
it is evident that in some situations the velocity degrades the perception. What happens if the targets are really "hidden", that is to say more precisely,   if searching and moving are  incompatible ? In recent years, many works have been devoted to answering this question. Most of them rely on the following simple two-state model, historically introduced to account for the search behavior of the "saltatory animals".

\subsection{A basic model of  intermittence}

\label{section_animaux_2005}

\subsubsection{Observations: the case of "saltatory animals"}

 Anyone who has ever lost his keys knows that instinctively we often adopt an intermittent behavior  combining local scanning phases and relocating phases.
Indeed, numerous studies of foraging behavior of a broad range of animal species show that such an  intermittent behavior is commonly observed and that the durations of search and displacement phases vary widely \cite{Bell,kramer,animauxObrien}. The spectrum, which goes from cruise strategy (for large fishes that swim continuously such as tuna), to ambush or sit-and-wait search, where the forager remains stationary for long periods (such as rattlesnake), has remained uninterpreted 
for a long time. As explained in the introduction, the interest of this type of intermittent strategies, often referred to as "saltatory"\cite{kramer,animauxObrien} in the context of foraging animals, can be understood intuitively when the targets are "difficult" to detect and sparsely distributed, as it is the case for many
foragers (such as ground foraging birds, lizards, planktivorous fish...\footnote{Note that there are counter-examples such as birds of prey which can detect targets even at large velocities}): since a fast movement is known to significantly degrade perception abilities \cite{kramer,animauxObrien}, the forager must search slowly. Then, it has to relocate as fast as possible in order to explore a previously unscanned space, and search slowly again. 

Even though numerous models based on optimization of the net energy gain \cite{OBrien:1989,r5,r6} predict an optimal strategy for foragers, the large number of unknown parameters used to model the complexity of the energetic constraint, renders a quantitative comparison with experimental data difficult. In the model presented in this section,  the search time is assumed to be the relevant quantity optimized by the forager in order to obtain a sufficient daily amount of food and to precede other competing foragers. The energy cost is treated only as an external constraint that sets the maximal speed of the animal. 
As explained in the next sections,  this purely kinetic model of target search captures the essential features of saltatory search behavior observed for foragers in experiments \cite{kramer}, when the predator has no information about the prey location. 

\subsubsection{The Model}

\imageal{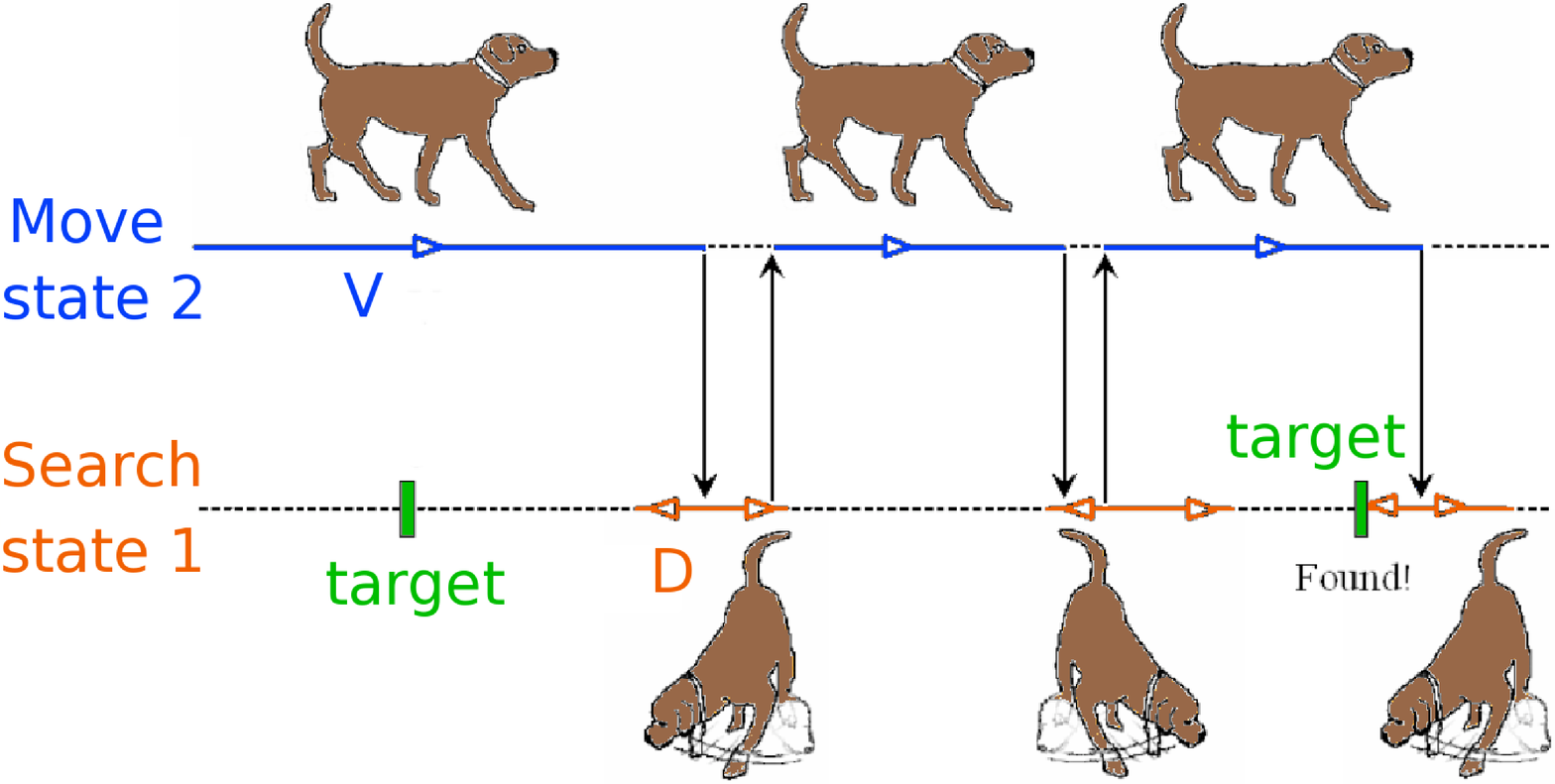}{Basic model for intermittent search}{chien}{0.35\linewidth}

The central point of this schematic model \cite{Animaux1D}  is
that it relies on the explicit description of searching trajectories as intermittent.
 In the following it is
assumed that the searcher displays alternatively two distinct
attitudes (see figure \ref{chien})~:
\begin{itemize}
 \item a scanning phase, named phase 1,
during which the sensory organs of the searcher explore its immediate vicinity. 
This phase is modeled as a ``slow'' diffusive movement (a
continuous random walk with diffusion coefficient $D$). 
The
target is found when this movement reaches the target
location for the first time. 
As
focusing and processing the information received by sensory
organs require a minimum time, the scan phase
cannot be too short, which implies a minimal mean time spent in this phase $\tau_1^{min}$. 
\item a motion phase, named phase 2, during which
the searcher moves ``fast'' and is unable to detect targets.
These relocating moves are characterized by a ballistic
motion (at constant velocity $V$).  
In the case of animals, there is usually correlations in 
the angles between two successive ballistic phases \cite{animauxObrien}. 
We limit ourselves here to the case of high correlations,
which allow us to consider an effective 
1-dimensional problem for both phases, with phase 2 always in the same direction.
\end{itemize}
Next, it is assumed that the searcher randomly switches
from phase 1 (respectively, 2) to phase 2 (respectively, 1)
with a fixed rate per unit time $\lambda_1$ (respectively, $\lambda_2$), that is with no temporal memory.
It leads to exponentially distributed phase durations, in agreement with  numerous experimental studies \cite{CelSensory,CelFunda,distri_exp_poisson,amibes_exp_etc_li_norrelykke_cox}, the mean duration of phase $i$ being
$\tau_i=1/\lambda_i$.
Last, the
preys are assumed to be immobile (see section \ref{pascal} for a discussion of moving vs immobile targets). 

\subsubsection{Equations}

We now evaluate the average time needed to find a
target. 
The chosen geometry is a single target in $x=0$ on a segment of size $L$ with periodic boundary conditions. 
This geometry is equivalent to the case of regularly spaced targets or 
to the case of one target centered in a finite domain with reflective boundaries. 
$L$ is thus the typical distance between targets, or the size of the search domain.
The instantaneous state of the searcher
can be described by its position $x$ on the segment and by an
index $i$ which specifies its motion: $1$ corresponds to the slow detection phase, and 
$2$ to the ballistic non-reactive phase.
The survival probability $p_i(t,x)$  that, 
when the searcher starts at time $t=0$ from $x$ and in state $i$, 
the target has not yet been found at time $t$, is known 
to satisfy the backward Chapman-Kolmogorov differential
equations  \cite{Redner,gardiner}~:
\begin{equation}\label{1Dcorrel1}
 D \frac{\partial^2  p_1}{\partial x^2} + \frac{1}{\tau_1}\left(p_2(t,x)-p_1(t,x) \right)=\frac{\partial p_1}{\partial t}
\end{equation}
\begin{equation}\label{1Dcorrel2}
 -V \frac{\partial  p_2}{\partial x} + \frac{1}{\tau_2}\left(p_1(t,x)-p_2(t,x) \right)=\frac{\partial p_2}{\partial t}
\end{equation}
Knowing that $t_i(x)$, the mean first passage
time at the target, starting from $x$ in phase $i$, is  given by
\begin{equation}
t_i(x)=-\int_0^\infty t \frac{\partial p_i(t,x)}{\partial t} dt=\int _0 ^\infty p_i(t,x) dt,
\end{equation}
it is easily found from Eqs.(\ref{1Dcorrel1}), (\ref{1Dcorrel2}) to satisfy
\begin{equation}
 D \frac{d^2  t_1}{d x^2} + \frac{1}{\tau_1}\left(t_2(x)-t_1(x) \right)=-1
\end{equation}
\begin{equation}
 -V \frac{d  t_2}{d x} + \frac{1}{\tau_2}\left(t_1(x)-t_2(x) \right)=-1
\end{equation}
These differential equations have to be completed by boundary conditions. As we have periodic boundary conditions and as the target  at $x=0$ can be found only in state 1, we get  
$t_1(0)=t_1(L)=0$, $t_2(0)=t_2(L)$. 

\subsubsection{Results}

The average search time $\langle t \rangle$
is defined as the average of $t_1(x)$ over the initial position
$x$ of the searcher, which is uniformly distributed
over the segment $[0,L]$, 
as the searcher initially does not know the target's location. 
It is found to be given by \cite{Animaux1D}~: 
\begin{equation}\label{animaux_tmoyen}
\langle t \rangle = \left( {\tau_2}+{\tau_1} \right) 
 \left( \frac{L}{2} 
\frac{ \left( 
{{\rm e}^{{ \alpha}+{ \beta}}}-1 \right) \sqrt {1+4r}+ \left( 1+2r\right)  \left( {{\rm e}^{{
 \beta}}}-{{\rm e}^{{ \alpha}}} \right)   } {\sqrt {
1+4r}
 \left( {{\rm e}^{{ \beta}}}-1 \right) \left( {{\rm e}^{{ 
\alpha}}}-1 \right) \tau_2 V}-{\frac {1}{r}}-1 \right) ,
\end{equation}
with~: 
\begin{equation}
r=\tau_2^2V^2/(D \tau_1),
\end{equation}
\begin{equation}
 \alpha = \frac{L}{2} \left( -{\frac {1}{{\tau_2}\,V}}+\sqrt {{\frac {1}{{{\tau_2}
}^{2}{V}^{2}}}+4\,{\frac {1}{{D}\,{\tau_1}}}} \right) ,
\end{equation}
\begin{equation}
 \beta=-\frac{L}{2} \left( {\frac {1}{{\tau_2}\,V}}+\sqrt {{\frac {1}{{{\tau_2}
}^{2}{V}^{2}}}+4\,{\frac {1}{{D}\,{\tau_1}}}} \right) .
\end{equation}
In the limit of $L \gg V\tau_2, \sqrt{D \tau_1}, D\tau_1/(V\tau_2)$, this expression simplifies~: 
\begin{equation}\label{1Dcorrel_tmsimp}
\langle t \rangle \simeq {\frac {L \left( {\tau_2}+{\tau_1} \right)  \left( {D}
\,{\tau_1}+2\,{{\tau_2}}^{2}{V}^{2} \right) }{2{\tau_2}\,V\,
\sqrt {{D}\,{\tau_1}}\sqrt {{D}\,{\tau_1}+4\,{{\tau_2}
}^{2}{V}^{2}}}}.
\end{equation}
Note that thanks to intermittence,  $\langle t \rangle \propto L$, whereas 
for diffusion alone the mean detection time is $t_{\rm diff} =L^2/(12D)$).
Intermittence is thus favorable  (meaning that the gain, defined as $t_{\rm diff}/\langle t \rangle$ is grater than 1), at least for $L$ large enough.

Intermittence is favorable and the strategy can even be optimized. 
The mean search time is minimized for   $\tau_1^{opt}=\tau_1^{min}$, and $\tau^{opt}_2$ satisfying the relation~: 
\begin{equation}\label{1Dcorrel_rel}
 \tau_1^3+6\frac{\tau_1^2\tau_2^2}{\tau}-8\frac{\tau_2^5}{\tau^2}=0,
\end{equation}
where $\tau=D/V^2$ is an extra characteristic time, 
depending on the searcher's characteristics. 
This minimum takes a simple form in two different
regimes.
\begin{itemize}
\item  If $\tau_1 \gg \tau$, the minimum of the search time 
is for $\tau_1=\tau_1^{min}$, and~:
\begin{equation} \label{1Dcorrel_to_t1gd}
\tau_2^{opt} = \left( \frac{3\tau \tau_1^2}{4}\right)^{1/3} .
\end{equation}
In this regime, denoted by S for "searching", one has $\tau_1 > \tau_2$~: the
searcher spends more time scanning than moving .
\item If $\tau_1 \ll \tau$, the minimum of the search time 
is for $\tau_1=\tau_1^{min}$, and~:
\begin{equation}  \label{1Dcorrel_to_t1petit}
\tau_2^{opt} = \left( \frac{\tau^2\tau_1^3}{8}\right)^{1/5}.
\end{equation} 
In this regime, denoted by "M" for moving, one has $\tau_1 < \tau_2$, which means 
that the searcher spends more time moving than scanning.
\end{itemize}

\subsubsection{Comparison with experimental data}

These results have been compared to experimental  data from  
\textcite{animauxObrien} and \textcite{kramer}, who
provide the average duration of detection and ballistic phases,  
characterizing the saltatory behavior of  18 different species, 
as various as planktivorous fish, ground
foraging birds, or lizards. 
The optimal strategy obtained above is shown to account reasonably well 
for these data (see fig. \ref{compareanimaux} and \textcite{Animaux1D, modele, physicA} for further details).

\imagea{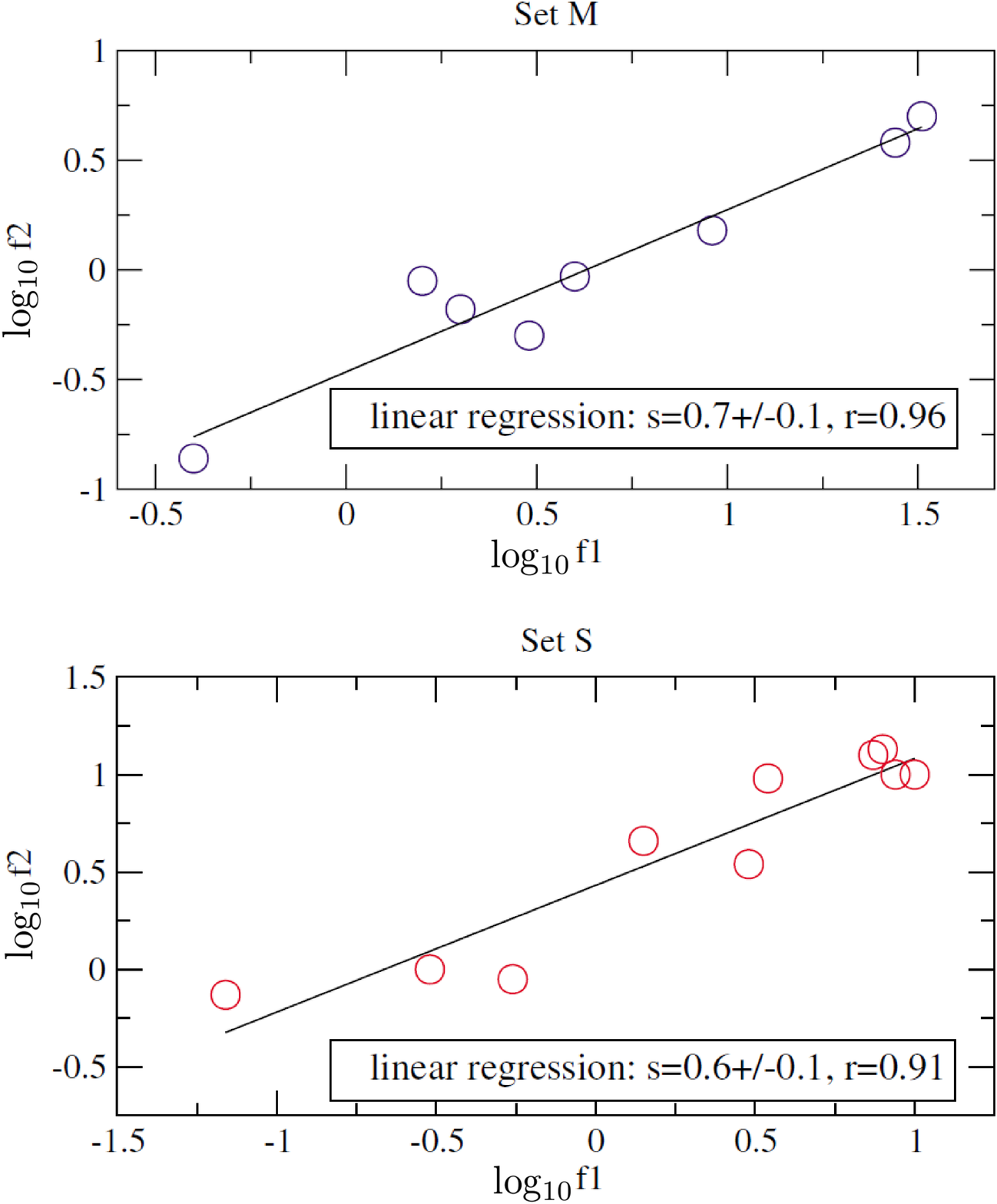}{Log-Log plot of experimental data \cite{animauxObrien,kramer,Animaux1D} 
of saltatory search behaviors and their linear regression. Here $f_i=1/\tau_i$ 
}{compareanimaux}{10}

These results show that the saltatory patterns observed 
are a way to optimize the search, 
and that it is probably a reason why this type of patterns is observed so often,
as it could have been favored by natural selection.

\bigskip


\subsection{Two-dimensional intermittent search processes: An alternative to L\'evy strategies }
\label{animaux_alternative}
\subsubsection{Motivation}

The model of intermittent search presented previously was one-dimensional, 
with ballistic phases infinitely correlated, in the sense that the direction 
taken is always the same. 
Here we present a model of
intermittent search strategies in dimension two \cite{PRE2006, SpecialIssue2006}  which encompasses a  much broader field
of applications, in particular for animal or human searchers.
It is shown that bidimensional intermittent search strategies
 {\it do optimize} the search time for  non revisitable targets, 
i.e. targets that are \new{destroyed} upon discovery (see section \ref{optimizing}). 
The optimal way to share the time between the phases of non reactive displacement and of reactive search is explicitly determined.
Technically, this  approach relies on  an approximate
analytical solution based on a decoupling hypothesis, which proves
to reproduce quantitatively  numerical simulations over a wide
range of parameters.

\subsubsection{The Model}

\doublimagem{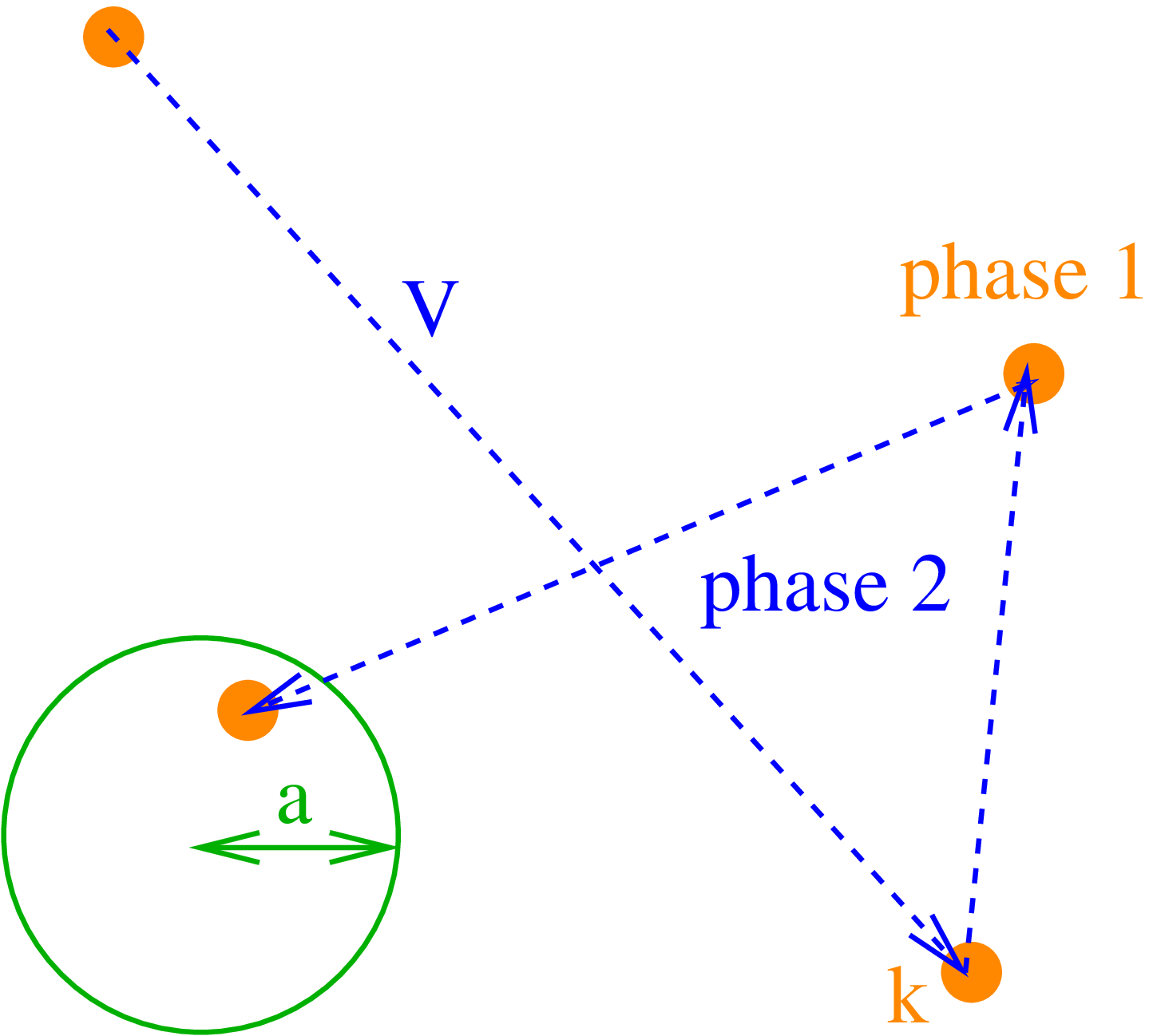}{Static mode. The slow reactive phase
 is static and detection takes place with
finite rate $k$.}{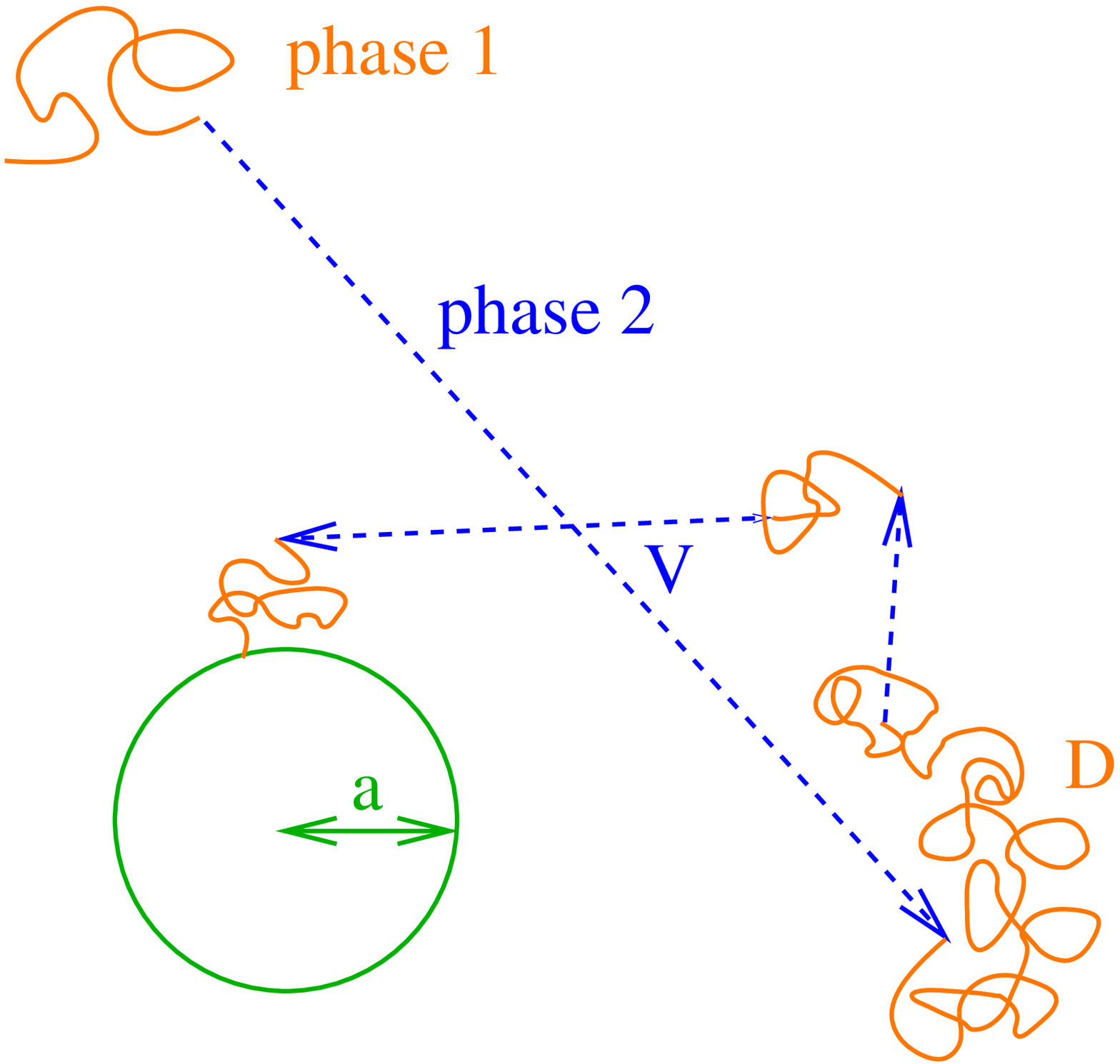}{Diffusive mode. The slow reactive phase is diffusive and detection is infinitely efficient.}{Two models of intermittent search: The searcher alternates slow reactive phases (regime 1)
of mean duration $\tau_1$, and fast non reactive ballistic phases (regime 2) of mean duration $\tau_2$}{figmodel}

Following the previous model, we consider a two state searcher \refi{figmodel} of
position $\bf r$ that performs slow reactive phases (denoted 1),
 randomly interrupted by fast relocating ballistic flights of
constant velocity $V$ and random direction (phases 2). We assume the
duration of each phase $i$ to be exponentially distributed with mean
$\tau_i$. As fast motion usually strongly degrades perception
abilities, we consider again that the searcher is able to find a target
only during  reactive  phases 1.  The detection phase involves
complex biological processes that we do not aim at modeling accurately here.
However, we put forward here two modes of detection. The
first one, referred to in the following as the "diffusive mode",
corresponds to a diffusive modeling (with diffusion coefficient $D$)
of the search phase like in the previous model, in
agreement with observations for vision \cite{huey}, tactile sense or
olfaction \cite{Bell}.  The detection is assumed to be  infinitely
efficient in this mode~: a target is found as soon as the
searcher-target distance is smaller than the reaction radius $a$. On
the contrary, in the second mode, denoted as the "static mode", the
reaction takes place with a finite rate $k$, but the searcher is
immobile during search phases. Note that this description is
commonly adopted in reaction-diffusion systems \cite{rice} or operational
research \cite{gardecote}. A more realistic description is obtained by combining both
modes and considering a diffusive searcher with diffusion
coefficient $D$ and finite reaction rate $k$. In order to reduce the
number of parameters and to extract the main features of each mode,
we study them separately by taking successively the limits
$k\to\infty$ and $D\to0$ of this general case. More precisely, in
these two limiting cases, we address  the following questions ~:
what is the mean time it takes the searcher to find a target?   Can
this search time be minimized ? And if so for  which values of the
average durations $\tau_i$ of each phase?

\subsubsection{Basic equations}

We now present the basic equations combining the two  search modes
introduced above  in the case of a  point-like target centered in a spherical
domain of radius $b$ with reflexive boundary.  Note that this
geometry mimics both relevant situations of a single target and of
infinitely many regularly spaced non revisitable targets.
As in the previous model,  the mean first passage time to a target
satisfies  backward equations \cite{Redner} (see section \ref{section_methods}) for derivation), which write here :
\begin{equation}\label{back1bis}
D\nabla^2_{\bf r}t_1+\frac{1}{2\pi\tau_1}\int_{0}^{2\pi}(t_2-t_1)d\theta_{\bf V}-k{\rm I}_a({\bf r})t_1=-1
\end{equation}
\begin{equation}\label{back2}
{\bf V}\cdot\nabla_{\bf r}t_2-\frac{1}{\tau_2}(t_2-t_1)=-1
\end{equation}
where $t_1$ stands for the mean first-passage time starting from state 1 at position ${\bf r}$, and $t_2$ for the mean first passage time starting from state 2 at
position ${\bf r}$ with velocity ${\bf V}$\new{, of direction characterized by the angle $\theta_{\bf V}$}. Here ${\rm I}_a({\bf r})=1$ if $|{\bf r}|\le a$ and  ${\rm I}_a({\bf r})=0$ if $|{\bf r}|> a$.
 In the present form, these integro-differential equations do not seem to allow for an exact resolution with standard methods. We thus resort to an approximate decoupling scheme, which relies on the following idea.
If the searcher initially starts in phase 2, and if the target is close, 
its initial direction matters. 
But as soon as the initial position is far from the target, there are numerous reorientations before finding the target, 
implying that the initial direction does not matter. 
Consequently, if $b \gg a$ and once the mean search time has been averaged over the starting position,  
the effect of the  initial direction can be neglected. 
It allows us to make an approximation and solve the system (for more technical details, see appendix and \textcite{PRE2006,SpecialIssue2006}).

\subsubsection{Results for the diffusive mode of detection}

For the diffusive mode of detection ($k\to\infty$), 
 an analytical approximation for the  search time can be obtained (see appendix \ref{section_generic_2DvD}).
In the case of low target density
($a\ll b$), which is the  most relevant for hidden target search
problems, three regimes  arise.
In the first regime $a\ll b\ll D/V$, the relocating phases are not efficient and intermittence is useless.
 In the second regime $a\ll D/V \ll b$,  it can be shown (see section \ref{section_generic_2DvD}) 
that the intermittence can significantly speed up the search
(typically by a factor 2), but that it does not change the order of magnitude
of the search time.  On the contrary, in the last regime $D/V\ll a\ll
b$, the optimal strategy, obtained for
\begin{equation}\label{grandv_a}
\tau^{opt}_{1}\sim \frac{D}{2V^2}\frac{\ln^2 (b/a)}{2\ln (b/a) -1}, \;\tau^{opt}_{2}\sim \frac{a}{V}(\ln(b/a)-1/2)^{1/2},
\end{equation}
leads to a  search time  arbitrarily smaller than the non intermittent
 search time when $V\to\infty$. 
Note that 
 this optimal strategy  corresponds to a scaling law
\begin{equation}
\frac{\tau_{1}^{opt}}{\tau^{opt}_{2}}\sim \frac{D}{a^2}\frac{1}{\left(2-1/\ln(b/a)\right)^2},
\end{equation}
which does not depend on $V$.

\subsubsection{Results for the static mode of detection}

We now turn to the static mode ($D\to 0$) \new{(see appendix \ref{section_generic_2Dvk})}.
In this case, intermittence is trivially necessary to find the target, and the optimization of the search time 
 leads for $b\gg a$ to~:
\begin{equation} \label{statique_a}
\tau_{1,{\rm min}}=\left(\frac{a}{Vk}\right)^{1/2}\left(\frac{2\ln(b/a)-1}{8}\right)^{1/4},\;
\end{equation}
\begin{equation}\label{statique2_a}
\tau_{2,{\rm min}}=\frac{a}{V}\left(\ln(b/a)-1/2\right)^{1/2},
\end{equation}
which corresponds to the scaling law $\tau_{2,{\rm min}}=2k\tau_{1,{\rm min}}^2$, which still does not 
depend on $V$.

\subsubsection{Conclusion}

This bidimensional   two-state model of search processes
for non-revisitable targets  closely relies on the
experimentally observed intermittent strategies adopted by foraging animals. Using a decoupling
approximation numerically validated,   it can be analytically solved, allowing us to draw 
several conclusions. 
 (i) The mean search time $\langle t\rangle$ presents a global minimum
for finite values of the $\tau_i$, which means that  intermittent strategies  constitute optimal strategies, as opposed to L\'evy  \new{walks} which are optimal only for revisitable targets. (ii)
The optimal  $\tau^{opt}_{1}$ obtained for two modes of detection are different and depend
explicitly on $D$ and $k$, leading to different scaling laws which
are susceptible to discriminate between the two search modes .
 (iii) A very striking and non intuitive feature is that both modes of
search studied lead to the {\it same optimal value} of $\tau^{opt}_{2}$. As this optimal time  does not depend on
the specific characteristics $D$ and $k$ of the search mode, it seems to constitute a general property of intermittent search strategies.
The robustness of these conclusions will be discussed further in section \ref{section_generic} in the framework of a more general model.

\subsection{Should foraging animals really adopt Levy strategies ?}

As seen before, 
intermittent strategies are an alternative to L\'evy walks (defined in section \ref{subsection_Levy}) for 
interpreting trajectories of foraging animals. 
However, the L\'evy walks are often thought as optimal and widespread in nature. Is it really true?

\subsubsection{The albatross story}

Many foraging animals, including albatrosses, deers and bumblebees to name a few, 
 have been thought for long 
to adopt L\'evy strategies described in the pioneering work of  \textcite{viswaNat}. 
These foraging behaviors were repeatedly accounted for  by a simple model stating in  the more general
framework of search processes that L\'evy walks 
are optimal search strategies, as they constitute the best way to explore space.  Recently, \textcite{revisitingViswaNat}  
reanalyzed these data, completed by newly gathered data on foraging albatrosses, and showed that 
in fact there was no experimental evidence for 
the L\'evy  \new{walk} 
behavior\footnote{\new{Albatrosses behavior was followed by a humidity sensor on the birds. 
Flights were taken as the ``dry`` phases, interspersed with humid phases, when the birds touches the ocean. 
Very long ''flights'' eventually proved to be rest time, when the bird was in its nest. 
Once these misinterpreted dry phases removed, the distribution of flights durations is no longer a power-law.}}. 
This study questions 
the interpretation of several experimental works, but also raises a new important and puzzling question~: why
animals do not adopt the L\'evy  \new{walk} strategy which  has however been reported to be an optimal search strategy? Here we clarify
this apparently paradoxical situation.

\subsubsection{Do animals really perform L\'evy walks?}

As the optimality of L\'evy strategies crucially requires 
conditions on the targets (regenerating at the same place, patched or hard to capture) 
and conditions on the searcher (no switch when a target is found, which is a very simple form of memory), 
it can not be taken as a general rule even if realistic for certain species. 
On the contrary,   we argue that the  general question of determining the best strategy 
for finding a single hidden target belongs to the situation of destructive search, where  
in the framework of the model of \textcite{viswaNat}, the most efficient way to find a randomly hidden target 
is simply  a  linear ballistic motion and {\it not} a L\'evy strategy (see section \ref{subsection_Levy}). 
As a consequence, there is
no paradox~:  the reason why  L\'evy walks are not observed in the work of \textcite{revisitingViswaNat}
 is probably because they do not constitute 
robust optimal search strategies.  

\begin{figure}[h!]
\begin{center}
\includegraphics[height=6 cm]{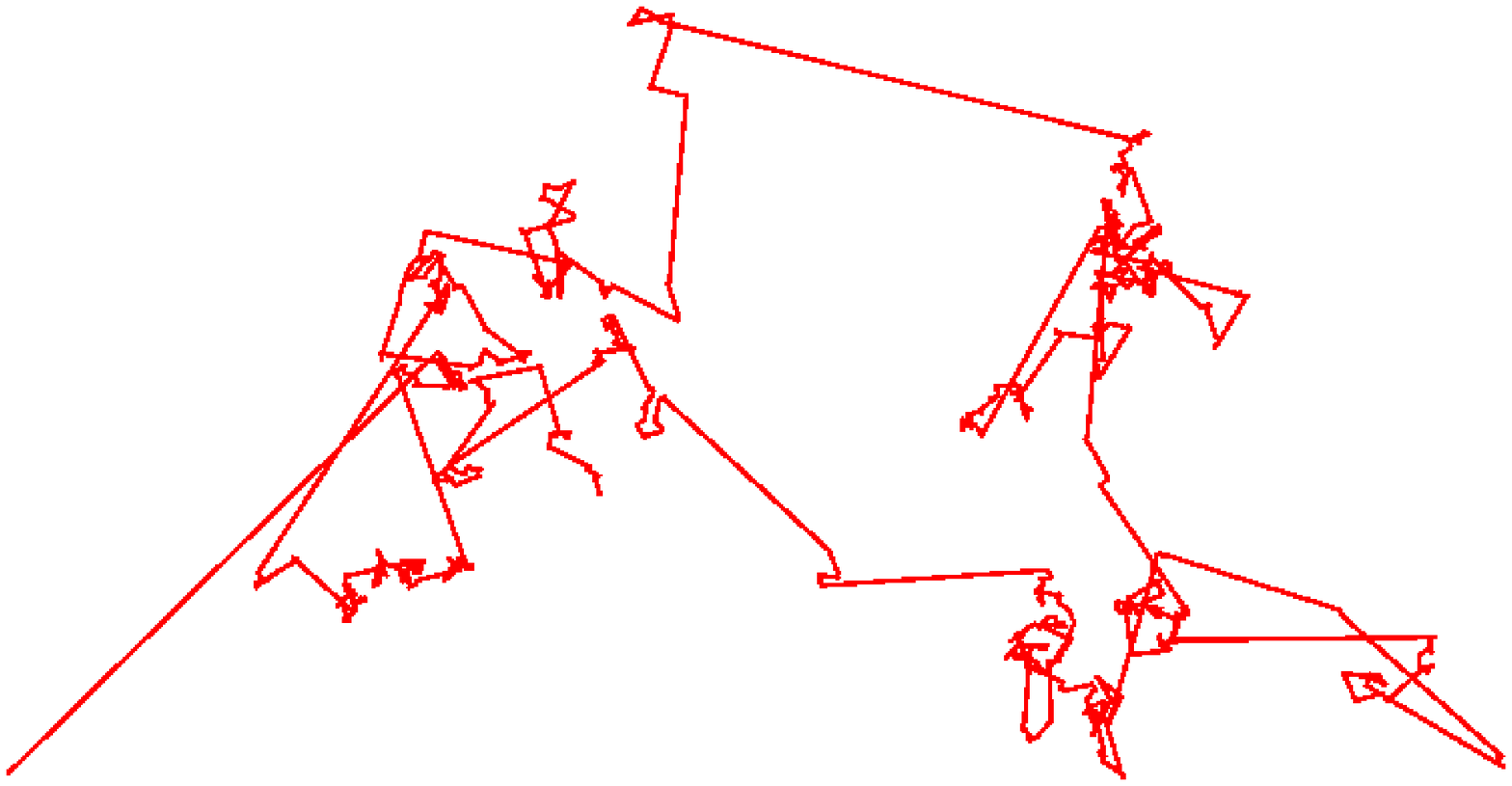}

L\'evy walk with $\mu=2$

\includegraphics[height=6 cm]{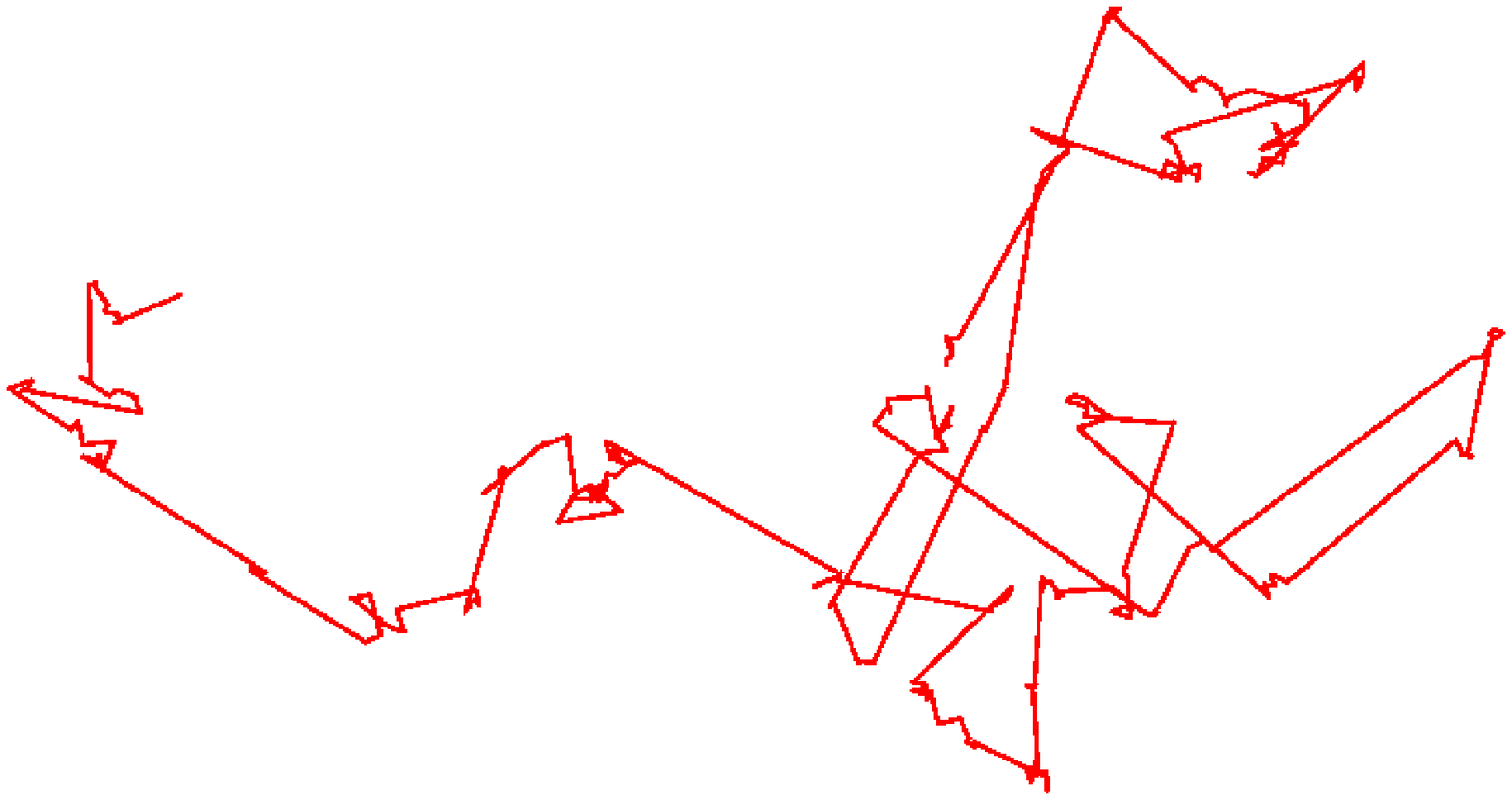}

Composite random walk~: alternation of about 10 short steps (mean 1, distributed exponentially), 
and  one large step (mean 10, distributed exponentially).

\caption{Comparison between a L\'evy walk and a composite random walk~: 
They are not easy to distinguish at short time scales}\label{Levy_compositeRW}
\end{center}
\end{figure}

And what about other experimental observations? 
Among  experimental studies analyzing organisms trajectories 
as a succession of segments interspersed with turns, 
an important proportion reports times between turns distributed exponentially 
(a list of examples far from exhaustive~: 
\textit{C.Elegans} worm \cite{CelSensory,CelFunda}, 
 fish \cite{distri_exp_poisson}, 
plankton in a part of the conditions studied by \textcite{BartumeusPlancton},  
amoebae \cite{amibes_exp_etc_li_norrelykke_cox}, etc.). 
However, apart from the controversial albatrosses study \cite{revisitingViswaNat}, 
there is a boom in articles claiming that L\'evy behavior is observed for some animal species. 
A part of them can be dismissed as evidence of L\'evy behavior. 
On the one hand, as explained in details in \textcite{revisitingViswaNat}, 
due to experimental limitations, 
most data cover only a very limited range, which makes difficult a reliable identification of power laws.
On the other hand, patterns and processes should not confused, 
as underlined by \textcite{BenhamouContreLevy07}. 
The same observed patterns, 
can often be explained by different models. 
It is not because a trajectory is similar to L\'evy walk trajectories 
that the underlying process is necessarily a L\'evy walk.   
For example, a composite classical random walk can look very similar to a L\'evy walk 
for a time short enough \refi{Levy_compositeRW}. 
Nonetheless, not all studies should be discarded , since limited studies neither prove nor rule out L\'evy strategies.
(see \textcite{revueViswa} for a review).
As underlined by \textcite{revueViswa}, 
other selection pressures could be predominant.
For example, when targets location is known,  exploitation 
could be optimized instead of search, 
and L\'evy walks could emerge from interactions between the environment and the searcher 
\cite{spidermonkeys, levyBande, levy_humain_car_rues}. 

\subsection{Conclusion on animal foraging}

L\'evy walks are a  fashionable model
for interpreting trajectories of foraging animals. 
However, on the one hand, there is controversy about at least a part of the experimental 
data which were thought to support L\'evy walks. 
On the other hand, 
the conditions in which L\'evy walks are optimal are very restrictive.  
However, it does not rule out any contribution of L\'evy statistics in the context of search processes. 
For example, as discussed by \textcite{intermittentlevy} 
and in section \ref{section_temporal_levy}, 
L\'evy statistics can be advantageously used in the context of intermittent trajectories.
Additionally, we argue that some 
animals cannot detect their target when they are moving ballistically, 
and in fact alternate these fast but blind phases with detection phases. 
Mean search time with intermittence can be smaller than with detection phase alone, 
and it can be minimized tuning the mean durations of each phase. 
Intermittent search strategies, because they rely on the experimental observation that 
speed degrades perception, and because they prove optimal an robust, 
are good candidates for interpreting animals trajectories.

\section{Intermittent search strategies at the microscopic scale }

\label{section_ADN}

The previous section has shown that  intermittent search strategies are 
observed at the macroscopic scale.
They are also observed at the microscopic scale. 
In the following, we will focus on two examples : 
the localization  by a protein of a specific DNA sequence, 
and  the active transport of vesicles  in cells. 

\subsection{Protein/DNA interactions}
\subsubsection{Biological context}

\begin{figuresmall}[h!]
\small
\centering\includegraphics[scale = 0.7]{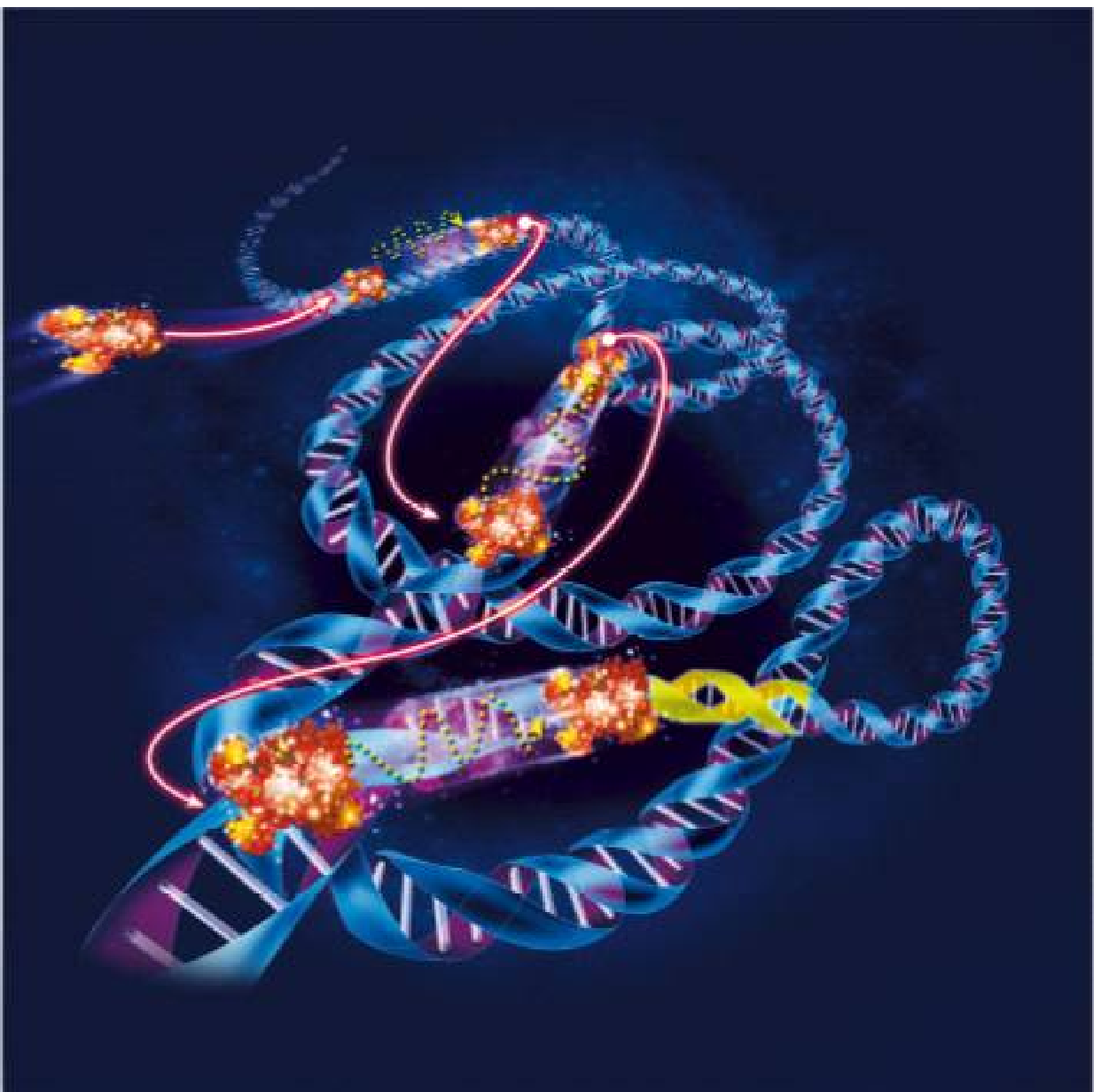}
\caption{\small Artist view of a DNA/protein interaction, which combines 
1--dimensional sliding phases and 3--dimensional relocation phases. 
Picture by Virginie Denis/Pour la Science 352, February 2007.}\label{DNA}
\normalsize
\end{figuresmall}

Various functions of living cells - and therefore at larger scales of living organisms - are regulated by coordinated chemical reactions 
between specific molecules, which are often present only in a few copy number.
The importance of the kinetics of such search processes between reaction partners   can be illustrated by 
the bacterial restriction and modification system \cite{ref_generale_restriction_modif_systs}, which involves 
couples of methyltransferase and restriction enzymes that recognize the same sequence on DNA 
(for example \textit{Eco}RV recognizes the sequence $GATATC$ \cite{halford_seq_reconnaissance_ecoRV}). 
Methyltransferase enzymes methylate this specific sequence on the bacterial DNA in order to protect it from restriction enzymes, whose function is oppositely to cut the DNA at this specific sequence. This function is first aimed at impairing 
any intruder  viral DNA which enters the cell and which 
is very likely to contain the target sequence. 
Indeed, this  sequence,   typically 4-8 base pairs, is very short as
compared to the viral genome, which, depending on the virus, 
can be made of $10^3-10^6$ base pairs (typically $5.10^4$ for bacteriophages). 
The infected bacteria then faces a vital search problem :
restriction enzymes must find their target sequence on the viral DNA reliably to inactivate the virus before it exploits the bacteria machinery and kills it.

More generally, it is well established that some sequence-specific proteins find their target site 
in a remarkable short time. 
For the lac repressor for example, 
\textcite{riggs}  measured association rates 
orders of magnitude larger than those expected for reactions limited by  
the classical three-dimensional diffusion 
(results confirmed by \textcite{HsiehBrenowitz}
at different salt concentrations, 
ruling out electrostatic effects  as the only explanation). 
\textcite{halford40ans} argues that in fact only a few enzymes react significantly faster 
than the 3-dimensional diffusion limit. However, this author underlines
that   many enzymes react at rates close to the diffusion limit, 
and that this observation  is still impressive.  
Indeed, classical experiments are performed with a considerable excess of DNA, which is very likely to contain sequences similar to the target sequence and therefore act as traps slowing down the enzymes in their search. In 
a series of seminal articles, 
\textcite{winter1,winter2,winter3}  
proposed that 3D diffusion 
(or ``hopping''/``jumping'') was not the only 
motion available to the protein, even if no energy is consumed 
(unlike some enzymes which consume energy to scan DNA processively). 
They suggested that in some cases, proteins could  bind non specifically to DNA due to a weak electrostatic interaction and diffuse along the chain in a process named sliding (see \textcite{revue_hippel} and \textcite{PRL_dahirel_paillusson} for more details on the weak electrostatic interaction). 
It was then  argued that the combination of sliding and 3D diffusion, \textit{i.e.} 
facilitated diffusion, can make the search for a sequence two orders of magnitude faster than 3D diffusion alone and henceforth sufficiently efficient (see also \citet{AdamDelbruck}). 

Actually, this search mechanism can be classified as intermittent, 
in the general meaning defined in introduction. 
Indeed, on the one hand,  3-dimensional diffusion off the DNA is fast, 
but does not allow for target detection. 
On the other hand,
sliding is a phase of motion along DNA, which therefore enables target detection, but which is 
much slower due to a higher effective friction.

The pioneering studies on facilitated diffusion \cite{riggs,winter1,winter2,winter3} are based on ensemble measurements, 
which were for a long time the only way to experimentally access to protein/DNA interaction. 
Recently developed techniques make possible the observation of
 this interaction at the level of a single molecule, 
with a resolution in space and time still improving  
(for a recent review on the experimental results of such techniques, see \textcite{revue_dna_tirfm}).
It is now confirmed directly that 
many proteins searching for a specific sequence on DNA combine 
 ``hopping/jumping'' and ``sliding'' (see figure \ref{DNA}).
Sliding phases have been clearly identified (both in vitro \cite{kabata_science_93} and in vivo \cite{elf_dans_cellule,RalfFEBS,WANG}), 
as well as hopping/jumping phases  \cite{Gowers,broek,hopUL42,NAR08}. 

\new{With these new single-molecule experiments, theoretical models have bloomed too.
Here, we first}
 present \old{below} a stochastic approach of a simplified version of the problem,  
which shows that it is the intermittent nature of the trajectories that makes possible such high reaction rates. 
This minimal model permits to calculate explicitly the mean search time for such intermittent reaction paths, 
and shows that reactivity can even be optimized by properly tuning simple dynamic parameters of intermittent trajectories. 
Next, we discuss 
\new{the different directions of extension of recent theoretical models. }.

\subsubsection{Minimal model of intermittent reactions paths}

\label{section_DNA_minimal_model}

We present here a simple model of intermittent reaction paths with minimal ingredients 
 \cite{DNA}.
We first define the model, 
then explain the main steps of the calculation, and
eventually give the results.

\begin{figure}[h!]
\centering\includegraphics[scale = 0.4]{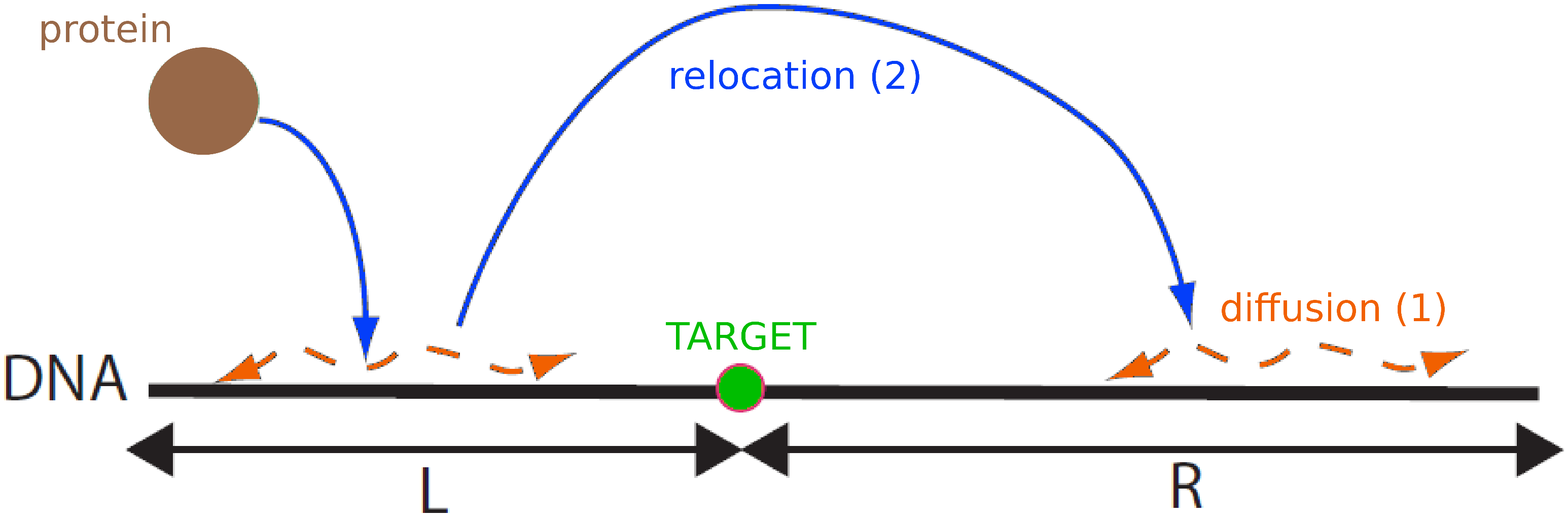}
\caption{\small A model of intermittent transport for  DNA/protein interactions.}\label{DNAmodel}
\end{figure}

\popetit{Definition of the model.}
We consider a generic protein searching
for its target site on a DNA molecule (see figure \ref{DNAmodel}). 
The pathway followed by the protein, considered as a
point-like particle,  is a succession of 1D diffusions  along the
DNA strand (sliding phases denoted phases 1) and 3D excursions in the surrounding solution (denoted phases 2, during which the target is not accessible).
In this minimal model we assume that the target site is a perfect reactive
point of the DNA molecule, which means that reaction occurs as soon as the target is reached by the protein. Note that in this case
the protein can find the target site only by diffusing along DNA, and therefore follows the scheme of intermittent search  presented in introduction. The key quantity that we evaluate in this section is the search time, or reaction time,  defined as the mean first-passage time \cite{Redner,Condamin:2005qr,Condamin:2005db,NatureSylvain, Condamin:2007yg, Condamin:2007rz, pnasSylvain,Benichou:2008}   of the protein at the target, here denoted by $\langle T \rangle$. This quantity gives directly access in a mean field approximation to   the first order reaction constant   $K=1/\langle T \rangle$ \cite{Berg76}.

We now introduce further ingredients of the model. The time spent by the protein on  DNA  during each sliding phase
 is assumed to follow an exponential law with dissociation
frequency $\lambda_1$. This law relies on a Markovian
description of the chemical bond which is commonly used for bimolecular association. The probability density that the protein
leaves the DNA at a random time $t$ is then given by
$\lambda_1 \exp(-\lambda_1 t)dt$, and the mean duration of a sliding event reads $\tau_1=1/\lambda_1$.

The one--dimensional motion on DNA during sliding phases 1 is  modeled by a
continuous Brownian motion  with diffusion coefficient $D$. We assume that the ends of the DNA chain act on the protein as
reflecting boundaries (see for instance \textcite{jeltsch:1998hy}), but in practice this assumption is unimportant for long DNA molecules. Moreover, the case of circular DNA, such as plasmids, is readily obtained by taking the particular case $L=R$.
We next assume that  the 3D excursions of phase 2 are 
uncorrelated in space. This means that after dissociation 
from DNA, the protein will rebind the DNA at a random position independently of its starting position. This is justified when the DNA is in a random coil conformation, as in this case even short 3D excursions can lead to a long effective translocation of
the linear position of the protein on DNA. We further assume that the probability density  $P_{3D}(t)$  of the duration     
$t$ of such 3D excursions is exponentially distributed, and write  $P_{3D}(t)=\lambda_2  \exp(-\lambda_2 t)$.  
This assumption is justified, at least for the tail of the distribution, as soon as the 3D excursions of the protein are confined in a closed volume, 
for instance  \new{an experimental volume in vitro, or in vivo} the cell or a cell compartment. The mean time $\tau_2=1/\lambda_2$ spent in the surrounding solution in phase 2 can then be shown to be proportional to the confining volume \cite{Kacbook,Blanco03, Olivier2005domaine,Condamin:2005qr,NatureSylvain,Benichou:2008a}.

We next introduce $P_{1D}(t|x)$, which is the conditional probability density that the
protein, being on the DNA at position $x$ and at time $t=0$, will
dissociate at time $t$ before any encounter with the
target site. We rewrite this quantity as :

\begin{eqnarray}
\displaystyle P_{1D}(t|x)=\lambda_1 \exp(-\lambda_1 t)Q(t|x)
\end{eqnarray}
where $Q(t|x)$ is the conditional probability density that the protein,
starting from the position $x$, does not meet the target site  during a single sliding event. The probability
density  $j(t|x)$ of the first passage to the target site position at time $t$
without dissociation is then related to $Q(t|x)$ according to $Q(t|x)=1-\int_0^tj(t'|x)dt'$.

Last, we introduce  $\bar{P}_{1D}(t|x)$, which is the conditional probability density that
the protein, being on DNA at position $x$  at time $t=0$, will find
the target site for the first time at time $t$ within a single sliding phase 1, without leaving the DNA:

\begin{eqnarray}
\displaystyle \bar{P}_{1D}(t|x)=\exp(-\lambda_1 t)j(t|x).
\end{eqnarray}

Given these quantities, we show below that the first-passage density of the protein to
the target site, and consequently the reaction constant,  can be calculated explicitly.

\popetit{ First-passage density.}   
By calculating the first-passage density, we
obtain the mean reaction time, as well as all associated moments. We assume that
the  protein  starts at $t=0$ in state 1 (bound to the DNA) at position
$x$. We consider a generic event  whose  number of 3D 
excursions is  $n-1$, and denote the duration of successive sliding phases   $t_1,\ldots,t_{n}$,  and  the duration of successive 3D excursions 
$\theta_1,\ldots,\theta_{n-1}$.  The probability density of such an
event, for which the protein finds the target site  for the first time at 
time
$t=\sum_{i=1}^{n}t_i+\sum_{i=1}^{n-1}\theta_i$
is:

\begin{eqnarray}
\displaystyle
P_n(t|x)=\bar{P}_{1D}(t_{n})P_{3D}(\theta_{n-1})P_{1D}(t_n)\ldots
P_{1D}(t_2)P_{3D}(\theta_1)P_{1D}(t_1|x)
\end{eqnarray}
where $P_{1D}(t)$ and $\bar{P}_{1D}(t)$ are averaged over the
initial position of the protein: $P_{1D}(t)=\left<P_{1D}(t|x)
\right>_x$ and ${\bar P}_{1D}(t)=\left<{\bar P}_{1D}(t|x)
\right>_x$.  We denote by $L$ the DNA length on the
``left'' side of the target site  and by $R$ the length on the
``right'' side of the target site. The average of a function $f$ over
the initial position $x$ is given by $\left<f(t|x) \right>_x\equiv
\frac{1}{L+R}\int_{-L}^Rf(t|x){\rm d}x$.

To obtain the density of first passage to the target site,
$F(t|x)$, we sum over all possible numbers of
excursions and we integrate over all intervals of time, ensuring that $t=\sum_i^{n}t_i+\sum_i^{n-1}\theta_i$. The average
over the initial position of the protein, $F(t)=\left<F(t|x)\right>_x$,
can be expressed as:
\begin{eqnarray}
\label{debut}
\displaystyle F(t)=\sum_{n=1}^{\infty}\int_0^{\infty}dt_1\ldots
dt_nd\theta_1\ldots d\theta_{n-1} \delta\left(
\sum_{i=1}^{n}t_i+\sum_{i=1}^{n-1}\theta_i-t \right) \\
\nonumber \left[ \prod_{i=1}^{n-1} P_{3D}(\theta_i)
\right]\left[\prod_{i=1}^{n-1}P_{1D}(t_i) \right]
\bar{P}_{1D}(t_{n}).
\end{eqnarray}
Taking the Laplace transform of $F(t)$,
$\widehat{F}(s)=\int_0^{\infty}dte^{-st}F(t)$,  we obtain:
\begin{eqnarray}
\label{densite} \displaystyle
\widehat{F}(s)=\left<\widehat{j}(\lambda_1+s|x)\right>_x\;
\left\{1-\frac{1-\left<\widehat{j}(\lambda_1+s|x)\right>_x}{\left(1+s/
\lambda_1\right)\left(1+s/ \lambda_2\right)}\right\}^{-1},
\end{eqnarray}
where $\widehat{j}(s|x)$ is the Laplace transform of
$j(t|x)$. This expression completely solves the problem {\it for any 1D motion}. We will see next  that  the
main quantities of physical interest can be extracted from this
formula.

\popetit{ Optimal search strategy.}  
The relevant quantity to describe the protein/DNA association reaction is the mean
time $\left<T\right>$ necessary for the protein to find the
target site (see above). This mean time is obtained from the derivative of the
first passage density by the following relation:

\begin{eqnarray}
\displaystyle
\left<T\right>=-\left(\frac{\partial\widehat{F}(s)}{\partial
s}\right)_{s=0}
\end{eqnarray}
which combined with equation (\ref{densite}) gives:

\begin{eqnarray}
\displaystyle
\left<T\right>=\frac{1-\left<\widehat{j}(\lambda_1|x)\right>_{x}}
{\left<\widehat{j}(\lambda_1|x)\right>_{x}}\left(\frac{1}{\lambda_1}+\frac{1}{\lambda_2}
\right).
\end{eqnarray}
This expression is  general and holds for any $1D$ motion in the slow phase 1.
Now, we calculate this quantity in the case where the phase 1 is  a free $1D$ diffusion.
The one--dimensional Laplace transform of the first passage
probability density is well known (see \textcite{Redner}), and leads to, after averaging over the starting position $x$:
\begin{eqnarray}
\label{MST} \displaystyle
\left<T\right>=\left(\frac{1}{\lambda_1}+\frac{1}{\lambda_2}\right)
\left\{\frac{\sqrt{\frac{\lambda_1}{D}}(L+R)}{\tanh\left(\sqrt{\frac{\lambda_1}{D}}L\right)
+\tanh\left(\sqrt{\frac{\lambda_1}{D}}R\right)}-1\right\},
\end{eqnarray}
where $D$ is the diffusion coefficient. This defines as a by product the association constant of the reaction as $K=1/\left<T\right>$. Two first  comments  are in order. (i) First, as soon as the length of the DNA strand is
large enough (more precisely as soon as
$\sqrt{\frac{\lambda_1}{D}}L\gg1$ or
$\sqrt{\frac{\lambda_1}{D}}R\gg1$), $\left<T\right>$
grows linearly with the length of the DNA. This mirrors the
efficiency of intermittent reaction paths, as compared to the
quadratic growth obtained in the case of pure sliding. In
particular, the boundary effects are negligible for this quantity
as soon as the overall length is large enough. (ii) Second, this expression is
valid for a very large class of 3D motions. More precisely, it holds
as soon as the mean first return time $\tau_{2}$ corresponding to the 
3D motion is finite and independent of the departure and arrival
points. 

We now come to  an important question  recently addressed in \textcite{MirnySlutsky,DNA}, which concerns the {\it optimization } of such
intermittent reaction paths. We here assume that the mean search time is a limiting quantity which might have been minimized in the course of evolution.  In this context, we consider $\lambda_1=1/\tau_1$, which characterizes the protein/DNA affinity,  as the adjustable parameter. Indeed, this quantity depends strongly both on
the structure of the protein and on physiological conditions such as the ionic strength, and therefore could widely vary from one protein to another. On the contrary $\lambda_2=1/\tau_2$ depends mostly  on the
properties of the environment, such as the DNA conformation, which is itself subject to very stringent constraints and therefore much less likely to be varied.  Another adjustable parameter is the 1D diffusion coefficient $D$.  Optimizing the search time
with respect to this parameter is trivial: it is found that $D$ should be as large
as possible (assuming that $D$ and $\lambda_1$ are  independent), but obviously one should keep in mind that $D$ is controlled by the hydrodynamic radius of the  protein which can not be too small. For these reasons we focus here on $\lambda_1$.


\begin{figure}[h!]
\centering\includegraphics[scale = 0.4]{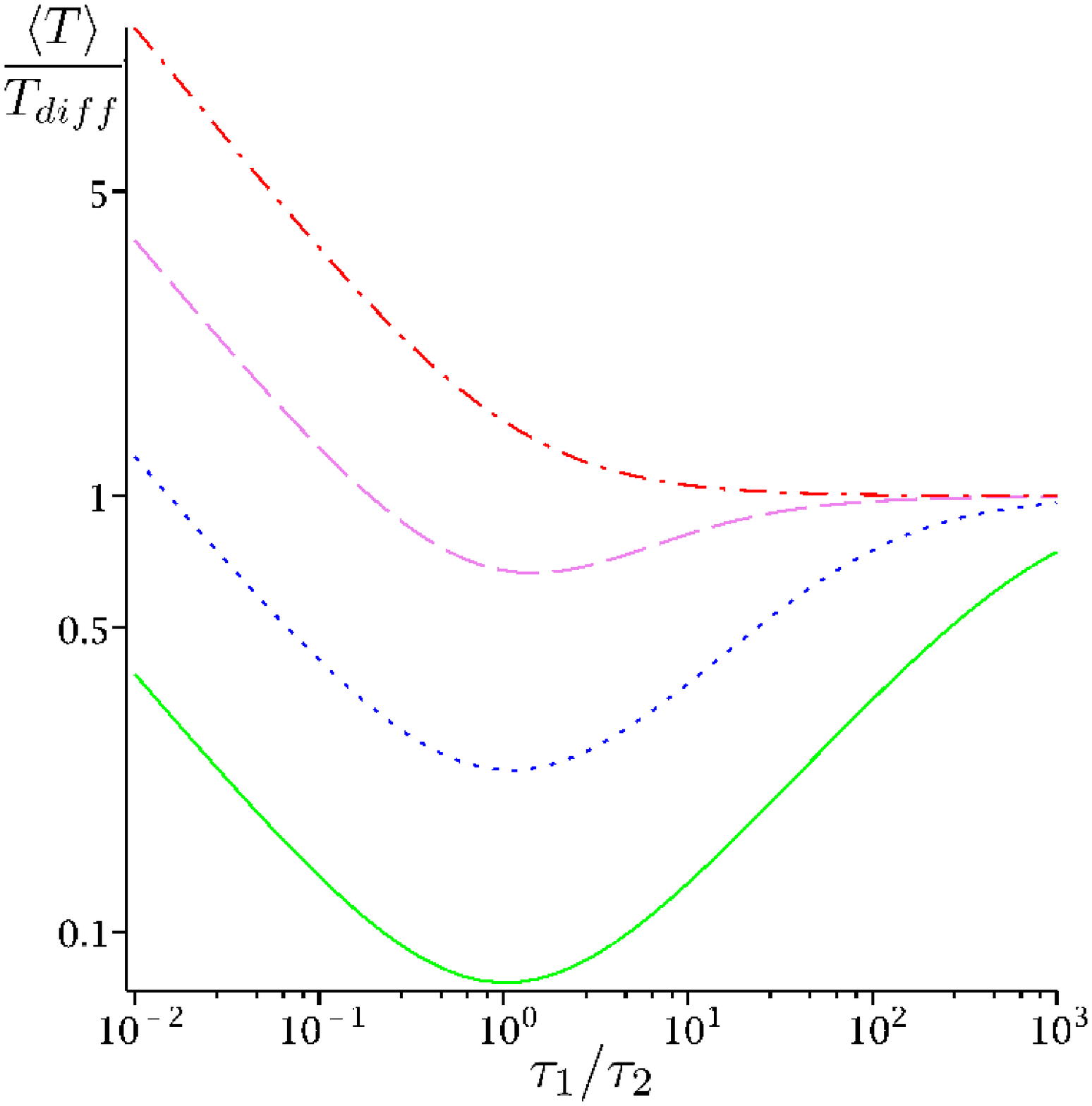}
\caption{\small Mean first passage time of a DNA--binding protein to its target site, 
renormalized by the mean first passage time by diffusion alone ($T_{diff}$), as a function of $\tau_1/\tau_2$. 
$T_{diff}/\tau_2=1$ (red dashed-dotted line), $T_{diff}/\tau_2=10$ (violet dashed line), $T_{diff}/\tau_2=100$ (blue dotted line), $T_{diff}/\tau_2=1000$ (green solid line).  For a small DNA length $\ell$ (and therefore small $T_{diff}$) (red dashed-dotted line), the reaction constant depends monotonously on $\tau_1$ and intermittent reaction paths are inefficient. For larger values of $l$ (other curves), the reaction rate can be optimized as a function of $\tau_1$. 
Here we averaged other the initial position of the target. }\label{KDNA}
\end{figure}

It can be seen qualitatively  that $\left<T\right>$ is large for both
$\lambda_1$ very large (in the limit $\lambda_1$ infinite, the protein
 is never on the DNA), and $\lambda_1$ very small (pure sliding limit which gives a quadratic growth with the DNA length), and could therefore be minimized for and intermediate value of $\lambda_1$. The sign of the derivative of the mean search time at $\lambda_1=0$ shows that it can indeed be minimized provided that 
\begin{equation} \label{condex}
\lambda_2>15D\;\frac{L^2+R^2-LR}{L^4+R^4+4LR(L^2+R^2)-9R^2L^2}.  \end{equation}
This condition means that bulk excursions, to be favorable, should be shorter than a fraction of the typical time needed to scan the full DNA molecule by 1D diffusion. In particular, it requires that the DNA length is long enough.  If this condition is fulfilled, the search time can indeed be minimized (see figure \ref{KDNA}).  A  careful analysis  of the
implicit equation satisfied by $\lambda_1$ at the optimum leads
to the following expansion for large $\ell=L+R$

\begin{eqnarray}
\label{lammin}
\displaystyle
\lambda_1=\lambda_2-4\frac{\sqrt{D\lambda_2}}{\ell}-\frac{8D}{\ell^2}-\frac{40D^{3/2}}{\sqrt{\
\lambda_2}\ell^3}+O\left(\frac{1}{\ell^4}\right).
\end{eqnarray}
Equations (\ref{condex}) and (\ref{lammin}) refine the result of \citet{MirnySlutsky}, which predicts than the optimal strategy is realized when $\lambda_1=\lambda_2$. This result actually 
holds  in the large $\ell$ limit, or more precisely for
$\sqrt{\frac{\lambda_1}{D}}\ell\gg1$. For intermediate values of
$\ell$
boundary effects become important and the minimum can be
significantly different.

The $\left<T\right>$ value at the minimum is
particularly interesting. We compare it to the case of pure sliding
where $\left<T_s\right>=\ell^2/(3D)$: 
\begin{equation}
\frac{\left<T\right>}{\left<T_s\right>}=\frac{6}{\ell}\sqrt{\frac{D}{\lambda_1}}
\end{equation}
 The efficiency of the 3D mediated strategy is therefore much more
important when the DNA chain is long. For example, using standard values for  $\lambda_1$ (a few $10^{-2}\ s.$)
and $D$ (typically $10^{-2}\ \mu^2/s.$) and for a DNA substrate of
length $10^6$ bp, the mean reaction time  is three orders of magnitude  smaller than for a pure sliding strategy. Beyond the importance of such results for understanding the kinetics of gene transcription, this first minimal model shows that intermittent reactive paths are indeed very efficient, and that  they can even allow to optimize the reaction kinetics.

\subsubsection{Towards a more realistic modeling}

The model introduced above provides a simple way to discuss the minimization of the search time. 
Further approaches have been developed to model target search by proteins.  We present below  the main models used in the literature, and   discuss their relevance to real target search problems by proteins in cellular conditions.

\po{Main approaches}
Generally speaking, theoretical models of facilitated diffusion rely on the basic assumption that the protein alternates phases of 1D diffusion along the DNA and phases of free diffusion when the protein is desorbed from the DNA. The existence of such two distinct states, whose dynamics is usually characterized by  association/dissociation rates, is supported by direct experimental observations as discussed above. Additionally, molecular dynamics simulations taking into account the electrostatic interaction between the negatively charged DNA and the locally positively charged protein (see for example \citet{florescu,PRL_dahirel_paillusson}) have shown that these two states naturally arise on the basis of the electrostatic interaction only, suggesting the robustness of the facilitated diffusion mechanism. Such studies at the molecular scale could serve as a tool to calculate the association/dissociation rates  used in the models of facilitated diffusion discussed in this section, which all take into account only effectively the molecular interactions.

\popetit{Stochastic modeling}

The minimal model presented in section \ref{section_DNA_minimal_model} relies on the statistical analysis of the trajectory  of a single protein, and can henceforth be qualified as a stochastic model. Similar stochastic methods have been  used and complemented  in
 \citet{Lomholt2005,LomholtZaidMetzler,Lomholt2009,adncirculaire,adncirculaireNgrand,adncirculaire_resume,ADNkafri}, ... 
and have the advantage, when solvable, of giving access to the full distribution of the search time, yielding a refined information on the search kinetics. Moreover they can be adapted in some cases to take into account anomalous transport both in the 1D and 3D phases as discussed below.

\popetit{Kinetic approach}  

The main alternative to the stochastic approach  is given by what can be called  kinetic models, which assume a steady-state homogeneous concentration of proteins in contrast with the single protein description of stochastic models. Such models therefore rely on a mean-field approximation which proves to be efficient to evaluate the mean search time thanks to scaling arguments.
A first example  is given by  \citet{HalfordMarko}, where  scaling arguments are used  to  roughly estimate the time for the protein to find the  DNA coil, and
then the time to find the target inside the coil, which eventually yields an optimal sliding length. More generally, 
the key ingredient of kinetic models, developed mainly in \citet{HuGrosbergShklovskii,HuGrosbergBruinsma08,HuShklovskii} , is that the system is assumed to be in a stationary state. Under this hypothesis, the flux of particles delivered by the 3D diffusion into the sphere of influence of the target, whose size is defined as  the "antenna length" $\xi_a$,  must be equal to the flux of particles delivered by 1D diffusion into the target.
Such  balance equation generically reads
\begin{equation}\label{grosberg}
J\sim D_3c_{free}\xi_a\sim D_1 c_{ads}/\lambda_a,
\end{equation}
 where the concentrations of free ($c_{free}$) and adsorbed ($c_{ads}$) proteins are assumed to be at equilibrium, {\it ie } satisfying $c_{free}/c_{ads}=K$ where $K$ is the equilibrium constant associated to the association/dissociation rates. Importantly, the  antenna length, defined as the typical scale below which the dominant transport is sliding instead of 3D diffusion, has size $\xi_a$ when measured in 3D space, but takes another value $\lambda_a$ when measured along the DNA. Making assumptions on the DNA conformation (for instance random coil or fractal globule), different scaling laws between  $\lambda_a$  and  $\xi_a$ can be proposed. Equation (\ref{grosberg}) then permits to determine $\xi_a$ and henceforth to give the scaling of the  mean search time $1/J$. The advantage of this method is that it permits, through the relation between    $\lambda_a$  and  $\xi_a$, to take into account various models of DNA conformations, which is much harder to achieve in the stochastic approach.
  Such models, whose results are compatible with the stochastic approach, provide in addition a useful picture of facilitated diffusion. Indeed, in these models the effect of sliding can be seen as  effectively making the target of the size of the antenna length, which is much larger than the real target size, and therefore speeding up the search.

\bigskip

~

\medskip

Finally, these two approaches are quite complementary and both require  as an input the modelling of 1D and 3D phases. The minimal model of section \ref{section_DNA_minimal_model}  describes the 1D phase as regular diffusion, while 3D phases are assumed to result in completely random relocations over the DNA. Beyond this minimal model, the specific description of these two phases has motivated numerous works and many refinements have been discussed in the literature. We review in the next sections the main models that have been proposed to provide a more realistic description of 1D and 3D phases.

\po{Descriptions of the 1D phase (sliding and recognition)}

\imagea{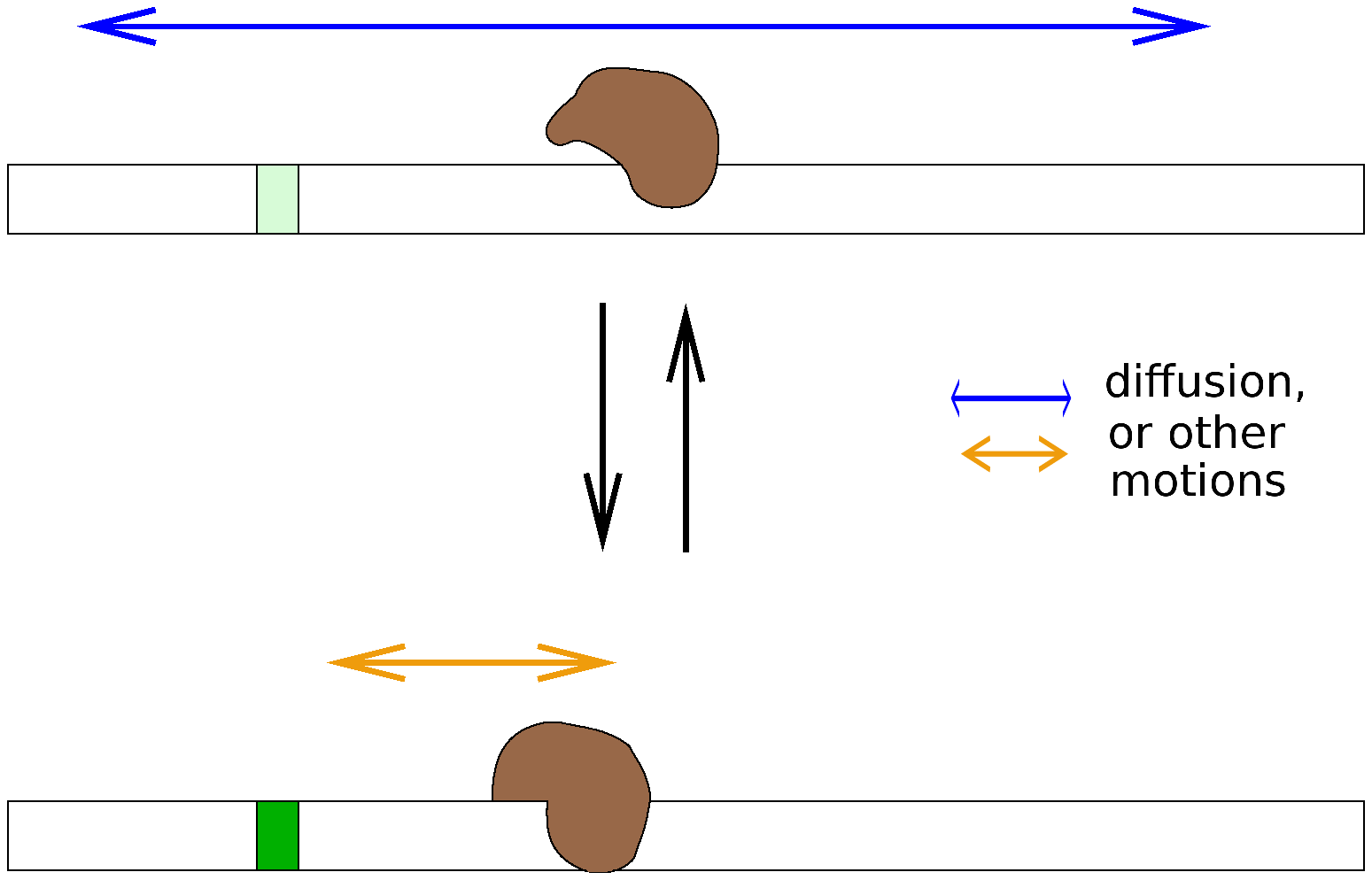}
{Sliding is often represented by diffusion with perfect reactivity on the target. 
This figure shows the two main directions for a more realistic description of the 1D phase~: on the one hand, 
the sliding is not necessarily diffusive, and on the other hand the 1D phase could be in fact a combination of 2 phases, 
one  fast but with low recognition, and another slow (or immobile), but with high recognition of the target. }
{model_ADN_1D}{5}

\popetit{Anomalous diffusion in the sliding phase} 

As stated above, the phase of one-dimensional non specific interaction of a protein with DNA, ``sliding'', is generally described as  Brownian diffusion as in the minimal model of section \ref{section_DNA_minimal_model}. If this hypothesis seems to be confirmed by in vitro experiments \cite{kabata_science_93,NAR08}, it could not always be the case, in particular in vivo.
A first limitation of this simple description  appears in the case of many proteins binding to DNA, as it is the case in vivo, which are likely to create  traffic jams \citep{SokolovMetzlerPantWilliamsPRE2005,looping_elf}. Such crowding effects in one dimension are known  to  potentially lead to a subdiffusive behavior. Additionally, even in  the case of a single protein, 
it should be kept in mind  that the DNA sequence is not homogeneous, and the disorder in the sequence can also impact on sliding.  
The heterogeneity in the sequence is often modelled by a disordered energy landscape, whose distribution is  Gaussian \citep{maria_seq_ADN,Wunderlich2008,HuShklovskii}. 
 \citet{maria_seq_ADN} show that in this case sliding is not purely diffusive~: 
at short times, the protein will be trapped in local minima, leading to a subdiffusive behavior. The diffusive behavior is recovered only at larger times, or equivalently for 
sliding lengths longer than  a hundred of  base pairs.

Anomalous diffusion in the sliding phase has been discussed at the theoretical level in \citet{adncirculaire} 
(see also \citet{adncirculaireNgrand,adncirculaire_resume}), which extend 
the minimal model of facilitated diffusion 
\citet{DNA} summarized in section \ref{section_DNA_minimal_model}. 
In particular, 
the Laplace transformed  search  time distribution is obtained  for several non Brownian sliding motions such as
ballistic, self-similar or halted motions  (in particular when halts durations 
are widely distributed, leading to a subdiffusive behavior), therefore covering standard models of anomalous diffusion.  
Importantly, \citet{adncirculaire}  find that whatever the model of sliding, there are always regimes in which intermittence is favorable, similarly to the case of Brownian sliding. They further  show that  in the case of 3D excursions with finite mean durations, 
the mean search time with an arbitrary sliding mechanism remains of order $\propto \ell$. This indicates that 
for long enough DNA, intermittence is favorable  
for a wide range of sliding motions, be it normal or anomalous, which supports the robustness of the facilitated diffusion mechanism.

\popetit{Target recognition}

 A simplification which is often used in the literature consists in assuming that the target is  perfectly reactive, \textit{i.e.} that reaction occurs with probability one at the very first passage of the protein to the target sequence. 
\citet{MirnySlutsky} however stress that if there is an activation barrier at the target site, the protein as a chance to pass on the target without entering the recognition process, and therefore to miss it. The roughness $\sigma$ of the  energy landscape of the sequence can then play a crucial role. If $\sigma$ is of order $1 k_B T$, \citet{MirnySlutsky} find that
the sliding  diffusion  is fast, but that the protein 
has a high chance to miss the target. On the contrary, 
if   $\sigma$ is of order $ 5k_B T$, the 
recognition probability is  high, 
but the sliding diffusion coefficient is very low, 
leading to a huge search time. Such result seems to set conflicting constraints on the search, since an efficient target search process requires both speed and reliability in target recognition. To overcome this paradox, \citet{MirnySlutsky}
propose on the basis of direct structural observations \cite{kalodimos_science2004} that the protein can perform conformational  changes in the sliding phase, and switches between a fast search state (with low $\sigma$) and a slow recognition state (with high $\sigma$). They  show that
if the two energy landscapes 
are strongly correlated, 
it is possible to conciliate high speed and efficient recognition. Similar models of 2-state proteins in the sliding phase have been studied more quantitatively both in the framework of the kinetic (see 
\citet{HuGrosbergBruinsma08}) and the stochastic approach (see   \citet{ADNkafri}), confirming that such mechanism indeed permits a fast search with reliable recognition.

\po{Descriptions of the 3D phase (jumping/hopping)}

We now review the different descriptions of the 3D  phase. Most of models of facilitated diffusion implicitly require the knowledge of the probability $\pi(x|x_0)$ that a protein which  desorbs the DNA at position $x_0$ will eventually rebind for the first time to the DNA  at position $x$, where the coordinate $x$ measures the distance along the chain. This quantity depends both on the dynamics of the protein and on the conformation of the DNA, and in practice can be determined explicitly only in the case of an ideal infinite cylindrical DNA (see below), which makes  assumptions necessary in realistic situations. Depending on the relocation length $|x-x_0|$, 3D excursions have been given in the literature different names, mainly either "hops" (referring to "small" $|x-x_0|$) and jumps (referring to "large" $|x-x_0|$). Since the definition of jumps and hops may vary according to authors, the limit between both being somehow arbitrary, we give below the one that  will be used  in this review.

\imagea{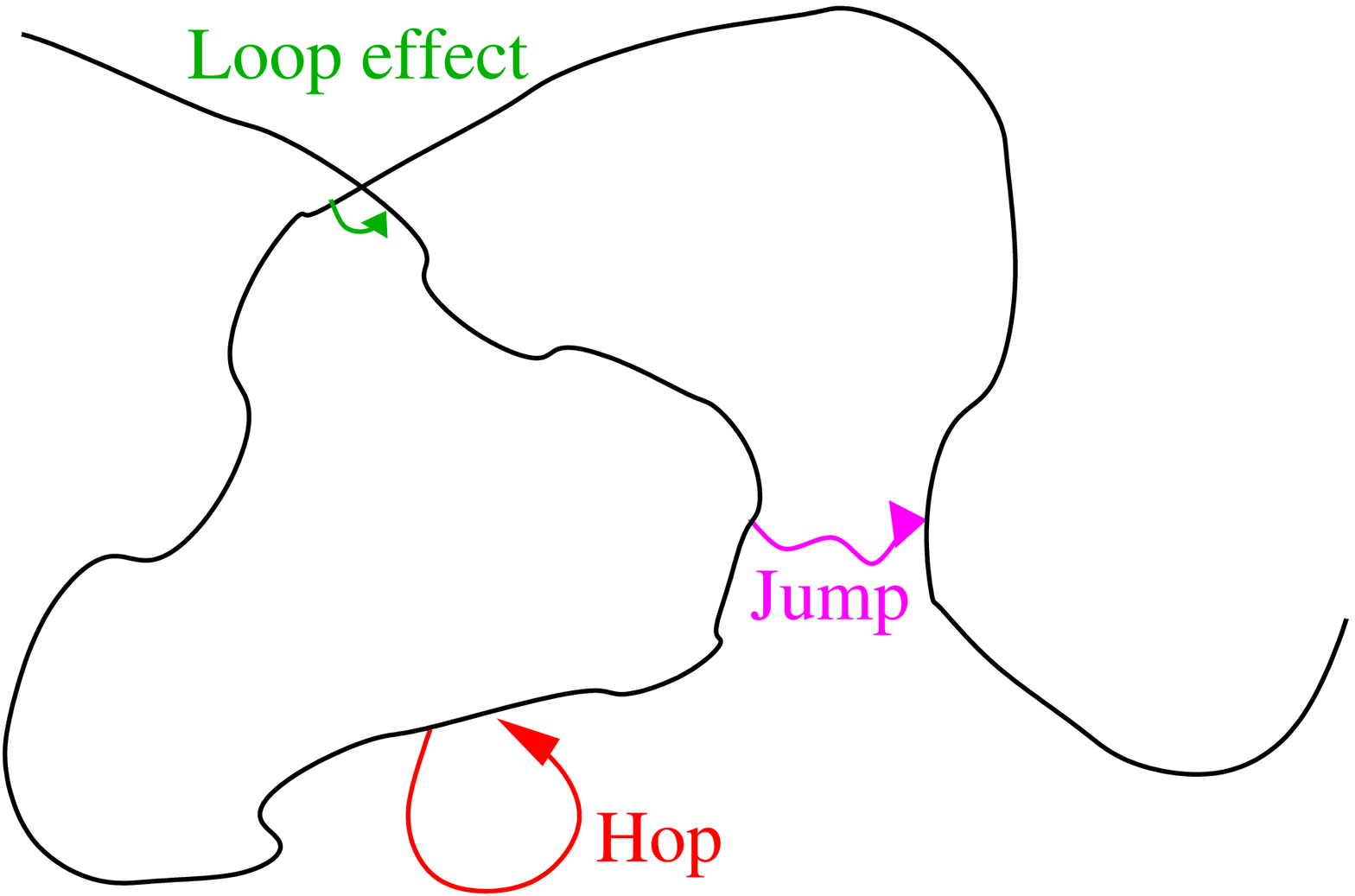}
{Description of 3D excursions.  Jumps are 3D excursions whose starting and ending points on the DNA sequence are uncorrelated. In practice, for confined DNA conformations as in cellular medium,  jumps have a span (measured along the DNA contour) larger than the density-density correlation length $\xi_c$ of DNA. Conversely, hops are 3D excursions whose starting and ending points on the DNA sequence are correlated, or equivalently  3D excursions whose relocation length is smaller than the correlation length $\xi_c$.}
{model_ADN_3D}{5}

\popetit{Jumps} 
We define as jumps the 3D excursions  whose starting and ending points on the DNA sequence are uncorrelated, {\it ie}  such that the relocation probability $\pi(x|x_0)$ is independent of $x$. This definition of course depends on the DNA conformation, and makes sense only for dense enough conformations such as a random coil (for instance for free DNA in solution) or even denser packings that can be  expected in vivo (such as fractal globule structures \cite{GrosbergNechaev} ). In such structures, sequences which are far apart along the DNA  chain can be actually very close in the 3D space. Hence, 3D excursions whose relocation length is larger than the typical distance between DNA segments  are likely to end at any remote location on the DNA sequence, and should therefore be considered as jumps according to our definition. The lower bound of the  relocation length of jumps therefore strongly depends on the DNA conformation and is in practice hard to evaluate. In the case of interest of a confined DNA (as in cellular conditions), the typical distance between DNA segments  can be estimated by  the DNA density-density correlation length $\xi_c$. This suggests an alternative and  equivalent definition of jumps as 3D excursions  whose relocation length is larger than  $\xi_c$.
A widespread assumption used for example in the minimal model of section \ref{section_DNA_minimal_model} 
(see also  \citet{HuGrosbergShklovskii} at scales larger then the antenna size or  \citet{DNA,adncirculaire,ADNkafri}) consists in taking all 3D excursions as jumps, {\it ie} as random uniform relocations over the DNA. Even if not exact, this assumption has been checked numerically on the example of a quenched self avoiding DNA and as proved to be very satisfactory \cite{kafri_intersegment} at high enough DNA concentration. 
Interestingly,   this definition of jumps highlights the importance of the local DNA concentration on the search efficiency, as observed in  \citet{broek,Lomholt2009} :
the more densely packed the DNA, the smaller the correlation length $\xi_c$, and therefore 
the higher the probability of jumps enabling the protein to explore previously  unscanned areas, 
and the less the time spent in 3D phases. 

The assumption that all excursions of relocation length larger  than  the typical distance between DNA segments lead to a completely uniform relocation over the DNA is however not exact. Indeed, large scale correlations in the 3D conformation of the DNA chain may exist, as in the model case  of a free random coil conformation. In particular in eukaryots, where DNA is packed in the nucleus, recent studies such as  \cite{science2009avecMirny} support a hierarchized structure of DNA which could induce long range correlations in the conformation and therefore a  non uniform relocation probability. The impact of DNA conformation on the relocation probability has been tested on the example of  an annealed worm-like chain polymer by \textcite{MarioDiaz} who found that
 at short times, 3-dimensional excursions  
of approximately the DNA persistence length  are actually less abundant than both shorter and a bit larger relocations : 
 by definition, closer sequences along the DNA are closer in 3D space,
 but sequences further than the persistence length can take advantage of loops and actually be even closer in 3D space. 
 More generally, the loop statistics impacts on the relocation probability and can lead to a non uniform relocation probability $\pi(x|x_0)$ . \citet{Lomholt2005} argue  that since polymers form loops whose linear size $x$ is distributed according to  $p(x)\sim|x|^{-1-\alpha}$ 
(for instance $\alpha=0.5$ for Gaussian chains, $\alpha \simeq 1.2$ for self-avoiding walks), relocations distributed according to the same law are favored and  should also be taken into account. In an annealed version of such a model, the authors show that depending on $\alpha$ the optimal strategy can widely vary. It should be added that loops can enable another relocation mechanism for proteins with multiple binding sites called intersegmental transfer 
(see \citet{Hu2007} and \citet{kafri_intersegment}) which can be shown  for modelling purposes  to be widely equivalent to 3D excursions. Finally,  the approximation of uniform random relocations proved to be useful and can be validated numerically for simple DNA conformations {\cite{kafri_intersegment}, but a better knowledge of the in vivo conformation of DNA would be necessary to assess more precisely the relocation probability.

\popetit{Hops} 
Echoing the definition of jumps, we define hops as 3D excursions whose starting and ending points on the DNA sequence are correlated, or equivalently as 3D excursions whose relocation length is smaller than the typical distance between DNA segments, which is given by  the correlation length $\xi_c$ for confined DNA. Due to their local character, hops are often effectively taken into account in the sliding mechanism (see \citet{HuGrosbergShklovskii} for example).  If this assumption is in practice useful, it also raises additional questions. (i) First, hops do not explore continuously the DNA, and a protein performing hops have higher chances to miss the target (note that hops permit as a counter part to bypass obstacles on the DNA), and this non perfect reactivity has to be taken into account. (ii)  Second, if hops are included in an effective sliding mechanism, then the effective diffusion coefficient as to be determined, as well as the effective transition rate from this effective sliding state to the jumping state. This last point amounts in practice in calculating the relocation probability $\pi(x|x_0)$ of hops, which gives as a byproduct the probability that after desorption from DNA, the protein performs a jump rather than a hop.

\begin{figure}[t]
\centering\includegraphics[width =1\linewidth,clip]{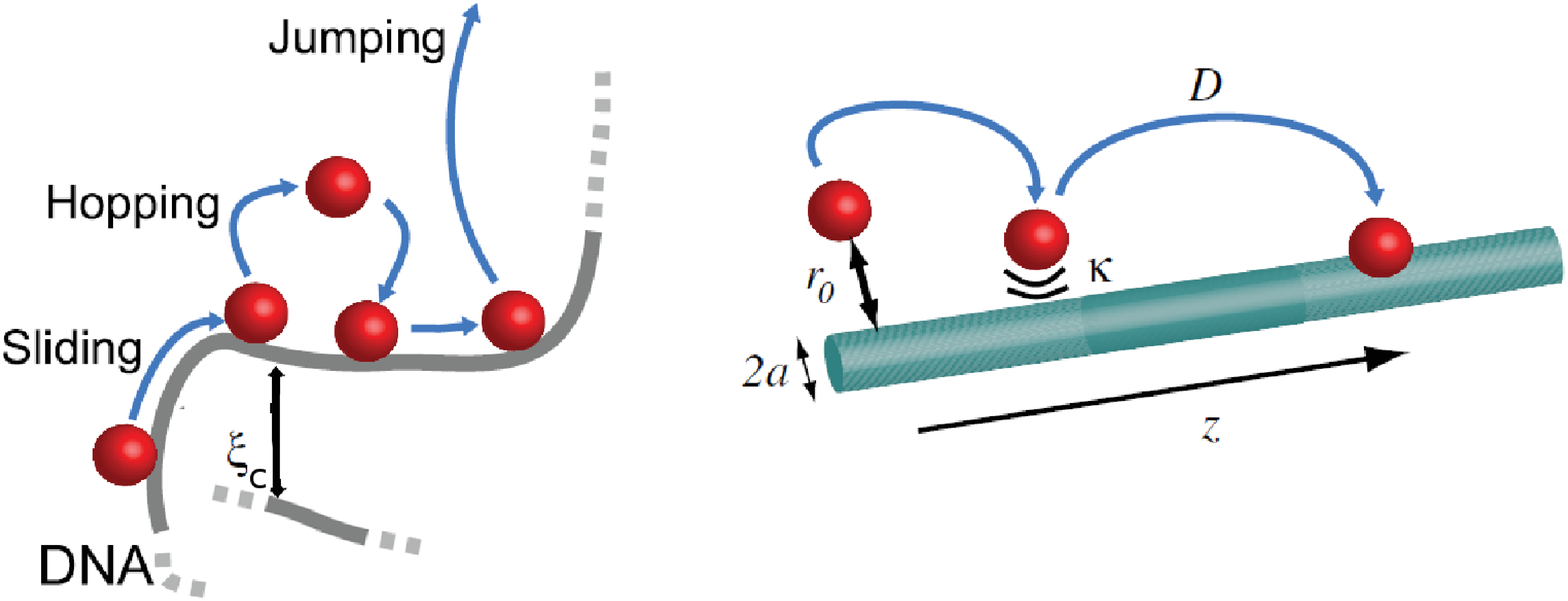}
 \caption{Facilitated diffusion of a protein on DNA. {\it Left:} schematic definition of sliding, hopping and jumping. Hops are 3D excursions whose starting and ending points on the DNA sequence are correlated, or equivalently  3D excursions whose relocation length is smaller than the typical distance between DNA segments. In the case of confined DNA, this distance is estimated by  the DNA density-density correlation length $\xi_c$.   {\it Right:} model parameters. }
\label{fig1}
\end{figure}

This problem, which has been studied numerically in \citet{Wunderlich2008},  can actually be studied analytically  \cite{postNAR08}, since at the scale of hops, which are smaller than the DNA persistence length $\xi_p$, the DNA can be simply modelled as a cylinder of effective radius  $a=R_{DNA}+R_{protein}$ (see figure \ref{fig1} left).
Denoting $D$ the 3D diffusion coefficient of the protein (assumed point-like). The density probability $\pi(x|{\bf r_0})$ of being adsorbed on the DNA  at the longitudinal abscissa  $x$, starting from the point ${\bf r_0}$ then satisfies   (for details on derivation of this equation and the following ones, see \textcite{postNAR08}, and  \textcite{metzler_cylindre,Levitz08,Berg76} for similar analysis):
\begin{equation}
 \Delta_{\bf r_0} \pi(x|{\bf r_0})=0
\end{equation}
Assuming radiative boundary condition on the cylinder surface to account for the adsorption rate on DNA,
 \begin{equation}
 \partial _{r_0}\pi(x|{\bf r_0})(r_0=a)=\kappa \pi(x|{\bf r_0}) (r_0=a),
\end{equation}
 one can show that:
 \begin{equation}\label{pi}
\pi(x|{\bf r_0})=\frac{1}{\pi}\int_0^\infty \cos(kx)\frac{K_0(kr_0)}{K_0(ka)+K_1(ka)k/\kappa}{\rm d} k,
\end{equation}
where $K_i$ are Bessel functions.
 This relocation distribution, in good agreement with experimental data  from \textcite{NAR08}
 obtained on an extended DNA, enables to address the questions of point (ii) above.
 First, Eq.(\ref{pi}) gives 
the analytical distribution of hops, since for $x<\xi_c$ the DNA can be well approximated by a cylinder. In turn, as shown in  \cite{postNAR08}, this gives access to the effective diffusion coefficient of a combine motion of sliding and hops.
Second, according to our definition,  all relocations  with $x > \xi_c$ will be jumps.
Thus, the probability that a 3D relocation is a jump rather that a hop is given by   the complementary cumulative distribution:  
\begin{equation}\label{cum}
C(x=\xi_c)=\int_{|x|>\xi_c} \pi(x|{\bf r_0})dx \sim  \frac{\ln(r_0/a)+1/\kappa a}{\ln (\xi_c/a)}.
\end{equation}}
Going back to the search problem, in regimes where jumps are favorable, 
 decreasing the correlation length speeds up the search process, as found by  \textcite{broek}.

\popetit{Crowding effects }

In both prokaryots and prokaryots, the cellular medium is very crowded, and 3D excursions of proteins are likely to be hindered. The normal or anomalous nature of transport in cellular medium is still debated. For instance,
\citet{DixVerkman} supports that 3D motion is mostly normal diffusion;  
whereas  \citet{MalchusWeiss} suggest that there is more and more evidence for subdiffusion. 
Experimentally, it is found that some tracers exhibit transient behavior 
(subdiffusive at small timescales, diffusive at larger time scales) : 
measures however depend on the size and nature of the tracer, on the time and length-scales covered and other experimental conditions, which may explain the lack of consensus on the problem.
The influence of subdiffusion on target search will depend on the microscopic mechanism at play. 
There are three main mechanisms leading to  subdiffusion, 
as underlined section \ref{resume_random_walks} : 
random walk on a fractal medium, random walk with long waiting times (CTRW), or random walk with long range correlations
such as  Fractional Brownian Motion (FBM). 
Which of these possibilities best describes transport in  crowded environments such as the cellular medium is however still unclear 
(see  \citet{CTRWweak_ergodicity_breaking_burov}, 
\citet{Bancaud:2009}, 
\citet{SzymanskiWeiss} or  \citet{FBMtejedor_benichou_voiturie_metzler} for various opinions on the subject).

\citet{LomholtZaidMetzler} explore the effect of  a crowded environment with 
subdiffusion $\langle r^2(t)\rangle \propto t^{\alpha}$ ($0<\alpha<1$) caused by 
waiting times distributed as $p(t)\sim \tau^{\alpha}/t^{1+\alpha}$. 
They argue that  
because of these waiting times, 
the probability that the protein has not yet left the DNA at time $t$ and the probability 
that an unbound protein has not yet bound to DNA after a time $t$ 
both scale as $1/t^{1+\alpha}$. 
Their results have two main practical implications. 
On the one hand, in an experiment, since proteins can remain stuck for very long times, 
ensemble averages 
do not lead to the same results 
as time averages as also highlighted by \citet{CTRWweak_ergodicity_breaking_burov}. 
On the other hand, since  proteins would slide for a longer time and 
as it would take them a very long time to return to DNA, 
the genes coding for transcription factors should be 
close to their target sequences, as also underlined by \citet{Wunderlich2008}. The analytical determination of the relocation  distribution above can be  extended  to a fractal medium \cite{postNAR08}. 
In this case, using the  \textcite{procaccia} formalism, the  large $x$ behavior  of the relocation distribution is obtained as 
$\pi(x|{\bf r_0})\sim r_0^{d_w-d_f^{\perp}}/x^{1+d_w-d_f^\perp},$ where     $d_f^\parallel$ is  
the dimension of the projection of the fractal on the axis parallel to the cylinder \cite{postNAR08}. Hence, 
 the relocation  distribution  always decays faster than in the case of normal diffusion ($\sim 1/z \ln^2 (z/a)$).   
As a consequence,  the proportion of jumps in the case of random conformation of DNA in the fractal  type crowding  scales like $C(\xi)\sim  \xi^{-d_w+d_f^\perp}$~: 
 it is much smaller than for regular diffusion, 
 which shows that fractal crowding favors hops and changes the overall intermittent search.

\po{Beyond the mean : variability of the search time}

Minimizing the mean search time is the optimization procedure the most often used (see for example \citet{DNA}). However,  the entire  distribution of the search time can be  needed  to assess the search kinetics on all time scales. Obviously in the case of simple exponential distributions of the search time, the mean is sufficient to describe the full dynamics. However, several models discussed below have shown that the search time distribution could not always be a single exponential. Such departure from an exponential distribution can have important consequences, such as large fluctuations of the search time which could be an extra source of variability in gene expression.

\popetit{Effect of trapping sequences}

A first possible  source of fluctuations in  the search time could come from the existence of trapping sequences along the DNA. Since a target sequence is typically 10 base pairs long, very similar sequences are statistically unavoidable and can be expected to be local minima in the protein/DNA interaction energy landscape. \citet{ADNkafri} propose a model in which the protein can be stuck on such sequences which are similar to the target. The corresponding trapping times naturally introduce new time scales in the problem, which potentially could be very long. In the framework of this model, it is shown that the search time distribution is best described by {\it two} exponentials. In particular, the mean search time is controlled by long trapping events even when they are very unlikely, and can be orders of magnitude larger than the median search time which is controlled by the trajectories that do not fall into the traps. The main outcome of such model is that it reconciles the possibility of having long lived stable complexes ({\it ie} deep traps), and very fast typical search times, which were casted so far as  paradoxical requirements.

\popetit{Effect of $n$ searchers} 

The influence of the number of searchers has been  discussed by several authors  
\citet{SokolovMetzlerPantWilliamsPRE2005,adncirculaire,adncirculaireNgrand,adncirculaire_resume}. 
Importantly, as stressed in \citet{ADNkafri}, 
the mean time for $n$ independent searchers 
  is simply given by the mean time divided by the number of searchers only if the search time distribution for a single searcher is a single exponential. In the case of non exponential distributions ({\it eg} a sum of weighted exponentials) it can be shown that the effect of the number of searchers can be much stronger since the weight of long time scales decreases exponentially with $n$, selecting for $n$ large enough only the shorter time scale of the problem (see \citet{ADNkafri}).  Moreover, if the concentration of searchers increases considerably, 
searchers cannot be considered as independent anymore and will act as "roadblocks" along the DNA.
These roadblocks will decrease the effective sliding length, and may also hide the target, overall slowing down the search process.
This results in a trade-off, as stressed in \citet{looping_elf}.
On the one hand, 
the more the proteins, the more the searchers for the target and the quicker the search.
On the other hand, the more the proteins, the more the crowding, the less efficient the search of a single protein.
These authors predict that the optimum is obtained for  $10^4 - 10^5$ DNA binding proteins for E.Coli, which is 
close to the experimental value of 30~000 proteins.

\popetit{Dependence on the starting point: colocalization} 

Another origin of fluctuations of the search time can be due to its dependence on the starting position of the protein. \citet{Wunderlich2008} show that
a  target which is close to the starting point of the protein
can be found within a single  sliding phase, which yields  a very short search time and a rather  low variability.
In contrast, if the target is far away from the starting point of the protein, 
it is found after numerous 3D excursions.
The mean search time is then much longer, and the spread of the distribution of the search time  
is larger. \citet{KolesovMirnyPNAS07,Wunderlich2008} then argue that for increasing the efficiency of the transcription factor, its coding sequence ({\it ie} its starting position) should be colocalized with 
its target sequence (see also \citet{epladn,Benichou:2009b} for a further optimization of this colocalization 
effect with respect to the diffusion coefficient of the protein).  Colocalization is indeed  observed in real prokaryotes genomes. Such mechanism can however be invoked only in prokaryotes, where there is no cell compartments separating protein production from 
DNA.

More generally, geometric effects on search kinetics have been discussed in \cite{Claire}, where it is shown that low dimensional effects, such as sliding or diffusion on fractals, can lead to non exponential distributions which depend strongly on the starting position of the searcher. Such mechanism could be important for eukaryots. Indeed, in eukaryotes, DNA is packed inside the nucleus in what is called the chromatin, and some  DNA regions could be more or less accessible depending on the chromatin configuration.
For example the DNA close to the nucleus pores is much more accessible to incoming proteins than the
DNA buried deep inside the nucleus, leading potentially to very different search times. \citet{kampmann_chromatine} argues qualitatively that proteins 
binding to DNA could take advantage of the heterogeneities of the chromatin, and,  
depending on the searched sequence, adopt different optimal strategies. The geometric effect can be particularly important in the case of genes which need to be activated simultaneously. Indeed, their colocalization in the nucleus permits to share the transcription material, since the search time for a transcription factor going from one to the other will be much smaller than in the case of a random localization  in the nucleus.



\subsubsection{Conclusion on protein/DNA interactions}

The mechanism of facilitated diffusion of proteins on DNA is intrinsically intermittent in the general meaning defined in this review~: 
it is a combination of one dimensional motion  in close interaction with DNA, called ``sliding'', which enables target detection,  
and faster 3D excursions. 
From the theoretical point of view, this further example of intermittent process has been shown to significantly speed up the search. 
Over the past few years, strong experimental evidences have been brought in,  showing that this mechanism is indeed at play.  
Interestingly, in this case in the "fast" phase the searcher is not able to detect the target not because of a lowering of its perception abilities, 
but simply because the motion takes place in a geometrical space which does not contain the target. 
Most of microscopic realizations of intermittence fall in  this case as we will see in next section on a further example.

\subsection{Active transport of vesicles in cells}
After this first microscopic example of intermittent search, 
we turn to another example~: active 
transport of vesicles reacting at specific locations in cells.

\label{section_generic_vesicles}

\subsubsection{Active transport in cells}

\begin{figuresmall}[h!]\small
\begin{center}
      \includegraphics[width=12cm]{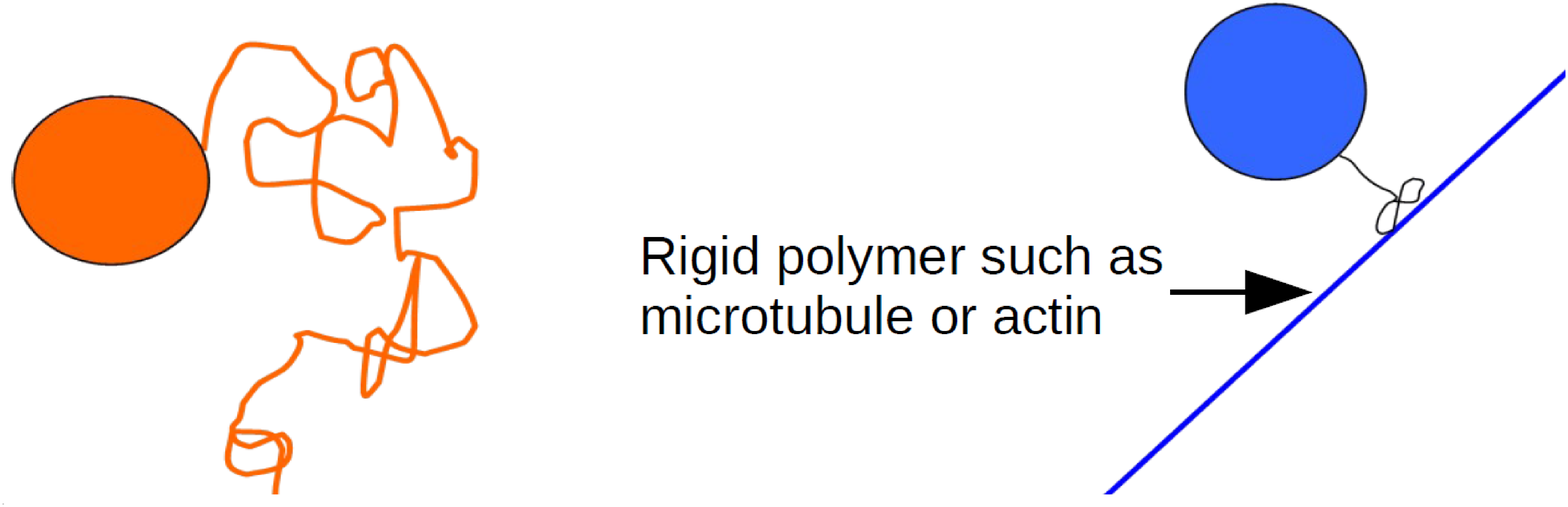}

\end{center}

   \begin{minipage}[c]{.46\linewidth}
\begin{center}

Does not need ATP

Distance$^2 \propto Dt$ (or less if subdiffusive)

Diffusion coefficient decreases when the size of the object increases

\end{center}
   \end{minipage} \hfill
   \begin{minipage}[c]{.46\linewidth}
\begin{center}

 Consumes ATP

Distance = $Vt$

Velocity $\sim 1~\upmu$m$.$s$^{-1}$, 
almost independent of the object size (see for example \textcite{refvmoteurfixe})

\end{center}
   \end{minipage}\hfill

\caption{Transport options for vesicles inside cells.}\label{comparaison_transport_vesicules}
\normalsize\end{figuresmall}

 Various motor proteins such as  kinesins or  myosins are able to convert the chemical fuel provided by ATP 
into mechanical work by interacting with the semiflexible oriented filaments (mainly F--actin and microtubules) of the cytoskeleton   \cite{alberts}.  
As many molecules or larger cellular organelles like vesicles, lysosomes or mitochondria, hereafter referred to as tracer particles, 
can randomly bind and unbind to motors, the overall  transport of a tracer in the cell can be described as alternating phases of standard diffusive transport (sometimes subdiffusive), 
and phases of active directed transport powered by motor proteins \cite{alberts,Salman2005} \refi{comparaison_transport_vesicules}. 
In particular, \textcite{karatekid} studied the rate of transitions between ballistic, diffusive 
and ``on the target'' states of vesicles, and found that 
the vesicles studied are much more likely to react in the free diffusive phase than when bound to motors.
Active transport \new{in this case} is therefore clearly a further example of intermittent behavior. 
Active transport in cells has been extensively studied both experimentally, for instance by single particle tracking methods \cite{myosine_rapide,Howard1989,ELBAUM1,ELBAUM2}, and theoretically by evaluating the mean displacement of a tracer  \cite{Shlesinger1989,Ajdari1995,Salman2005}, or stationary concentration profiles \cite{Nedelec2001}.
This transport is important for example for dynamically regulating the 
distribution of proteins such as membrane receptors.

Most of cell functions are regulated by coordinated chemical reactions which involve  low concentrations of reactants (such as ribosomes or vesicles carrying targeted proteins), and which are therefore limited by transport. 
An analytical model 
based on the idea of intermittence has been introduced in  \textcite{natphys08} (see also \textcite{natphys08comment, proceedingSigmaphi}),  and   enables the determination of the kinetic constant of transport limited reactions  in active media,
and further shows that the kinetic constant can be optimized.
We give below the main results of the model, and further details can be found in appendix.

\subsubsection{Model}

The model relies on the idea of intermittent search strategies 
and has important similarities with the model of section \ref{section_animaux}, 
which are discussed in section \ref{section_generic}. 
We consider a tracer particle evolving in a $d$--dimensional space (in practice $d=1,2,3$) which performs thermal diffusion phases of diffusion coefficient $D$ (denoted phases 1),
 randomly interrupted by  ballistic excursions bound to motors (referred to as phases 2) of
constant velocity $V$ and direction pointing in the solid angle $\omega_{\bf V}$ \refi{micro3D}. The distribution of the filaments orientation is denoted by $\rho(\omega_{\bf V})$, and will be taken as either  disordered or polarized (see figures \ref{micro3D}, \ref{micro2D}, \ref{micro1D}), which schematically reproduces the different states of the cytoskeleton \cite{alberts}. The random
duration of each phase $i$ is assumed to be exponentially distributed with mean
$\tau_i$.
The tracer $T$ can react with reactants $R$ (supposed immobile) during  free diffusion phases 1 only, as $T$ is assumed to be inactive when bound to motors. %
Reaction occurs with a finite probability per unit of time $k$ when the tracer-reactant distance is smaller than a given reaction radius $a$. In what follows  the kinetic constant $K$ of the reaction $T+R\to R$ is explicitly determined.

\subsubsection{Methods}

We now present the basic equations  in the case of a   reactant centered in a spherical
domain of radius $b$ with reflecting boundary.  This
geometry both mimics  the relevant situation of a single target and provides a mean field approximation of the general case of randomly located reactants with concentration $c=a^d/b^d$, where $b$ is the typical distance between reactants.  We start from a mean field approximation of  the first order reaction constant \cite{Berg76}  and write   $K=1/\langle t \rangle$, where the key quantity of our approach is the reaction time $\langle t \rangle$ which is defined as the mean first passage time   \cite{Redner,NatureSylvain} of the tracer at a reactant position uniformly  averaged over its initial
position. $t_1$ is defined as the mean reaction time if the tracer starts in phase 1 at position ${\bf r}$, 
and $t_2$ is defined as the mean reaction time if the tracer starts in phase 2 at
position ${\bf r}$ with velocity ${\bf v}$. For the active intermittent dynamics defined above, $t_1$ and $t_2$ 
satisfy the following backward equations \cite{Redner} (see section \ref{section_generic} for derivation)~:
\begin{equation}\label{back1}
\left\{
\begin{array}{l}
\displaystyle D\Delta_{\bf r}t_1+\frac{1}{\tau_1}\int(t_2-t_1)\rho(\omega_{\bf V} )d\omega_{\bf V}-k{\rm I}_a({\bf r})t_1=-1\\
\displaystyle {\bf V}\cdot\nabla_{\bf r}t_2-\frac{1}{\tau_2}(t_2-t_1)=-1\end{array}\right.
\end{equation}
where \new{$\Delta_{\bf r}$ and $\nabla_{\bf r}$ are the Laplacian and the gradient on the initial position,} ${\rm I}_a$ is the indicator function of the sphere of radius $a$. As these equations (\ref{back1}) are of integro-differential type, standard methods of resolution are not available for a general distribution $\rho$.

However, in the case of a \textit{disordered} distribution of filaments ($\rho(\omega_v)=1/\Omega_d$, where $\Omega_d$ is the solid angle of the d--dimensional sphere), 
these equations can be solved exactly in dimension 1.
In  dimension 2 and 3, an approximate scheme has to be introduced; the details of the calculation are given in appendix. 
We present here  simplified expressions of the resulting kinetic constant by taking alternatively the limit $k\to\infty$, which corresponds to the ideal case of perfect reaction, and the limit  $D\to0$ which allows us to isolate the $k$ dependence.

\subsubsection{Active transport in the cytoplasm}

\imagea{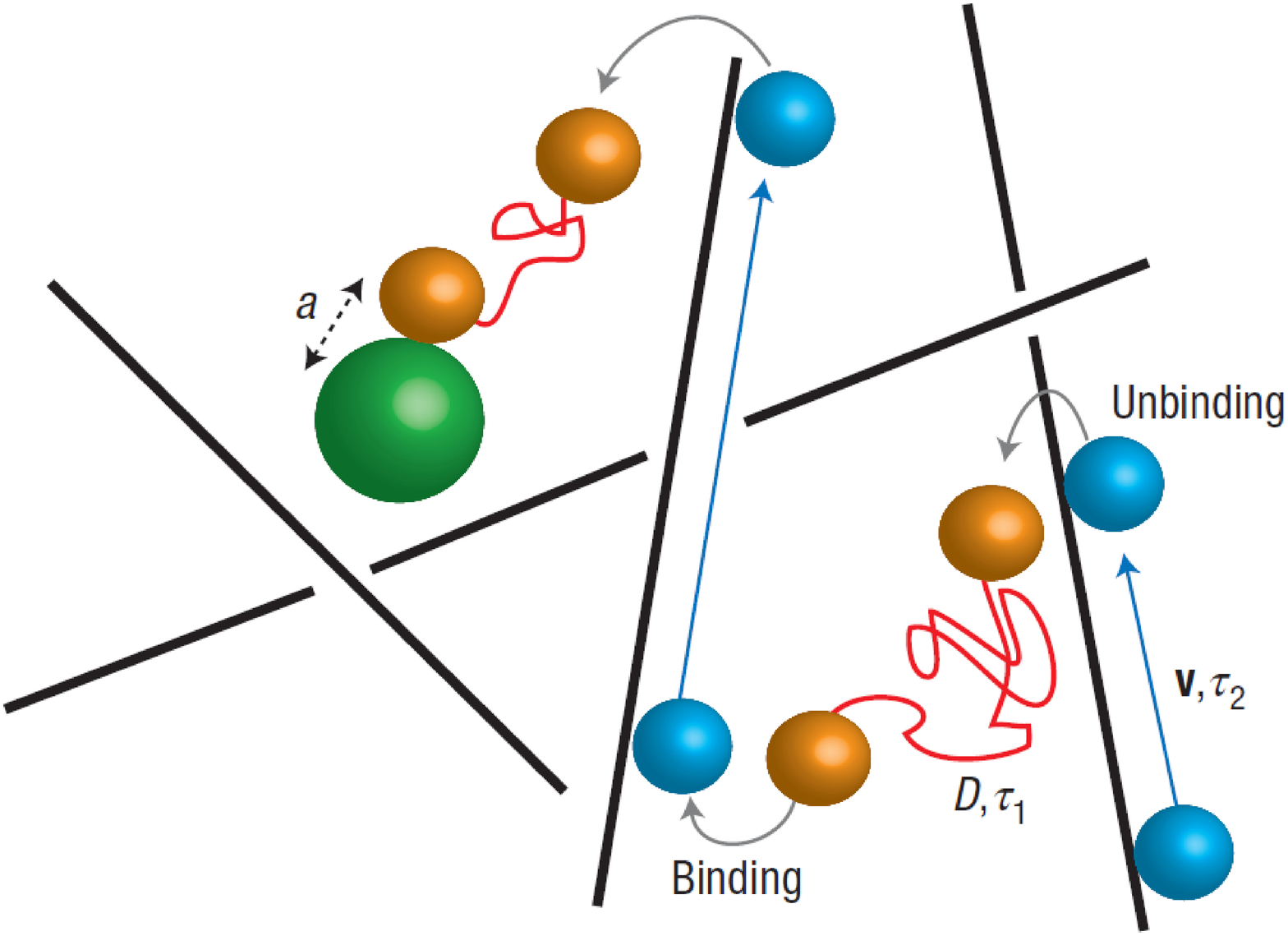}{Vesicle transport in the bulk (3 dimensions).}{micro3D}{7}

We first  discuss the  $d=3$ disordered case \refi{micro3D}, which provides a general description of the actin cytoskeleton of a cell in non polarized conditions, or of a generic in vitro active solution.
An analytical form of the mean first passage time  $\langle t \rangle =1/K_{3d}$ is given in appendix, and plotted in figure \ref{K3d}. Strikingly, $K_{3d}$ can be maximized  as soon as the reaction radius exceeds a threshold $a_c\simeq 6 D/V$ for the following value of the mean interaction time with motors:
\begin{equation}
\tau_{2,3d}^{\rm opt} =\frac{\sqrt{3}a}{Vx_0}\simeq 1.078 \frac{a}{V},
\end{equation}
where  $x_0$ is the solution of $ 2 \tanh(x)-2 x+x \tanh(x)^2=0$.
The $\tau_1$ dependence is very weak, but one can roughly estimate the optimal value by $\tau_{1,3d}^{\rm opt}\simeq 6 D/V^2$.
This gives in turn the maximal reaction rate
\begin{equation}
K^m_{3d} \simeq \frac{c V}{a}\,\frac{\sqrt{3}\left(x_0 -\tanh(x_0)\right)}{ x_0^2  },
\end{equation}
so that the gain with respect to the reaction rate $K^{p}_{3d}$ in a passive medium is
 $G_{3d}=K^m_{3d}/K^{p}_{3d}\simeq CaV/D$ with $C\simeq 0.26$.

\begin{figuresmall}[h!]\small
   \begin{minipage}[c]{.46\linewidth}
\begin{center}
      \includegraphics[scale=0.4]{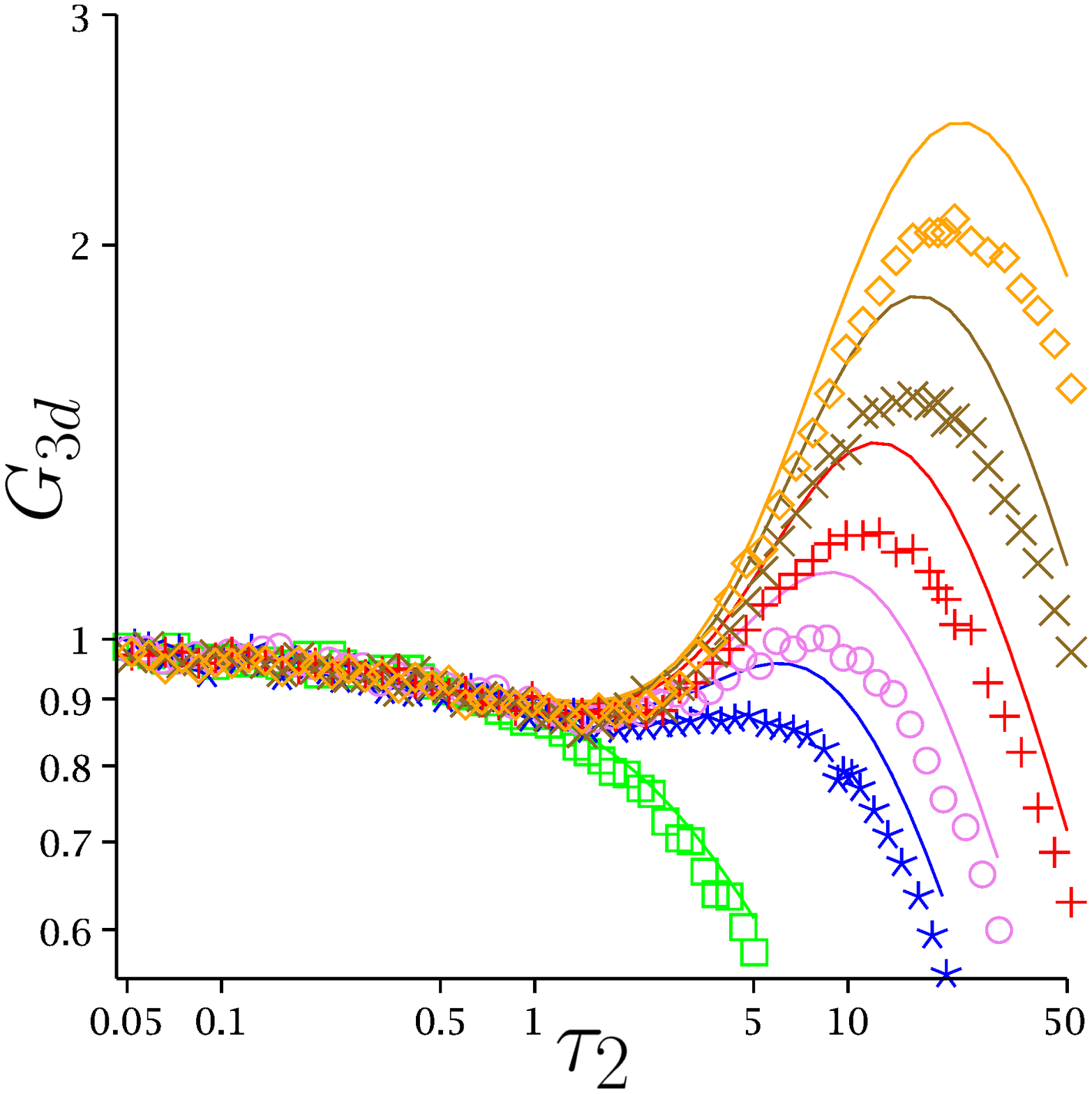}

b/a=5
\end{center}
   \end{minipage} \hfill
   \begin{minipage}[c]{.46\linewidth}
\begin{center}
      \includegraphics[scale=0.4]{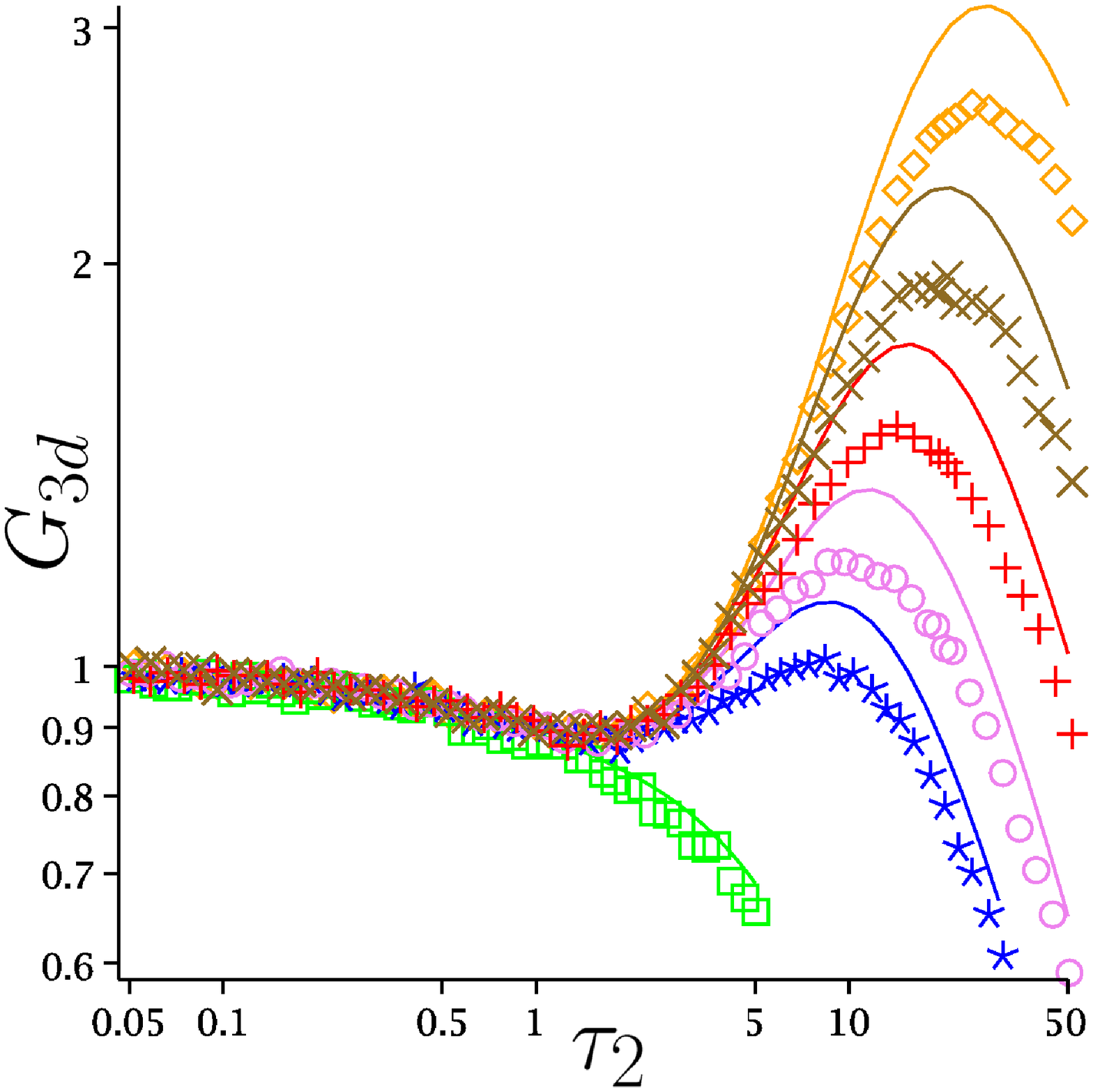}

b/a=40
\end{center}
   \end{minipage}\hfill

\caption{  Optimization of the reaction rate for intermittent active transport. Gain of reactivity due to active transport  in 3 dimensions  as a function of  $\tau_2$ for different values of the ratio $b/a$ (logarithmic scale). The analytical form (the mean detection time with diffusion alone  divided by the mean detection time with intermittence) (plain lines) is plotted against numerical simulations (symbols) for the following values of the parameters (arbitrary units): $a=1$ (green, $\square$), $a=5$ (blue, $\star$), $a=7$ (purple, $\circ$), $a=10$ (red, $+$), $a=14$ (brown, $\times$), $a=20$ (orange, $\diamond$), with $\tau_1=6$, $V=1$, $D=1$. $G_{3d}$ presents a maximum only for $a>a_c\simeq 4$. 
}\label{K3d}\normalsize
\end{figuresmall}

Several comments are in order. (i) First,  $\tau_{2,3d}^{\rm opt}$ neither depends on $D$, nor on the reactant concentration. A similar analysis for $k$ finite (in the $D\to 0$ limit) shows that this optimal value does not depend on $k$ either \refs{section_generic}, which proves that the optimal mean interaction time with motors is widely independent of the parameters characterizing the diffusion phase 1. (ii) Second, the value $a_c$ should be discussed. In standard cellular conditions $D$ ranges from $\simeq 10^{-2}~\upmu$m$^2.$s$^{-1}$ for vesicles to $\simeq 10~\upmu$m$^2.$s$^{-1}$ for small proteins, whereas the  typical velocity of a motor protein is $V\simeq 1~\upmu$m$.$s$^{-1}$, value which is widely independent of the size of the cargo  \cite{alberts}. This gives  a critical reaction radius $a_c$ ranging from $\simeq 10$~nm for vesicles, which is smaller than any cellular organelle,  to $\simeq 10~\upmu$m for single molecules, which is comparable to the whole cell dimension. Hence, this shows that in such 3--dimensional disordered case, active transport can optimize reactivity for sufficiently large tracers like vesicles, as motor mediated motion permits a fast relocation to unexplored regions, whereas it is inefficient for standard molecular reaction kinetics, mainly because at the cell scale molecular free diffusion is faster than motor mediated motion. This could help justifying  that many molecular species in cells are transported in vesicles. Interestingly, in standard cellular conditions $\tau_{2,3d}^{\rm opt}$ is of order $0.1~$s for a typical reaction radius of order $0.1~\upmu$m. This value is compatible with experimental observations \cite{alberts}, and suggests that cellular transport is close to optimum. (iii) Last, the typical gain for a vesicle of reaction radius $a \gtrsim 0.1~\upmu$m in standard cellular 
conditions is  $G_{3d}\gtrsim2.5$ \refi{K3d} and can reach $G_{3d}\simeq 10$ for the fastest types of molecular motors ($V\simeq 4~\upmu$m$.$s$^{-1}$, see  \textcite{alberts,myosine_rapide}), independently of the reactant concentration $c$. As we shall see below the gain will be significantly higher in lower dimensional structures such as axons.

\subsubsection{Active transport at membranes}

\imagea{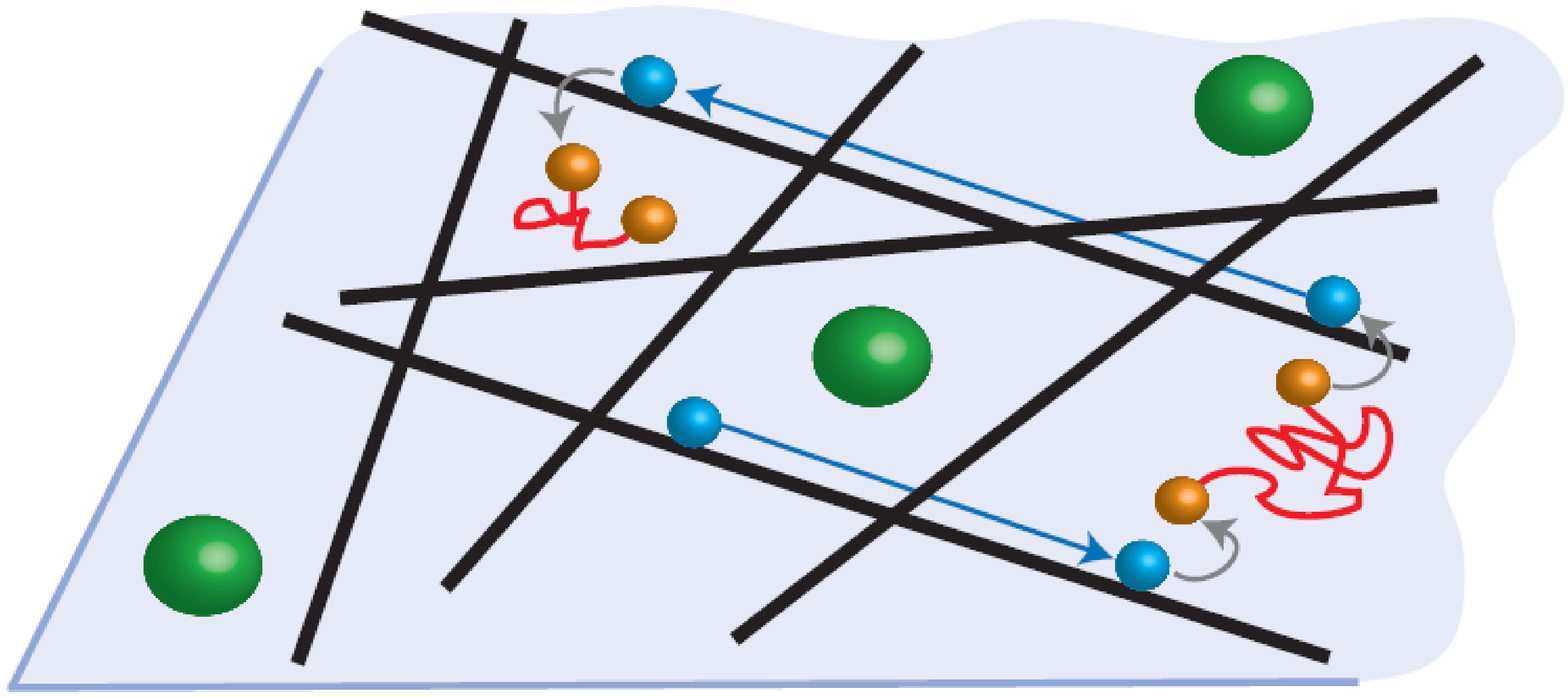}{Planar structures such as membranes and lamellipodia ($d=2$).}{micro2D}{7}

We now come to the $d=2$ disordered case \refi{micro2D}. Striking examples in cells are given by  the cytoplasmic membrane, which is closely coupled to the network of cortical actin filaments, or the lamellipodium of adhering cells \cite{alberts}.  In many cases the orientation of filaments can be assumed to be random. 
It can be shown that as for $d=3$ \refs{section_generic}, the reaction rate $K_{2d}$ can be  optimized in the regime $D/V\ll a\ll b$. Remarkably, the optimal interaction time $\tau_{2,2d}^{\rm opt}$ takes one and the same value in the two limits $k \to \infty$  and  $D \to 0$~:
\begin{equation}
\tau_{2,2d}^{\rm opt}\simeq \frac{a}{V\sqrt{2}}(\ln(1/c)-1)^{1/2},
\end{equation}
which indicates that again $\tau_{2,2d}^{\rm opt}$ does not depend on the parameters of the thermal diffusion phase, neither through $D$ nor $k$. In the limit $k \to \infty$  one has $\tau_{1,2d}^{\rm opt} = \frac{D}{8V^2}\frac{\ln^2(1/c)}{\ln(1/c)-1}$, and the maximal reaction rate can then be obtained~:
\begin{equation}
K^m_{2d} \simeq \frac{c V}{a\sqrt{2\ln(1/c)}}.
\end{equation}
Comparing this expression to the case of passive transport yields a gain  $G_{2d}=K^m_{2d}/K^{p}_{2d}\simeq av\sqrt{\ln(1/c)}/(4D\sqrt{2})$. As in the  $d=3$ case, this proves that active transport  enhances reactivity for large enough tracers (with a critical reaction radius $a_c\simeq D/V$ of the same order as in the $d=3$ case) such as vesicles. However, here the gain $G_{2d}$ depends on the reactant concentration $c$, and can be more significant~: with the same  values of $D$, $V$ and $a$ as given above in standard cellular conditions, and for low concentrated reactants (like specific membrane receptors) with a typical distance between reactants $b\gtrsim 10~\upmu$m,  the typical gain is $G_{2d}\gtrsim 8$, and reaches $10$ for single reactants (like examples of signaling molecules) \refi{G2d}.

\begin{figuresmall}[h!]\small
   \begin{minipage}[c]{.46\linewidth}
\begin{center}
  \includegraphics[scale=0.4]{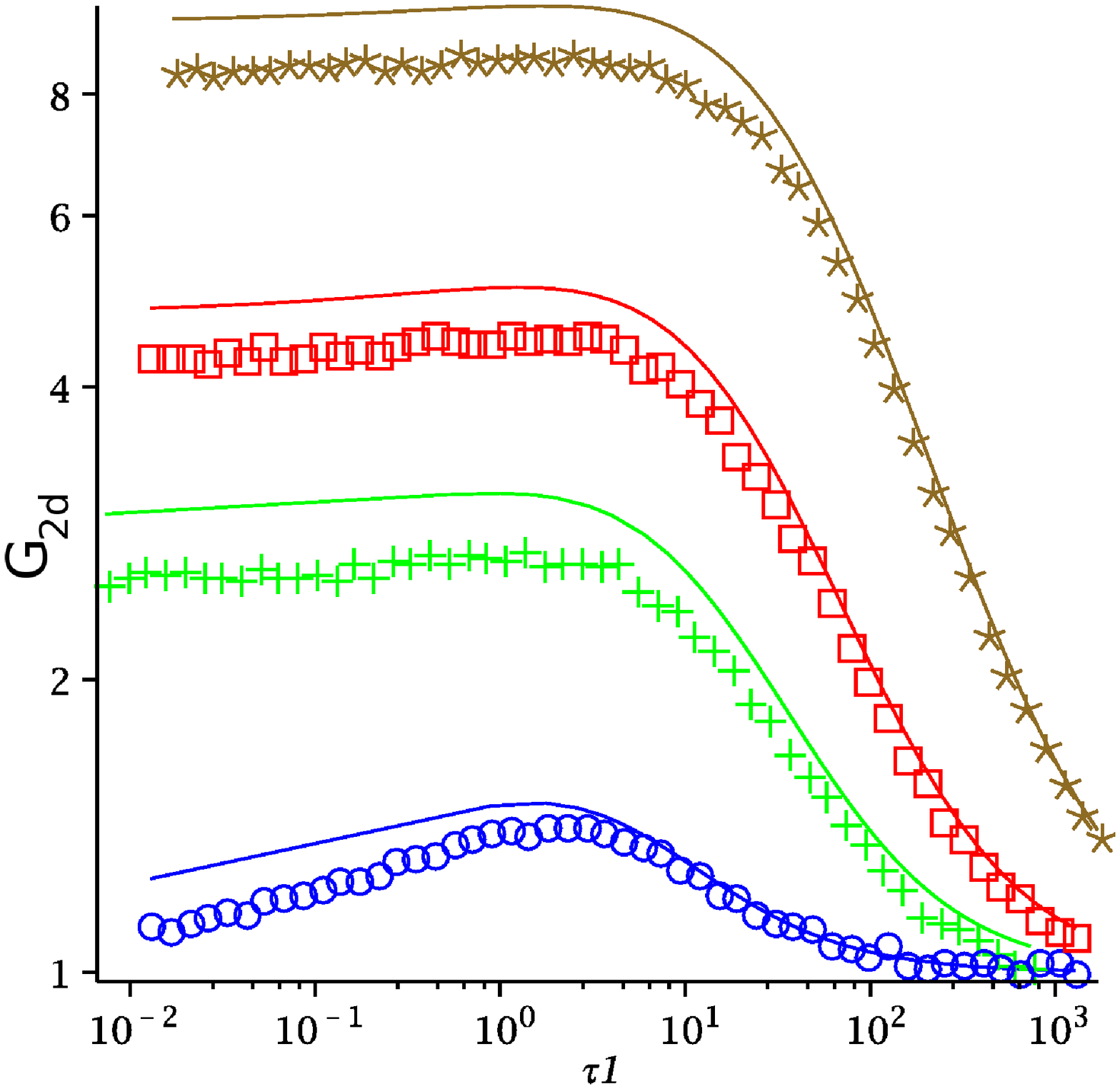}

$\tau_2 \simeq \tau_2^{opt}$.
\end{center}
   \end{minipage} \hfill
   \begin{minipage}[c]{.46\linewidth}
\begin{center}
      \includegraphics[scale=0.4]{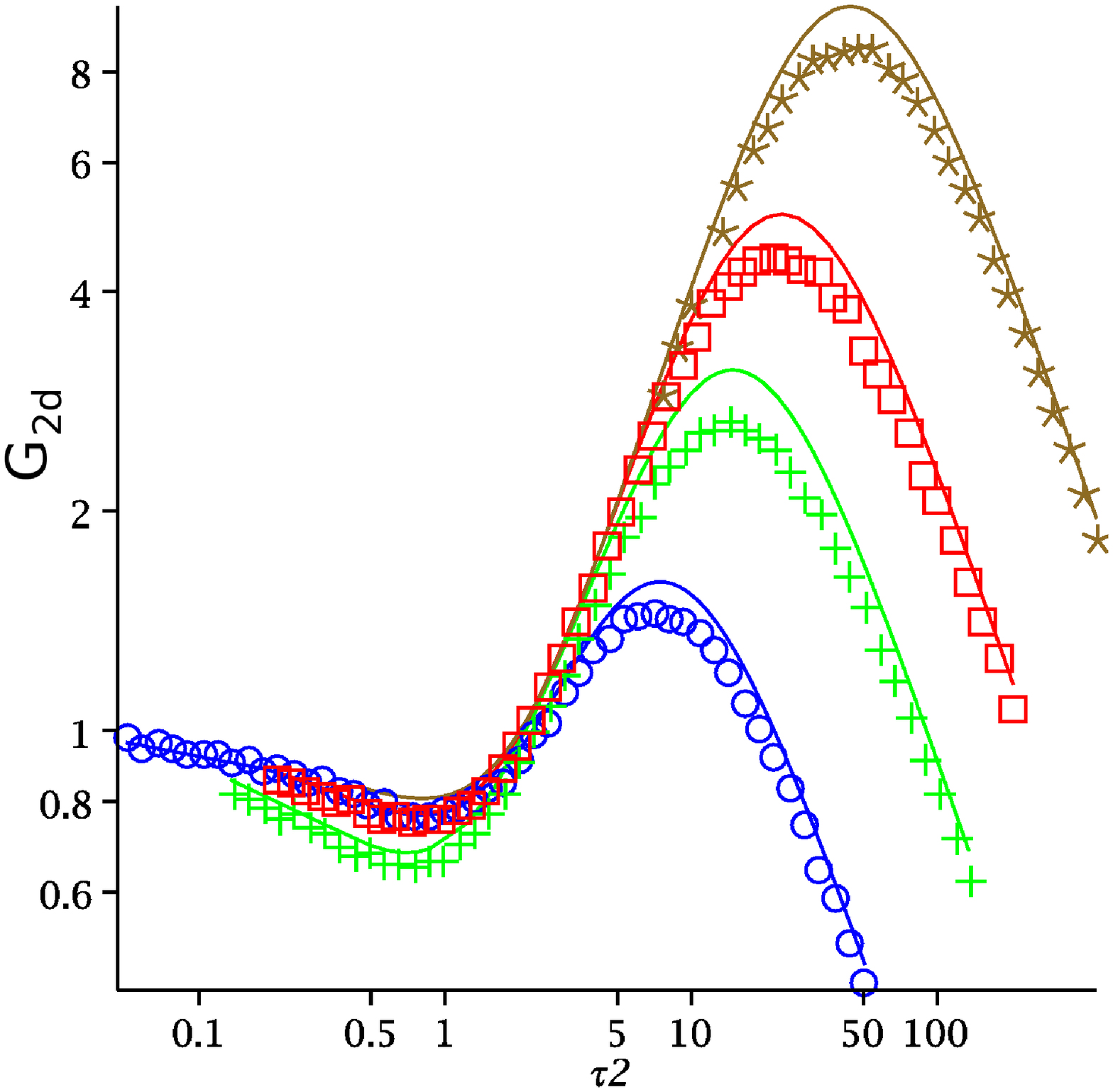}

 $\tau_1 \simeq \tau_1^{opt}$.
\end{center}
   \end{minipage}\hfill

\caption{  Optimization of the reaction rate for intermittent active transport. Gain of reactivity due to active transport $G_{2d}$ in two dimensions  as a function of  $\tau_1$ or $\tau_2$ (logarithmic scale). The analytical form (the mean detection time with diffusion alone  divided by the mean detection time with intermittence \refm{tmap}) (plain lines) is plotted against numerical simulations (symbols) for the following values of the parameters (arbitrary units): $a=20$, $b=2000$ (brown, $\star$), $a=10$, $b=1000$ (red, $\square$), $a=10$, $b=100$ (green, $+$), $a=2.5$, $b=250$ (blue, $\circ$) with $V=1$, $D=1$. 
 These curves represent standard cellular conditions (as discussed in the text).  
}\label{G2d}\normalsize
\end{figuresmall}

\subsubsection{Active transport in tubular structures}

\imagea{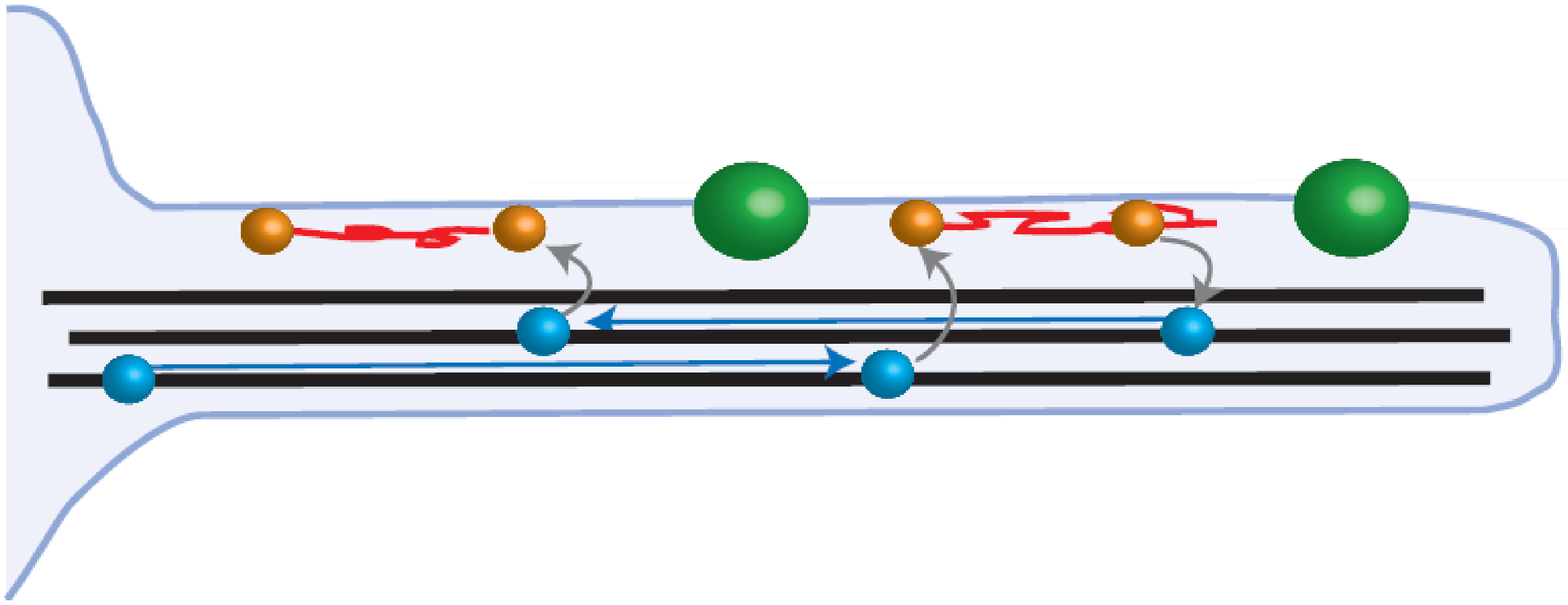}{Tubular structures in cells such as axons and dendrites ($d=1$).}{micro1D}{5}

The case of \textit{nematic order} of the cytoskeletal filaments, which depicts for instance the situation of a polarized cell \cite{alberts}, can be shown to be equivalent in a first approximation to the   1--dimensional case, which is exactly solvable \refi{micro1D} 
(for calculations, see section \ref{section_generic_1DvD}).
The $d=1$ case is also important on its own in cell biology as many 1--dimensional active structures such as axons, dendrites, or stress fibers  are present in living cells \cite{alberts}. As an illustration, we take the example of an axon,  filled with parallel  microtubules pointing their plus end in a direction  ${\bf e}$. We  consider  a tracer particle interacting with both kinesins (``+'' end directed motors, of average velocity  $V {\bf e}$ ) and dyneins (``-'' end directed motors,  of average velocity $-V {\bf e}$) with the same characteristic interaction time $\tau_2$  (see figure 1b). For this type of tracer, the mean first passage time satisfies equations (\ref{back1}) with an effective nematic distribution of filaments  $\rho(\omega_{\bf V})=\frac{1}{2}(\delta({\bf V}-{\bf e})+\delta({\bf V}+{\bf e}))$.
The reaction rate $K_{1d}$ is maximized in the regime $D/V\ll a\ll b$  for the following values of the characteristic times \refi{G1d}~: 
\begin{equation}\label{opt1d}
\tau_{1,1d}^{\rm opt}=\frac{1}{48} \frac{D}{V^2c}, \;\tau_{2,1d}^{\rm opt}=\frac{1}{\sqrt{3}}\frac{a}{V c^{1/2}},
\end{equation}
for $k\to \infty$. The maximal reaction rate $K_{1d}^{m}$ is then given by
\begin{equation}
 K_{1d}^{m}\simeq \frac{\sqrt{3}Vc^{3/2}}{2a},
\end{equation}
 and  the gain  is $G_{1d}=K_{1d}^{m}/K^{p}_{1d}\simeq aV/(2\sqrt{3}Dc^{1/2})$, which  proves that active transport can optimize reactivity as in higher dimensions. Very interestingly the $c$ dependence of the gain is much more important than for $d=2,3$, which shows that the efficiency  of active transport is strongly enhanced in 1-dimensional or nematic structures at low concentration. Indeed, with the same values of $D$, $V$ and $a$  as given above in standard cellular conditions, and for a typical distance between reactants $b\gtrsim 100~\upmu $m (like low concentrated axonal receptors), one obtains a typical gain $G_{1d}\gtrsim 100$ \refi{G1d}. In the limit of finite reactivity ($k$ finite and $D\to 0$) one has $\tau_{1,1d}^{\rm opt} = \sqrt{\frac{a}{Vk}} \left(\frac{2\ln(1/c)-1}{8} \right)^{1/4}$ and the same optimal value (\ref{opt1d}) of $\tau_{2,1d}^{\rm opt}$. As in higher dimensions  $\tau_{2,1d}^{\rm opt}$  depends neither on the thermal diffusion coefficient $D$ of phases 1, nor on the association constant $k$, which shows that the optimal interaction time with motors  $\tau_2^{\rm opt}$  presents remarkable universal features. Furthermore, this approach permits an estimate of $\tau_2^{\rm opt}$  compatible with observations in standard cellular conditions, which suggests that cellular transport could be close to optimum.

\begin{figuresmall}[h!]\small
   \begin{minipage}[c]{.46\linewidth}
\begin{center}
      \includegraphics[scale=0.4]{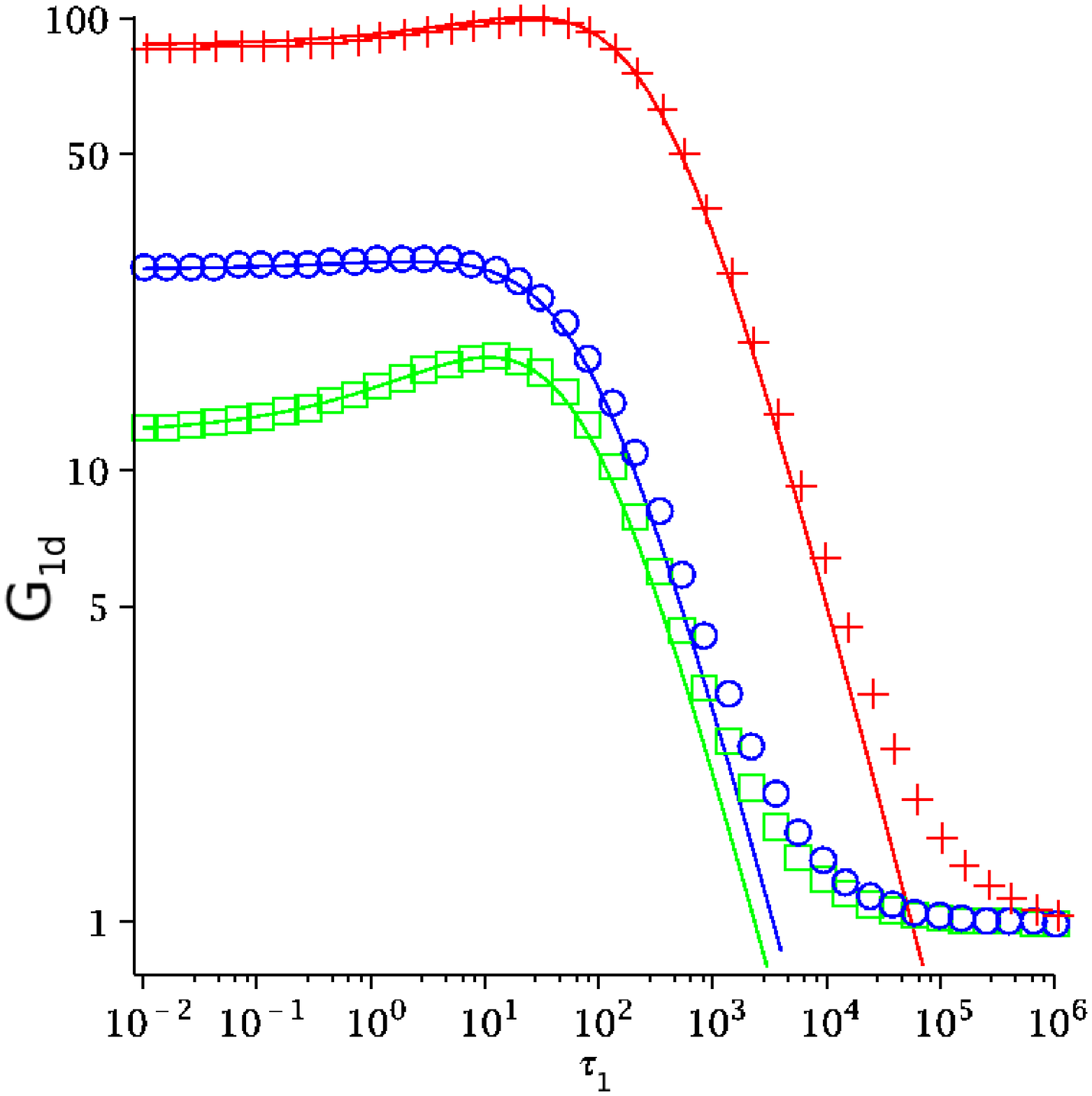}

$\tau_2 \simeq \tau_2^{opt}$.
\end{center}
   \end{minipage} \hfill
   \begin{minipage}[c]{.46\linewidth}
\begin{center}
      \includegraphics[scale=0.4]{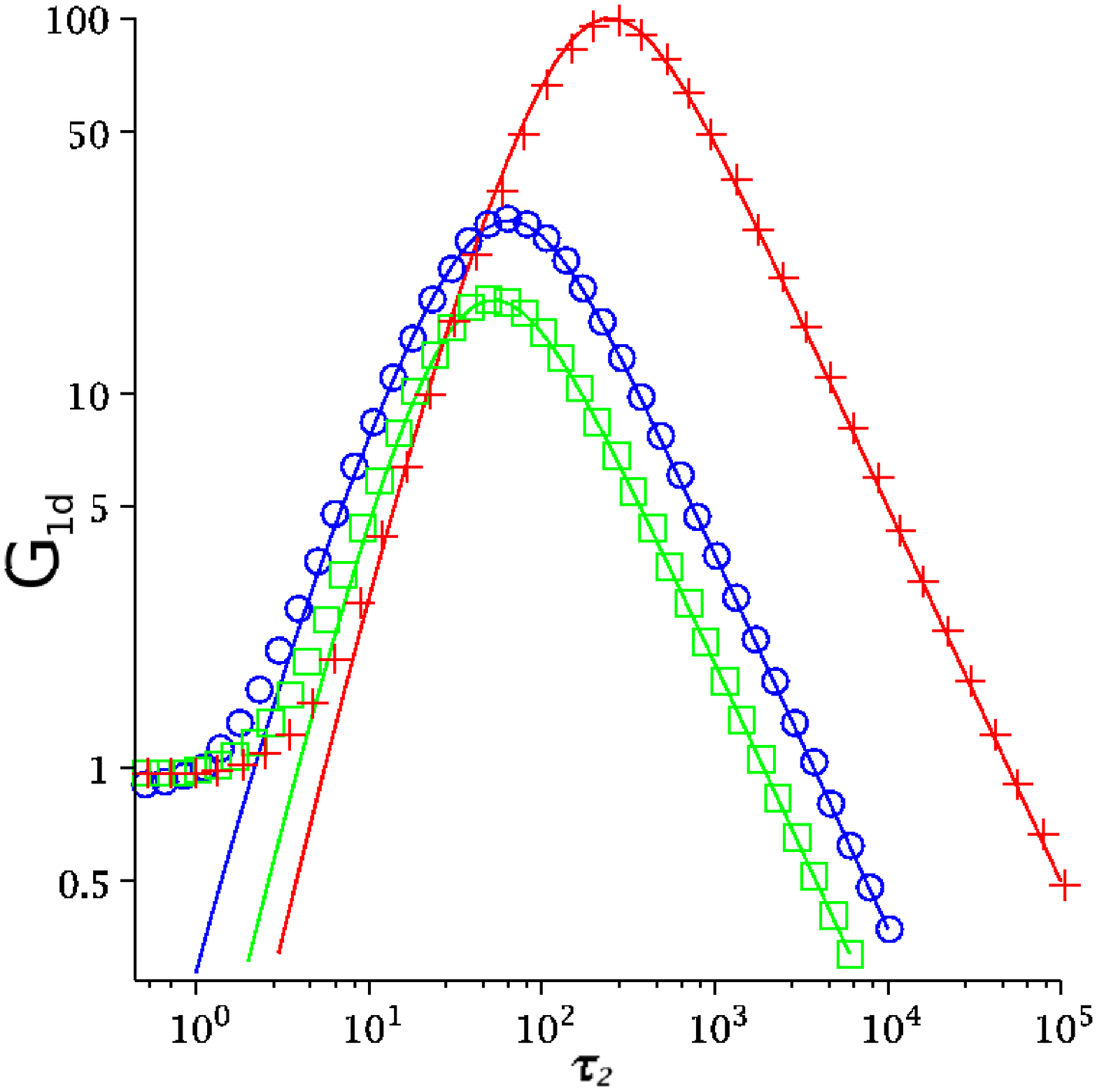}

 $\tau_1 \simeq \tau_1^{opt}$.
\end{center}
   \end{minipage}\hfill

\caption{  Optimization of the reaction rate for intermittent active transport. Gain of reactivity due to active transport $G_{1d}$ in one dimension  as a function of  $\tau_1$ or $\tau_2$ (logarithmic scale).  The analytical form (the mean detection time with diffusion alone  divided by the mean detection time with intermittence \refm{tmn})  is plotted against the exact solution  (symbols), for the following values of the parameters (arbitrary units): $D=1,V=1$ for all curves and $a=10,b=10^4$ (red, $+$), $a=10,b=10^3$ (blue, $\circ$), $a=2.5,b=10^3$ (green, $\square$). Standard cellular conditions (as discussed in the text) correspond to blue and red curves.
}\label{G1d}\normalsize
\end{figuresmall}

\subsubsection{Conclusion on intermittent active transport}

Starting from the observation of vesicles 
alternating free diffusion and phases bound to motors performing ballistic motion, 
and from the observation 
that (at least in some cases), vesicles can only react in the free phase, 
 a model for intermittent active transport has been proposed.
 The reaction rate, which can be approximated by the inverse of the mean first passage time, can be  explicitly calculated in this model for various cellular geometries (bulk cytoplasm, membranes, tubular structures).
This shows that  intermittent transport can indeed 
increase reaction rates, 
in particular for large objects such as vesicles,
and in particular in low dimensions. 
The model for the reactive phase is either diffusive or static (with a reaction rate), 
and both lead to the same optimal duration of the ballistic phase.
The latter point is investigated in more details in next section.  
%
%


\section{Intermittent search : a robust strategy}
\label{section_generic}

As shown in previous sections, intermittent search strategies are observed at the macroscopic scale (foraging animals)
as well as at the microscopic scale (localization of a DNA sequence by a protein, vesicle transport in cells).
The models we have used to interpret these findings, in particular in sections  \ref{animaux_alternative} and \ref{section_generic_vesicles}, 
present similar general features. 

In \textcite{LeGros}, a generic model of intermittent search 
based on these   general features has been introduced, and studied  
systematically in 1, 2, and 3 dimensions, 
and for three different modelings of the detection phase.  This rather technical  section (completed by appendix)  gathers the main tools usually involved  in the calculation of first-passage properties of intermittent random-walks, and utilized throughout this review. Finally, general conclusions on
intermittent random walks can be drawn from this systematic study, and are summarized in table \ref{recapgeneral}.

\subsection{Introduction}

The generic model presented in this section follows the original definition of intermittence given in introduction and 
relies  on a succession of slow phases with detection, and ballistic phases without detection,
without direction correlation between ballistic phases.  
This model is minimal in the sense that the searcher has low memory skills. 
Indeed, without correlations between ballistic phases, there is no spatial memory. 
We also assume a Markovian searcher, \textit{i.e.} with no temporal memory. 
As previously we address the following main questions: is it beneficial for the search to perform such fast but non reactive phases?
Is it possible, by properly tuning the kinetic parameters  of trajectories (such as the durations of each of the two phases) to minimize the search time?
We develop in what follows a systematic analytical study of intermittent random walks in one, two and three dimensions and fully  characterize the optimal regimes. 
Overall, this systematic approach allows us to identify  robust features of intermittent search strategies. 
In particular, the slow  phase that enables detection 
is often hard to characterize experimentally. 
Here we propose and study three distinct  modelings  for this phase, which allows us 
to assess to which extent our results are robust and model independent.
Our analysis covers in details  intermittent search problems in one, two and three dimensions and is aimed at giving a quantitative basis -- as complete as possible -- to model real search problems involving intermittent searchers.

We first define the model and introduce the methods.  
Then we summarize the results for  the search problem in dimension one, two and three, 
 for 
 different types of motion in the slow phase.  
Eventually we synthesize the results 
in the table \ref{recapgeneral} where all cases, their differences and similarities are gathered.
This table finally leads us  
 to draw general conclusions.

\subsection{Model and notations }

\subsubsection{Model}

We consider an intermittent searcher that switches  between two phases.
The switching rate $\lambda_1$ (resp. $\lambda_2$) from phase 1 to phase 2 (resp. from phase 2 to phase 1)  is  time-independent, which assumes that the searcher has no temporal memory and  implies an  exponential distribution of durations of each phase $i$ 
of mean $\tau_i=1/\lambda_i$. 

\begin{figuresmall}[h!]\small
   \begin{minipage}[c]{.3\linewidth}
\begin{center}
      \includegraphics[width=4cm]{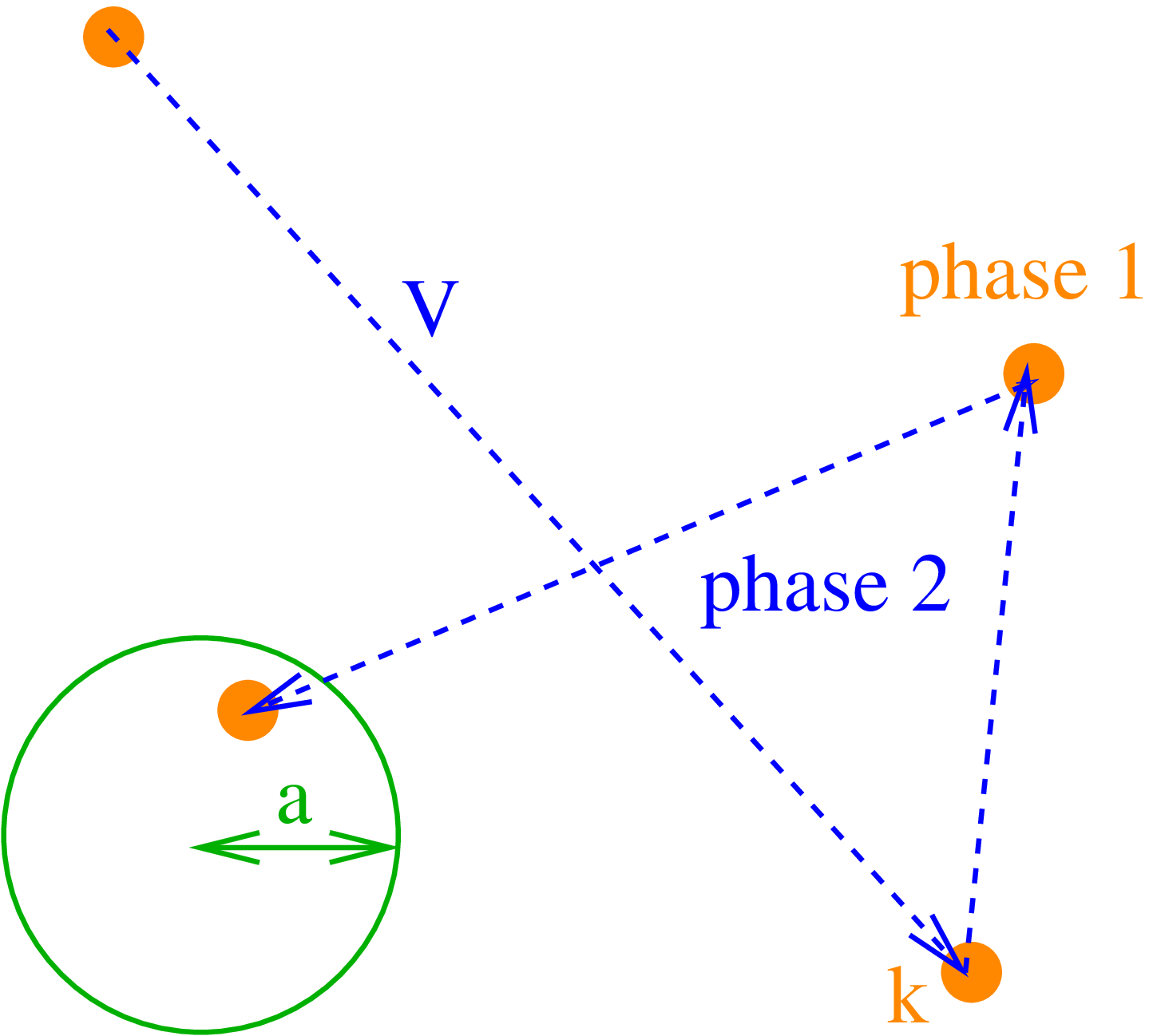}

\end{center}
   \end{minipage} \hfill
   \begin{minipage}[c]{.3\linewidth}
\begin{center}
      \includegraphics[width=4cm]{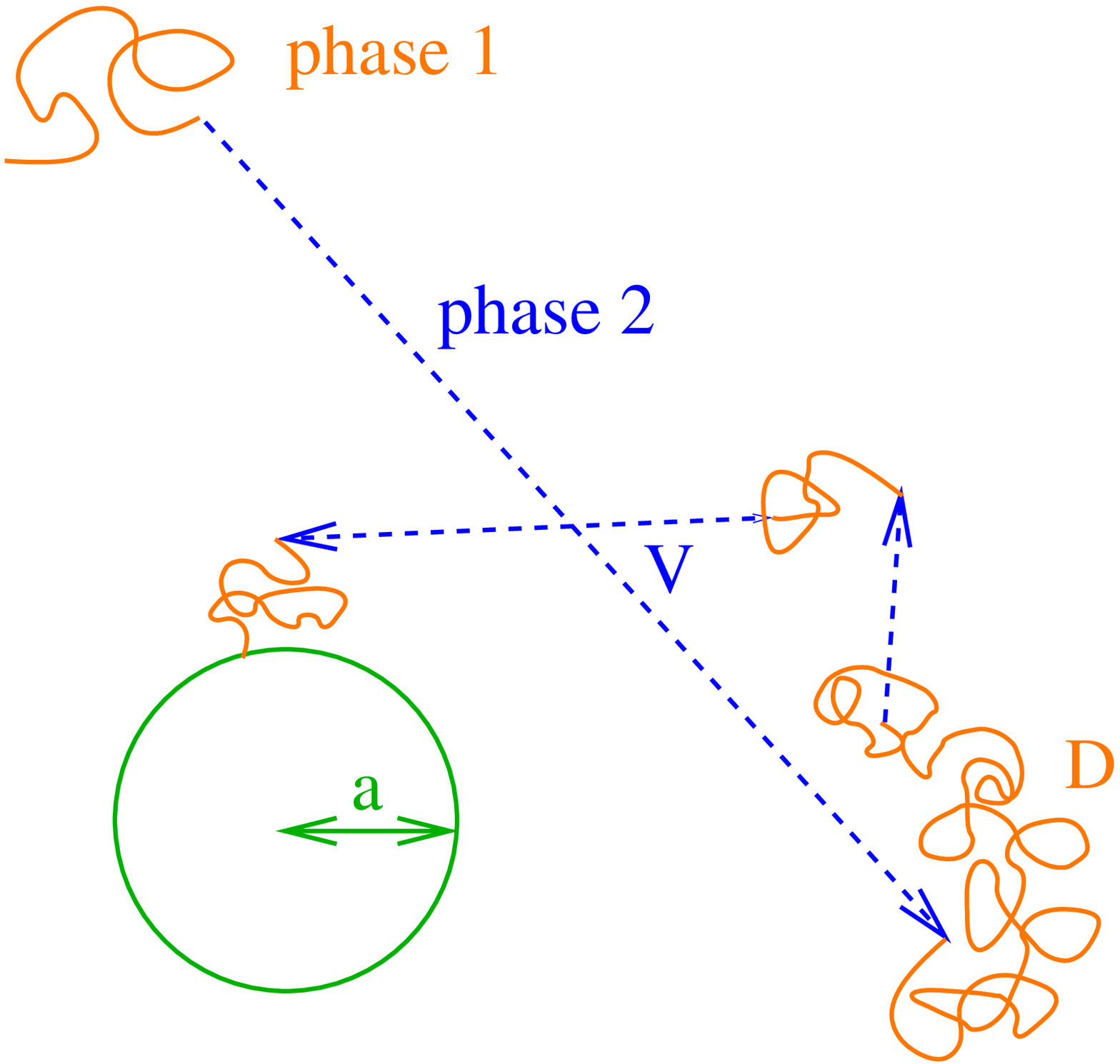}

\end{center}
   \end{minipage}\hfill
   \begin{minipage}[c]{.3\linewidth}
\begin{center}
      \includegraphics[width=4cm]{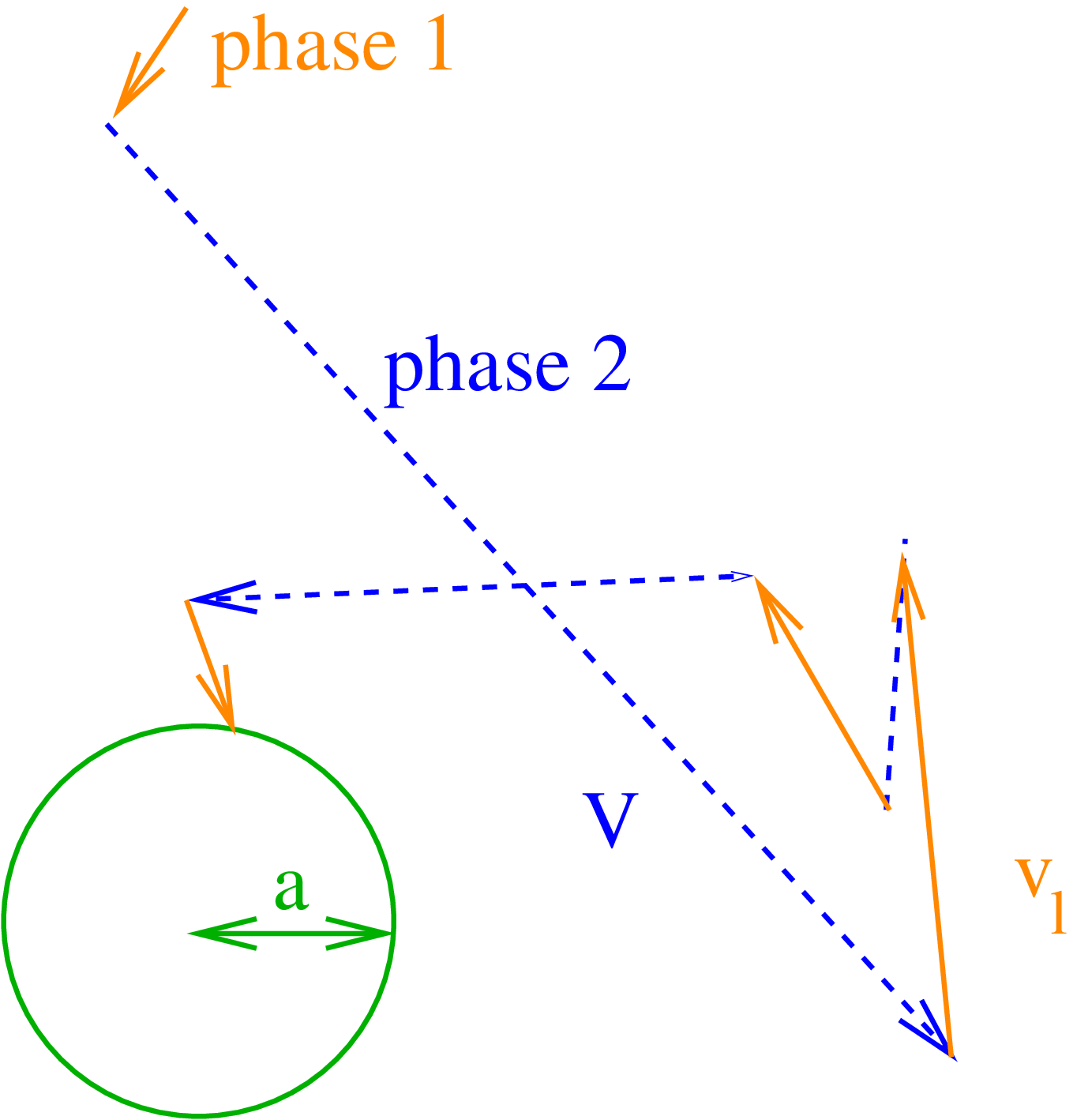} 

\end{center}
   \end{minipage}

   \begin{minipage}[c]{.3\linewidth}
\begin{center}

Static mode

\end{center}
   \end{minipage} \hfill
   \begin{minipage}[c]{.3\linewidth}
\begin{center}

Diffusive mode

\end{center}
   \end{minipage}\hfill
   \begin{minipage}[c]{.3\linewidth}
\begin{center}

Ballistic mode

\end{center}
   \end{minipage}

\caption{The three different descriptions of phase 1 (the phase with detection), 
here represented in two dimensions.}\label{modes_generic}\normalsize
\end{figuresmall}

Phase 1 denotes the phase of slow motion, 
 during which the target can be detected if it lies within a distance from the searcher which is smaller than a given detection radius  $a$, 
which is the maximum distance within which the searcher can get information about target location. 
We propose 3 different modelings of this phase, 
in order to cover various real life situations \refi{modes_generic}.
\begin{itemize}
\item In the  ``static mode'',  the searcher is immobile, and detects the target with probability per unit time $k$ if it lies at a distance  less than $a$.
\item In the second modeling, called the ``diffusive mode'', the searcher performs a continuous diffusive motion, with  diffusion coefficient $D$,
and finds immediately the target if it lies at a distance  less than $a$. 
\item In the last modeling,  called the ``ballistic mode'', the searcher  moves ballistically in a random direction with constant speed $v_l$ and reacts immediately with  the target if it lies at a distance  less than $a$. 
We note that this mode is equivalent to the model of L\'evy walks searches proposed in \textcite{viswaNat},
except for the law of the time between reorientations \refs{subsection_Levy}. 
It was shown that for destructive search, \textit{i.e.} targets that cannot be revisited, 
the optimal strategy is obtained for a straight ballistic motion, without reorientations \refs{subsection_Levy}. 
In what follows it is shown  that if another motion, ``blind'' (\textit{i.e.} without detection) but with higher velocity is available, 
there are regimes outperforming  the straight line strategy. 
\end{itemize}
Some comments on these different modelings 
of the slow phase 1 are to be made.  
First, these three modes  schematically cover experimental observations of the behavior of animals searching for food \cite{Bell,animauxObrien}, where the slow phases of detection are often described as  static, random or with slow velocity.
Several real situations 
are also likely to involve a combination of two modes.  For instance the motion of a reactive particle in a cell not bound to motors can be described by a combination of the diffusive and static modes. For the sake of simplicity,  these modes are treated independently, and this approach can therefore be considered as a limit of more realistic models. 
Finally, combining  these three schematic modes covers a wide range of possible motions, from subdiffusive (even static), 
diffusive, to superdiffusive (even ballistic). 
Beyond the modeling of real-life systems, 
studying different detection modes enables us to assess the robustness of the results. 

The phase 2 denotes the fast phase during which the target cannot be found. 
In this phase, the searcher performs  a ballistic motion at a constant speed $V$ in a random direction, 
redrawn for each new phase 2, independently of previous phases. 
In real examples correlations between successive ballistic phases could exist,  
as observed for foraging animals \cite{animauxObrien}.
If correlations are very high, it is close to a one dimensional problem 
with all the phases 2 in the same direction, a different problem already treated in section \ref{section_animaux_2005}. 
We consider here the limit of low correlation, that is of a searcher with no memory skills.

We assume that the searcher evolves  in a $d$-dimensional spherical domain of radius $b$, with reflective boundaries and with one centered immobile target 
of radius $a$.
 As the searcher does not initially know the target's location, 
 we start the walk from a random point of the $d$-dimensional sphere,
and average the mean target detection time over the initial position.
 This geometry models the case of a single target in a finite domain, and  also provides a good approximation
of an infinite space with regularly spaced targets. Such regular array of targets  corresponds to a mean-field approximation of random distributions of targets, which can be more realistic in some experimental situations. 

\subsubsection{Methods}\label{section_methods}

We explain here the general methods, and introduce the notations. 

We define $s_i(\mathbf{r},t) $ the probability that the searcher has not yet found the target at $t$, 
 {\it starting} from $\mathbf{r}$ in state $i$, 
where state $i=1$ is the slow motion phase with detection and 
state $i=2$ is the fast motion phase without target detection. 
Note that in dimension one, the space coordinate will be denoted by $x$, 
and in the case of a  ballistic mode for phase 1, the upper index in $t_i^\pm $ stands for  
ballistic motion  with direction $\pm x$. 
The survival probability  $s_i(\mathbf{r},t) $ is solution of the following standard backward differential Chapman-Kolmogorov equations \cite{gardiner} :
\begin{equation}
 L_i^\dagger s_i( \mathbf{r},t) + \frac{1}{\tau_i}\left(s_j( \mathbf{r},t) - s_i( \mathbf{r},t)\right)-k s_i(\mathbf{r},t) I_a(\mathbf{r})\delta(i-1)=\frac{\partial s_i(\mathbf{r},t) }{\partial t},
\end{equation}
with  $I_a(\mathbf{r})=1$ when $r<a$, and 0 else, and $L_i^\dagger$ is the adjoint operator of the transport operator. For example $L_i^\dagger=D\Delta$ for diffusion and $L_i^\dagger= {\bf v}\cdot\nabla$ for a ballistic motion of velocity $ {\bf v}$ . The mean first passage time to the target $t_i(\textbf{r})$ 
for a  searcher {\it starting} in the phase $i$ from point $\textbf{r}$ is then given by~: 
\begin{equation}
 t_i(\textbf{r}) =-\int_0^{\infty} t \frac{\partial s_i(\textbf{r},t)}{\partial t} dt=\int_0^{\infty}  s_i(\textbf{r},t) dt
\end{equation}

Consequently, for each phase $i$,  $t_i(\textbf{r})$ is solution of : 
\begin{equation}
 L_i^\dagger t_i( \mathbf{r}) + \frac{1}{\tau_i}\left(t_j( \mathbf{r}) - t_i( \mathbf{r})\right)-k t_i(\mathbf{r}) I_a(\mathbf{r})\delta(i-1)=-1 .
\end{equation}
We assume that the searcher starts in phase 1, 
and to take into account the fact that it does not initially know the target's location, 
we average 
the mean  detection time  over the starting point, leading to the following definition of the mean search time~: 
\begin{equation}
t_m=\frac{1}{V(\Omega_d)}\int_{\Omega_d} t_1(\textbf{r})d\textbf{r},
\end{equation}
with $\Omega_d$ the d-dimensional sphere of radius $b$ and $V(\Omega_d)$ its volume. Unless specified, we will consider the low target density limit $a\ll b$.

Our general aim is to minimize $t_m$ as a function of the mean durations $\tau_1,\tau_2$ of each phase, and in particular to determine under which conditions an intermittent strategy (with finite $\tau_2$) is faster than a usual single state search in the phase 1 only, which is given by the limit $\tau_1\to\infty$.  In the static mode, intermittence is necessary for the searcher to move, 
and is therefore always favorable. 
In the diffusive mode, we will compare 
the mean search time with intermittence $t_m$ to the mean search time for a single state diffusive searcher  $t_{\rm diff}$, 
and define the gain as $gain=t_{\rm diff}/t_m$.
Similarly in the ballistic mode, we will compare $t_m$ to the mean search time for a single state ballistic searcher  $t_{bal}$ and 
define the gain  as $gain=t_{bal}/t_m$. 

The upper index 
``opt'' is used to denote the value of a parameter or variable at the minimum of $t_m$ .

Calculations are exact in dimension one, whereas  approximation schemes (which can be checked numerically) are needed
 in dimension two and three. The main calculation steps are given in appendix and further technical details  can be found in   \cite{LeGros}. We now summarize the main results for each dimension.

\subsection{Dimension one}
Besides the fact that it involves more tractable calculations, the one-dimensional case is also interesting to model real search problems. 
As discussed before, at the microscopic scale,  tubular structures of cells such as axons or dendrites in neurons can be considered as one-dimensional \cite{alberts}. 
The active transport of reactive particles, which alternate between diffusion phases and ballistic phases when bound to molecular motors, can be schematically captured by this generic model with diffusive mode \cite{natphys08}.
At the macroscopic scale, one could cite animals like ants \cite{fourcassieWallFollowingAnts} which 
tend to follow tracks or one-dimensional boundaries.  
\new{More generally, borderlines between different habitats, such as a shoreline, 
can be considered as one-dimensional.}

It is showed in appendix that intermittent search strategies  in dimension one share  similar features for the static, diffusive and ballistic detection modes. 
In particular, all modes show regimes where intermittence is favorable and lead to a minimization of the search time.
Strikingly, the optimal duration of the non-reactive relocation phase 2 is quite independent of the modeling of the 
reactive phase~: $\tau_2^{opt} = \frac{a}{3V} \sqrt{\frac{b}{a}}$ for the static mode, 
for  the ballistic mode (in the regime $v_l < v_l^c \simeq \frac{V}{2}\sqrt{\frac{3a}{b}}$), and for the diffusive mode (in the regime $b>\frac{D}{V}$ and $a\gg \frac{D}{V}\sqrt{\frac{b}{a}}$). This shows the robustness of the optimal value  $\tau_2^{opt}$.

\subsection{Dimension two}

As discussed previously, the two-dimensional problem is particularly well suited to model animal behaviors;   
it is also relevant to  the microscopic scale, since it  mimics for example the case of  
cellular traffic on membranes\cite{alberts}. 
While in dimension  one  the mean search time can be calculated analytically, 
 in dimension two (and later in dimension three) approximation schemes are necessary, and can be  
  checked by numerical simulations.

Remarkably, for the   three different modes of detection
(static, diffusive and ballistic), there is  a regime where intermittence
 minimizes the search time for one and the same $\tau_2^{opt}$, given by 
 $\tau_2^{opt}= \frac{a}{V}\sqrt{\ln\left( \frac{b}{a} \right) -\frac{1}{2}}$. 
As in dimension one, this indicates that  optimal intermittent strategies are robust and widely  
independent of the details of the description of the detection mechanism.

\subsection{Dimension three}
The  three dimensional case is also  relevant to biology. 
At the microscopic scale, it corresponds for example to 
intracellular traffic in the bulk cytoplasm of cells, 
or at larger scales to animals living in  dimension three, such as 
 plankton \cite{BartumeusPlancton}, or \textit{C.elegans} in its natural habitat (soil) \cite{CelHabitat}. 
As it was the case in dimension two, 
different assumptions have to be made to obtain analytical expressions of the search time. 
Such assumptions can be checked by
numerical simulations using the same algorithms as in dimension two.

For the three possible modelings of 
the detection mode (static, diffusive and ballistic) in dimension three, 
there is a regime where the optimal strategy is  intermittent.  
Remarkably, and as was the case in  dimensions one and two,  
the optimal time to spend in the fast non-reactive phase 2 is independent 
of the modeling of the detection mode and reads $\tau_2^{opt} \simeq 1.1 \frac{a}{V}$. 
Additionally, while the mean first passage time to the target scales as $b^3$, 
 the optimal values of the durations of the two phases do
not depend on the target density $a/b$.

\subsection{Discussion and conclusion}


 To summarize, the  methods of calculation  developed in this section allow one to show 
that  the mean search time of intermittent random walks  can be minimized under broad conditions. 
The table \ref{recapgeneral} summarizes the main results of this minimization. 
This study shows  that the optimal durations of the two phases and the gain of intermittent search 
(as compared to a single state search) 
do depend on the target density in dimension one. 
In particular,  the gain can be very high at low target concentration. 
Interestingly, this dependence is smaller in
 dimension two,  
and vanishes in dimension three. 
The fact that intermittent search is more advantageous in low dimensions (1 and 2) 
can be understood as follows.  
At large scale, the intermittent searcher  effectively performs  a random walk, and therefore scans a space of dimension two. 
In an environment of dimension one (and marginally of dimension two), 
the searcher therefore oversamples the space, and it is favorable to perform  large jumps 
to go to previously unexplored areas. On the contrary, 
 in dimension three, 
the random walk is transient, and the searcher on average always scans previously unexplored areas, 
which makes large jumps less beneficial.

Additionally, these results show 
 that, for various modeling choices of the slow reactive phase, 
there is one and the same optimal duration of the fast non reactive phase, 
which depends only on the space  
 dimension. This further supports the robustness of optimal intermittent search strategies. 
Such robustness and efficiency  could explain why intermittent 
trajectories are observed so often, and in various forms.

\begin{table*} 
\begin{center}
\includegraphics[height=20 cm]{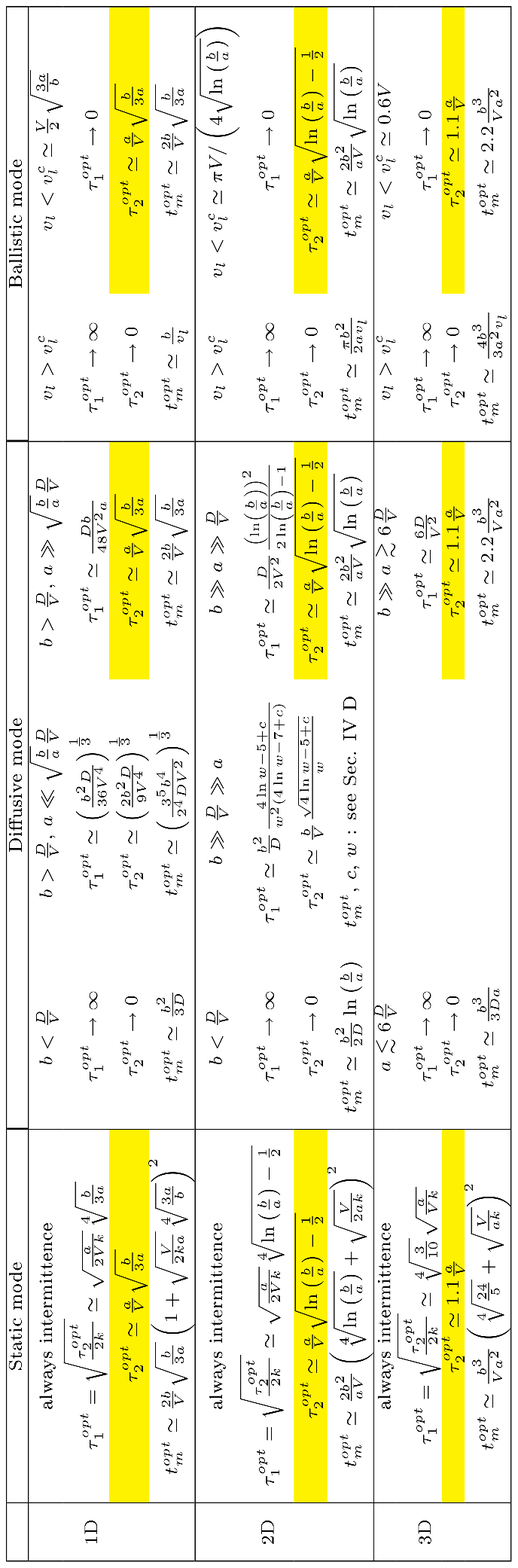}
\end{center}
\caption{Recapitulation of main results of the generic intermittent search model~: strategies minimizing the mean first passage time on the target. In each cell, validity of the regime, optimal $\tau_1$, optimal $\tau_2$, minimal $t_m$ ($t_m$ with $\tau_i=\tau_i^{opt}$). Yellow background highlight the value of $\tau_2^{opt}$ independent from the description of the slow detection phase 1. Results are given in the limit $b \gg a$. 
Complements to the intermediate regime for the diffusive mode in 2 dimensions~: 
$c=4 \left(\gamma-\ln(2)\right)$ with $\gamma$ the Euler constant; 
$w$ solution of $\frac{2Vb}{wD} \ln\left(4\ln(w)-5+c \right)=-8(\ln w)^2+\left(6+8\ln(b/a)\right)\ln(w)-10\ln(b/a)+11-c\left(c/2+2\ln(a/b)-3/2 \right)$; 
in this regime we have $t_m^{opt} \simeq \frac{b^2}{D} \ln\left(\frac{b}{a}\right) \frac{1}{4\ln (w)-5}\left[1+\frac{wD(4\ln(w)-7)}{bV \sqrt{5\ln(w)-5}} \right] \left[ 1+2\ln(w)\ln\left(\frac{b}{aw} \right)\right]$. Adapted  from \textcite{LeGros}}  \label{recapgeneral}
\end{table*} 

\finsection

\section{Extensions and perspectives}
\label{section_perspectives}

Far from closing the problem, 
the generic model presented  previously  opens interesting perspectives. 
In this section, we highlight a few promising directions : (i) Influence of the targets distribution  ;  (ii) Taking into account a more involved searcher, enjoying now some orientational and temporal memory. 
Indeed, in the generic model of section \ref{section_generic}, the searcher has minimal memory skills. 
On the one hand, the phase duration distribution is exponential, which means that there is no 
temporal memory :  the effect of other duration distributions is studied in section \ref{section_autres_temps}.
On the other hand, the direction of each new ballistic ``blind'' phase is taken at random, independently of the previous phases,  
meaning that there is no orientational  memory : 
we study the effect of correlations in section \ref{section_correlations}. (iii) Moving targets, which can be more realistic both at the microscopic and macroscopic scale. Next, we briefly  review in section \ref{section_autres} similar models of intermittent search which have been proposed recently in other  contexts and finally 
discuss how 
further models could also be applied to design efficient searches instead of interpreting biological systems  \refs{section_desing}.

\subsection{Influence of the target distribution on the search time}

\label{section_poisson}

We first study the influence of target distribution on the previous results. For the sake of simplicity we study the one dimensional model of section  \ref{section_animaux_2005}.

\subsubsection{How real targets are distributed?}

In the context of foraging animals, target distributions  are often described as regular, random or patched \cite{Bell} 
(see figure \ref{repartition_cibles_fig}). 
In the models presented in previous sections, 
the chosen geometry can be interpreted as one target in a finite domain, 
or as an infinite array of regularly spaced targets. 
The regular distribution is representative of the real-life case of targets that repel each other, 
thus being as far  from each other as possible. 
This distribution is also a mean-field approximation of other distributions. 
As the regular distribution has already been studied, 
let us discuss the other representative distributions. 

\begin{figuresmall}[h!]
\small
\begin{center}
\includegraphics[width=12cm]{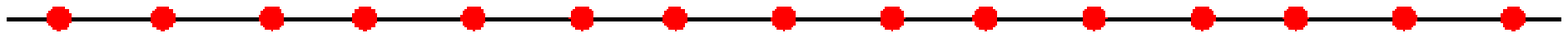}

regular distribution

~

\includegraphics[width=12cm]{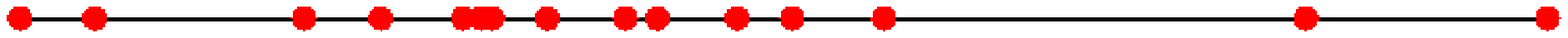}

Poissonian distribution

~

\includegraphics[width=12cm]{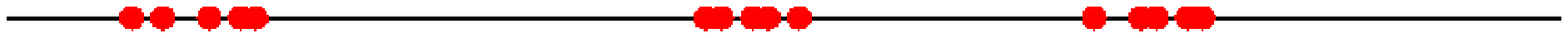}

patched distribution

\caption{Examples of target distributions.}\label{repartition_cibles_fig}
\end{center}\normalsize
\end{figuresmall}

If targets are in patches, for example when they attract each other,  when a target is found  
it is likely that  other targets are present in the immediate surroundings. 
Thus a simple strategy is to switch behavior
when a target is encountered, as proposed for example by \textcite{BenhamouPatch}.
The search is then in two steps : finding a patch, and exploiting it. 
For the first step, previous results   are still valid, 
except for the density of targets which has to be replaced by the density of patches. 

In the following we focus on the  last case of Poissonian targets, which corresponds to situations 
of non-interacting targets.

\subsubsection{Analytical results in the case of a Poissonian distribution of targets}

In the case of a 1 dimensional Poissonian distribution of targets, the distance between two consecutive targets is exponentially distributed. 
Except for this change, the other parameters remain as defined in the model of section \ref{section_animaux_2005}. 

The mean search time is in general hard to calculate for a Poissonian target distribution, 
which can be seen as frozen disorder. 
However, estimates (for $L \gg D/V$) can be given in 3 regimes (see \textcite{Europhys2006,desordonnegros} for details)~:
\begin{itemize}
\item In the large ballistic displacements limit (when $V\tau_2 \gg \sqrt{D\tau_1}$), 
two successive diffusive phases can be considered as non-overlapping. It can be shown that in this regime~: 
\begin{equation}\label{poisson_t_large_ball}
\langle t \rangle \simeq L \frac{\tau_1+\tau_2}{2 \sqrt{D\tau_1}}.
\end{equation}
\item In the small ballistic displacements limit (when $V\tau_2 \ll \sqrt{D\tau_1}$), 
 successive diffusive phases often overlap. It leads to~: 
\begin{equation}\label{poisson_t_small_ball}
\langle t \rangle \simeq L \frac{\tau_1+\tau_2}{ V\tau_2}.
\end{equation}
\item The most interesting situation is the intermediary regime. 
Indeed, in the first case (large ballistic displacements), relocations are too long and 
overshoot the target; and in the second case (small ballistic displacements), 
there are often repetitive scans of the same areas. 
In the intermediary regime, the mean first passage time to the target can be approximated by~: 
\begin{equation}\label{poisson_t_inter_ball}
\langle t \rangle \simeq  L \frac{\tau_1+\tau_2}{ V \tau_2}\frac{(1+\theta)^2(1+\epsilon\theta)}{(1+4\theta+2\epsilon \theta^2)},
\end{equation}
with $\theta=V\tau_2/\sqrt{D\tau_1}$ and $\epsilon =\sqrt{D\tau_1}/L$. 
\end{itemize}
This last regime enables a discussion of the efficiency of the intermittent search. 
The efficiency can be quantified  by comparing $\langle t \rangle^{opt}$,  the mean search time with intermittence at the minimum with 
 $\tau_{\rm diff}=L^2/(2D)$, the mean search time with diffusion alone.
It can be shown that intermittence  decreases the search time in the limit of low target density, and 
that the mean search time is minimized for $\tau_1$  as small as possible.  
The optimization with respect to $\tau_2$ leads to two regimes, depending on the minimal  value of $\tau_1$ as compared to the 
previously introduced timescale $\tau=D/V^2$, characteristic of the searcher.~: 
\begin{itemize}
 \item When  $\tau_1 \gg \tau$~: 
\begin{equation}\label{poisson_tau2_tau1gd}
\frac{\tau_2^{opt}}{\tau} \sim \sqrt{\frac{7}{4}}\left(\frac{\tau_1}{\tau} \right)^{3/4}.
\end{equation}
 At the optimum, the mean search time is~: 
\begin{equation}
\langle t \rangle^{opt} \sim \frac{L}{2V}\sqrt{\frac{\tau_1}{\tau}},
\end{equation}
and the gain is~: 
\begin{equation}\label{poisson_G_tau1gd}
G \sim \frac{L}{\sqrt{D\tau_1}} =\sqrt{\frac{2\tau_{\rm diff}}{\tau_1}},
\end{equation}
where $\tau_{bal}=L/V$ is the typical time needed to travel in the ballistic mode the distance between two consecutive targets. 
As we shall see in the following, 
in this regime the approximations are very accurate. 
 \item When  $\tau_1 \ll \tau$~: 
\begin{equation}\label{poisson_tau2_tau1pt}
\frac{\tau_2^{opt}}{\tau} \sim \frac{1}{2}\sqrt{\frac{\tau_1}{\tau}}.
\end{equation}
 At the optimum, the mean search time is~: 
\begin{equation}
\langle t \rangle^{opt} \sim \frac{3L}{4V}= \frac{3}{4}\tau_{bal},
\end{equation}
and the gain is~: 
\begin{equation}\label{poisson_G_tau1pt}
G \sim \frac{2LV}{3D} =\frac{4 \tau_{\rm diff}}{3\tau_{bal}}.
\end{equation}
As we shall see in the following, the approximations are qualitatively good in this regime, 
but not as precise as in the other regime. 
Indeed, the gain obtained here would mean that the mean first passage time to the target is
smaller than $\tau_{bal}$, which is the minimal mean time to travel to the target 
(except if $\tau_{bal}>\tau_{\rm diff}$).
In fact, as can be seen in figure \ref{gain}, simulations show that $ \langle t \rangle^{opt} \to \tau_{bal}$. 
It means that  
very fast intermittence enables the searcher to retain the best of the two phases~: 
reactivity of phase 1 and motion of phase 2. 
\end{itemize}

\doublimagem{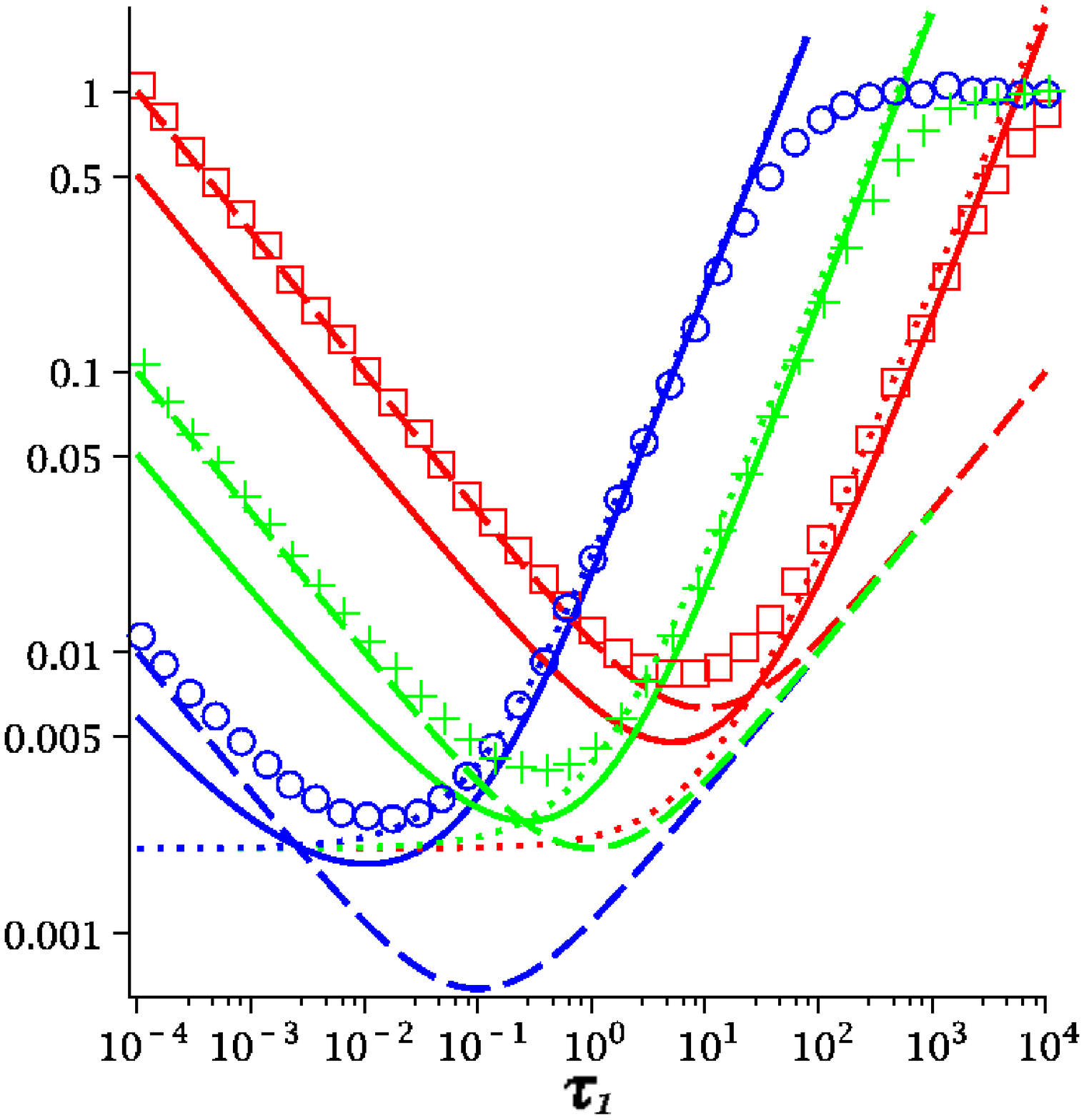}
{$ln(\langle T \rangle /\tau_{\rm diff})$ as a function of $ln(\tau_1)$. $\tau_2=10$ (red, $\square$),  $\tau_2=1$ (green, $+$),  $\tau_2=0.1$ (blue, $\circ$).  }
{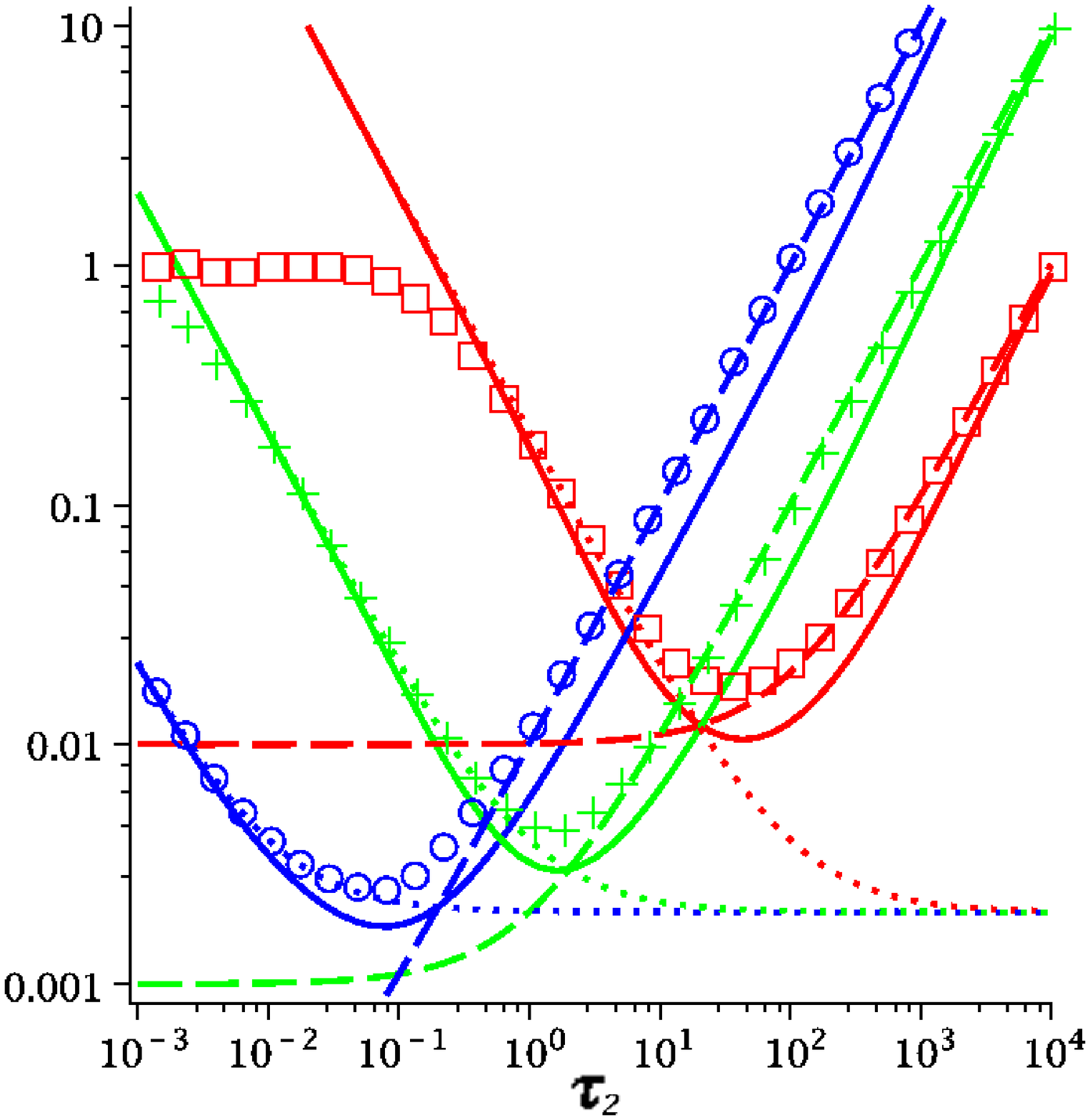}
{$ln(\langle T \rangle /\tau_{\rm diff})$ as a function of $ln(\tau_2)$. $\tau_1=100$ (red, $\square$),  $\tau_1=1$ (green, $+$),  $\tau_1=0.01$ (blue, $\circ$).  }
{Validity of the approximations. Mean first passage time to the target, renormalized by the mean first passage time without intermittence. 
Small ballistic displacements approximation \refm{poisson_t_small_ball} (dashed line). Large ballistic displacements approximation  \refm{poisson_t_large_ball}  (dotted line). Intermediary approximation  \refm{poisson_t_inter_ball} (line). Numerical simulations (symbols). $D=1$, $V=1$, $L = 10^{3}$. }{approx}

\imagea{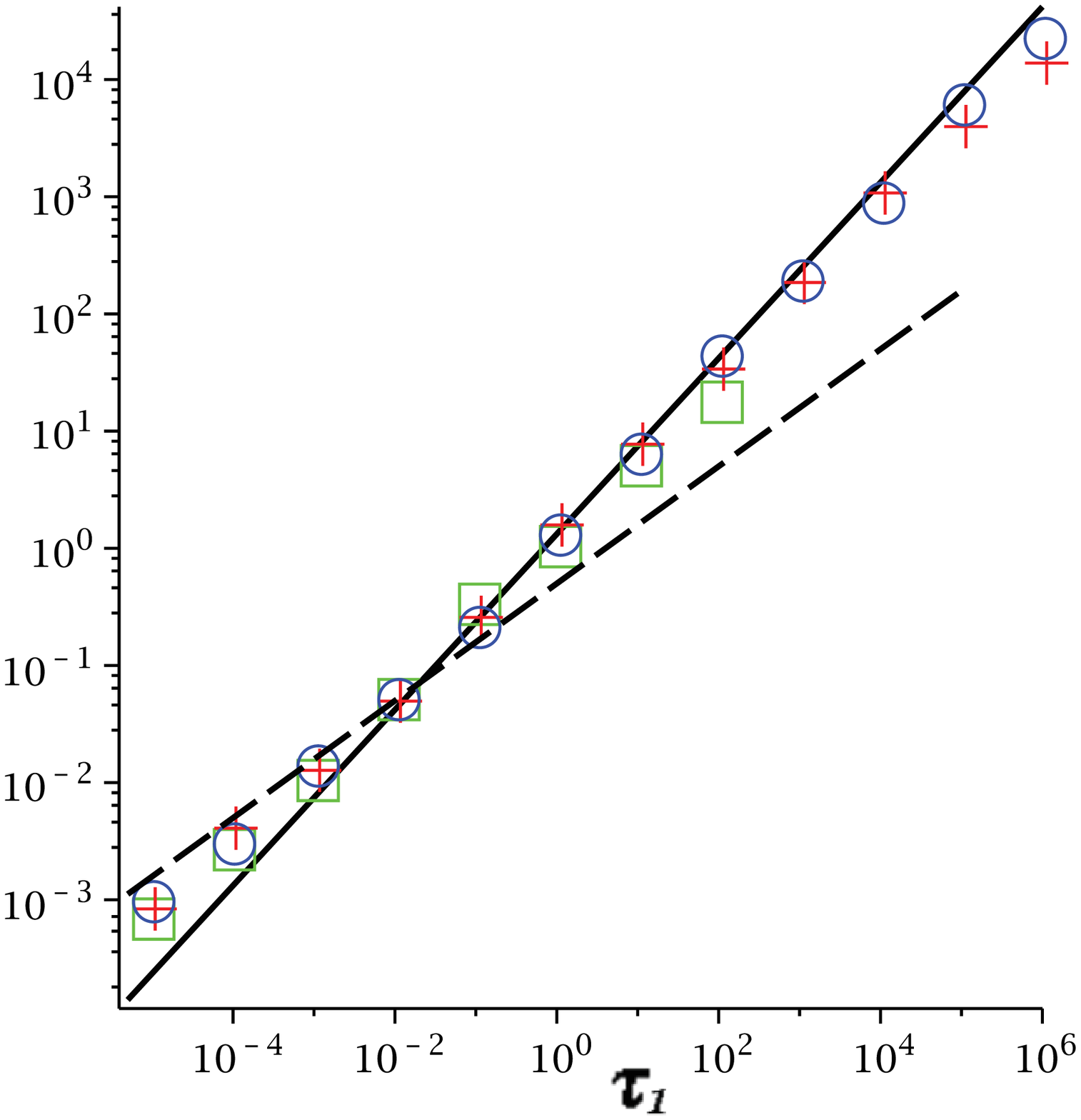}
{$ln(\tau_2^{opt})$ as a function of $ln(\tau_1)$. Small $\tau_1$ analytical prediction \refm{poisson_tau2_tau1pt} (dashed black line). Large $\tau_1$ analytical prediction \refm{poisson_tau2_tau1gd} (solid black line). Numerical values (symbols), for  $L=10$ (green $\square$),  $L=10^{3}$ (red $+$), $L=10^{5}$ (blue $\circ$). $D=1$, $V=1$.}{opt}{8}

\imagea{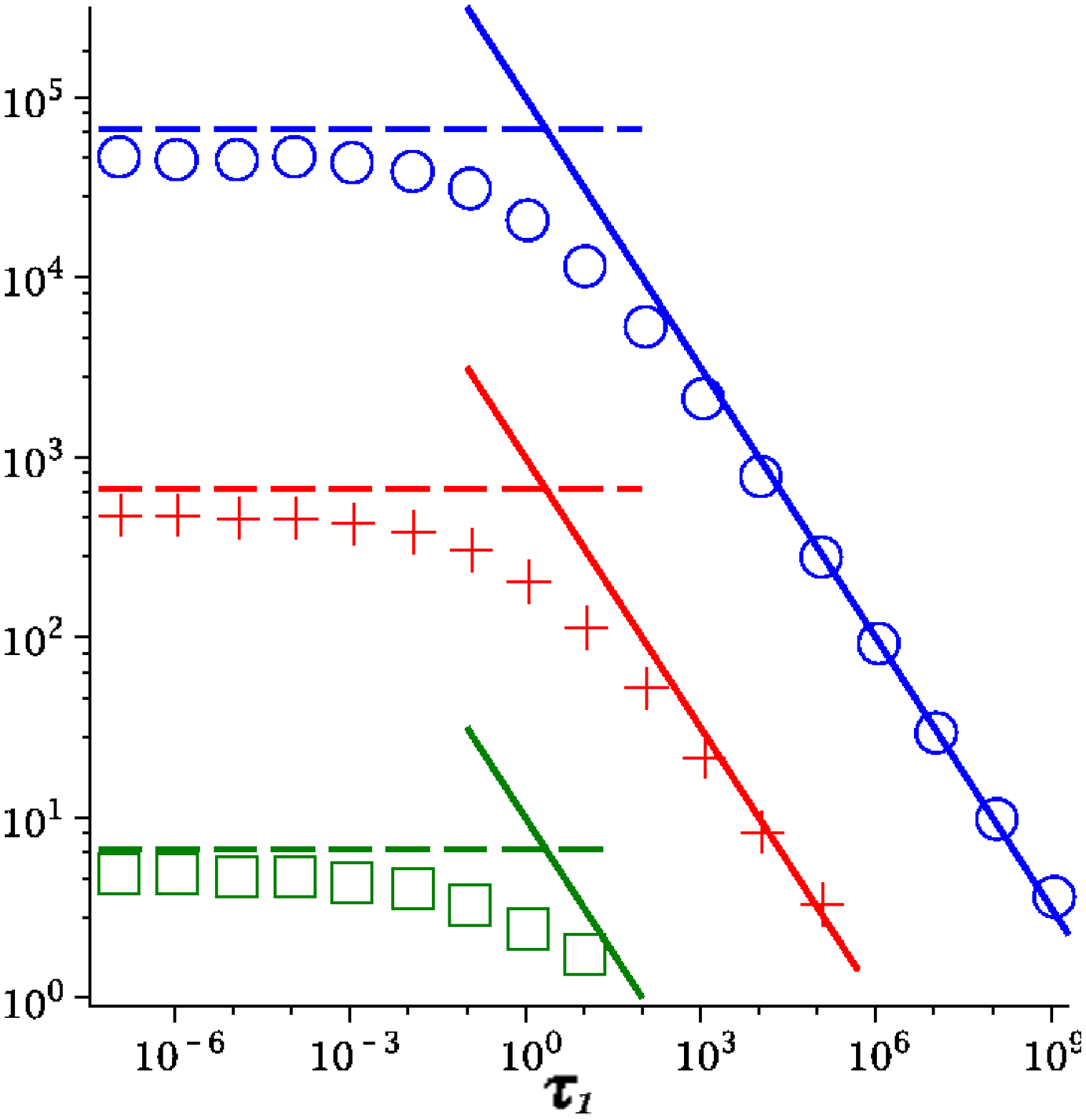}{$ln(G)$ as a function of $ln(\tau_1)$ ($\tau_2$ taken optimal). Small $\tau_1$ analytical prediction \refm{poisson_G_tau1pt} (dotted line). Large $\tau_1$ analytical prediction \refm{poisson_G_tau1gd} (line). Numerical simulations (points). $L=10^{1}$ (green, $\square$),  $L=10^{3}$ (red, $+$), $L=10^{5}$ (blue, $\circ$). $D=1$, $V=1$. }{gain}{8}

Figure \ref{approx} represents the mean search time $\langle t \rangle$ as a  function of $\tau_1$ and $\tau_2$ for typical values
of the other parameters. It enables to compare the numerical results with the approximations
\refm{poisson_t_large_ball} and \refm{poisson_t_small_ball}, and with the intermediary approximation \refm{poisson_t_inter_ball}. 
It shows that the approximations
of large and small ballistic displacements are valid in the expected conditions, and  
 the intermediary approximation \refm{poisson_t_inter_ball}
 correctly reproduces the existence and the position of the minimum of $\langle t \rangle$. 
Figure \ref{opt}  supports the scaling laws relating $\tau_1$ and the corresponding optimal
waiting time $\tau_2$ at the optimum. The exponent 3/4 of the theoretical scaling law \refm{poisson_tau2_tau1gd} 
for $\tau \ll \tau_1$ is very well
confirmed by the simulations. This is not the case for the law \refm{poisson_tau2_tau1pt} for $\tau \gg \tau_1$, which 
indicates that the approximations should be handled with care for short waiting times $\tau_1$, $\tau_2$,
although their results are qualitatively correct.
Figure \ref{gain} shows the gain as a function of $\tau_1$ in different possible conditions. It supports
the conclusions of the theoretical study, and indeed confirms that the gain due to intermittence
can be very important if  $\tau_{\rm diff} \gg \tau_{bal}$.

\subsubsection{Conclusion}

In the case of a Poissonian distribution of targets,  
 intermittence remains valid as a strategy minimizing the search time. 
The
optimal strategy still consists in taking $\tau_1$ as small as possible. 
However, $\tau_2^{opt}$ is different from the case of regularly spaced targets. 
The optimal
mean duration of ballistic flights scales as $\sqrt{\frac{7}{4}}(\tau_1^3 \tau)^{1/4}$ in
the limit $\tau_1 \gg \tau = D/V^2$. 
In this regime, at the optimum,  $\langle t \rangle \simeq \frac{1}{2} \frac{L}{V}\sqrt{\frac{\tau_1}{\tau_2}} $,  
with a gain compared to diffusion alone $\propto L/\sqrt{D\tau_1}$.

\subsection{Taking into account partial correlations in ballistic phases}
\label{section_correlations}

\subsubsection{Motivation}

The models developed in sections \ref{section_animaux_2005} and  \ref{section_generic}, 
are very similar. 
The searcher alternates between a slow reactive phase, 
and a fast ballistic blind phase. 
The main difference is that in section \ref{section_animaux_2005}, ballistic phases
are always in the same direction, 
whereas on the contrary, in section \ref{section_generic}, 
the direction of  each new ballistic phase,  is  random and independent of the 
 previous ballistic phase. 


In the case of animal trajectories, the successive directions of ballistic phases are usually correlated  \cite{animauxObrien}. 
We have considered so far two extremes cases~: no correlation or infinite range correlations. In both cases, there are regimes where intermittence is favorable. 
However, in the case of infinite range  correlations, 
 the shorter the duration of each phase, the smaller the search time. 
In contrast, in the case without any correlation, 
the minimal search time is obtained for finite values of $\tau_1$ and $\tau_2$, which even diverge with the system size.
In the intermediate case of finite range correlations, determining the nature 
 of the minimum  is an interesting theoretical question. 
Besides, as real biological systems often present correlations, 
it is an important issue to take into account correlations in the  generic model of intermittence. 
We present in what follows  the simplest case of the static mode of detection  in dimension one . 

\subsubsection{Model}

The searcher is either in the reactive phase 1 (where it is immobile,  and finds the target with probability per unit time  $k$ if the target is at a distance smaller than $a$),
or in the ballistic phase 2, of velocity $V$. For  each new ballistic phase, the direction of $V$ is  the same  as in the previous ballistic phase with probability $p$, 
and in the opposite direction with probability $1-p$.
The distribution of the duration of the phases is exponential, of mean $\tau_i$, and the distance between two targets is $2b$.

In the case of no correlations ($p=1/2$) the mean search time has been calculated  in the section \ref{section_generic_statique1D} (see equation 
 \refm{tm1Dvk}), where it was shown that the optimum is obtained for $\tau_1^{opt}=\sqrt{\frac{a}{Vk}}\left(\frac{b}{12a} \right)^{1/4}$
and $\tau_2^{opt}=\frac{a}{V}\sqrt{\frac{b}{3a}}$.

In the general case, the methods of section \ref{section_generic} can be adapted to calculate analytically the 
mean search time starting from a random position in state 1, which can be written as~:
 \footnotesize
\begin{equation}\label{tm_complet}
t_m=\left( \tau_2+\tau_1 \right)  \left( {\frac {1}{k\tau_1}}+
 \frac{ b-a}{b}  \left(\frac{2}{3}{\frac { \left( 1-p \right)  \left( b
-a \right) ^{2}}{{\tau_2}^{2}{V}^{2}}}+1+{\frac {1}{k\tau_1}}+\frac{
u \left( b-a \right)}{\sqrt {k\tau_1} \tau_2 V} \coth \left( {\frac {
\sqrt {k\tau_1}ua}{\tau_2\,V \left( 1+k
\,\tau_1 \right) }} \right)\right)\right),
\end{equation}\normalsize
with $u=\sqrt {2(1-p)+k\tau_1}$.

\subsubsection{Minimization of the mean search time}
\po{Case of infinite range correlation $p=1$}

\imagea{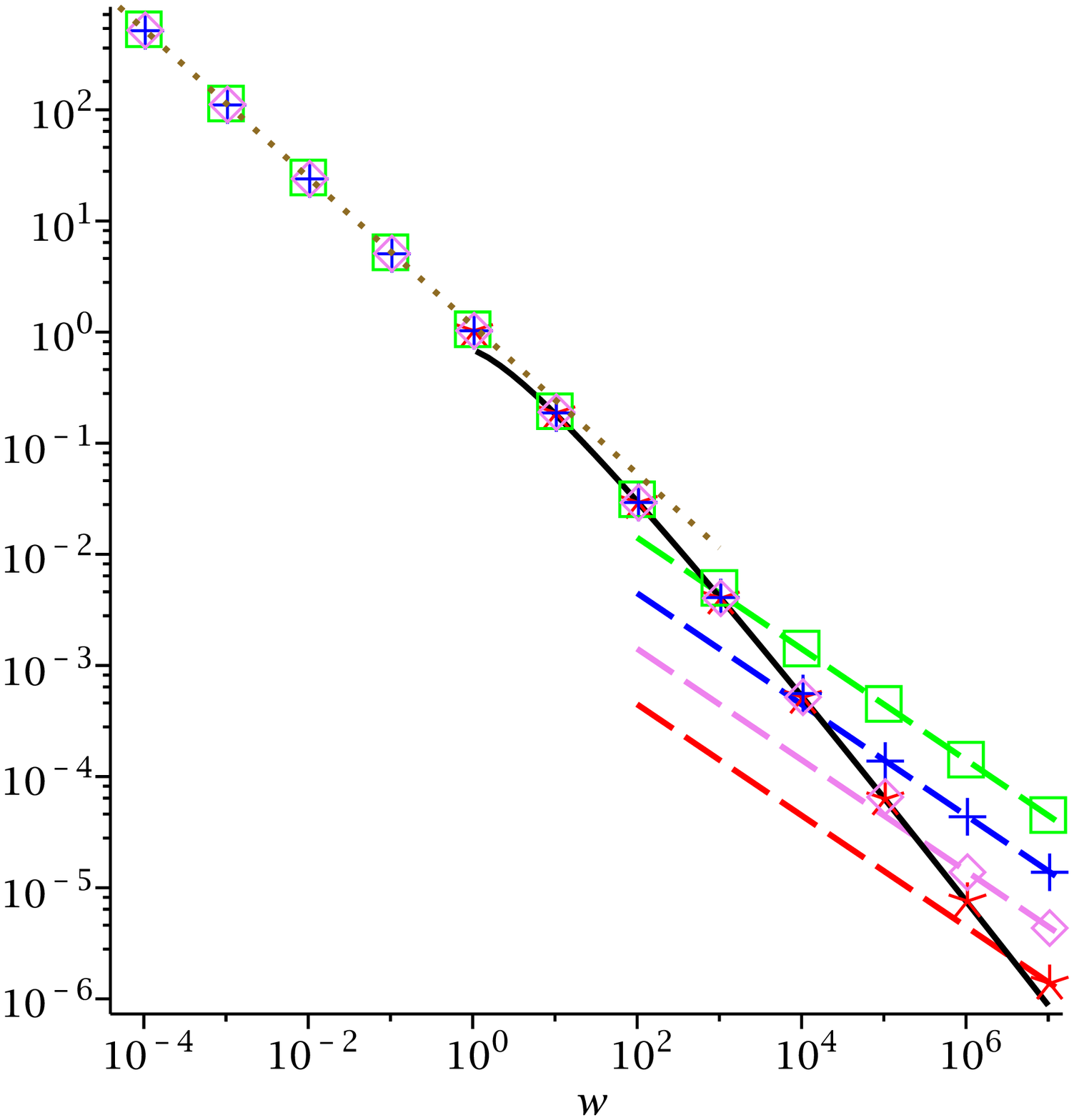}{Minimization of the mean search time for the static mode with infinite correlation ($p=1$). $\alpha^{opt}$ as a function of $w$.
Theoretical expression for $w$ small (brown dots), 
for $w$ intermediate (black solid line), 
for $w$ large (colored dashed lines). 
Optimization of the exact expression (with $\tau_2 \to 0$ and $\tau_1=\alpha \tau_2$) (symbols).  
$b=100$ (green, $\square$), $b=10^3$ (blue, $+$), $b=10^4$ (violet, $\diamond$), $b=10^5$ (red, $\star$). $a=1$, $V=1$.}
{p1_optalpha_wresume}{6}

It can be shown that the mean search time  is minimized  for  $\tau_1$ and $\tau_2$ tending to  0, with  $\tau_1=\alpha \tau_2$.
We define $w=ak/v$, and depending on this parameter~: 
\begin{itemize}
 \item $w<1$~: $\alpha^{opt} \simeq \left(\frac{3}{2w^2} \right)^{\frac{1}{3}}$,
\item $1<w<w^*$~: $\alpha^{opt} \simeq \frac{\ln(4w)}{2w}$,
\item $w>w^*$~: $\alpha^{opt} \simeq \sqrt{\frac{2a}{wb}}$.
\end{itemize}
$w^*$ is defined as the solution of $\frac{\ln(4w^*)}{2w^*} = \sqrt{\frac{2a}{w^*b}}$. 
These expressions are in good agreement with the numerical minimization of the exact expression 
of the mean search time (see figure \ref{p1_optalpha_wresume}).

\po{Case of intermediate correlations}

The mean search time obtained in \refm{tm_complet} is difficult to optimize. 
An important question raised is to determine whether the mean search time is minimized for finite $\tau_1$ and $\tau_2$ (as in the case $p=0.5$), 
or for $\tau_1$ and $\tau_2$ tending to 0 (as in the case $p=1$).
An answer can be obtained by noticing that 
a lower bound of  the mean search time is given by~:
\begin{equation}
 t_m \geq (\tau_1+\tau_2)\left(\frac{1}{k\tau_1}+\frac{(b-a)^3}{b} \frac{2(1-p) }{3\tau_2^2 V^2}\right).
\end{equation}
Supposing  that the minimum is realized for at least one of the $\tau_i \to 0$, three cases arise. 
\begin{itemize}
 \item $\tau_1 \to 0$ with $\tau_1 \ll \tau_2$. In this case $t_m \geq \tau_2 /(k\tau_1) \to \infty $.
\item $\tau_2 \to 0$ with $\tau_2 \ll \tau_1$.  In this case $t_m \geq \tau_1 \frac{(b-a)^3}{b} \frac{2(1-p) }{3\tau_2^2 V^2} \to \infty$.
\item $\tau_1 \sim \tau_2$ and both $\to 0$.  In this case $t_m \geq \tau_2 \frac{(b-a)^3}{b} \frac{2(1-p) }{3\tau_2^2 V^2} \sim 1/\tau_2 \to \infty$.
\end{itemize}
Finally, this shows that the minimum is realized for finite values of $\tau_1$ and $\tau_2$ as soon as $p<1$. Actually, it can be shown that, except for $p$ close to 1, 
the minimum of the search time is obtained for~: 
\begin{equation}\label{t1corr}
\tau_1^{opt}=\sqrt{\frac{a}{Vk}}\left(\frac{b}{a}\frac{(1-p)}{6} \right)^{1/4} \mathrm{~and}
\end{equation}
\begin{equation}\label{t2corr}
\tau_2^{opt}=\frac{a}{V}\sqrt{\frac{b}{a}\frac{2(1-p)}{3}}.
\end{equation}
Interestingly, note that the relation  $\tau_2^{opt}=2k\left(\tau_1^{opt}\right)^2$ obtained initially
in the case of the absence of correlations (see \reft{recapgeneral}) still holds in this case.  

\doublimagem{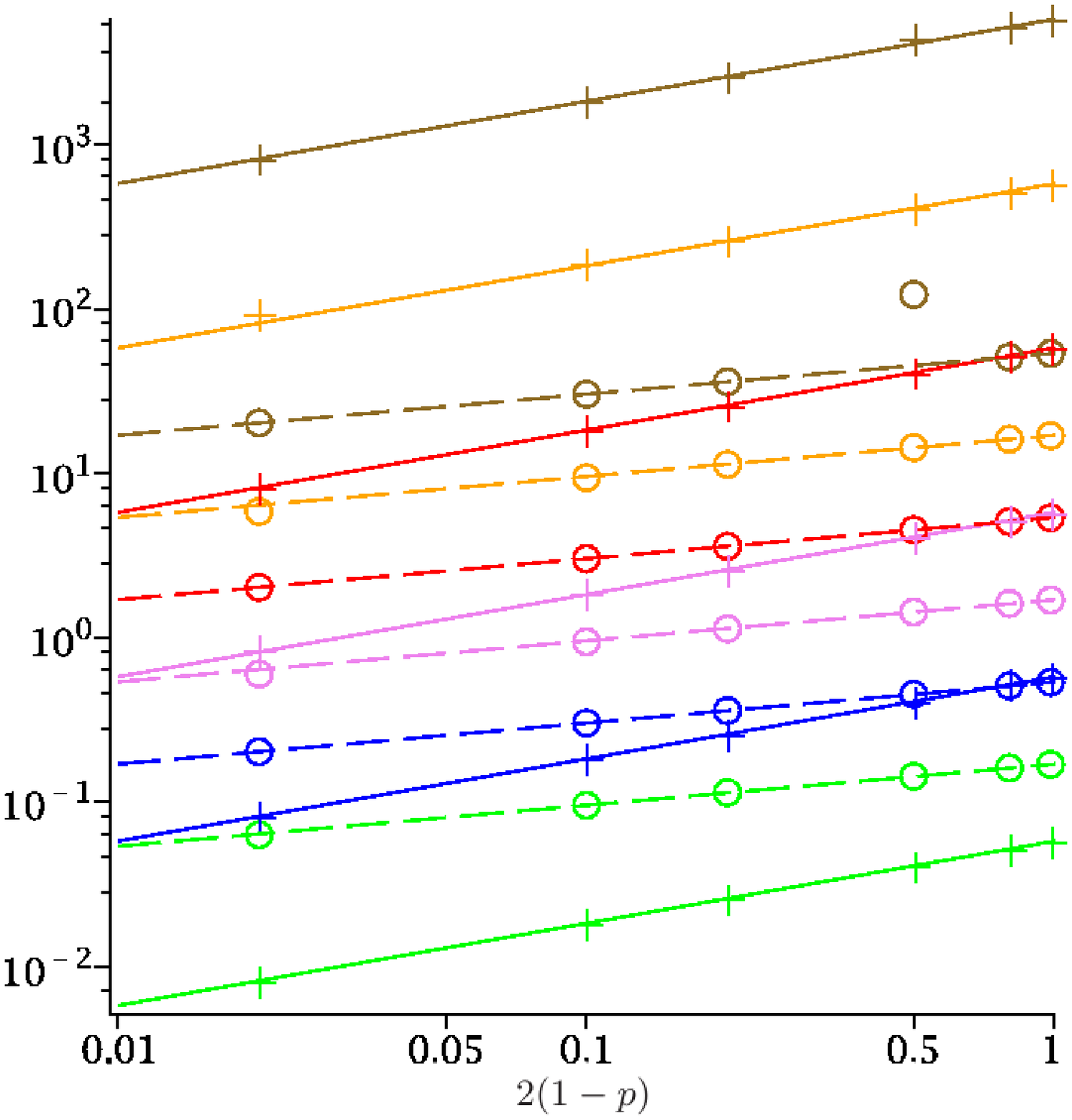}{}{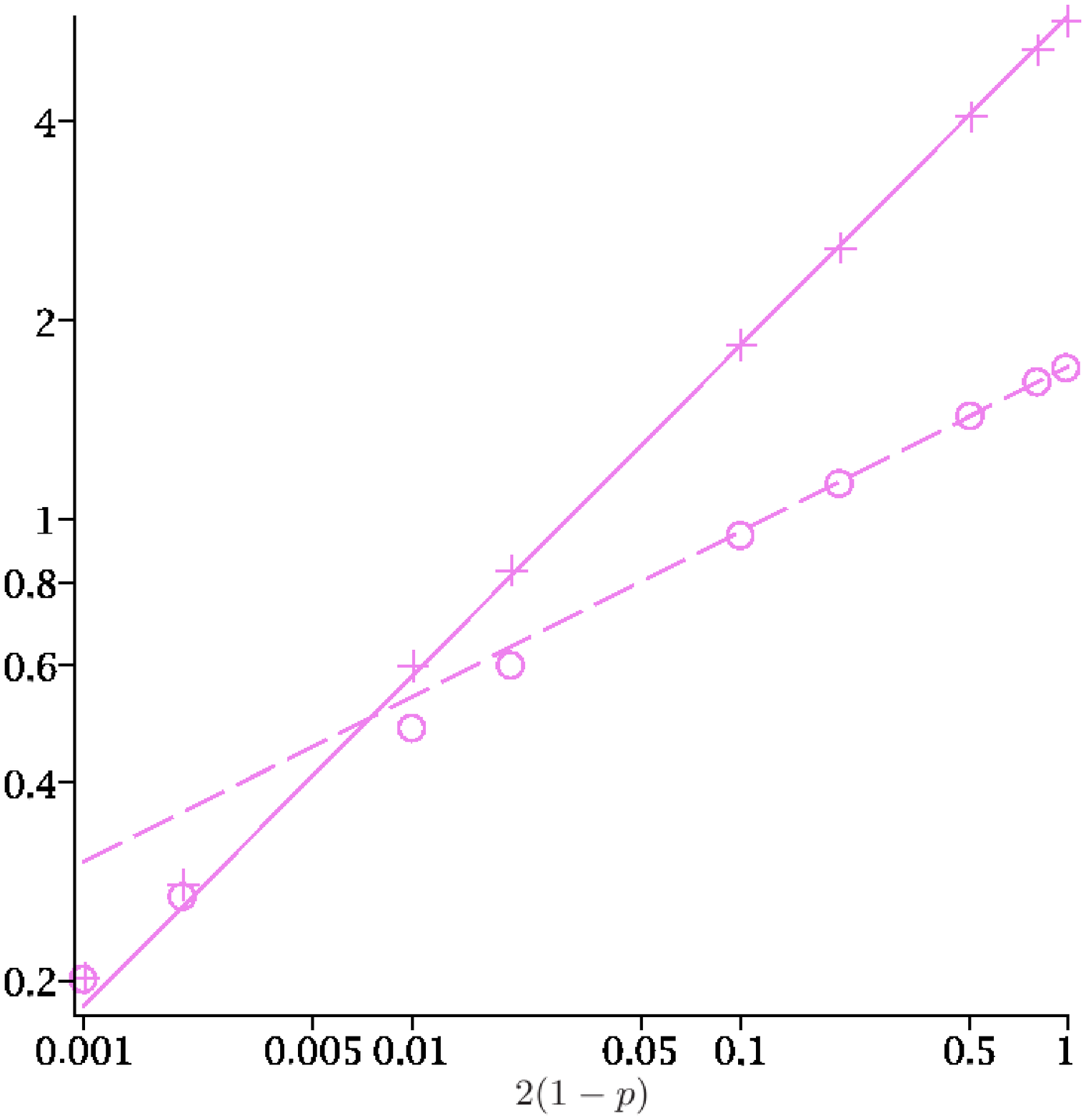}{}{$\tau^{opt}$ as a function of $2(1-p)$. 
Theoretical value of $\tau_1^{opt}$ \refm{t1corr} (dashed line)
and theoretical value of $\tau_2^{opt}$ \refm{t2corr} (solid lines), 
compared to the numerical minimization of the full exact mean search time, 
leading to  $\tau_1^{opt}$ ($\circ$) and $\tau_2^{opt}$ ($\square$). 
$a=0.01$, $b=1$ (green), $a=0.01$, $b=100$ (blue), $a=1$, $b=100$ (violet),  $a=1$, $b=10^4$ (red),  
$a=100$, $b=10^4$ (orange),  $a=100$, $b=10^6$ (brown).  
$k=1$, $V=1$.}{test_corr}

These expressions are in agreement with  the numerical 
minimization of the exact expression of the mean search time \refi{test_corr}, 
except when $1-p$ is very small.

\subsubsection{Conclusion}

In the simple case of the static mode in one dimension, 
the influence of correlations on the mean search time and its minimization can be studied. 
An exact expression of the mean search time shows that it is minimized 
 for finite values of $\tau_1$ and $\tau_2$ as soon as $p<1$. 
When $(1-p) \gg a/b$, the optimal durations $\tau_1^{opt}$ and $\tau_2^{opt}$ can be explicitly given,
and they are in continuity with the case without correlation $p=0.5$.

\subsection{Other distributions of phases duration}
\label{section_autres_temps}

 
The model presented in section \ref{section_generic} is minimal in the 
sense that the searcher has no memory. As seen in the previous section, 
a possibility is to add orientational memory. Another possibility is to 
add temporal memory. In the generic model, we have considered a ``Markovian'' 
searcher, in the sense that the rate of switching from one phase to the 
other is constant. It leads to an exponential distribution of the durations 
of the phases. In the following, we study the influence of the 
distribution of the duration of the phases (see also \cite{ctrwCorrelee1,ctrwCorrelee2,ctrwCorrelee3} for other types of correlations). A first possibility is to 
study the effect of distributions which are peaked around the mean duration, or even deterministic
\cite{SpecialIssue2006}.
A second possibility is to  study the case of L\'evy-distributed ``blind'' phases as in  \textcite{intermittentlevy}.

\subsubsection{Deterministic durations of the phases}

\label{section_temporal_memory}

\imagea{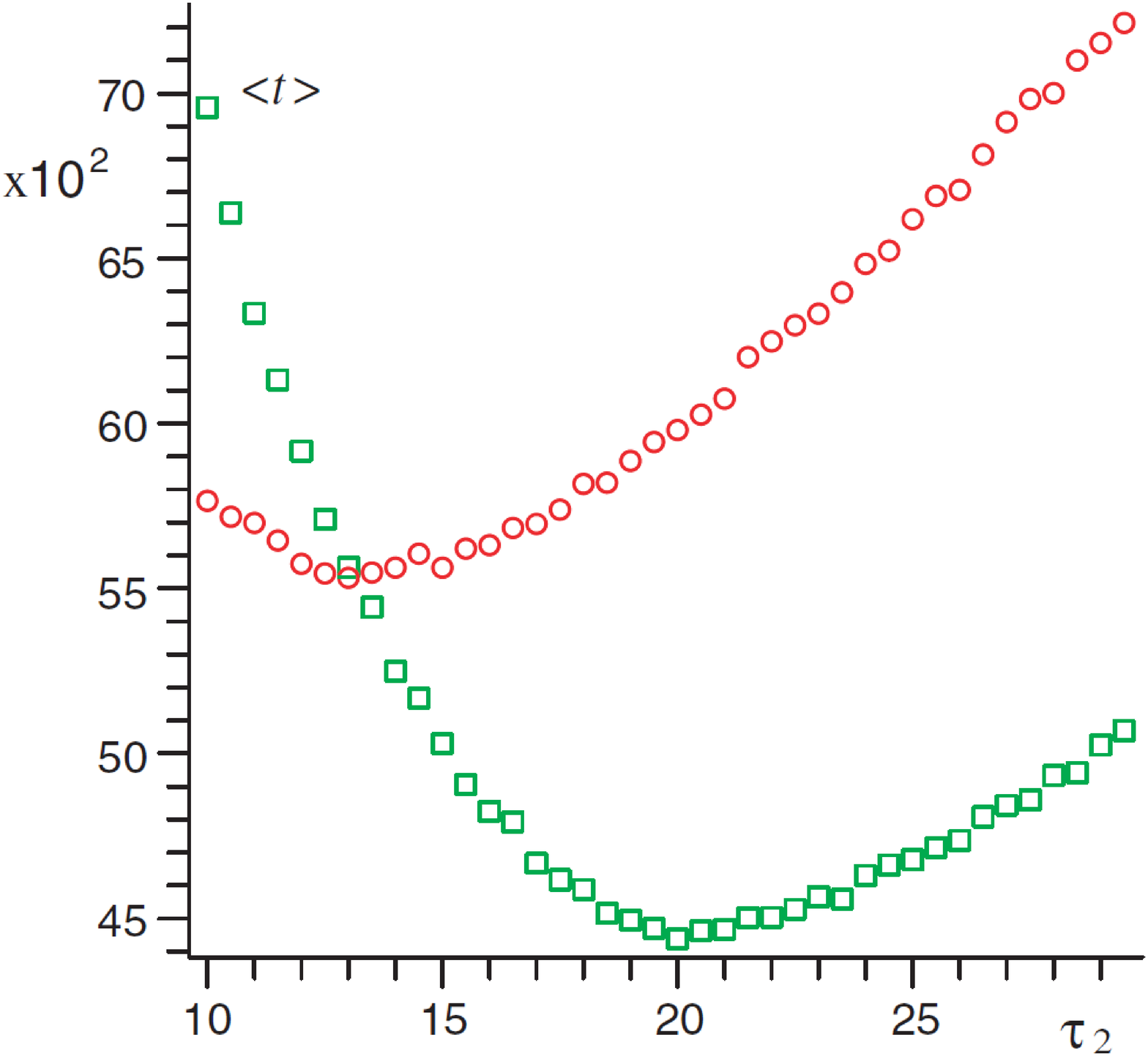}{Comparison between the search without temporal memory (\textcolor{red}{$\circ$}) and the search
with temporal memory (\textcolor{green}{$\square$}). Static mode in two dimensions. 
$t_m$ as a function of $\tau_2$.  $k = 1$, $V = 1$,
$b=113$, $a = 10$, $\tau_1 = 2.6$.}{generic_temps_fixe}{6}

In the generic model previously described, we have considered exponential durations of phases, which correspond to searchers with no temporal memory. 
In the  opposite case, a searcher with full  memory skills  could for example switch from one phase to the other 
at deterministic times instead of exponentially distributed times. The corresponding problem is not Markovian any more, which makes its analytical resolution much more 
 complex.  
We present here a summary of a numerical study  
of the effect of such 
temporal memory 
for a searcher with the static mode of detection  in dimension 2 (see  \textcite{SpecialIssue2006} for details and  \textcite{LosingTime} for a semi-analytical treatment in 1 dimension). 
First, this study shows that the optimal $\tau_1^{opt}$ and $\tau_2^{opt}$ are larger than  in the case without memory, 
but are of the same order of magnitude \refi{generic_temps_fixe}. 
Second, such temporal memory decreases the mean search time. 
Indeed, a deterministic duration of the relocation phase avoids both the very short and very long relocations, that are inefficient.
Third, and importantly,  the gain from this temporal memory is quite low (less than 40\% 
in 
an extended range of parameters, and decreasing with $b/a$ increasing) as compared to the case with no memory (see figure \ref{generic_temps_fixe}). 

%
%
%

\subsubsection{L\'evy distribution of the fast phase durations}

\label{section_temporal_levy}

\textcite{intermittentlevy} study analytically and numerically a 1--dimensional  intermittent random walk whose duration of relocation phases is taken from a L\'evy law ($p(l) \propto l^{-\alpha-1}$, with $1<\alpha <2$).  Apart from this distribution of the duration of ballistic phases, 
this model is identical to the generic model presented in section \ref{section_generic} (see also \textcite{LosingTime}), in the case of a  diffusive mode of detection in one dimension (in the particular case of a point-like target $a \to 0$).

The mean search time is evaluated with the exact formula (equation (9) of \textcite{intermittentlevy})~:
\begin{equation}
\langle t \rangle = \sum_{n=1}^{\infty} \frac{2(\tau_1+\tau_2)}{D\tau_1k_n^2+1-\lambda(k_n)}
\end{equation}
with $k_n=2\pi/L$, where $L$ is the distance between targets , and $\lambda$ is the characteristic function of the distribution~: $\lambda(k)=\exp(-\sigma^\alpha |k|^\alpha)$ 
($p(k)=\int_{-\infty}^{\infty}e^{ikx}p(x)dx $). The relation between $\sigma$, $\alpha$, the velocity $V$ and 
$\tau_2$ (the mean duration of phase 2, which is defined since $\alpha>1$)
 is given by (equation (10) of \textcite{intermittentlevy})~: 
\begin{equation}
 \sigma=\frac{\pi V \tau_2}{2 \Gamma\left(1-1/\alpha \right)}.
\end{equation}
A more tractable approximate expression of the mean search time can be derived (equation (14) of \textcite{intermittentlevy})~: 
\begin{equation}\label{tap}
\langle t \rangle =2(\tau_1+\tau_2)\left(\frac{L}{4\sqrt{D\tau_1}}+\left(\frac{L}{2\pi \sigma}\right)^\alpha \zeta(\alpha) \right) 
\end{equation}
with $\zeta(\alpha)=\sum_{n=1}^{\infty}n^{-\alpha}$ is the Riemann $\zeta$  function. 

As compared with the generic model of section  \ref{section_generic}, this model introduces an extra parameter $\alpha$, which as could be expected enables a further minimization of the mean search time. \textcite{intermittentlevy} claim that L\'evy laws are more efficient than exponential laws because they have no second moment and therefore are not bound to the central limit theorem.


\imagea{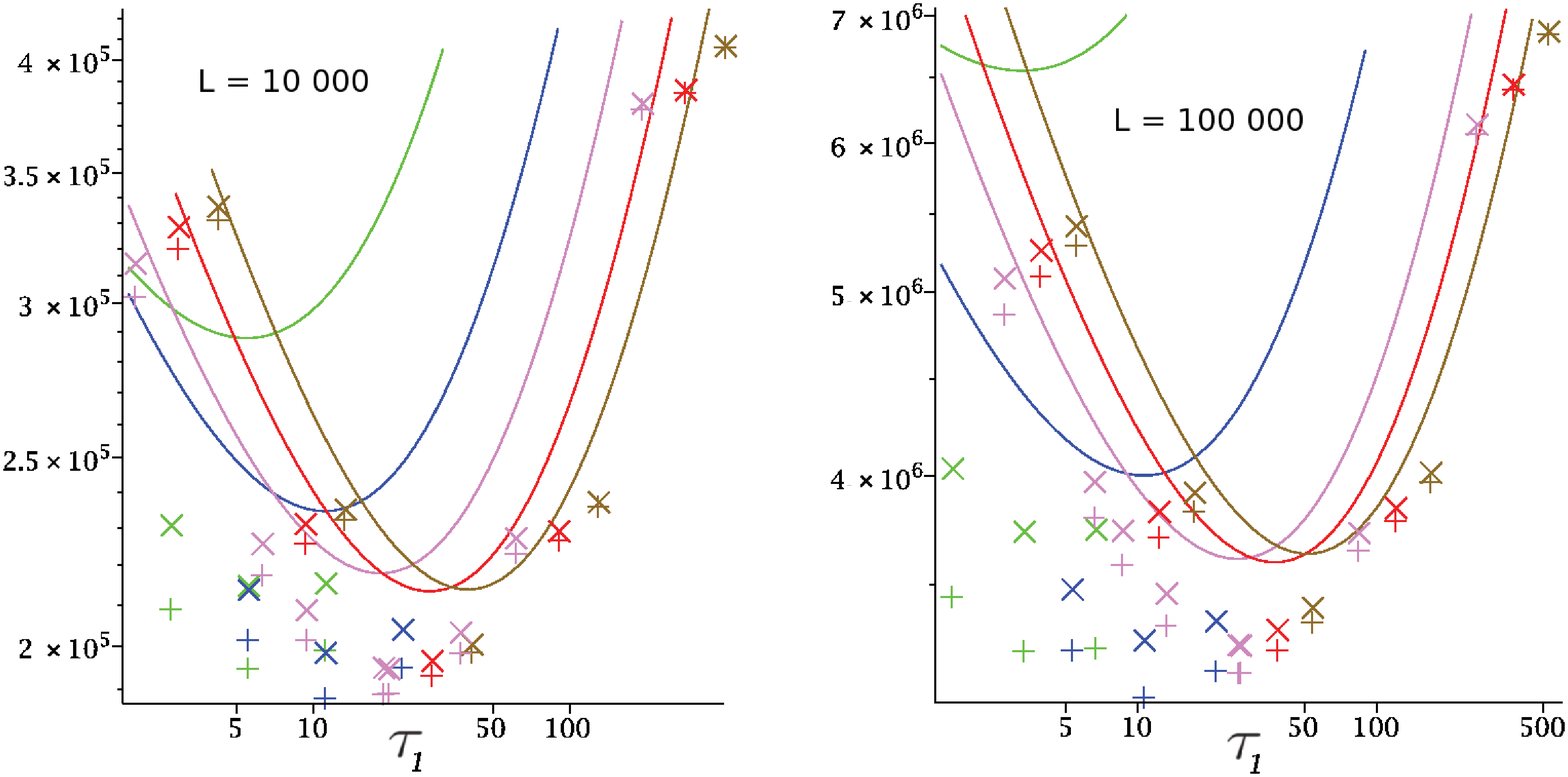}{$t_m$ as a function $\tau_1$, $\sigma$ being at the theoretical minimum (numerical minimization of equation \ref{tap}). 
lines~: approached analytical formula \ref{tap}. $\times$~: simulations without cut-off. $+$~: simulations with cut-off at $L$. $D=1$, 
$V=1$. \textit{Left} $L=10^4$, 
sienna~: $\alpha = 1.6$; red~: $\alpha = 1.5$; violet~: $\alpha = 1.4$; 
blue~: $\alpha = 1.3$; green~: $\alpha = 1.2$.  \textit{Right} $L=10^5$, sienna~: $\alpha = 1.6$; red~: $\alpha = 1.5$; violet~: $\alpha = 1.4$; 
blue~: $\alpha = 1.3$; green~: $\alpha = 1.2$.}{Lalpha}{6}
%
%
%

However, the L\'evy distribution is not the optimal distribution of the duration of ballistic phases, and distributions with a finite second moment can perform even better, as opposed to what is claimed in  \textcite{intermittentlevy}. Indeed, relocations larger than the distance between two targets $L$ cannot be profitable, and power law distributions are therefore inefficient in the regime of long times $t>L/V$. 
A simple example is given by a L\'evy distribution with an upper cut-off at $L$ \new{\refi{Lalpha}}.  
For $L=10^4$, 
numerical simulations show that the optimum without cut-off is realized for approximately  $\alpha \simeq 1.4$, 
with $t_m \simeq 195~000$; 
and 
the optimum with a cut-off at  $L$ is realized for  $\alpha \simeq 1.3$, with  $t_m 
\simeq 188~000$, that is $\simeq 3.7\%$ lower. 
For $L=10^5$, 
the optimum without cut-off is realized  for $\alpha \simeq 1.3$, $t_m \simeq 
3~260~000$; 
the optimum with a cut-off of $L$ is realized for $\alpha \simeq 1.2$, $t_m 
\simeq 
3~060~000$,   that is $\simeq 6.5\%$ lower. 
 Truncated distributions with well defined second moment  therefore outperform  L\'evy distributions.
 Hence, intermittent random walks with L\'evy-distributed relocations decrease the mean search time more 
efficiently  than in the case of exponentially distributed relocations, 
but it is not because of their infinite variance, and other distributions with second moment can perform even better. 

\new{Another point discussed by \textcite{intermittentlevy} is the robustness of the strategy : 
if $L$ is misevaluated, the efficiency of the intermittent search with exponential relocation durations 
decreases more than for the L\'evy distribution. Truncated L\'evy is probably of intermediate robustness.}

\subsection{The point of view of the target : Pascal principle} \label{pascal}
In this review,  we have addressed the question of determining optimal search strategies. One could also consider the opposite point of view and try to determine optimal survival strategies of targets. In the case where the target's motion is independent of the searcher's motion, the response is actually    given very simply by the so called Pascal principle \cite{Pascal_principe1,Pascal_principe2} for a broad class of  situations . More precisely, suppose that the motion of the searcher is time and space homogeneous and that it satisfies the following property: starting from a position $x$ (different from the target position), the transition probability to be at position $y$ at time $t>0$ is always maximum for $y=x$. Assume that the target can perform any stochastic motion, independently of the searcher which is assumed to perform a stochastic  motion. Then Pascal principle states that the survival probability of the target is maximum if the target remains immobile at its initial position. Of course, the validity of Pascal principle is restricted to special motions of the searcher: it holds for  a diffusive motion, but not, for instance, if the searcher undergoes a ballistic motion with constant velocity. However, the validity conditions are satisfied if the searcher  undergoes a ballistic motions with symmetrically distributed stochastic velocities, or if the displacements consist in teleportations, which are distributed symmetrically with respect to the initial position.
In these cases, the best strategy for the target is to remain immobile.

\subsection{Other models of intermittent search}
\label{section_autres_autres}
\label{section_autres}

In the past few years, several models relying on the mechanism of intermittent search have been developed in different contexts. We here briefly review these models which are in essence similar to the generic case discussed in section \ref{section_generic}, and which broadens the field of application of intermittent search.


\po{\textcite{glebpareildiscret,glebpareildiscret_suite}}

\imagea{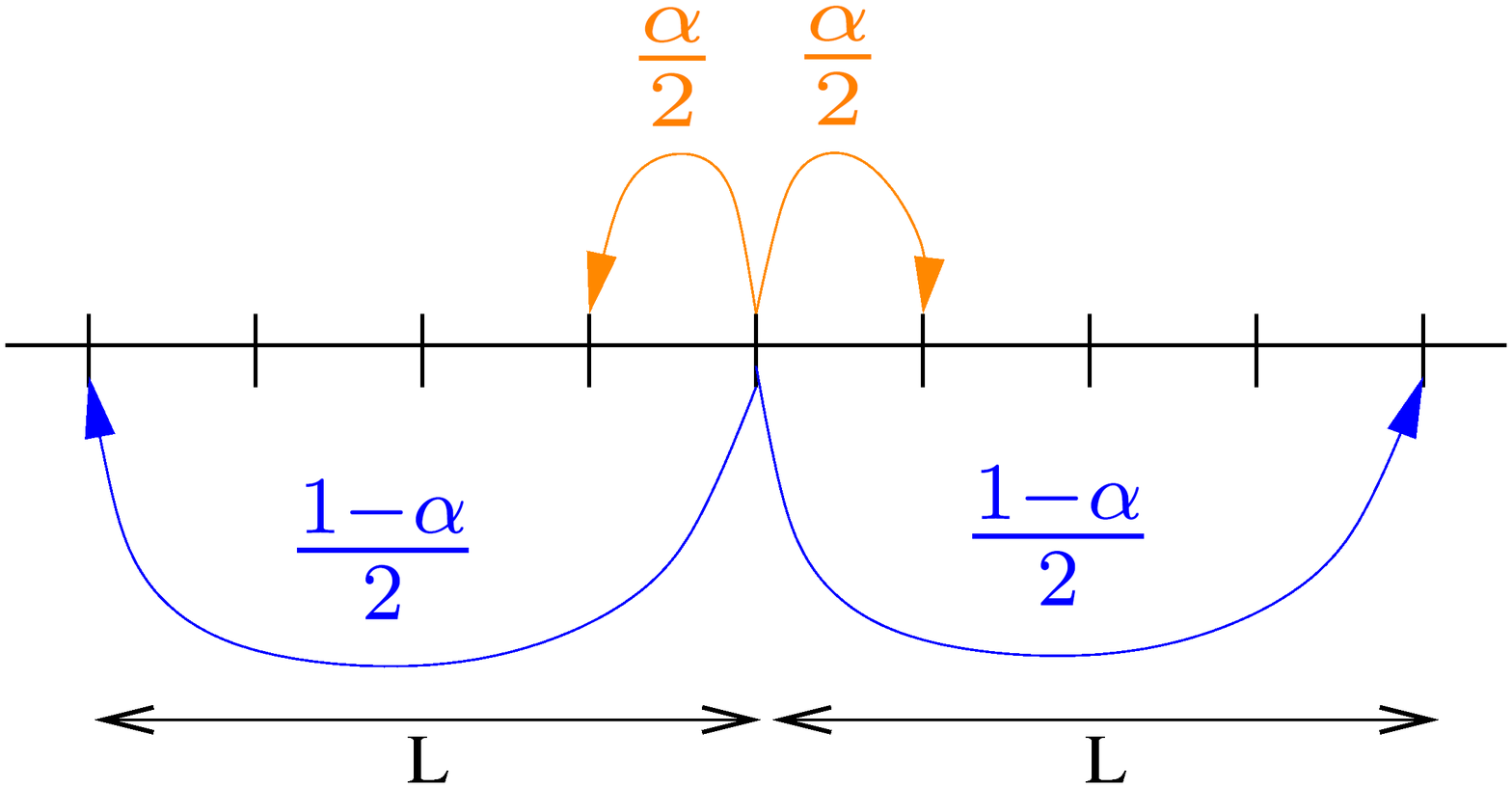}{Model used by \textcite{glebpareildiscret,glebpareildiscret_suite}.}{Gleb_schema}{5}

\textcite{glebpareildiscret,glebpareildiscret_suite} propose a model  very similar 
to the diffusive mode in one dimension of the generic model, 
but in discrete space, on a infinite lattice. 
At each time step, with probability $\alpha$, 
the searcher jumps to the neighboring node of the line 
(with equal probabilities for each side, which corresponds to diffusion).
With probability $1-\alpha$, it stays off-lattice during a time $T$ 
and after this time, it lands at a distance $L$ 
from its initial position (once again, with equal probabilities for each 
side) (see figure \ref{Gleb_schema}). 
This phase is equivalent to a ballistic non-reactive phase. 
Its duration is exactly $T$, 
whereas the duration of the diffusive phase with target detection is 
exponentially distributed, with mean duration $1/(1-\alpha)$. 
There is one target, but an infinite set of searchers, 
initially randomly distributed. 
The quantity maximized is the probability that at a given time $t$, 
the target has already been found by any of the searchers.
Oshanin et al. do find an optimal $\alpha$, but dependent on $t$. 
\new{If $L$ and $T$ are fixed, then the optimization with respect to $\alpha$ leads to $\alpha^{opt} \propto t^{1/3}$, 
which can be very small. If $T$ only  is fixed , the optimization with respect to both $\alpha$ and $L$ gives :  $\alpha^{opt}=0.5$ and $L^{opt} \sim \sqrt{t}/\ln(t)$.
Note that in the case  $T=1$, at the optimum the time has to be  equally shared between the two phases, which is 
reminiscent of the result obtained in the framework of the simple model of facilitated diffusion (section \ref{section_DNA_minimal_model}). }
Last, if $V=L/T$ is fixed, it is found that $L^{opt} \sim \sqrt{t}/\ln(t)$, $\alpha^{opt} \sim 1-4V\ln(t)/(3\sqrt{t})$, 
which means than the mean duration of the nearest-neighbors phase is 3/4 the duration of a large move. This model can actually be seen as the lattice version of the model given in \cite{LosingTime}, where similar results were obtained : (i) existence of a global minimum of the search time  ; (ii) the  ratio of times spent in both phases  at the minimum is given by a numerical constant ; (iii) this numerical constant is equal to $1/2$ in  \cite{LosingTime} (where the time spent off the lattice is exponentially distributed) instead of the $3/4$ given above (for deterministic times off lattice).  

\po{\textcite{RojoBuddeWio}}

\imagea{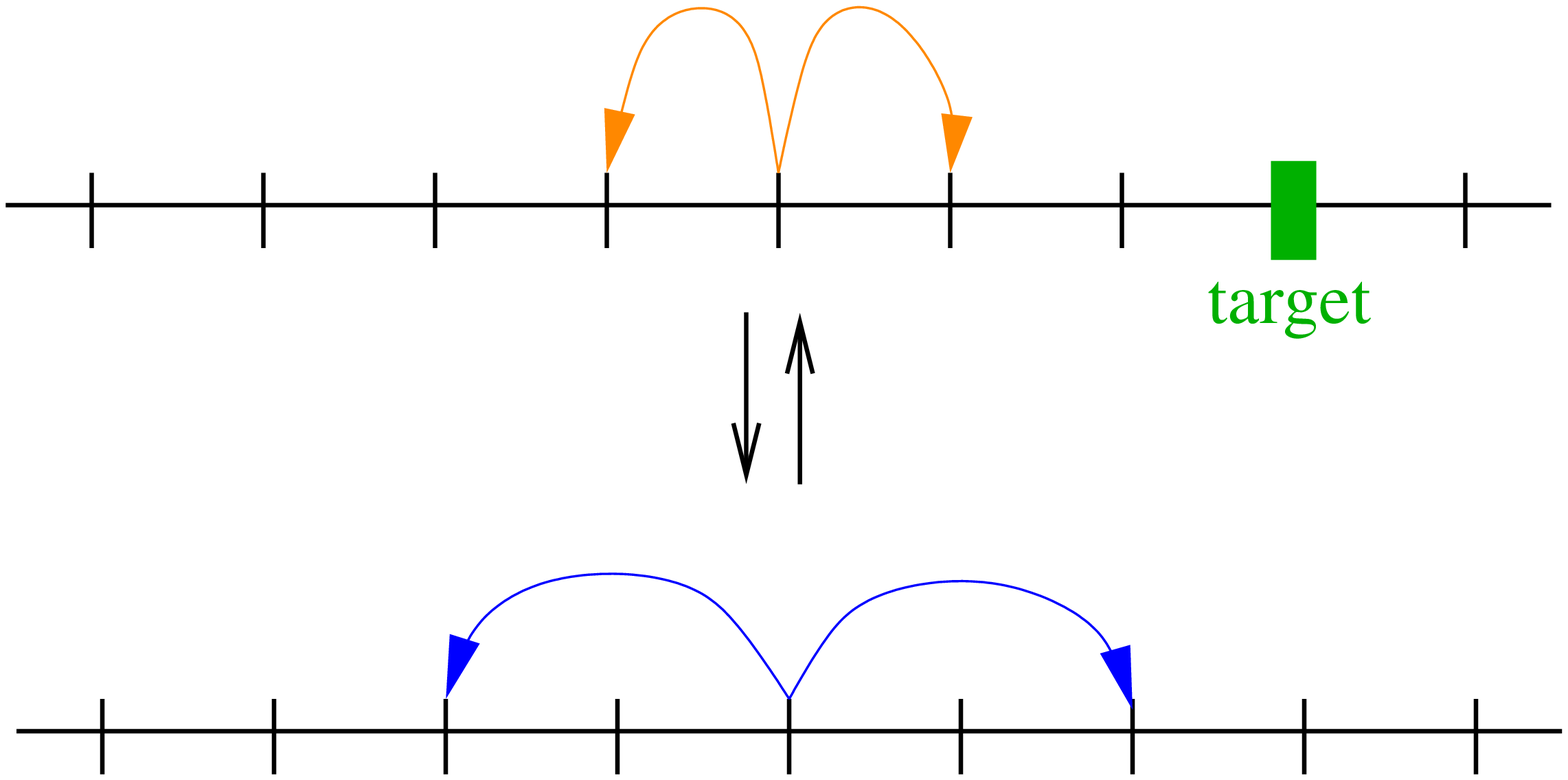}{Model used by \textcite{RojoBuddeWio}.}{rojo_schema}{4}

\textcite{RojoBuddeWio} propose a model which displays some similarities with the previous model (see figure \ref{rojo_schema}).  
The search domain  is also 
a one-dimensional discrete infinite lattice with one target, 
there are also an infinite set of searchers, 
and the quantity optimized is also the probability that the target is found by any of the searchers at a given $t$. 
The detection phase consists of jumps to the nearest-neighbors, 
with a given frequency. 
Such a rule is equivalent to diffusion. 
The non-reactive phase consists of jumps to the next nearest neighbors. 
It is again diffusion, but 
if the jump frequency is the same as in the other phase, 
it is a faster diffusion. 
In both phases, there is a fixed rate of switch to the other phase, 
leading to exponentially distributed  durations
of the phases. 
If one of the mean durations is fixed, 
the probability that the target is already found at $t$ 
is minimized for a finite duration of the other phase.  
But the optimum is for infinitely short phases, 
enabling the searcher to combine the faster diffusion of one phase 
and the detection capacities of the other phase. 

\po{\textcite{Holcman_diff_diff}}

\begin{figuresmall}[h!]\small
\begin{center}
   \begin{minipage}[c]{.68\linewidth}
\begin{center}

1D  

~

\includegraphics[width= 7.5 cm]{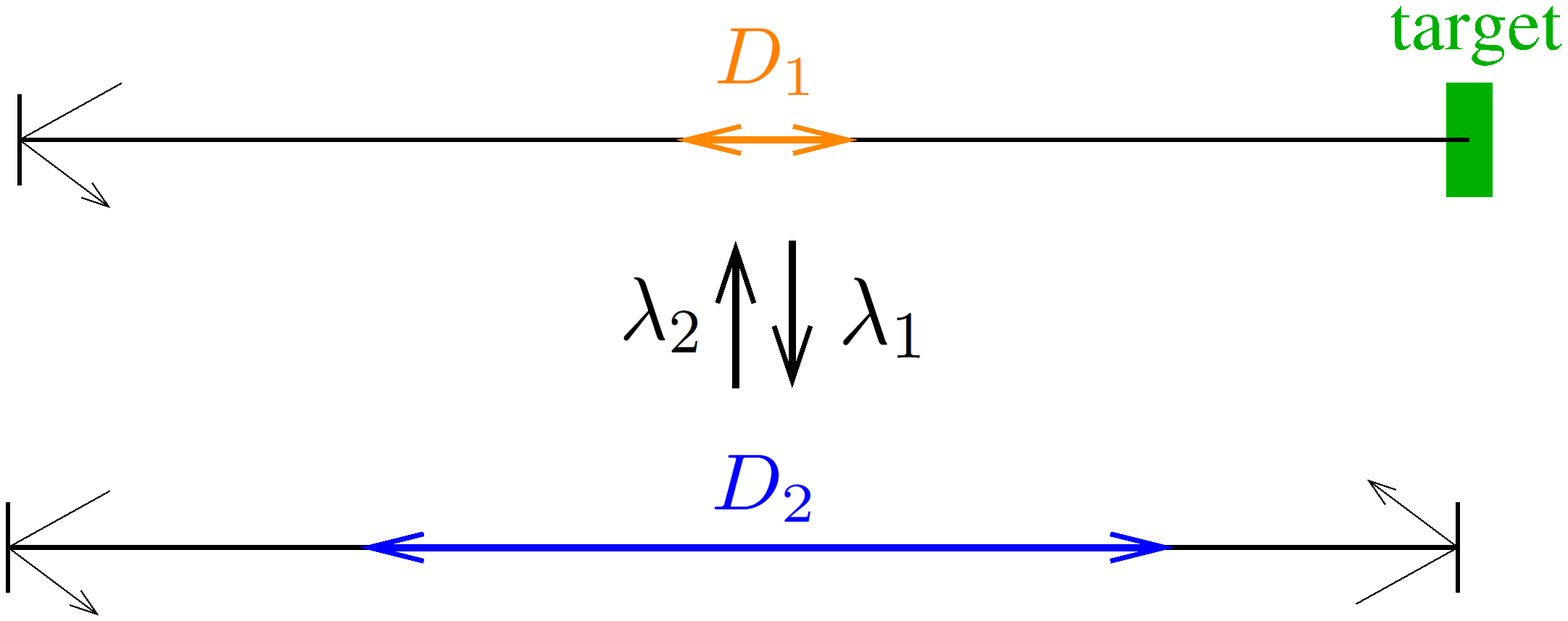}

\end{center}
   \end{minipage}\hfill
   \begin{minipage}[c]{.3\linewidth}

3D  

\begin{center}

\includegraphics[width=5cm]{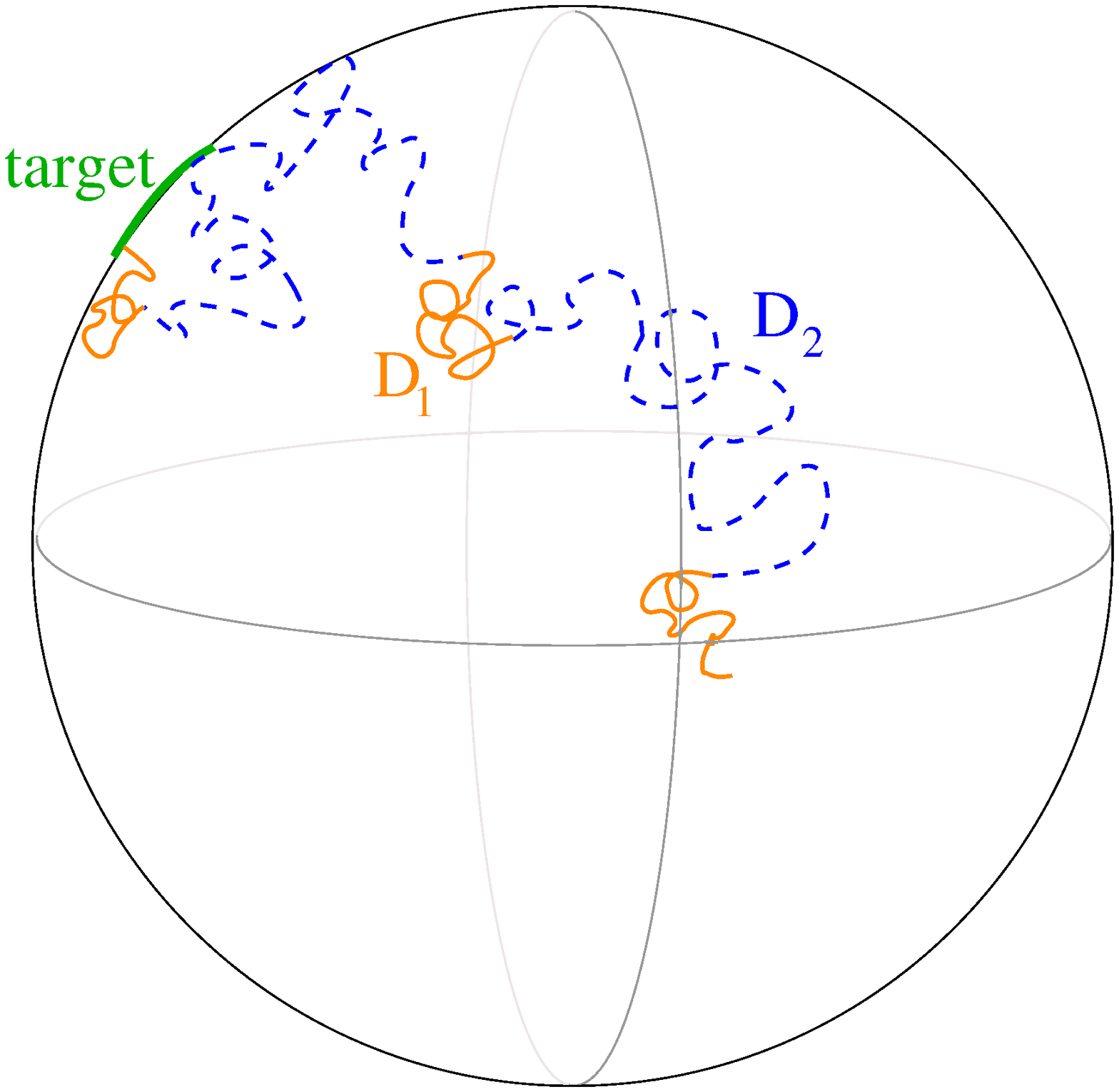}

\end{center}
   \end{minipage}
\caption{Model used by \textcite{Holcman_diff_diff}.}
\label{holcman_schema}
\end{center}\normalsize
\end{figuresmall}

\textcite{Holcman_diff_diff} propose a model which is also diffusive/diffusive (see figure \ref{holcman_schema}). 
They study this model first in one dimension~: 
the searcher's starting point is at one extremity of a segment, 
a reflecting boundary. 
The target is at the other end of the segment. 
However, in phase 1 (diffusion of coefficient $D_1$), 
the target can be  found, 
whereas in phase 2 (diffusion of coefficient $D_2$), 
both extremities are  reflecting. 
There are fixed rates of switching from one phase to another. 
The results show that there are two regimes~: 
if $D_1>D_2$, straightforwardly, the optimum for the searcher is to be in phase 1 only; 
if $D_2>D_1$, the optimum is to switch very rapidly between the two phases, 
such as to spend almost all the time in the faster phase 2, but not to miss the target. 
This model is extended to a 3-dimensional ball 
(the initial position is almost without importance in this geometry), 
but with a target of
radius $a$ on the border (which is reflecting everywhere else). 
The two phases are defined like in one dimension. 
\textcite{Holcman_diff_diff} give two limits in this case. 
It can be noticed that the expression we have obtained in the generic 
model for the diffusive mode in 3 dimensions 
(see \textcite{LeGros}) could be used, with $3V^2\tau_2^2 = D_{2}\tau_2$. 
In fact, our calculations use a  ``diffusive/diffusive'' approximation, with an effective $D_{2}^{\rm eff} = 3V^2 \tau_2$. 
The optimization will be quite different, because the dependence in $\tau_2$ is dramatically changed if instead of a fixed $D_2$,
$D_2$ is a function of $\tau_2$.  Indeed, the optimum for our generic model is for finite $\tau_1$ and $\tau_2$, 
whereas, even if not explicitly calculated, it is probable that the optimum for diffusion/diffusion in three dimensions 
is similar to the one-dimensional case, 
\textit{i.e.} for phases durations as small as possible. 
The goal of this model is to study  cellular signaling, 
with a ligand binding to a target which will transmit a signal. 

\po{ \textcite{Bressloff_1D,Bressloff_tree}}

\begin{figuresmall}[h!]\small
\begin{center}
   \begin{minipage}[c]{.73\linewidth}
\begin{center}

\includegraphics[width= 8.5 cm]{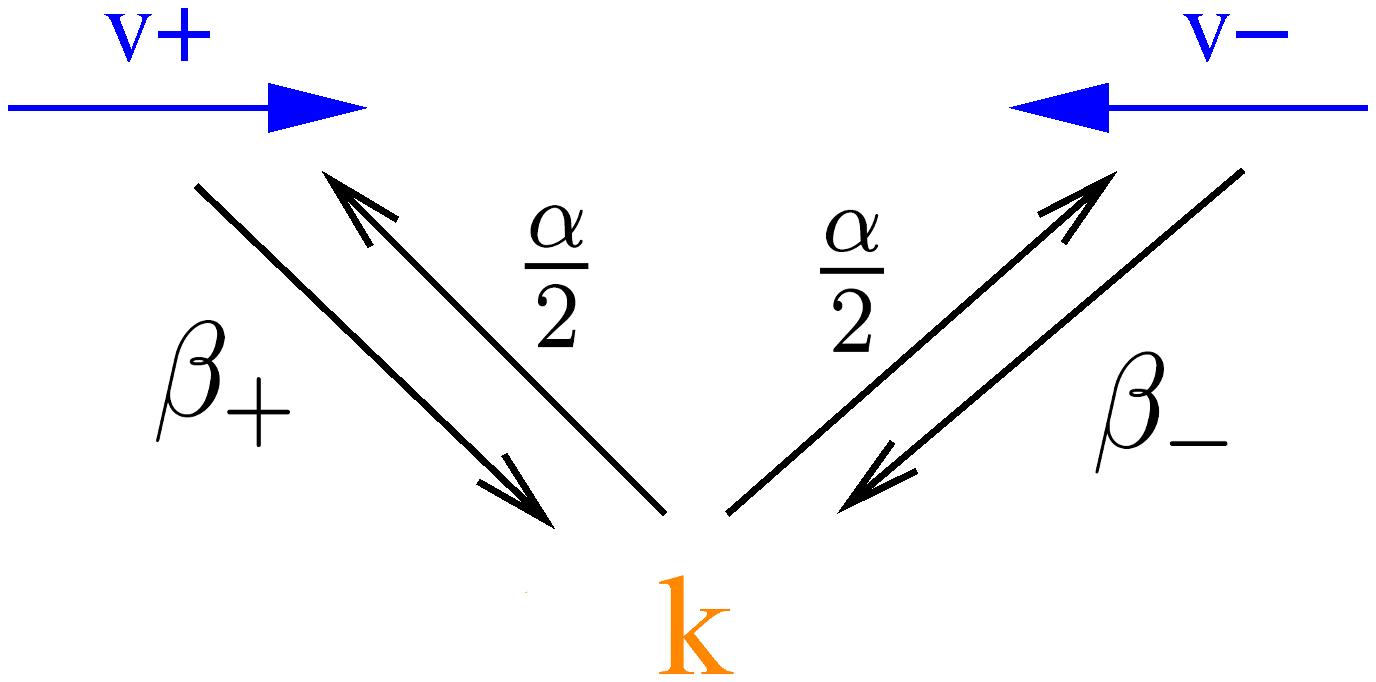}

\end{center}
   \end{minipage}\hfill
   \begin{minipage}[c]{.25\linewidth}

\begin{center}

\includegraphics[width=4cm]{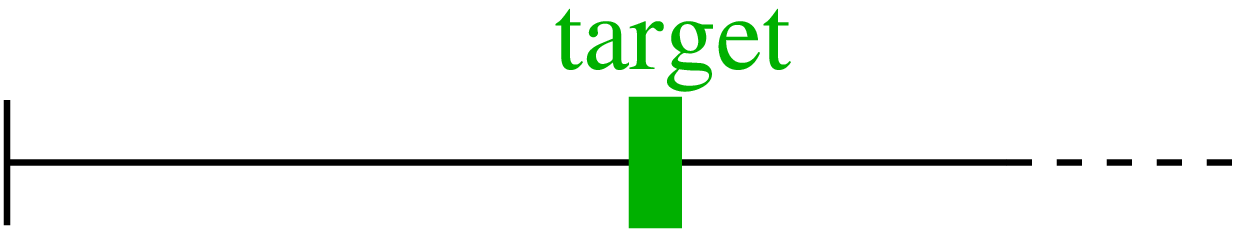}

\end{center}

\begin{center}
\includegraphics[width=4cm]{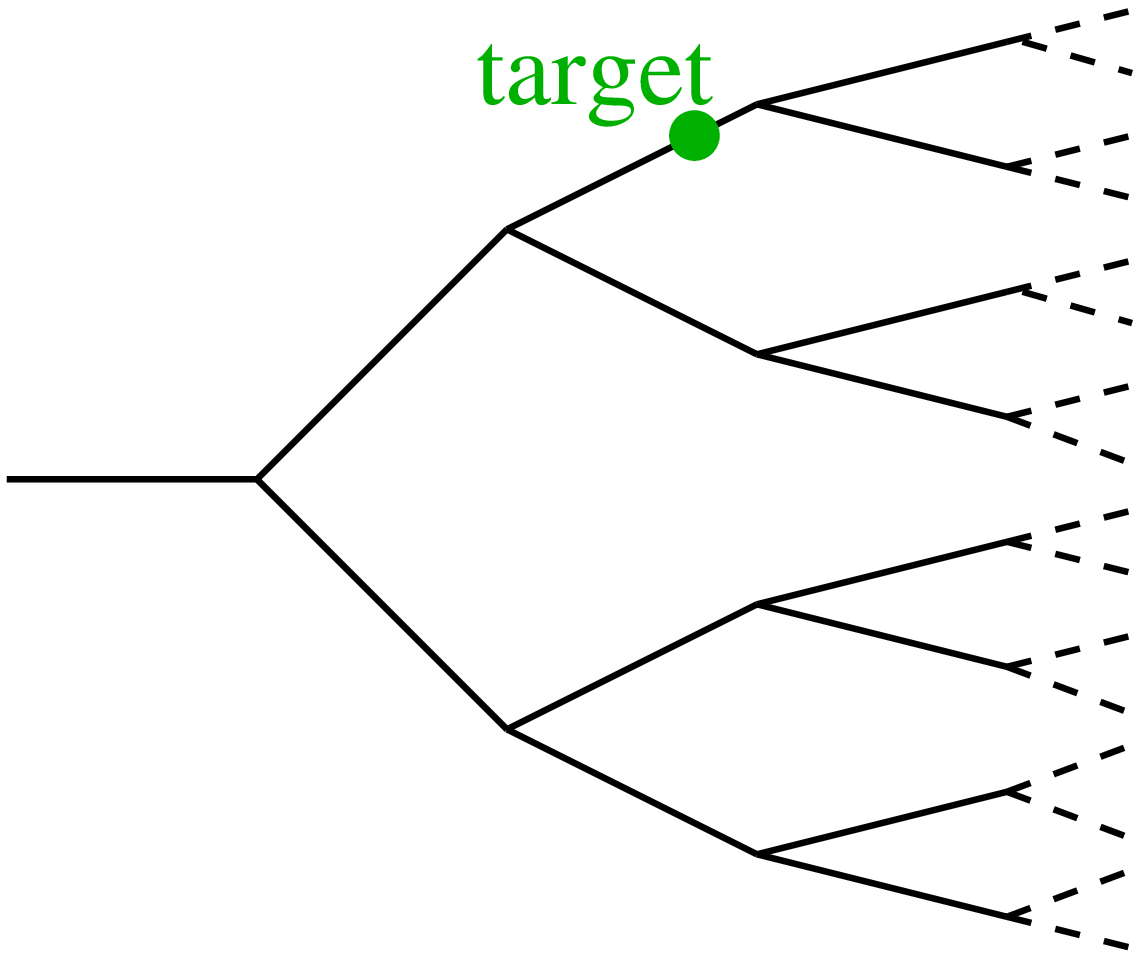}
\end{center}

   \end{minipage}
\caption{Model used by \textcite{Bressloff_1D,Bressloff_tree}.}
\label{bressloff_schema}\normalsize
\end{center}
\end{figuresmall}

\textcite{Bressloff_1D} present another model applied to intracellular transport, 
more precisely here to the transport  of mRNA granules inside neurons. 
They 
present a model in one dimension, standing 
for example for an axon with little branching. The starting point is at one extremity,  
which is reflecting~: it models granules produced in the soma of the neuron and that have to be exported to the axon. 
The target, a synapse, is somewhere in the segment. 
The other end of the segment is an absorbing boundary, representing 
that the vesicles containing the mRNA can be degraded, or that there can be other targets further away in the 
axon that can absorb the searcher. 
To complete this idea that there are several targets that are not equivalent, 
and that these targets are in competition, they also
calculate explicitly the probability that the searcher finds a target more often than the others. 
In this model, 
there are 3 states (see figure \ref{bressloff_schema})~: 
an immobile detection phase, similar to the static mode, 
switching to ballistic modes with probability $\alpha$ per unit time; 
a ballistic phase in direction $+$, with speed $v_+$, 
and with a transition rate to the detection mode $\beta_+$; 
a ballistic phase in direction $-$, with speed $v_-$, 
and with a transition rate to the detection mode $\beta_-$.
During the  two ballistic phases, the searcher cannot detect the target. 
Movement is biased to the direction $+$ if $v_+/\beta_+ >v_-/\beta_-$.

The results are based on the fact that on the segment, 
there are two contradicting constraints~: 
maximizing the hitting probability 
(as the searcher can be degraded before finding the target), 
and minimizing the time to find the target when the target is found. 
Indeed, if there is more bias, the target will be missed more often, but when found, the search time  will be smaller. 
With a fixed hitting probability, 
the mean first passage time to the target 
(on the condition that the target is found)  
is minimized when there is more bias. 
In other words, 
unidirectional motion 
is better than bi-directional motion in this case. 

\textcite{Bressloff_tree} 
extend this problem 
to the case of a directed tree. 
In this case, unidirectional motion has a drawback~: 
a wrong branch can be taken, 
annihilating any possibility to find the target. 
Biased bidirectional motion can be seen as an effective combination 
of a ballistic and a diffusive motion. 
It exists a critical hitting probability $p^*$. 
If the mean first passage time to the target 
is minimized given that the probability of finding the target is a given $p<p^*$,
unidirectional motion is better; 
but if the given probability is $p>p^*$, 
there is an optimal finite bias which minimizes the mean search time in case of success. 

\po{\textcite{intermittent_reseau}}

\imagea{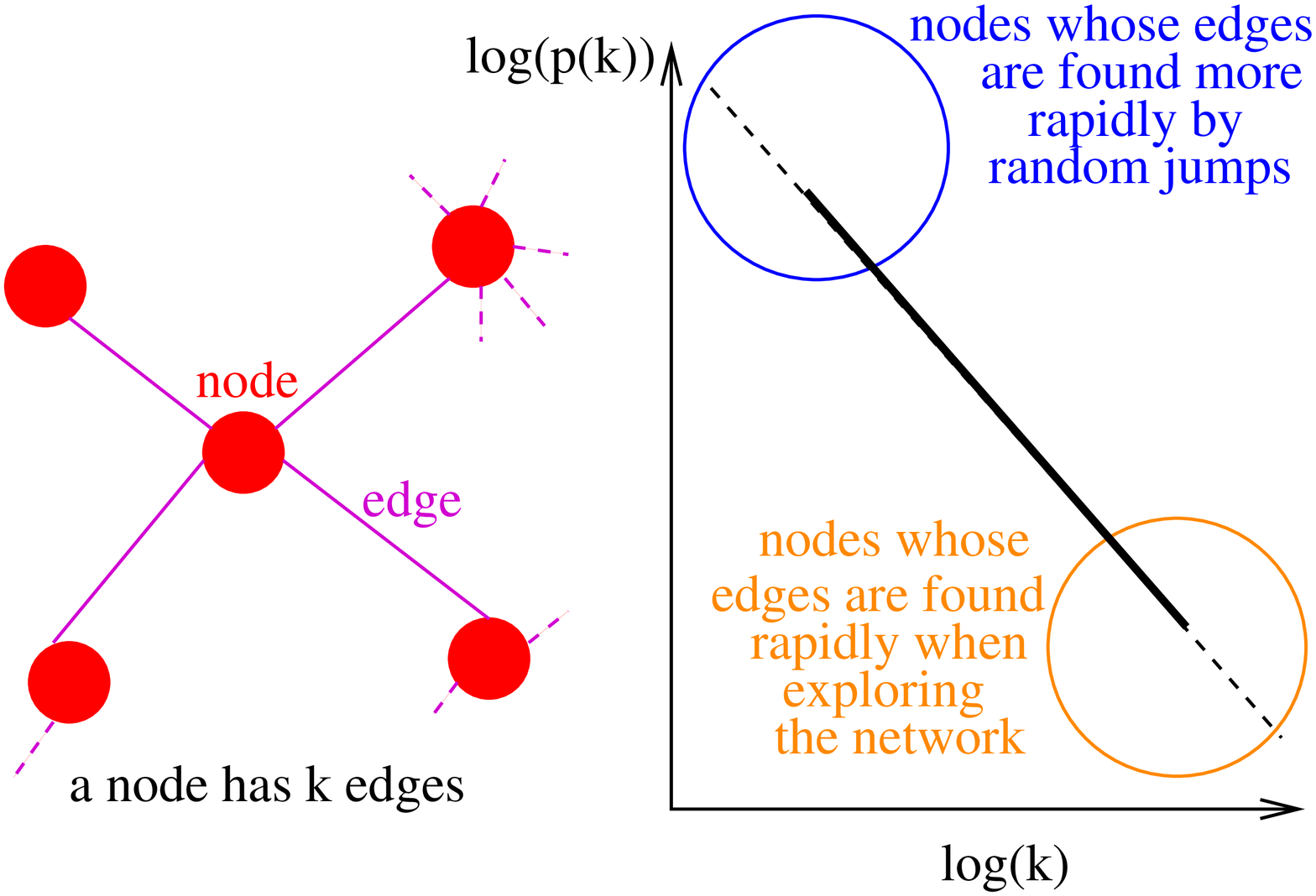}{Model presented by 
\textcite{intermittent_reseau}.}{schema_reseau_intermittent}{8}

Intermittence in networks such as the tree we have just seen  is an interesting extension. 
\textcite{intermittent_reseau} proposes (see figure \ref{schema_reseau_intermittent}) to explore a network  in which 
the degree (= number of neighbors) distribution is $p(k) \propto 
k^{-3}$, 
constructed as proposed by \textcite{barabasi_reseau}, or with some modifications. 
On this finite network, at each time step, the searcher 
chooses randomly one of the edges connected to the node where it is, 
and goes to the node connected by this edge. 
Every $t_w$, the searcher jumps to a completely random node. 
The question is whether the mean time to cover the nodes and the edges of the network can be optimized as a function of $t_w$.
For the nodes, the random jumping is a way to visit all the nodes with 
equal probability, thus $t_w$ should be as small as possible. 
For the edges, there is an optimal finite $t_w$. 
Indeed, if $t_w$ is small, most edges visited will emanate from low-connected nodes 
(as the low connected nodes are the more numerous nodes, such edges are more likely to be visited after a random jump), 
but if $t_w$ is large, the searcher would spend most of its time on the edges connecting 
high degree nodes, and  will take time to explore the whole network, 
especially for remote edges connecting nodes of low degree.

\subsection{Designing efficient searches}
\label{section_desing}

As seen previously, intermittent reaction paths are involved in various search problems 
involving biomolecules at the microscopic scale, 
as well as biological organisms at the macroscopic scale. 
Simple analytical models  show that intermittent transport can  actually  minimize the search time. 
A reason why such intermittent trajectories are widely observed 
could be simply that they constitute very  generic optimal search strategies, 
and consequently they could have been selected by evolution. 

Beyond   {\it modeling} what is observed in real-life biological examples, 
such intermittent strategies could also be used to 
{\it design} searches, at the microscopic and macroscopic scales. We briefly discuss here potential applications at the microscopic scale (for more details see \textcite{PCCP}).

\doublimagem{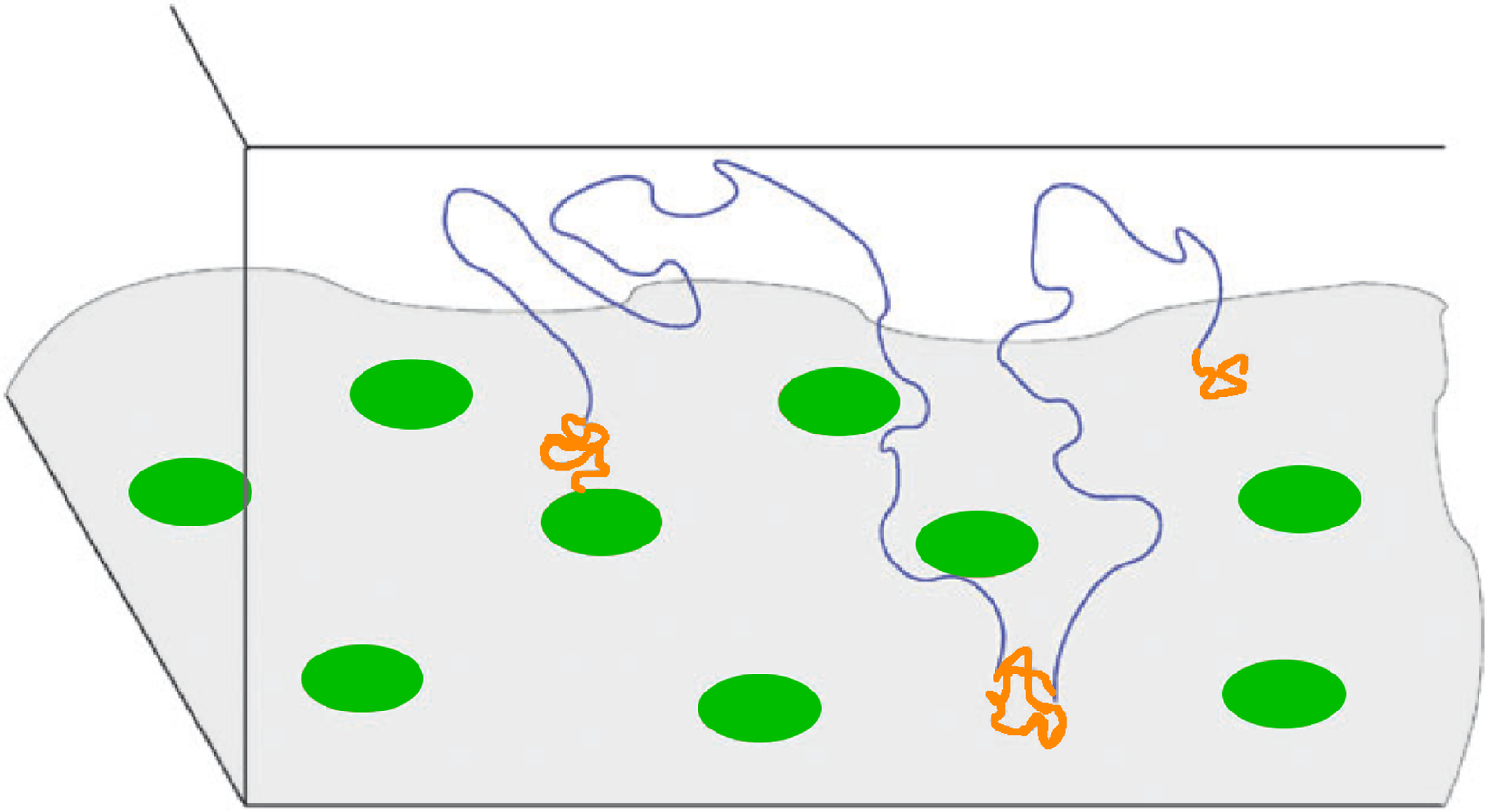}{Without flow}{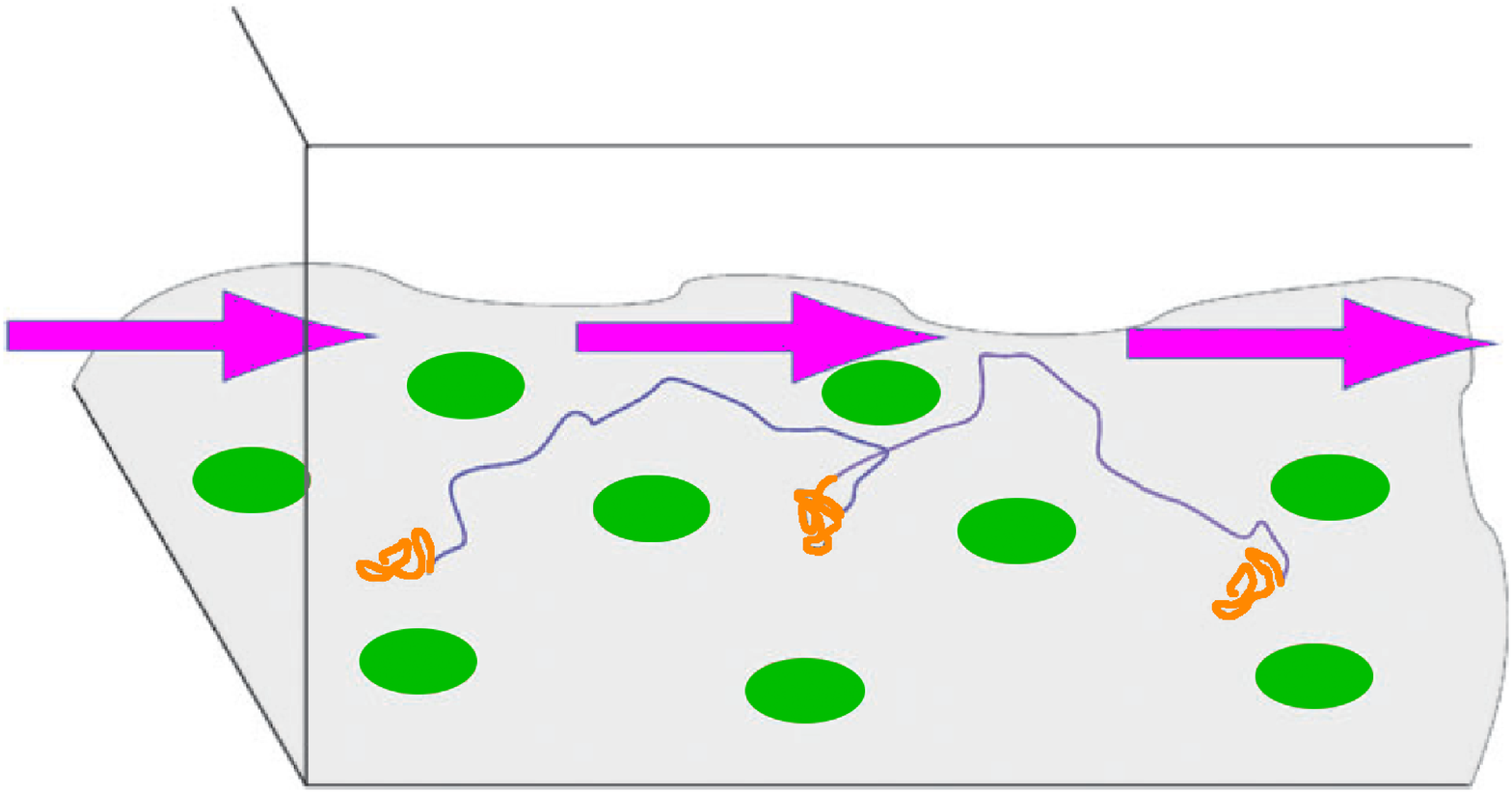}{With a flow}{Design of heterogeneous chemical reactions, 
 with targets (green disks) here fixed on a 2-dimensional surface. The reactant either diffuses in the volume (blue), 
or diffuses on the surface (orange thick line). The flow is represented by magenta arrows.}{desing_micro}

Heterogeneous chemical reactions, where the reactive  targets are located  at an interface, either 1-dimensional  (polymer) or 2-dimensional (surface) are intrinsically intermittent. 
Indeed, the reactants can either diffuse in the bulk volume, where target  cannot be found, 
or bind to the interface and diffuse more slowly (see figure \ref{desing_micro} left).

Beyond obvious optimizations 
(increasing the target and the reactant concentrations, 
increasing the diffusion coefficients of the reactant in the bulk or at the interface, etc.), 
 the mean durations of the  phases (free or bound to the interface)  are  the main adjustable parameters enabling to minimize the mean search time, and therefore to maximize the reaction rate. 
The main idea is that in the "teleportation" approximation (see section \ref{section_DNA_minimal_model}), after a bulk excursion, 
the distance between the reactant landing point on the surface 
and its starting point is larger than the typical distance between targets. 
In such a regime, each new bound phase is independent from the previous one, and the trajectory overlap is limited, which enhances the reactivity.

The mean durations of the phases can actually be tuned in real systems. 
First, the mean time a reactant remains bound to the interface depends on its affinity with the interface, which could be tuned. Second, 
the mean time spent in the bulk is mainly controlled by the confinement volume \citep{Blanco03,Olivier2005domaine,NatureSylvain}.
The confinement volume has therefore to be as small as possible, but it should   be large enough to make the  ``teleportation''  approximation
 valid. This constraint defines a critical volume, and can actually be by-passed by  applying a hydrodynamic flow parallel to the surface, 
which makes the teleportation approximation valid even for a very short bulk excursions, 
provided that the velocity of the flow is high enough (see figure \ref{desing_micro} right). 
In this  regime it can be shown \cite{PCCP} that the reaction rate can be optimized by tuning the affinity of the reactant for the interface in a similar way as in section \ref{section_DNA_minimal_model}.

 At the molecular level, we stress that intermittent transport could also be useful for in vitro chemistry. Indeed, we have shown that intermittent transport naturally pops up in the context of reaction at interfaces, where reactants combine surface diffusion phases and bulk excursions, and could permit to enhance reactivity. In this case, adjusting chemically the typical association time of the reactants with the interface, makes possible to optimize the reaction rate.

\finsection

\section{Conclusion}

Intermittent search strategies rely on a simple mechanism~: 
the searcher alternates between  two phases, 
one during which the target can be detected, but with slow motion, and 
another of faster motion but without target detection.

This mechanism of intermittence has emerged from the observation of real-life biological searches at various scales. 
At the macroscopic scale, an example is given by animals searching for hidden food, 
which alternate between fast ballistic relocation phases with no target detection, and phases of slower motion  aimed at detecting the target.  
A simple model based on this observation 
permits to show analytically that the mean search time can be minimized 
as a function of the phases mean duration. 
There is one single way to share time between the two phases to find the target as fast as possible. 
This intermittent search is then an optimal search strategy. With this respect, this model 
is an alternative to the famous L\'evy walks model which is optimal only in restrictive conditions.

Intermittence is also observed at the microscopic scale.  
Indeed, for some biochemical reactions in cells, which involve a very low concentration of reactants, reaction pathways are not always simple Brownian trajectories. They can rather be qualified as intermittent, since they combine slow diffusion phases on the one hand, and a second mode of faster transport on the other hand, which can be either a faster diffusion mode as in the case of DNA-binding proteins, or a ballistic mode powered by molecular motors in the case of intracellular transport. Analytical models actually  show that such intermittent trajectories are very efficient, since they  significantly reduce reaction times. Interestingly, it is shown that reaction rates can even be maximized by adjusting simple biochemical parameters. The gain is small in dimension three, but for lower dimensional structures such as  membranes (2D) or polymers or  tubular structures (1D), the gain can be very large at low  target concentration. Such efficiency -- and optimality -- could explain why intermittent transport is observed  in various forms in the context of reactions in cells.

Since these intermittent search strategies are observed at various scales, 
one could suggest that they constitute a generically efficient search mechanism. 
The systematic analysis of a generic model in the framework of intermittent random walks,  
in 1, 2 and 3 dimensions,  and for three different descriptions of the slow reactive phase,
permits  to assess quantitatively the robustness of this mechanism. 
In fact, this study shows that the optimality of these search strategies is a widely robust result. 
Finally, if intermittent random walks are observed in real biological systems at various scales, 
it is probably because they do constitute an efficient search strategy. 
Beyond these modeling aspects, one can  suggest that such intermittent strategies could also be used to design  optimizable search strategies.

\finsection

\begin{acknowledgements} 
Support of ANR Grant DYOPTRI is acknowledged.
\end{acknowledgements}

\appendix\section{Recap on random walks and L\'evy processes}

\label{resume_random_walks}

Regular random walks obey Gaussian statistics, and have a mean square displacement growing linearly with time : $\langle r^2(t) \rangle \sim t^\alpha$, with $\alpha=1$. Inversely,  transport processes characterized by non-linear scalings  with time of the mean square displacement are  termed "anomalous", either subdiffusive if $\alpha<1$ or superdiffusive if $\alpha>1$. In this review, we shall make use of  standard models of subdiffusion (Continuous Time Random Walks, Diffusion on Fractals and Fractional Brownian Motion) and superdiffusion (L\'evy flights and L\'evy walks), whose definitions are briefly reminded here for consistence (see for instance  \citet{perco} for a  more complete discussion).


\subsection{{Subdiffusion}}

\subsubsection{Continuous time random walks} 

A first class of models leading to subdiffusion  stems from continuous time random walks (CTRWs)  and their continuous space limit described by fractional diffusion equations. The anomalous behavior in these models originates from a heavy-tailed distribution of waiting times : at each step the walker lands on a trap, where it can be trapped for extended periods of time. 
Technically, the CTRW is a standard random walk with random waiting times, drawn from a probability density function  $\psi(t)$. The CTRW model has a normal diffusive behavior if the mean waiting time is finite. For heavy-tailed distributions such that
\begin{equation}
\psi(t)\propto \frac{1}{t^{1+\beta}},\;\;{\rm at \;large\; times,}
\end{equation}
the mean waiting time diverges for $\beta<1$ and the walk is subdiffusive with $\alpha=\beta$.

When dealing with a tracer particle, traps can be out-of-equilibrium chemical binding configurations, and the waiting times are then the dissociation times; traps can also be realized by the free cages around the tracer in a dense hard sphere-like crowded environment, and the waiting times are the life times of the cages \cite{pnasSylvain,saxton96,saxton07}.

\subsubsection{Diffusion on fractals}

Another kind of model for subdiffusion relies on spatial inhomogeneities as exemplified by diffusion in deterministic (such as Sierpinski gasket)  or random fractals (such as critical percolation clusters). The subdiffusive behavior is in this case caused by the presence of fixed obstacles  that create numerous dead ends, as illustrated by De Gennes's ant in a labyrinth \cite{saxton94,degennes}. This results in an effective subdiffusion in the embedding space, with an exponent $\alpha<1$, whose value depends on the fractal structure \cite{perco,BundeHavlin,dauriac}. 

\subsubsection{Fractional Brownian motion}

Fractional Brownian motion (FBM) is a third model of subdiffusion \cite{mandelbrot}, usually defined for systems in dimension 1. It  was introduced to take into account correlations in a random walk: the state of the system at time $t$ is influenced by the state at time $t' < t$. More precisely, it is a Gaussian process of autocorrelation function of the form
\begin{equation}
\langle X(t_1) X(t_2) \rangle \propto t_1^{2H} + t_2^{2H} -|t_1-t_2|^{2H},
\end{equation}
with $0<H<1/2$, so that $\alpha=2H<1$ (FBM can also be defined for $1/2<H<1$, but in this case it leads to superdiffusion). Note that Brownian diffusion is recovered for $H=1/2$.
 FBM is used to describe the motion of a monomer in a polymer chain or single file diffusion. Recently, it has also been proposed to underline the diffusion in a crowded environment \cite{SzymanskiWeiss}.

\subsection{Superdiffusion}

\subsubsection{L\'evy flights}

L\'evy flights are random walks such that, at each step $t$, the walker jumps in some random uniformly distributed direction, to a distance $r$ drawn from a probability density function 
\begin{equation}
p(r)\propto \frac{1}{r^{1+\beta}}. 
\end{equation}
 It can be shown that, if $\beta<2$,  superdiffusion emerges,  with $\alpha=2/\beta$, while, if $\beta>2$, regular diffusion is recovered.
 
 \subsubsection{L\'evy walks}

L\'evy walks differ from L\'evy flights in that, now, the time to make a step of size $r$ is taken proportional to $r$ \cite{ShlesingerKlafterLevyFlightWalk}.  Physically, it can be seen as a random walker performing  jumps still drawn from the probability density function 
\begin{equation}
p(r)\propto \frac{1}{r^{1+\beta}},
\end{equation}
but this time at a constant velocity. The resulting mean square displacement can be shown to be given by
\begin{equation}
\langle r^2\rangle \propto \begin{cases} t^2  & \text{if $0<\beta<1$,}
\\
t^2/\ln t &\text{if $\beta=1$,}\\
 t^{3-\beta}  & \text{if $1<\beta<2$,}
\\
 t\ln t  & \text{if $\beta=2$,}
\\
 t  & \text{if $\beta>2$.}
\end{cases}
\end{equation}

Applications of L\'evy walks are given in the main text, in the context of random search problems. Note that, in this context, the terms "L\'evy walks" and ''L\'evy flight" are often used indifferently to design   ''L\'evy walks".

\appendix\section{Mean first-passage times of intermittent random walks}
\label{appendix}

\subsection{Dimension one}

\subsubsection{Static mode}
\label{section_generic_statique1D}


In this section we assume that the detection phase is modeled by  the static mode. Hence the searcher does not move during the reactive phase 1, 
and has a fixed reaction  rate $k$ per unit time  with the target if it lies within its detection radius  $a$ \refi{modes_generic}. 
It is the limit of a very slow searcher in  the reactive phase.

\po{Equations}
Outside the target (for $x>a$), we have the following backward equations for the mean first-passage time~: 
\begin{equation}
 V \frac{d t_2^+}{dx}+\frac{1}{\tau_2}(t_1-t_2^+)=-1,
\end{equation}
\begin{equation}
 - V \frac{d t_2^-}{dx}+\frac{1}{\tau_2}(t_1-t_2^-)=-1, \mathrm{~and}
\end{equation}
\begin{equation}
 \frac{1}{\tau_1}\left( \frac{t_2^++t_2^-}{2} -t_1 \right)=-1.
\end{equation}
Inside the target ($x\le a$), the first two equations are identical, but the third one is written~: 
\begin{equation}
  \frac{1}{\tau_1} \frac{t_2^++t_2^-}{2} -\left(\frac{1}{\tau_1}+k  \right)t_1=-1.
\end{equation}
We introduce $t_2=\frac{t_2^++t_2^-}{2}$ and $t_2^d=\frac{t_2^+-t_2^-}{2}$. 
Then outside the target we have the following equations~: 
\begin{equation}
  V \frac{d t_2}{dx}-\frac{1}{\tau_2}t_2^d=0,
\end{equation}
\begin{equation}
  V^2 \tau_2 \frac{d^2 t_2}{dx^2}+\frac{1}{\tau_2}(t_1-t_2)=0,
\end{equation}
\begin{equation}
 \frac{1}{\tau_1}(t_2-t_1)=-1.
\end{equation}
Inside the target the first two  equations are identical, but the last one writes~: 
\begin{equation}
  \frac{1}{\tau_1}t_2-\left(\frac{1}{\tau_1}+k  \right)t_1=-1.
\end{equation}
Due to the symmetry $x \leftrightarrow -x$, we can restrict the study to the part $x \in [0,a]$ and the part $x\in [a,b]$. 
This symmetry also implies ~: 
\begin{equation}
\left. \frac{dt_2^{in}}{dx}\right|_{x=0}=0,
\end{equation}
\begin{equation}
\left. \frac{dt_2^{out}}{dx}\right|_{x=b}=0.
\end{equation}
In addition,  continuity at $x=a$ for $t_2^+$ and $t_2^-$ gives: 
\begin{equation}
t_2^{in}(x=a)=t_2^{out}(x=a),
\end{equation}
\begin{equation}
t_2^{d,in}(x=a)=t_2^{d,out}(x=a).
\end{equation}
This set of linear equations enables us to explicitly determine  $t_1$, $t_2$, $t_2^d$  inside and outside the target.

\begin{figuresmall}[h!]\small
   \begin{minipage}[c]{.46\linewidth}
\begin{center}
      \includegraphics[width=6cm]{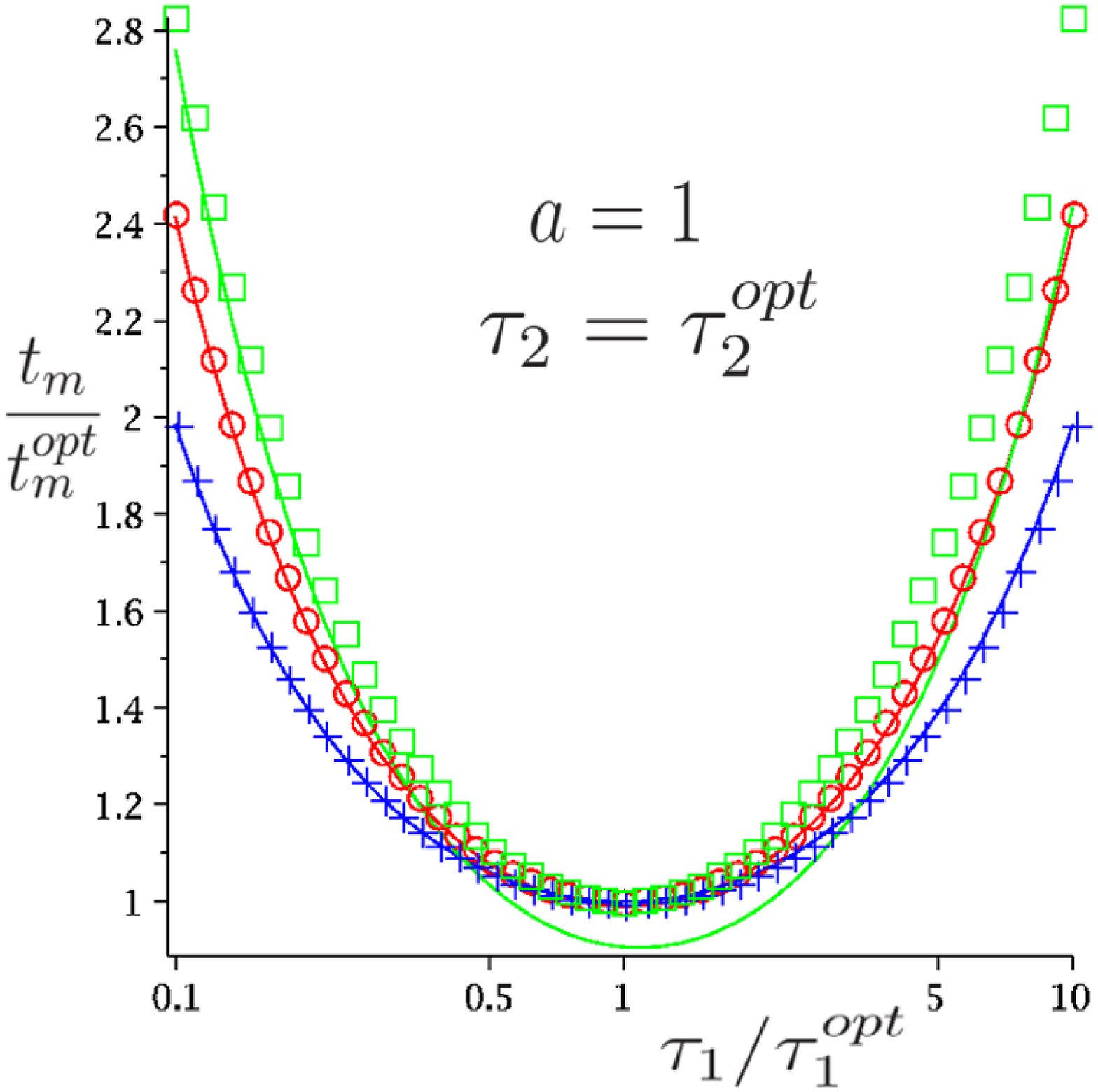}

\end{center}
   \end{minipage} \hfill
   \begin{minipage}[c]{.46\linewidth}
\begin{center}
      \includegraphics[width=6cm]{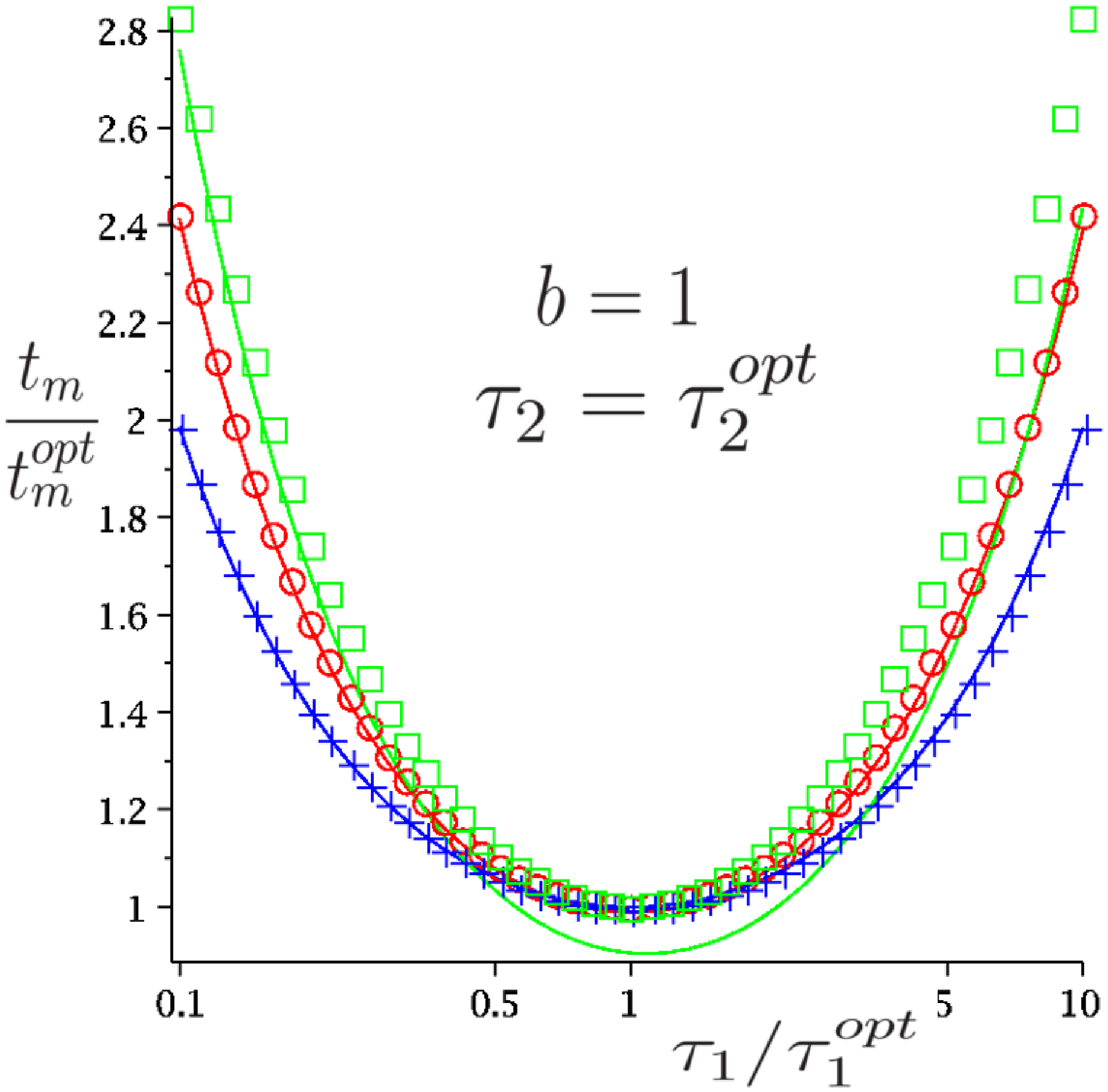}

\end{center}
   \end{minipage}\hfill

   \begin{minipage}[c]{.46\linewidth}
\begin{center}
      \includegraphics[width=6cm]{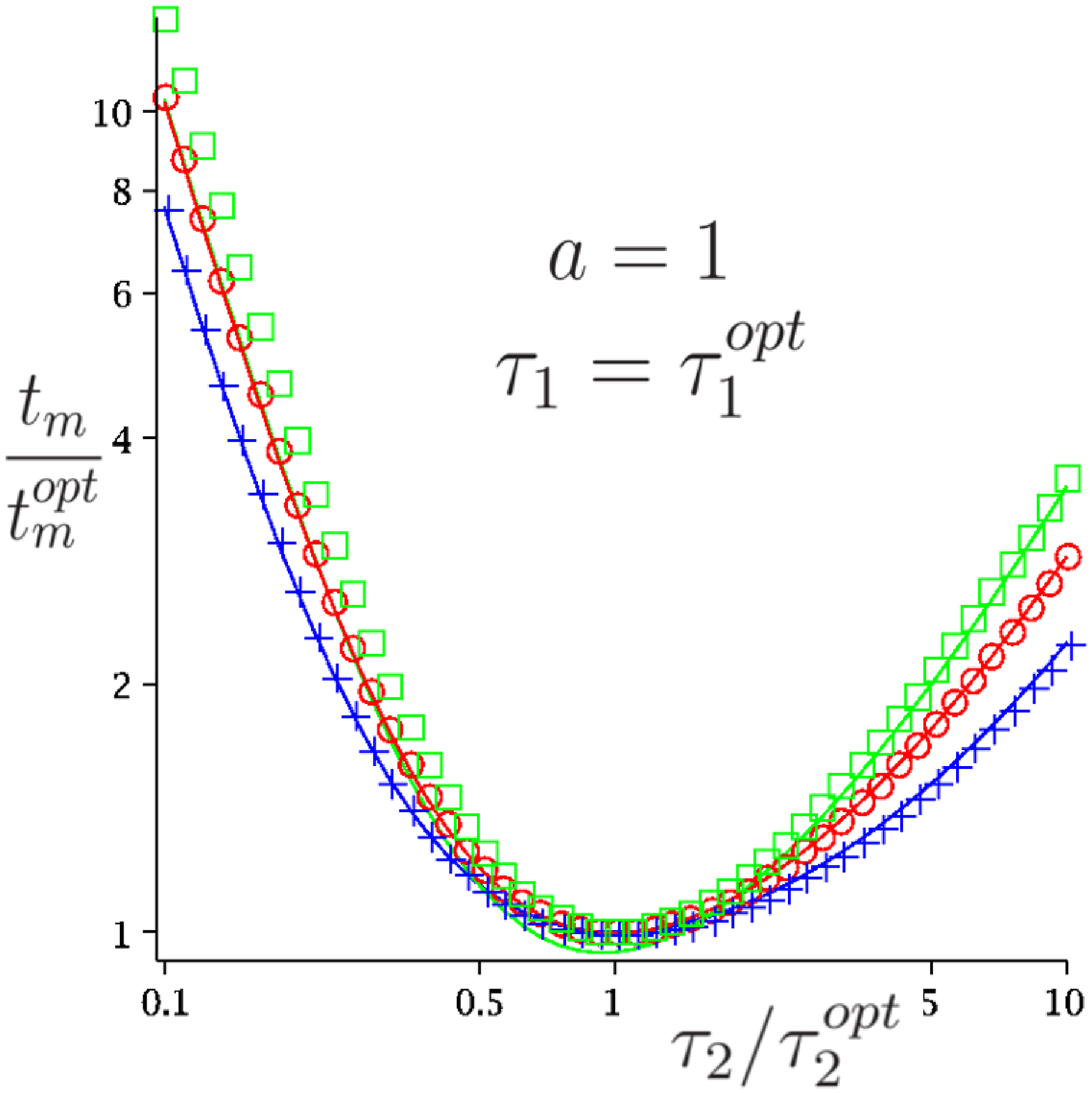}

\end{center}
   \end{minipage} \hfill
   \begin{minipage}[c]{.46\linewidth}
\begin{center}
      \includegraphics[width=6cm]{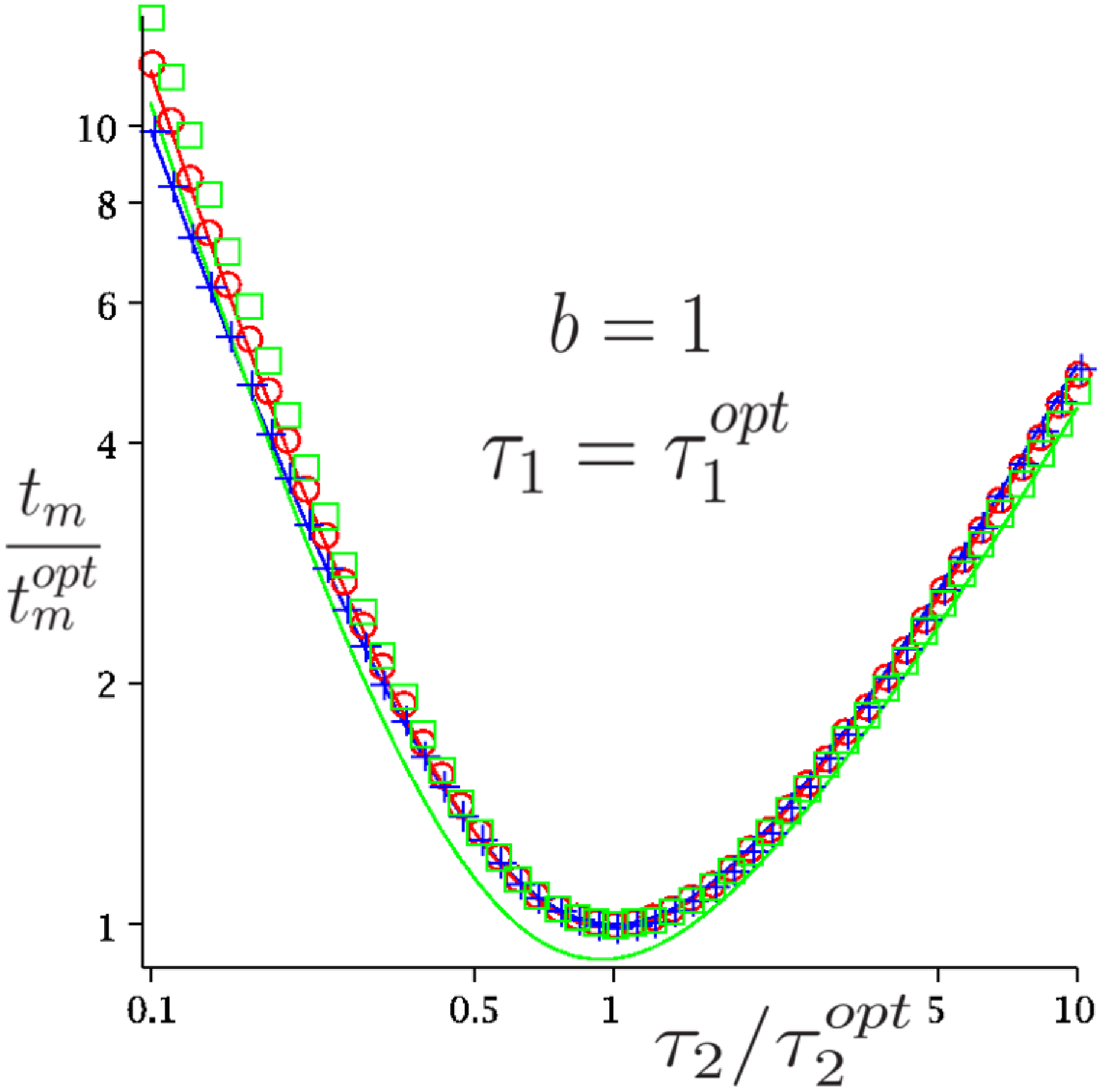}

\end{center}
   \end{minipage}\hfill

\caption{Static mode in one dimension. Exact expression of $t_m$ \refm{tm1Dvk} (lines) compared to the approximation of $t_m$  \refm{tm1Dvksimp} (symbols), 
both rescaled by $t_m^{opt}$. $\tau_1^{opt}$ from \refm{1Dvk_t1},  $\tau_2^{opt}$ from \refm{1Dvk_t2}. $V=1$, $k=1$. $b/a=10$ (green, $\square$), $b/a=100$ (red, $\circ$), $b/a=1000$ (blue, $+$).}\label{1Dvk_approx}\normalsize
\end{figuresmall}

\po{Results}
An exact analytical expression of the mean first passage time to the target is then given by ~: 
\begin{equation}\label{tm1Dvk}
 t_m = \frac{\tau_1+\tau_2}{b}\left( \frac{b}{k\tau_1} + \frac{(b-a)^3}{3V^2 \tau_2^2}+\frac{\beta (b-a)^2}{V\tau_2}coth\left(\frac{a}{V\tau_2 \beta} \right) + (b-a)\beta^2\right),
\end{equation}
where $\beta=\sqrt{(k\tau_1)^{-1} +1}$.

We obtain in the limit  $b\gg a$ ~: 
\begin{equation}\label{tm1Dvksimp}
 t_m = (\tau_1+\tau_2)\left( \frac{b^2}{3V^2 \tau_2^2}+\left(\frac{1}{k\tau_1}+1\right) \frac{ b}{a}\right).
\end{equation}

We use this approximation \refm{tm1Dvksimp} to 
find $\tau_1$ and $\tau_2$ values minimizing $t_m$~:
\begin{equation}\label{1Dvk_t1}
 \tau_1^{opt}=\sqrt{\frac{a}{Vk}}\left(\frac{b}{12a} \right)^{1/4},
\end{equation}
\begin{equation}\label{1Dvk_t2}
 \tau_2^{opt}=\frac{a}{V}\sqrt{\frac{b}{3a}}.
\end{equation}
 Importantly,  the optimal duration of the relocation phase 
does not depend on $k$, \textit{i.e.} on the description of the detection phase.

\subsubsection{Diffusive mode}
\label{section_generic_1DvD}

We now turn to the diffusive modeling of the detection phase. 
The detection phase 1 is now diffusive, with immediate detection
of the target if it is within a radius $a$ from the searcher \refi{modes_generic}. 

\po{Equations}

Along the same lines, the backward equations for the mean first-passage time read outside the target ($x>a$) ~:
\begin{equation}
 V \frac{d t_{2}^+}{dx} + \frac{1}{\tau_2} (t_1-t_2^+)=-1,
\end{equation}
\begin{equation}
-V \frac{d t_{2}^-}{dx} + \frac{1}{\tau_2} (t_1-t_2^-)=-1, 
\end{equation}
\begin{equation}
 D \frac{d^2 t_{1}}{dx^2} + \frac{1}{\tau_1} \left(\frac{t_2^+}{2}+\frac{t_2^-}{2}-t_1 \right)=-1,
\end{equation}
and inside the target ($x\le a$)~: 
\begin{equation}
  V \frac{d t_{2}^+}{dx} - \frac{1}{\tau_2} t_2^+=-1,
\end{equation}
\begin{equation}
-V \frac{d t_{2}^-}{dx} - \frac{1}{\tau_2} t_2^-=-1, \mathrm{~and}
\end{equation}
\begin{equation}
 t_{1}=0.
\end{equation}
 Boundary conditions result as previously from   continuity and symmetry.

\po{Results}\label{r1dvd1}

Standard but lengthy calculations lead to an exact expression of the mean first detection time of the target $t_m$ given in \cite{LeGros}.
 Three  regimes can be identified. 
 \begin{itemize}
 \item In the first regime ($b<\frac{D}{V}$) intermittence is not favorable. 
 \item For $b >\frac{D}{V}$ and  $ \frac{bD^2}{a^3 V^2}<1$  intermittence is favorable. In the  low target density ($b \gg a$) limit, 
 the following  approximation of the mean first passage time around its minimum can be obtained~: 
\begin{equation}\label{tmn}
 t_m=  \left(\tau_1+\tau_2\right)b\left(\frac{b}{3V^2\tau_2^2}+\frac{1}{\sqrt{D\tau_1}} \right).
\end{equation}
 This expression gives a good approximation of $t_m$ 
in this regime, in particular around the optimum (\refi{tcasn}).
The simplified $t_m$ expression \refm{tmn} is minimized for~:
\begin{equation}\label{l11dvd1}
 \tau_1^{opt}=\frac{1}{2} \sqrt [3]{\frac{2b^2D}{9V^4}},
\end{equation}
\begin{equation}\label{l21dvd1}
 \tau_2^{opt}=\sqrt [3]{\frac{2b^2D}{9V^4}},
\end{equation}
\begin{equation}\label{toptn}
 t_m^{opt} \simeq  \sqrt[3]{\frac{3^5}{2^4} \frac{b^4}{DV^2}}.
\end{equation}
This compares to the case without  intermittence according to~:
\begin{equation}\label{gainn}
 gain^{opt}=\frac{t_{\rm diff}}{t_m^{opt}}\simeq \sqrt[3]{\frac{2^4}{3^8}} \left(\frac{bV}{D} \right)^{\frac{2}{3} } \simeq 0.13 \left(\frac{bV}{D} \right)^{\frac{2}{3}} .
\end{equation}
\item For $b >\frac{D}{V}$ and $1 \gg \frac{bD^2}{a^3 V^2}$  intermittence is favorable.

In this regime, the mean search time can be approximated by:
\begin{equation}\label{tsimp1DvD}
 t_m \simeq \frac{b}{a}(\tau_1+\tau_2)\left(\frac{a}{a+\sqrt{D\tau_1}}+\frac{ab}{3V^2\tau_2^2} \right).
\end{equation}
This expression gives a good approximation of $t_m$, at least around the optimum \refi{tsimpfig}, which is characterized by:
\begin{equation}\label{l11dvd2}
 \tau_1^{opt}=\frac{D b}{48V^2 a},
\end{equation}
\begin{equation}\label{l21dvd2}
 \tau_2^{opt} = \frac{a}{V} \sqrt{\frac{b}{3a}},
\end{equation}
\begin{equation}
  t_m^{opt}\simeq \frac{2a}{V\sqrt{3}}\left(\frac{b}{a} \right)^{3/2},
\end{equation}
\begin{equation}\label{gain1dvd2}
 gain\simeq \frac{1}{2\sqrt{3}}\frac{aV}{D}\sqrt{\frac{b}{a}}.
\end{equation}
Note that the gain can be very large at low target density, and that the value obtained for $\tau_2^{opt}$ 
is the same as in the static mode.  
\end{itemize}

\doublimagem{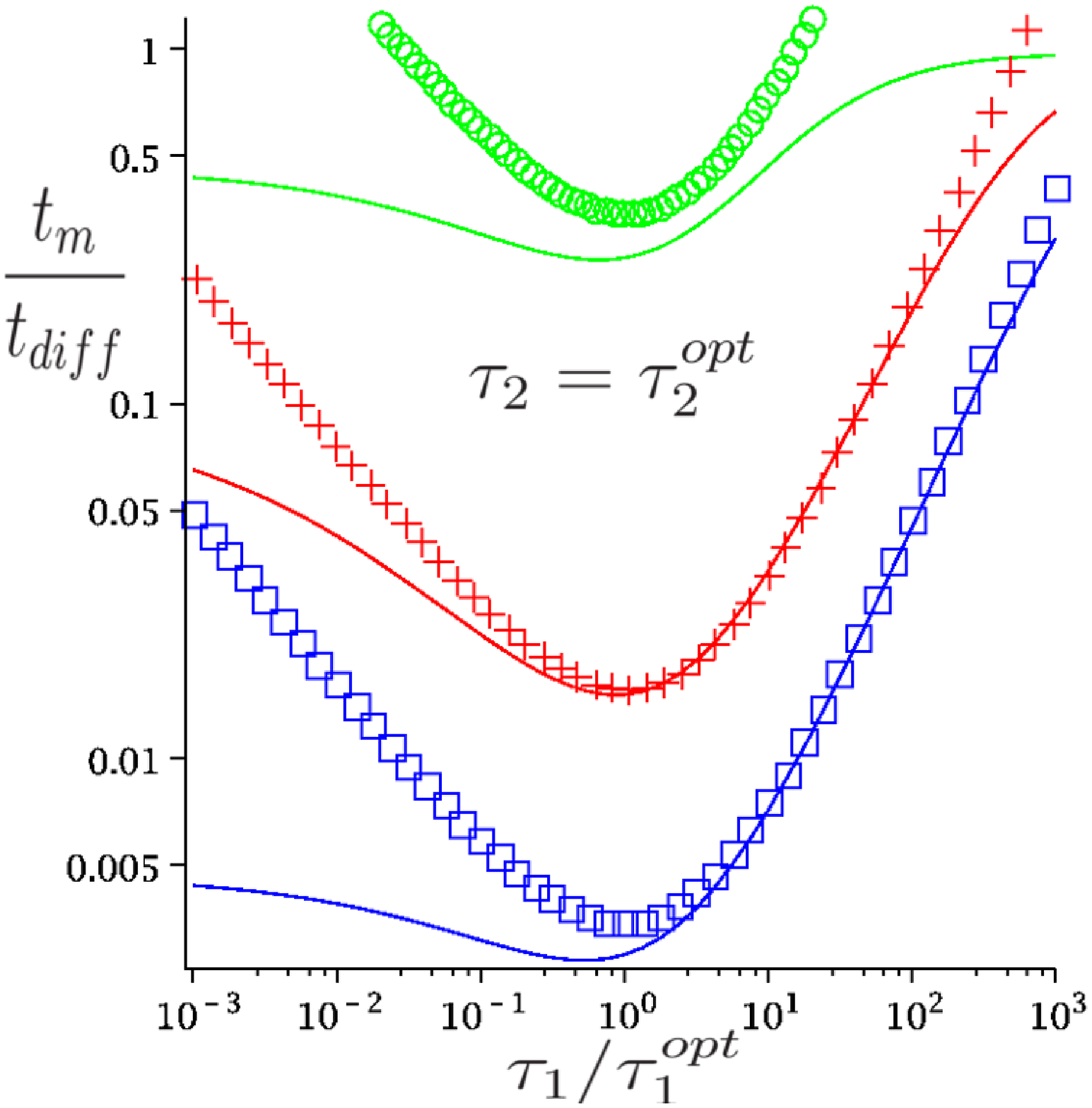}
{}
{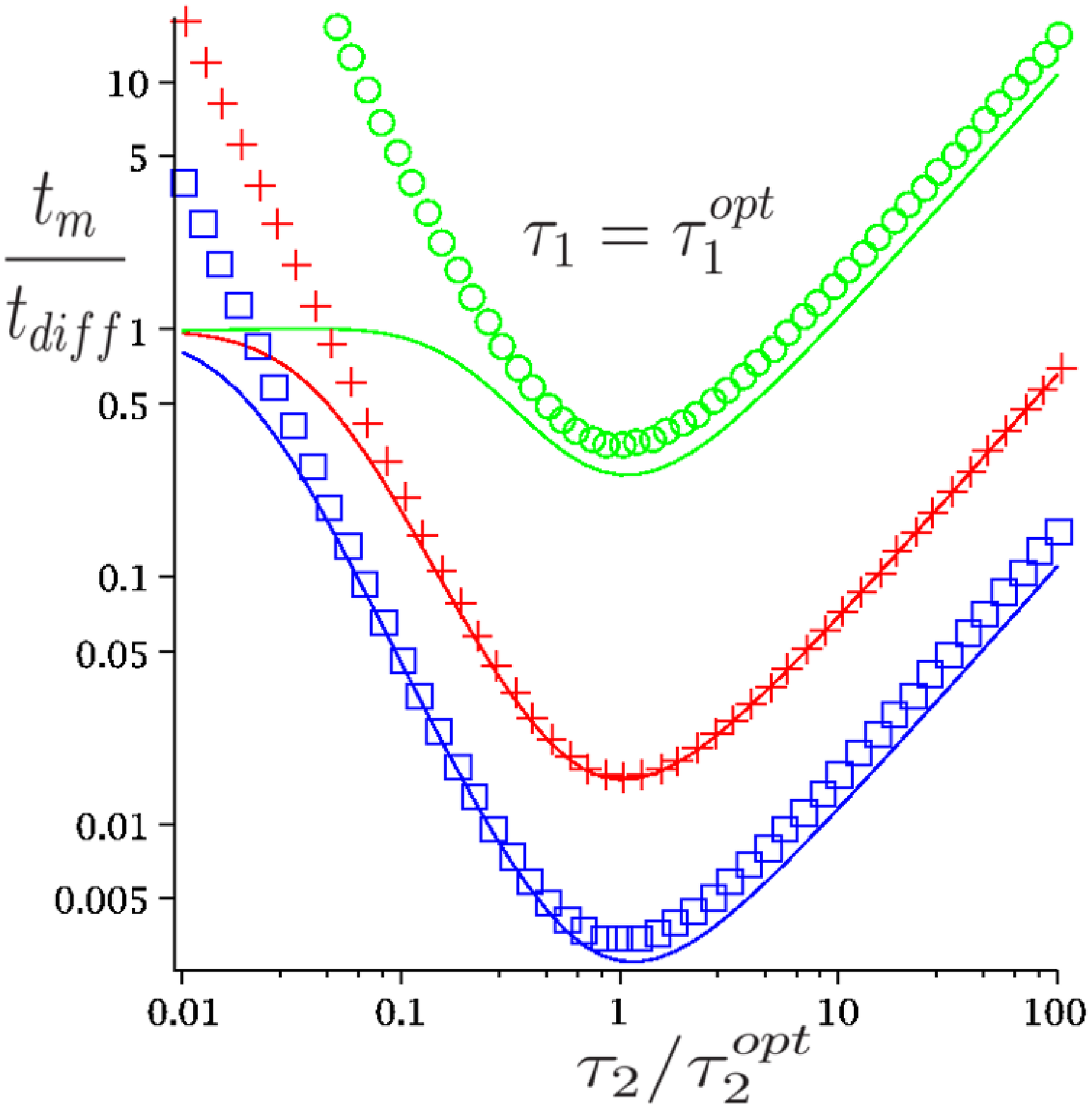}
{}
{Diffusive mode in 1 dimension. $\frac{t_m}{t_{\rm diff}}$,   exact expression  (line), and approximation in the regime of favorable intermittence and $ \frac{bD^2}{a^3 V^2} \gg 1$ \refm{tmn} (symbols). $a=1$ and $b=100$ (green, $\circ$), $a=1$, $b=10^4$ (red, $+$), $a=10$, $b=10^5$ (blue, $\square$). $D=1$, $V=1$. $\tau_1^{opt}$ is from expression \refm{l11dvd1}, $\tau_2^{opt}$ is obtained from expression \refm{l21dvd1}.}{tcasn}

\doublimagem{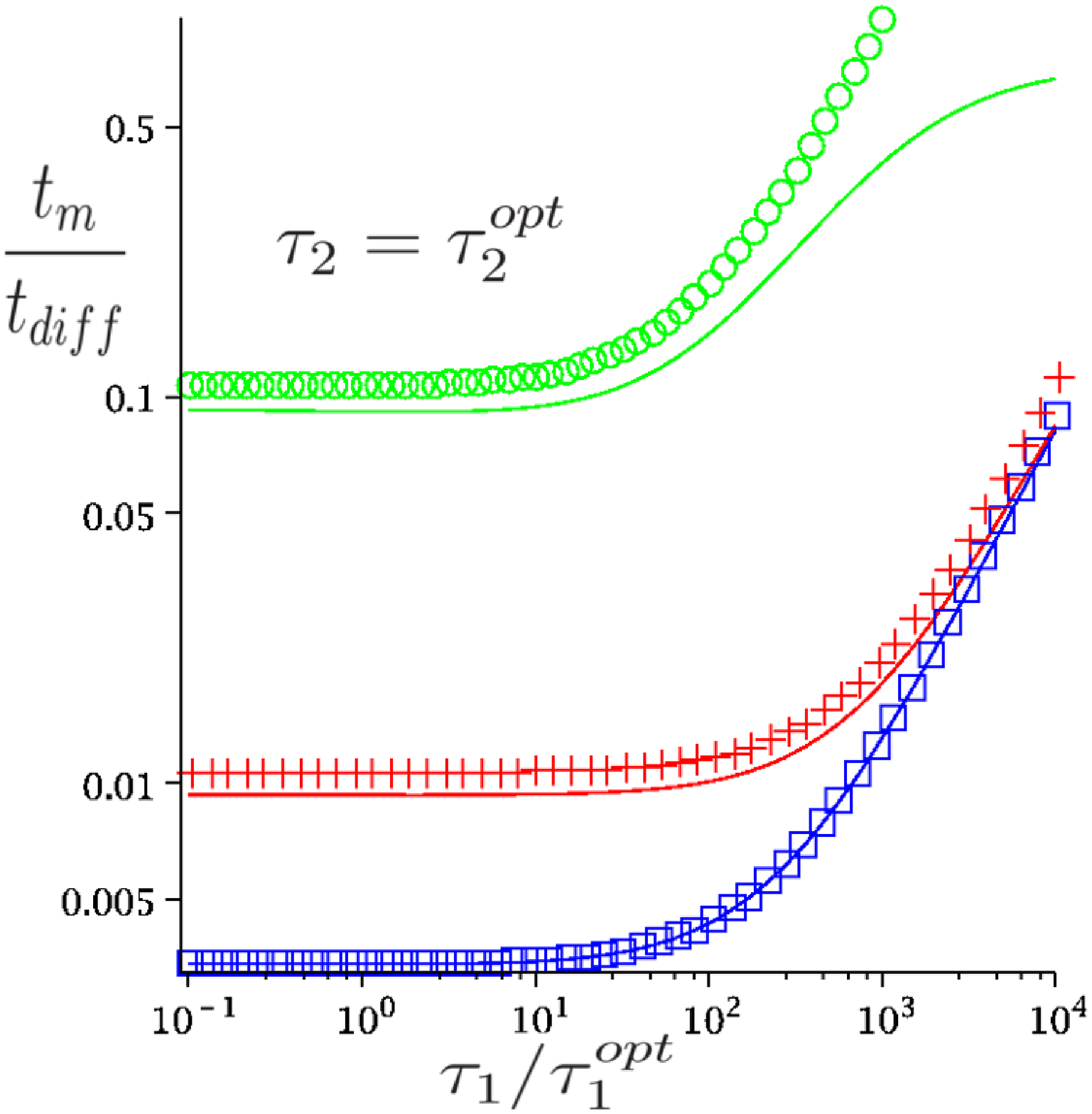}
{ }
{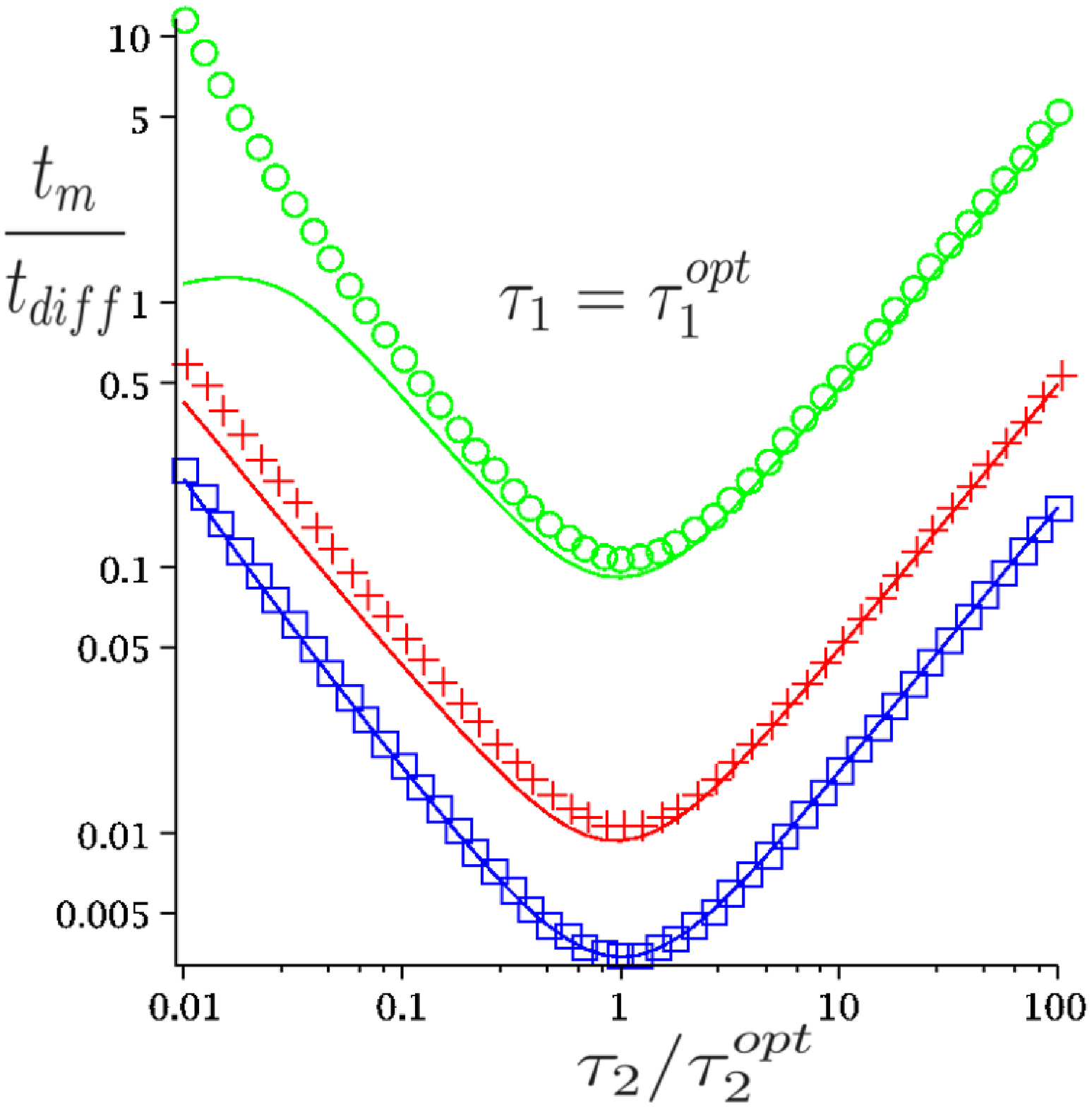}
{}
{Diffusive mode in 1 dimension. $\frac{t_m}{t_{\rm diff}}$,  exact expression (line), and approximation in the regime of favorable intermittence and $ \frac{bD^2}{a^3 V^2} \ll 1$ \refm{tsimp1DvD} (symbols). $a=10$ and $b=100$ (green, $\circ$), $a=10$, $b=1000$ (red, $+$), $a=100$, $b=10^4$ (blue, $\square$). $D=1$, $V=1$. $\tau_1^{opt}$ is from expression \refm{l11dvd2}, $\tau_2^{opt}$ is obtained from expression \refm{l21dvd2}.}{tsimpfig}

\subsubsection{Ballistic mode}

We now treat the case where the detection  phase 1 is modeled by the  ballistic mode \refi{modes_generic}. This model schematically accounts for the general observation that  speed often degrades perception abilities.  Our model corresponds to the extreme case where only two modes are available~: either the motion is slow and the target can be found, 
or the motion is fast and the target cannot be found. Note that this model  can be compared to the model of \textcite{viswaNat}, 
where there is only the detection phase.

\po{Equations}

The backward equations read outside the target ($x>a$)~:
\begin{equation}
v_l \frac{dt_1^+}{dx} +\frac{1}{\tau_1} \left( \frac{t_2^+}{2}+ \frac{t_2^-}{2} -t_1^+ \right) =-1,
\end{equation}
\begin{equation}
-v_l \frac{dt_1^-}{dx} +\frac{1}{\tau_1} \left( \frac{t_2^+}{2}+ \frac{t_2^-}{2} -t_1^- \right) =-1,
\end{equation}
\begin{equation}
V \frac{dt_2^+}{dx} +\frac{1}{\tau_2} \left( \frac{t_1^+}{2}+ \frac{t_1^-}{2} -t_2^+ \right) =-1 \mathrm{~and}
\end{equation}
\begin{equation}
-V \frac{dt_2^-}{dx} +\frac{1}{\tau_2} \left( \frac{t_1^+}{2}+ \frac{t_1^-}{2} -t_2^- \right) =-1.
\end{equation}
Inside the target ($x\le a$), one has 
 $t_1^{+,in}(x)=t_1^{-,in}(x)=0$, and~:
\begin{equation}
V \frac{dt_2^{+,in}}{dx} - \frac{1}{\tau_2} t_2^{+,in} =-1,
\end{equation}
\begin{equation}
-V \frac{dt_2^{-,in}}{dx} - \frac{1}{\tau_2} t_2^{-,in} =-1.
\end{equation}

\po{Results}\label{r1dvv1}


In the case where the phase 1 is modeled by  the ballistic mode in one dimension, 
we have calculated the exact mean first passage time $t_m$ at the target.
$t_m$ can be minimized  as a function of $\tau_1$ and $\tau_2$, yielding  
 two possible optimal strategies~: 
\begin{itemize}
\item for  $v_l>v_l^c=V\frac{\sqrt{3}}{2}\sqrt{\frac{a}{b}}$, intermittence is not favorable ~:  $\tau_1^{opt} \to \infty$,  $\tau_2^{opt} \to 0$,
\item for  $v_l<v_l^c=V\frac{\sqrt{3}}{2}\sqrt{\frac{a}{b}}$,   intermittence is favorable, with $\tau_1^{opt} \to 0$ and  $\tau_2^{opt}\approx \frac{a}{3V} \sqrt{\frac{b}{a}} $ . The gain reads 
 \begin{equation}
 gain\approx \frac{\sqrt{3}}{2}\frac{V}{v_l}  \sqrt{\frac{a}{b}}.
\end{equation}
This shows that 
the gain is larger than  1 for  $v_l<v_l^c=V\frac{\sqrt{3}}{2}\sqrt{\frac{a}{b}}$, which defines the regime where intermittence is favorable.
\end{itemize}

Note that the model studied by \textcite{viswaNat} shows that when targets   are not revisitable,  the optimal strategy for a single state searcher is to perform a straight ballistic motion. This strategy corresponds to $\tau_1\to\infty$ in our model. 
Our results show that if a faster phase without detection is allowed, this 
 straight line strategy can be outperformed.

\subsubsection{Conclusion in one dimension}

Intermittent search strategies  in one dimension share  similar features for the static, diffusive and ballistic detection modes. 
In particular, all modes show regimes where intermittence is favorable and lead to a minimization of the search time.
Strikingly, the optimal duration of the non-reactive relocation phase 2 is quite independent of the modeling of the 
reactive phase~: $\tau_2^{opt} = \frac{a}{3V} \sqrt{\frac{b}{a}}$ for the static mode, 
for  the ballistic mode (in the regime $v_l < v_l^c \simeq \frac{V}{2}\sqrt{\frac{3a}{b}}$), and for the diffusive mode (in the regime $b>\frac{D}{V}$ and $a\gg \frac{D}{V}\sqrt{\frac{b}{a}}$). This shows the robustness of the optimal value  $\tau_2^{opt}$.

\subsection{Dimension two}

\subsubsection{Static mode}
\label{section_generic_2Dvk}

We study here the case where the detection phase is modeled by the static mode~: 
the searcher does not move during the detection phase 
and has a finite reaction rate with the target if it is within its detection radius $a$ \refi{modes_generic}.

\po{Equations and results}

\imagea{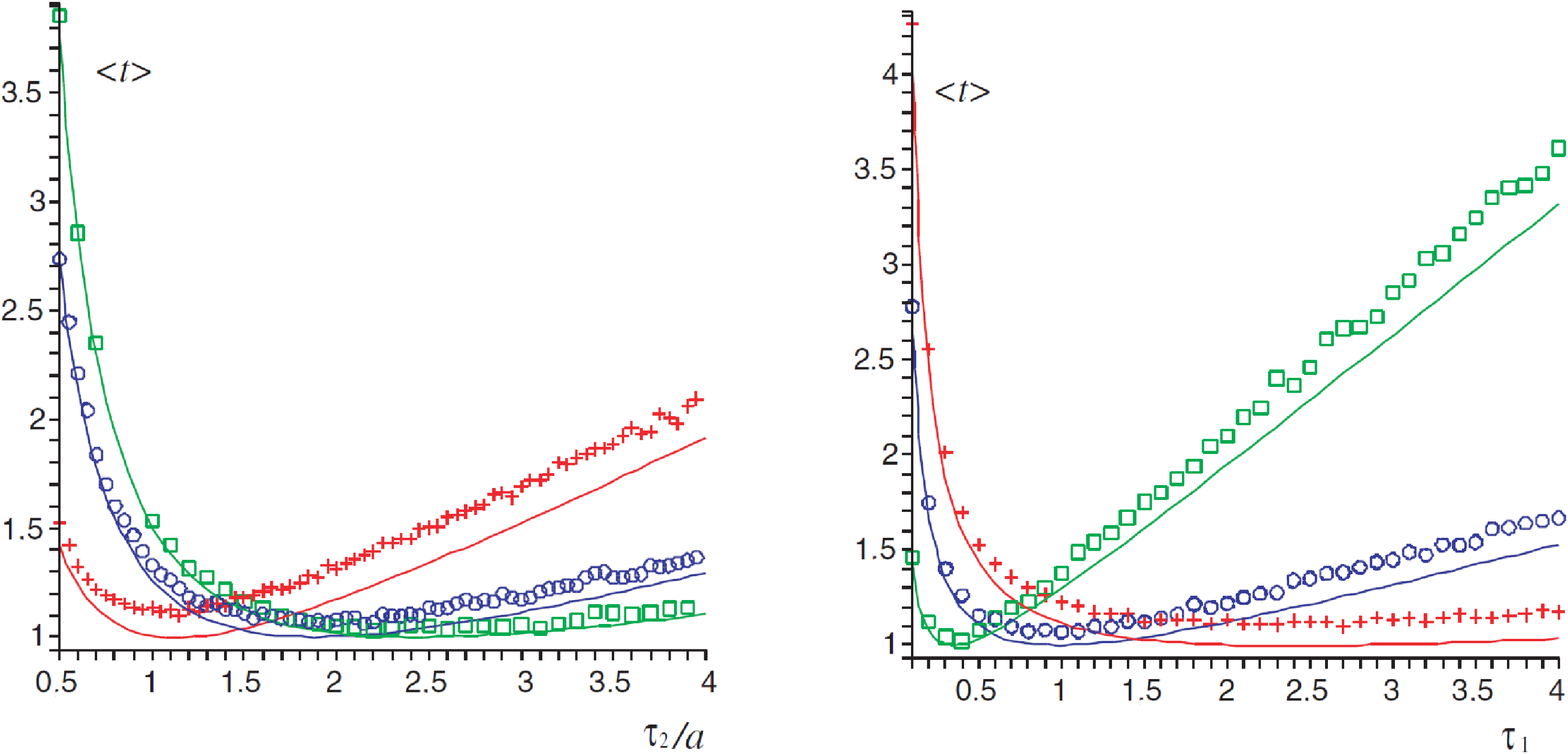}{Static mode in two dimensions. Simulations (symbols) and analytical approximate \refm{searchtime2} (lines). 
$k = 1$, $V = 1$,
$b = 56$; $a = 10$ (red, $+$) ($\tau_1^{opt}=2.41$, $\tau_2^{opt}=11.2$), $a = 1$ (blue, $\circ$) ($\tau_1^{opt}=0.969$, $\tau_2^{opt}=1.88$), $a = 0.1$ (green, $\square$) ($\tau_1^{opt}=0.348$, $\tau_2^{opt}=0.242$). \textit{Left}~: mean search time $t_m$ as a
function of $\tau_2/a$, with $\tau_1=\tau_1^{opt}$. \textit{Right}~: mean search time $t_m$ as a
function of $\tau_1$, with $\tau_2=\tau_2^{opt}$.}{generic_2Dk_tm}{6}

 The mean first passage time  at a target
satisfies the following backward equations \cite{Redner}:
\begin{equation}\label{back1k}
\frac{1}{2\pi\tau_1}\int_{0}^{2\pi}(t_2(\overrightarrow{r})-t_1(\overrightarrow{r}))d\theta_{\overrightarrow{V}}-k{\rm I}_a(\overrightarrow{r})t_1(\overrightarrow{r})=-1.
\end{equation}
\begin{equation}\label{back2k}
\overrightarrow{V}\cdot\nabla_{\bf r}t_2(\overrightarrow{r})-\frac{1}{\tau_2}(t_2(\overrightarrow{r})-t_1(\overrightarrow{r}))=-1
\end{equation}
The function ${\rm I}_a$ is defined by ${\rm I}_a(\overrightarrow{r})=1$ inside the target (if $|\overrightarrow{r}|\le a$) and  ${\rm I}_a(\overrightarrow{r})=0$ outside the target (if $|\overrightarrow{r}|> a$).
In the present form, these integro-differential equations (completed with boundary conditions)  do not seem to allow for an exact resolution with standard methods.
$t_2$ is the mean first passage time to the target, 
starting from $\overrightarrow{r}$ in phase 2, with speed $\overrightarrow{V}$, 
of angle $\theta_{\bf V}$, 
and with projections on the axes $V_{x}$, $V_{y}$. $i$ and $j$ can take either $x$ or $y$ as a value. 
The following decoupling assumption is introduced~:
\begin{equation}\label{deck}
\langle V_i V_j t_2\rangle_{\theta_{\bf V}}\simeq\langle V_i V_j\rangle_{\theta_{\bf V}}\langle t_2\rangle_{\theta_{\bf V}}
\end{equation}
and leads to the following approximation of the mean search time, which can be checked by numerical simulations \refi{generic_2Dk_tm}~:  \footnotesize
\begin{equation}\label{searchtime2}
t_m = \frac{\tau_1+\tau_2}{2k\tau_1 y^2}\left\{\frac{1}{x}(1+k\tau_1)(y^2-x^2)^2\frac{{\rm I}_0(x)}{{{\rm I}_1(x)}}
+\frac{1}{4}\left[8y^2+(1+k\tau_1)\left(4y^4\ln(y/x)+(y^2-x^2)(x^2-3y^2+8)\right)\right]\right\},
\end{equation}
\normalsize
\begin{equation}
{\rm where}\;x=\sqrt{\frac{2k\tau_1}{1+k\tau_1}}\frac{a}{V\tau_2} \;{\rm and}\;y=\sqrt{\frac{2k\tau_1}{1+k\tau_1}}\frac{b}{V\tau_2}.
\end{equation}
In that case, intermittence is trivially necessary to find the target.
In the regime  $b\gg a$, the optimization of the search time (\ref{searchtime2})
 leads to~: 
\begin{equation} \label{statique}
\tau_{1}^{opt}=\left(\frac{a}{Vk}\right)^{1/2}\left(\frac{2\ln(b/a)-1}{8}\right)^{1/4},\;
\end{equation}
\begin{equation}\label{statique2}
\tau_{2}^{opt}=\frac{a}{V}\left(\ln(b/a)-1/2\right)^{1/2},
\end{equation}
and the minimum search time is given in the large volume limit by~: 
\begin{equation}
\begin{split}
t_m^{opt}&={\frac {{b}^{2}}{{a}^{2}k}}- \frac{2^{1/4}}{\sqrt{Vka^3}}\,\frac{ ({a}^{2}-4b^2)\ln(b/a)+2b^2-a^2}{( 2\ln(b/a) -1 ) ^{3/4}}\\
&-\frac{\sqrt{2}}{48ab^2V}\,\frac{(96a^2b^2-192b^4)\ln^2(b/a)+(192b^4-144a^2b^2)\ln(b/a)+46a^2b^2-47b^4+a^4}{( 2\ln(b/a) -1 ) ^{3/2}}.\\
\end{split}
\end{equation}

\subsubsection{Diffusive mode}
\label{section_generic_2DvD}

We now assume that the searcher  diffuses during the detection phase \refi{modes_generic}. 
For this process, the mean first passage time to the  target
satisfies the following backward equation \cite{Redner}:
\begin{equation}\label{back1d}
D\nabla^2_{\bf r}t_1(\overrightarrow{r})+\frac{1}{2\pi\tau_1}\int_{0}^{2\pi}(t_2(\overrightarrow{r})-t_1(\overrightarrow{r}))d\theta_{\bf V}=-1,
\end{equation}
\begin{equation}\label{back2d}
\overrightarrow{V}\cdot\nabla_{\bf r}t_2(\overrightarrow{r})-\frac{1}{\tau_2}(t_2(\overrightarrow{r})-t_1(\overrightarrow{r}))=-1,
\end{equation}
with $t_1(\overrightarrow{r})=0$ inside the target ($r\le a$).  The same decoupling assumption as for the static case sis used \refm{deck}. 
It eventually leads to the following approximation of the mean search time, which can be  checked by numerical simulations \refi{generic_2ddiff_tm}~: 
\begin{equation}\disp\label{tmap}
t_m =(\tau_1+\tau_2)\frac{\disp1-a^2/b^2}{\disp(\alpha^2 D\tau_1)^2}\left\{\disp a\alpha(b^2/a^2-1)\frac{\disp M}{\disp 2L_+}-\frac{\disp L_-}{\disp L_+}-\frac{\disp \alpha^2D\tau_1}{\disp 8{\widetilde D}\tau_2}\frac{\disp(3-4\ln(b/a))b^4-4a^2b^2+a^4}{\disp b^2-a^2}\right\},
\end{equation}
with~: 
\begin{equation}
\begin{split}
{\rm with}\  L_\pm=&{\rm I}_0\left(\frac{a}{\sqrt{{\widetilde D}\tau_2}}\right)\left({\rm I}_1(b\alpha){\rm K}_1(a\alpha)- {\rm I}_1(a\alpha){\rm K}_1(b\alpha)  \right)\\
&\pm \alpha\sqrt{{\widetilde D}\tau_2}\;{\rm I}_1\left(\frac{a}{\sqrt{{\widetilde D}\tau_2}}\right)\left({\rm I}_1(b\alpha){\rm K}_0(a\alpha)+ {\rm I}_0(a\alpha){\rm K}_1(b\alpha)  \right),\\
\end{split}
\end{equation}
and~: 
\begin{equation}
\begin{split}
  M=&{\rm I}_0\left(\frac{a}{\sqrt{{\widetilde D}\tau_2}}\right)\left({\rm I}_1(b\alpha){\rm K}_0(a\alpha)+ {\rm I}_0(a\alpha){\rm K}_1(b\alpha)  \right)\\
&-4\frac{a^2\sqrt{{\widetilde D}\tau_2}}{\alpha(b^2-a^2)^2}{\rm I}_1\left(\frac{a}{\sqrt{{\widetilde D}\tau_2}}\right
)\left({\rm I}_1(b\alpha){\rm K}_1(a\alpha)- {\rm I}_1(a\alpha){\rm K}_1(b\alpha)  \right),\\
\end{split}
\end{equation}
where $\alpha=(1/(D\tau_1)+1/({\widetilde D}\tau_2))^{1/2}$ and ${\widetilde D} = v^2 \tau_2$.
The minimization as a function of $\tau_1$ and $\tau_2$ is as follows.

\imagea{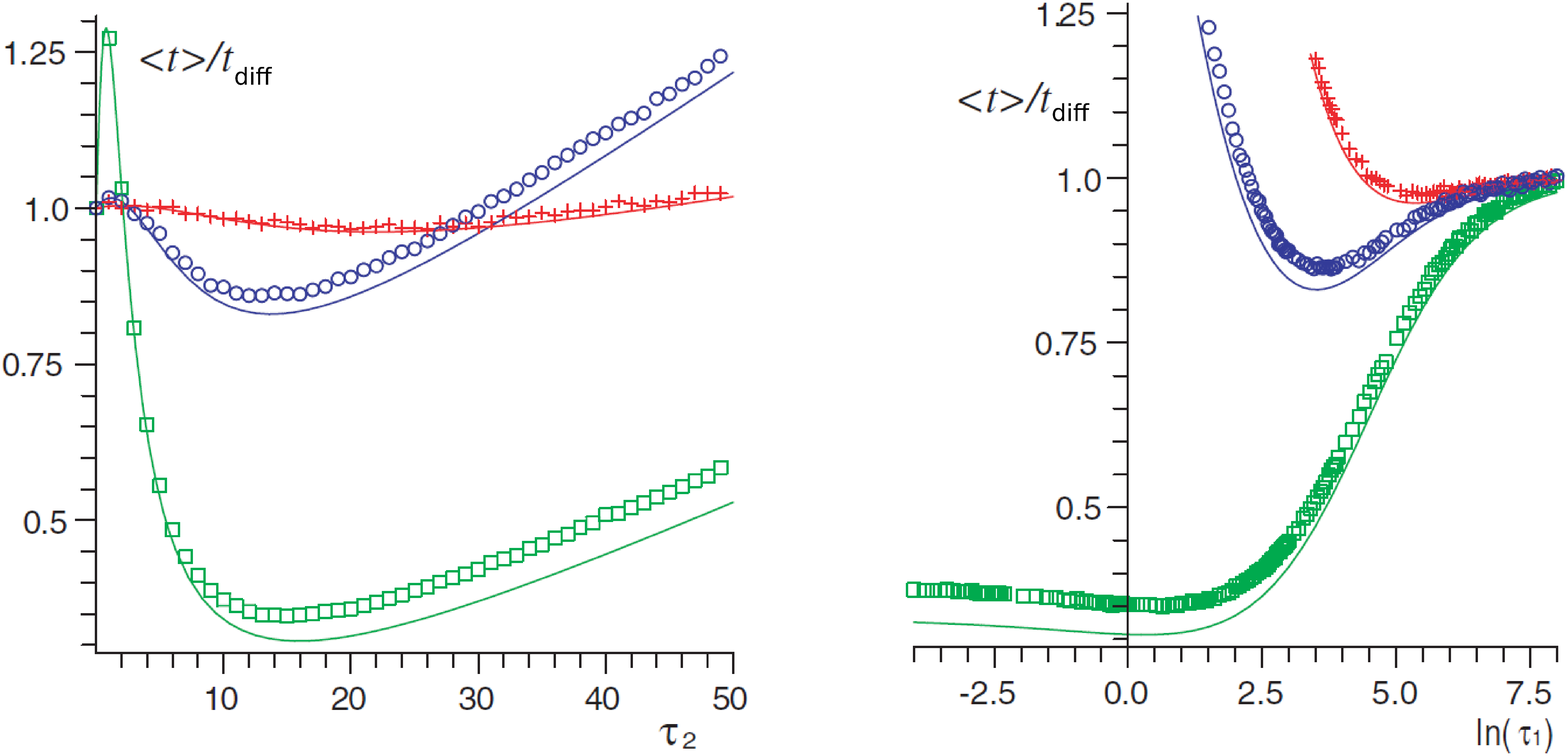}{Diffusive mode in two dimensions. 
Simulations (symbols) versus analytical approximate \refm{tmap} (line) of the search time, 
rescaled by the value in the absence of intermittence $t_{\rm diff}$ as a
function of $\tau_2$ (left) and $\ln(\tau_1)=$ (right), for $D = 1$, $V = 1$, $b = 226$. 
\textit{Left}~: $a = 10$, $\tau_1=1.37$ (green $\square$); $a = 1$, $\tau_1=33.6$ 
(blue $\circ$); $a = 0.1$, $\tau_1=213$ (red $+$). \textit{Right}: $a = 10$,  $\tau_2 = 15.9$ (green, $\square$); $a = 1$,
$\tau_2= 13.7$ (blue, $\circ$); $a = 0.1$, $\tau_2 = 22$ (red, $+$).}{generic_2ddiff_tm}{6}

\po{$a<b\ll D/V$~: intermittence is not favorable}

In that regime, intermittence is not favorable. Indeed, the typical time required to explore the whole domain of radius $b$ is of order 
$b^2/D$ with diffusive motion, which is shorter than the corresponding time $b/V$ with ballistic motion. As a consequence, it is never useful to interrupt the 
diffusive phases by mere relocating ballistic phases.
The mean first passage time to the target in this optimal regime of diffusion only is obtained using standard methods \cite{Redner} and reads  in the limit $b \gg a$~: 
\begin{equation}\label{tdiff2Db}
 t_{\rm diff} = \frac{b^2}{8 D_{\rm eff}}\left(-3+4 \ln \frac{b}{a} \right).
\end{equation}

\po{$a\ll D/V\ll b$~: first regime of intermittence}\label{r2dvd}

In this second regime, one can use the following approximate formula for the search time:
\begin{equation}\label{tapprox}
t_m =\frac{b^2}{4DV^2\alpha^2}\frac{\tau_1+\tau_2}{\tau_1\tau_2^2}
\left\{4\ln(b/a)-3-2\frac{(V\tau_2)^2}{D\tau_1}(\ln(\alpha a)+\gamma-\ln 2)\right\},
\end{equation}
$\gamma$ being the Euler constant. An approximate  criterion to determine if intermittence is useful can be obtained by expanding $t_m$ in powers of $1/\tau_1$ when $\tau_1\to\infty$ ($\tau_1\to\infty$ corresponds to the absence of intermittence), and requiring that the coefficient of the term $1/\tau_1$ is negative for all values of $\tau_2$. Using this  criterion, we find that  
intermittence is useful if 
\begin{equation}
\sqrt{2}\exp(-7/4+\gamma)Vb/D-4\ln(b/a)+3>0,
\end{equation}

In this regime, using Eq(\ref{tapprox}), the optimization of the search time  leads to~:
\begin{equation}\label{intermediare}
\tau_{1}^{opt}=\frac{b^2}{D}\frac{4\ln w-5+c}{w^2(4\ln w -7+c)}, \ \tau^{opt}_{2}=\frac{b}{V}\frac{\sqrt{4\ln w -5+c}}{w},
\end{equation}
where $w$ is the solution of the implicit  equation $w=2Vbf(w)/D$ with~:
\begin{equation}
\frac{\sqrt{4\ln w-5+c}}{f(w)}=-8(\ln w)^2+(6+8\ln(b/a))\ln w-10\ln(b/a)+11
-c(c/2+2\ln(a/b)-3/2),
\end{equation}
and  $c=4(\gamma -\ln(2))$,  $\gamma$ being the Euler constant.
An useful approximation for   $w$ is given by~:
\begin{equation}
w\simeq \frac{2Vb}{D}f\left(\frac{Vb}{2D\ln(b/a)}\right).
\end{equation}
The gain for this optimal strategy reads~: 
\begin{equation}
 gain=\frac{t_{\rm diff}}{t_m^{opt}}\simeq \frac{1}{2}\frac{4 \ln b/a -3 +4a^2/b^2 -a^4/b^4}{4\ln b/a -3+2(4\ln w)\ln (b/aw) }\left(\frac{1}{4\ln w -5}+\frac{wD}{bV}\frac{4\ln w -7}{(4\ln w -5)^{3/2}}\right)^{-1}.
\end{equation}
If intermittence significantly speeds up the search in this regime (typically by a factor 2), it does not change the order of magnitude of the search time.

\po{$ D/V\ll a\ll b$~: ``universal'' regime of intermittence}

 In the last regime $D/V\ll a\ll
b$, the optimal strategy is obtained for~:
\begin{equation}\label{grandv}
\tau_{1}^{opt}\simeq \frac{D}{2V^2}\frac{\ln^2 (b/a)}{2\ln (b/a) -1}, \;\tau_{2}^{opt}\simeq \frac{a}{V}(\ln(b/a)-1/2)^{1/2}.
\end{equation}
and the gain reads~:
\begin{equation}\label{gain2dvd}
 gain=\frac{t_{\rm diff}}{t_m^{opt}}\simeq \frac{\sqrt{2}aV}{8D}\left(\frac{1}{4\ln(b/a)-3}\frac{{\rm I}_0\left(2/\sqrt{2\ln(b/a)-1}\right)}
{{\rm I}_1\left(2/\sqrt{2\ln(b/a)-1}\right)}+\frac{1}{2\sqrt{2\ln(b/a)-1}}\right)^{-1}.
\end{equation}
Here, the optimal strategy leads to a significant decrease of the search time which can be rendered arbitrarily smaller than the 
 search time in absence of intermittence.

\subsubsection{Ballistic mode}

In this case, the searcher has access to two different speeds:
one ($V$) is fast but prevents the searcher from finding its target, 
and the other one ($v_l$) is slower but enables the searcher to detect the target \refi{modes_generic}.  

\po{Simulations}

\begin{figuresmall}
\begin{center}
      \includegraphics[width=\linewidth]{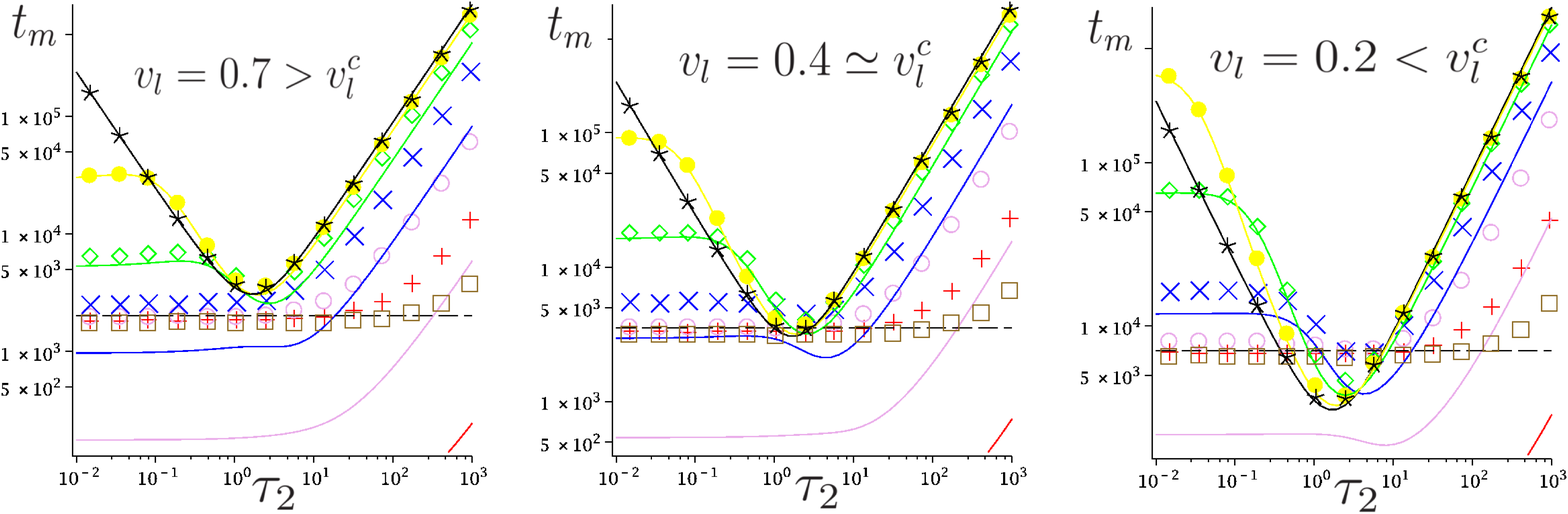}
\end{center}
\caption{ Ballistic mode
in two dimensions. $\ln(t_m)$ as a function of $\ln(\tau_2)$.
Simulations (symbols), diffusive/diffusive approximation \refm{tmap}
with \refm{d2dvv} (colored lines), $\tau_1 \to 0$ limit \refm{Pearson2D}
(black line), $\tau_1 \to \infty$ (no intermittence) \refm{topt2}
(dotted black line). $b=30$, $a=1$, $V=1$. $\tau_1=0$ (black, $\star$),
$\tau_1=0.17$ (yellow, $\bullet$), $\tau_1=0.92$ (green, $\diamond$),
$\tau_1=5.0$ (blue, $\times$), $\tau_1=28$ (purple, $\circ$),
$\tau_1=150$ (red, $+$), $\tau_1=820$ (brown, $\square$).}
\label{2dvvb30}\normalsize \end{figuresmall}

Since an explicit  expression of the mean search time is not available, a numerical study is performed. Exploring the parameter space numerically enables to identify the regimes where the mean search time is minimized. Then, for each regime, approximation schemes are developed to provide analytical expressions of the mean search time.
The numerical results presented in  figure \ref{2dvvb30} suggest  two regimes defined according to a threshold value  $v_l^{c}$ of  $v_l$ to be determined later on ~:  
\begin{itemize}
 \item for $v_l > v_l^{c}$,  $t_m$ is minimized for $\tau_2 \to 0$
\item for $v_l < v_l^{c}$, $t_m$ is minimized for $\tau_1 \to 0$~.
\end{itemize}

\po{Regime without intermittence ($\tau_2 \to 0$, $\tau_1 \to \infty$)}

Qualitatively, it is rather intuitive that for $v_l$ large enough 
(the precise threshold value $v_l^{c}$ will be determined next),  
phase 2 is inefficient since it does not allow for target detection. 
The optimal strategy is therefore  $\tau_2 \to 0$ in this case. 
In this regime, the searcher performs a ballistic motion, 
which is randomly reoriented with frequency $1/\tau_1$. 
Along the same line as in \textcite{viswaNat} 
(where however the times between successive reorientations are L\'evy distributed), 
it can be shown that the optimal strategy to find a target 
(which is assumed to disappear after the first encounter) 
is to minimize oversampling and therefore to perform a purely ballistic motion. 
In our case this means that in the regime $\tau_2 \to 0$, 
the optimal $\tau_1$ is given by  $\tau_1^{opt} \to \infty$.

In this regime, we can propose an estimate of the optimal search time  $t_{bal}$. 
The surface scanned during $\delta t$ is $2 a v_l \delta t$. 
$p(t)$ is the proportion of the total area which has not yet been scanned at $t$.
If we neglect correlations in the trajectory, $p(t)$ is solution of~: 
\begin{equation}
 \frac{d p }{dt} = -\frac{2a v_l p(t)}{\pi b^2}.
\end{equation}
Then, given that  $p(t=0)=1$,  we obtain~:
\begin{equation}
 p(t)=\exp\left(-\frac{2av_l}{\pi b^2}\right),
\end{equation}
and the mean first passage time to the   target in these conditions is~: 
\begin{equation}\label{topt2}
 t_{bal} = - \int_0^{\infty} t \frac{dp}{dt} dt = \frac{\pi b^2}{2av_l} .
\end{equation}
This expression yields  a good agreement with numerical simulations.
Note in particular that   $t_{bal} \propto \frac{1}{v_l}$.

\po{Regime with intermittence $\tau_1 \to 0$}
In this regime where $v_l < v_l^{c}$, the numerical study shows that the search time is minimized for  $\tau_1 \to 0$ \refi{2dvvb30}. 
We here determine the optimal value of $\tau_2$ in  this regime. To proceed we  approximate the problem by  the case of a diffusive mode previously studied  \refm{tmap}, with an effective diffusion coefficient~: 
\begin{equation}\label{d2dvv}
 D = \frac{v_l^2 \tau_1}{2}.
\end{equation}
This approximation is very satisfactory in the regime  $\tau_1 \to 0$   \refi{2dvvb30}. 

We can then use the results of the previous section for the diffusive mode in the  $\tau_1 \to 0$ regime and obtain:
\begin{equation}\label{Pearson2D}
t_m=\tau_2 \left(1-\frac{a^2}{b^2}\right) \left(1-\frac{1}{4} \frac{\left(3+4 \ln\left(\frac{a}{b}\right)\right) b^4-4 a^2 b^2+a^4}{\tau_2^2  V^2 \left(b^2-a^2\right)}+\frac{a}{V \tau_2 \sqrt{2}} \left(\frac{b^2}{a^2}-1\right) \frac{I_0\left(\frac{a \sqrt{2}}{\tau_2  V}\right)}{ I_1\left(\frac{a \sqrt{2}}{\tau_2  V}\right)}\right).
\end{equation}
The calculation of  $\tau_2^{opt}$ minimizing $t_m$ 
 then gives:
\begin{equation} \label{tau2opt}
 \tau_2^{opt} = \frac{a}{V}\sqrt{\ln\left( \frac{b}{a} \right) -\frac{1}{2}}.
\end{equation}
Finally the gain reads :
\begin{equation}\label{2dvvgain}
gain=\frac{t_{bal}}{t_m^{opt}}\simeq \frac{\pi V}{4 v_l}\left(\ln\left(\frac{b}{a}\right)\right)^{-0.5}.
\end{equation}

\po{Determination of $v_l^c$}

Note that  an estimate of $v_l^c$ can be obtained 
from (\ref{2dvvgain}) as   the value of $v_l$ for which $gain=1$~: 
\begin{equation}\label{2dvvvlc}
v_l^c \simeq \frac{\pi V}{4}\left(\ln\left(\frac{b}{a}\right)\right)^{-0.5} \propto \frac{V}{\sqrt{\ln(b/a)}}.
\end{equation}
It is noteworthy that intermittence is less favorable with increasing $b$. 
This effect is similar to the 1 dimensional case, even though it is less important here. 
It can be understood as follows: 
at very large scales the intermittent trajectory is reoriented many times and 
therefore scales as diffusion, which is less favorable than the non intermittent ballistic motion.

\subsubsection{Conclusion in dimension two}

Remarkably, for the   three different modes of detection
(static, diffusive and ballistic), we find a regime where intermittence
 minimizes the search time for one and the same $\tau_2^{opt}$, given by 
 $\tau_2^{opt}= \frac{a}{V}\sqrt{\ln\left( \frac{b}{a} \right) -\frac{1}{2}}$. 
As in one dimension, this indicates that  optimal intermittent strategies are robust and widely  
independent of the details of the description of the detection mechanism.

\subsection{Dimension three}

\subsubsection{Static mode}
\label{section_generic_3Dvk}

We study in this section the case where the detection phase is modeled by the static mode, for which  
the searcher does not move during the detection phase and has 
a finite reaction rate with the target if it is within a detection radius $a$ \refi{modes_generic}.

\po{Equations}

Denoting $t_1 (r)$ the mean first passage time to the target starting from a distance $r$ from the target  in phase 1 (detection phase),
and $t_{2,\theta,\phi}(r)$ the mean first passage time to the target starting from a distance $r$ from the target in phase 2 (relocation phase) 
with a ballistic motion in a direction characterized by $\theta$ and $\phi$, we obtain~: 
\begin{equation}
 \overrightarrow{V}.\overrightarrow{\bigtriangledown}t_{2,\theta,\phi} +\frac{1}{\tau_2}\left( t_1- t_{2,\theta,\phi}\right)=-1.
\end{equation}
Then outside the target  ($r>a$)~:
\begin{equation}
\frac{1}{\tau_1}\left( \frac{1}{4\pi}\int_0^{\pi} d\theta \sin\theta\int_0^{2\pi}d\phi t_{2,\theta,\phi} - t_1 \right) = -1, 
\end{equation}
and inside the target ($r\le a$)~: 
\begin{equation}
\frac{1}{\tau_1}\frac{1}{4\pi}\int_0^{\pi} d\theta \sin\theta\int_0^{2\pi}d\phi t_{2,\theta,\phi} - \left(\frac{1}{\tau_1} + k \right) t_1  = -1. 
\end{equation}
Defining $t_2=\frac{1}{4\pi}\int_0^{\pi} d\theta \sin\theta\int_0^{2\pi}d\phi t_{2,\theta,\phi}$, one obtains outside the target ($r>a$)~:  
\begin{equation}
\frac{1}{\tau_1}\left(t_2 - t_1 \right) = -1 ,
\end{equation}
and inside the target ($r<a$)~: 
\begin{equation}
\frac{1}{\tau_1} t_2 - \left(\frac{1}{\tau_1} + k \right) t_1  = -1. 
\end{equation}
Making a similar decoupling approximation as  in two dimensions, leads finally to~: 
\begin{equation}
\frac{ V^2 \tau_2}{3} \bigtriangleup t_2  -\frac{1}{ \tau_2}(t_1-t_2)=-1.
\end{equation}
These equations are solved inside and outside the target, using the following  boundary conditions~:
\begin{equation}
 \left.\frac{d t_2^{out}}{dr}\right|_{r=b}=0,
\end{equation}
\begin{equation}
t_2^{out}(a)=t_2^{in}(a),
\end{equation}
\begin{equation}
 \left.\frac{d t_2^{out}}{dr}\right|_{r=a}=\left.\frac{d t_2^{in}}{dr}\right|_{r=a},
\end{equation}
and the condition that 
$t_2^{in}(0)$ should be finite.

\po{Results}

The following  approximate expression of the mean search time is found in the low density limit~:

\begin{equation}\label{tm3dvksimp}
  t_m = \frac{b^3( \tau_2+ \tau_1)}{a} \left(\frac{(1+k  \tau_1) }{ \tau_1 k a^2}+\frac{6 }{5\tau_2^2 V^2} \right).
\end{equation}
This expression of $t_m$ can be minimized for~: 
\begin{equation}\label{tau13dvk}
 \tau_1^{opt}=\left(\frac{3}{10}\right)^{\frac{1}{4}} \sqrt{\frac{a}{Vk}},
\end{equation}
\begin{equation}\label{tau23dvk}
 \tau_2^{opt}=\sqrt{1.2}\frac{a}{V},
\end{equation}
and the minimum mean search time finally reads~: 
\begin{equation}\label{topt3dvk}
   t_m^{opt}=\frac{1}{\sqrt{5}}\frac{1}{k}\frac{b^3}{a^3}\left( \sqrt{\frac{ak}{V}}24^{1/4}+5^{1/4} \right)^2.
\end{equation}

\begin{figuresmall}
\begin{center}
      \includegraphics[width=\linewidth]{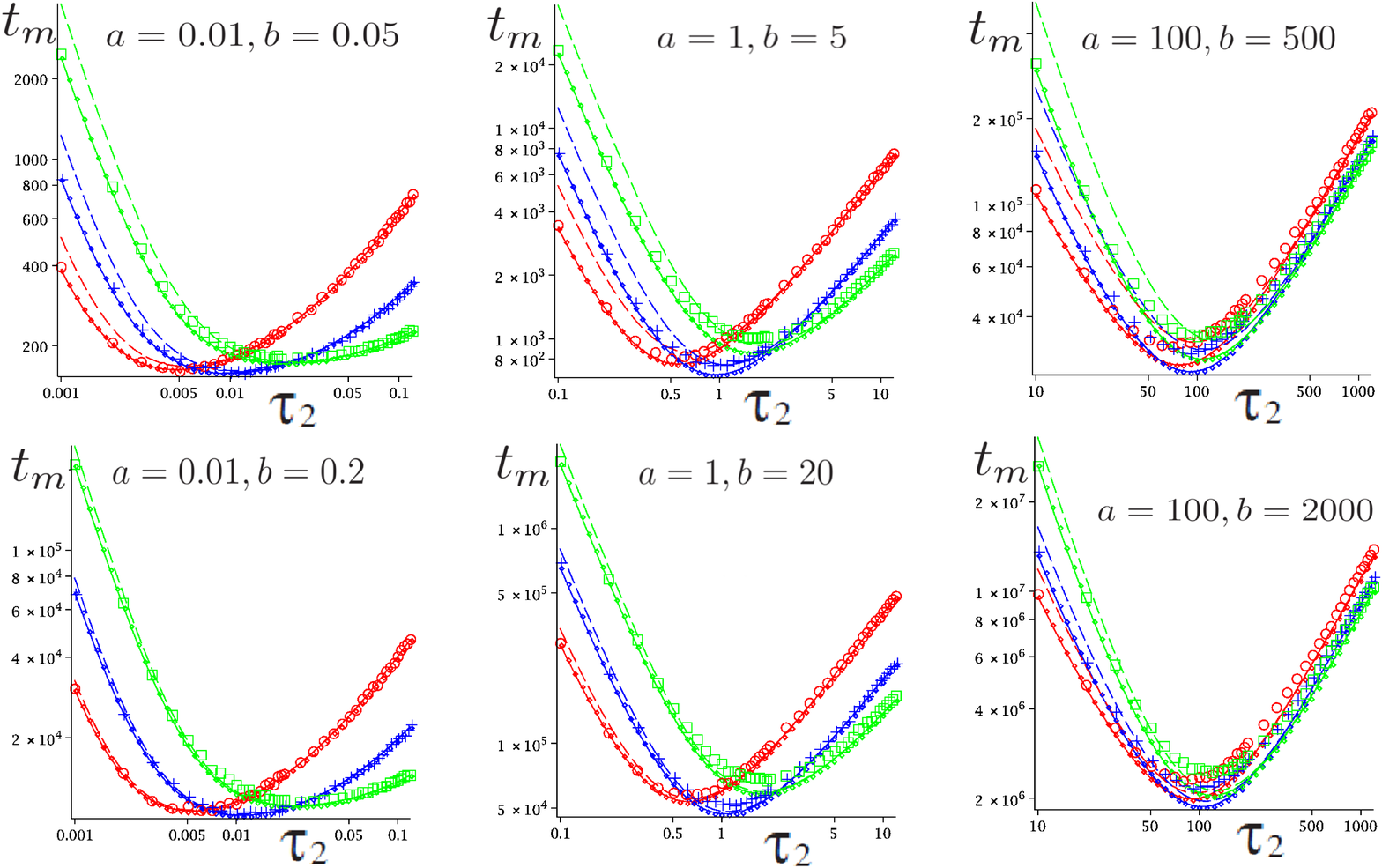}
\end{center}
\caption{ Static mode in 3 dimensions. $\ln(t_m)$ as a function of $\ln(\tau_2)$ for different values of $\tau_1$, $a$ and $b/a$. 
Comparison between simulations (symbols), analytical expression  (line) and its asymptotics   for $b \gg a$ obtained in \cite{LeGros} (small dots), 
and simple expression for $b\gg a$ and $\alpha$ small \refm{tm3dvksimp} (dashed line).
$\tau_1 \simeq \tau_1^{opt}  \simeq 0.74 \sqrt{\frac{aV}{k}}$ \refm{tau13dvk} (blue, $+$), 
$\tau_1 = 0.25 \sqrt{\frac{aV}{k}}$ (red, $\circ$), $\tau_1 = 2.5 \sqrt{\frac{aV}{k}}$ (green, $\square$).
$V=1$, $k=1$.}
\label{3Dvkcomp}\normalsize 
\end{figuresmall}

Data obtained by numerical simulations (figure \ref{3Dvkcomp}) are 
in good agreement with  the analytical expression \refm{tm3dvksimp} except for small $\tau_2$ or small  $b$, where a refined analytical expression can be obtained (see \cite{LeGros}).
In particular,  the position of the minimum is very well approximated, 
and the error on the value of the mean search time at  the minimum is close to 10\%.

With the static  detection mode, intermittence is always favorable and leads 
to a single  optimal intermittent strategy. 
As in one and two dimensions, the optimal duration of the relocation phase 
does not depend on $k$, \textit{i.e.} on the description of the detection phase. 
In addition, this optimal strategy does not depend on the typical distance between targets $b$.

One can notice than for the static mode in the three cases studied (1, 2, and 3 dimensions), 
we have the relation~: $\tau_1^{opt}=\sqrt{\tau_2^{opt}/(2k)}$. 
This relation between the optimal durations of the two phases is independent from the dimension.

\subsubsection{Diffusive mode}
\label{section_generic_3DvD}

We now study the case where the detection phase is modeled by a diffusive mode.
During the detection phase, 
the searcher diffuses and detects  
the target as soon as their respective
distance is less than $a$ \refi{modes_generic}.

\po{Equations}

Outside the target ($r>a$), one has~:
\begin{equation}
 \overrightarrow{V}.\overrightarrow{\bigtriangledown}t_{2,\theta,\phi} +\frac{1}{\tau_2}\left( t_1- t_{2,\theta,\phi}\right)=-1,
\end{equation}
\begin{equation}
D \bigtriangleup t_1  +\frac{1}{\tau_1}\left( \frac{1}{4\pi}\int_0^{\pi} d\theta \sin \theta\int_0^{2\pi}d\phi t_{2,\theta,\phi} - t_1 \right) = -1 ,
\end{equation}
and inside the target ($r\le a$)~: 
\begin{equation}
 \overrightarrow{V}.\overrightarrow{\bigtriangledown}t_{2,\theta,\phi} -\frac{1}{\tau_2} t_{2,\theta,\phi}=-1,
\end{equation}
\begin{equation}
t_1=0.
\end{equation}
With $t_2=\frac{1}{4\pi}\int_0^{\pi} d\theta sin\theta\int_0^{2\pi}d\phi t_{2,\theta,\phi}$, one obtains outside the target ($r>a$)~:  
\begin{equation}
D \bigtriangleup t_1^{out}+\frac{1}{\tau_1}\left(t_2^{out} - t_1^{out} \right) = -1 .
\end{equation}
The decoupling approximation described in previous sections then yields outside the target~:  
\begin{equation}
\frac{ V^2 \tau_2}{3} \bigtriangleup t_2^{out}  + \frac{1}{ \tau_2}(t_1^{out}-t_2^{out})=-1,
\end{equation}
and inside the target ($r\le a$)~: 
\begin{equation}
\frac{ V^2 \tau_2}{3} \bigtriangleup t_2^{int}  -\frac{1}{ \tau_2}t_2^{int}=-1.
\end{equation}
These equations are completed by the following boundary conditions~: 
\begin{equation}
 \left.\frac{d t_2^{out}}{dr}\right|_{r=b}=0,
\end{equation}
\begin{equation}
t_2^{out}(a)=t_2^{int}(a),
\end{equation}
\begin{equation}
 \left.\frac{d t_2^{out}}{dr}\right|_{r=a}=\left.\frac{d t_2^{int}}{dr}\right|_{r=a}.
\end{equation}

\po{Results in the general case}\label{r3dvd1}

Through standard but lengthy calculations  the above system can be solved and leads to  
an analytical approximation of $t_m$ (see \cite{LeGros}.
In the regime  $b\gg a$, and $b \sqrt{(\tau_1 D)^{-1}+3(\tau_2v)^{-2}} \gg 1$; one  obtains~:
\begin{equation}\label{tmb3dvd}
 t_m = \frac{b^3 \kappa_2^4(\tau_1+\tau_2)}{\kappa_1}\frac{\tanh(\kappa_2a)+\frac{\kappa_1}{\kappa_2}}{\kappa_1\kappa_2^2\tau_1Da\left(\tanh(\kappa_2a)+\frac{\kappa_1}{\kappa_2}  \right)-\tanh(\kappa_2 a)}
\end{equation}
with  $ \kappa_1=\frac{\sqrt{\tau_2^2V^2+3\tau_1D}}{\tau_2V\sqrt{D\tau_1}}$ and 
$\kappa_2=\frac{\sqrt{3}}{V \tau_2 }$.
It can be shown that $t_m$ only weakly depends on $\tau_1$, which indicates that  
this variable will be less important than $\tau_2$ in the minimization of the search time.
The  relevant order of magnitude for $\tau_1^{opt}$ can be evaluated by comparing the  
typical diffusion length $ L_{\rm diff} = \sqrt{6 D t}$ and the typical ballistic length $ L_{bal}=V t$. 
An  estimate of the optimal time $\tau_1^{opt}$  can be given   
by the time scale for which those lengths are of same order, which gives~:
\begin{equation}\label{tau13dvd}
\tau_1^{opt} \sim  \frac{6 D}{V^2}.
\end{equation}
In turn, the minimization of $t_m$ leads to~:
\begin{equation}\label{tau23dvd}
\tau_2^{opt} =\frac{\sqrt{3}a}{Vx},
\end{equation}
with $x$ solution of~:
\begin{equation}
 2 \tanh(x)-2 x+x \tanh(x)^2=0.
\end{equation}
This finally yields~:
\begin{equation}\label{3dvdtau2opt}
 \tau_2^{opt} \simeq 1.078 \frac{a}{V}.
\end{equation}
Importantly this approximate expression  is very close to the expression obtained for the static mode 
($\tau_2^{opt} = \sqrt{\frac{6}{5}}\frac{a}{V} \simeq 1.095 \frac{a}{V}$) \refm{tau23dvk}, and 
there is no dependence with the typical distance between targets $b$. 
The simplified expression of the minimal $t_m$ can then be obtained as: 
\begin{equation}\label{tmopt3dvd}
 t_m^{opt} = \frac{b^3 x^2}{\sqrt{3}a^2 V} \left(x-\tanh(x) \right)^{-1} \simeq 2.18 \frac{b^3}{a^2V},
\end{equation}
and the gain reads: 
\begin{equation}\label{3dvdgain}
 gain=\frac{t_{\rm diff}}{t_m^{opt}} \simeq 0.15 \frac{aV}{D},
\end{equation}
where the search time  without intermittence is given by $t_{\rm diff}\simeq b^3/(3Da)$ for $b\gg a$ \cite{Redner}.

\doublimage{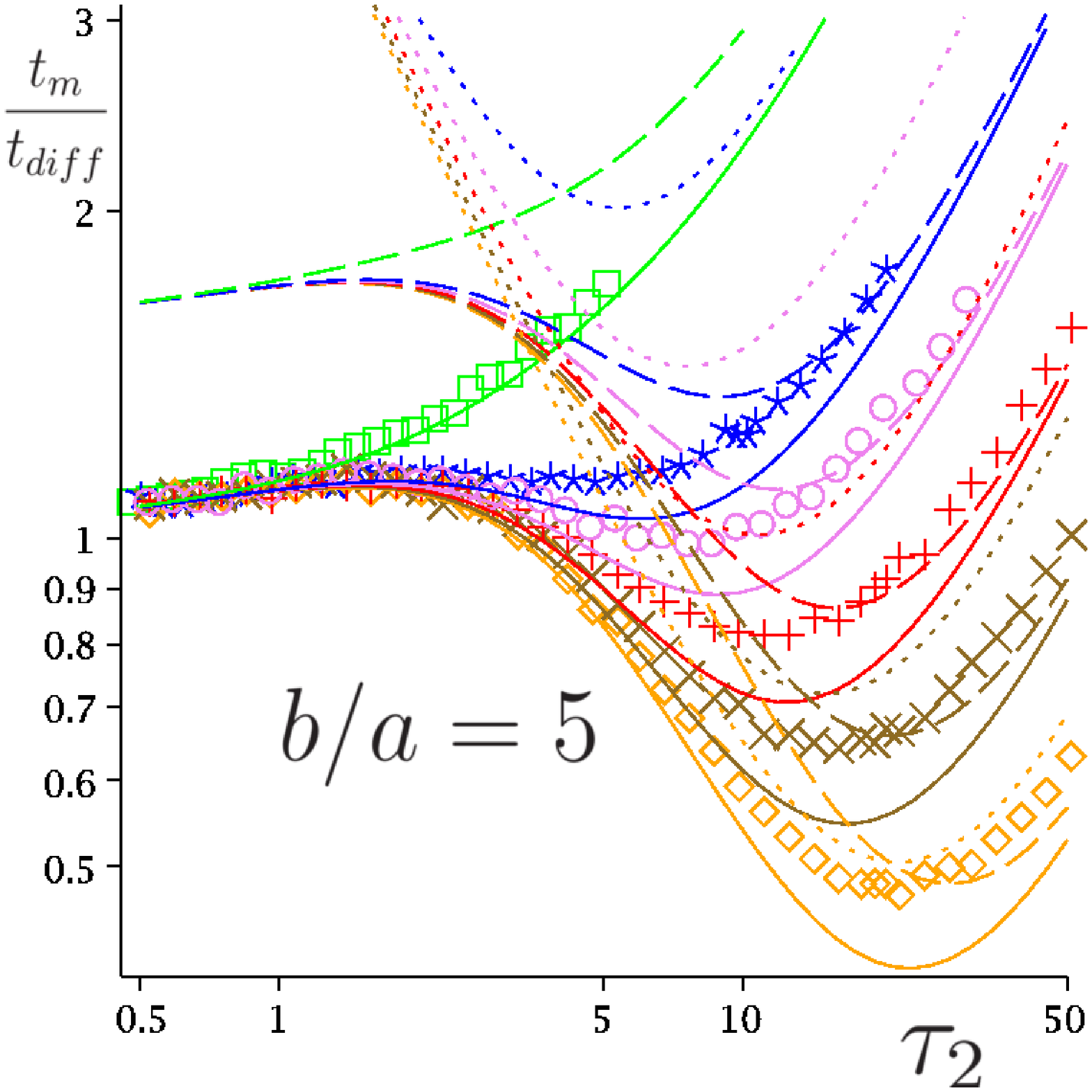}{}
{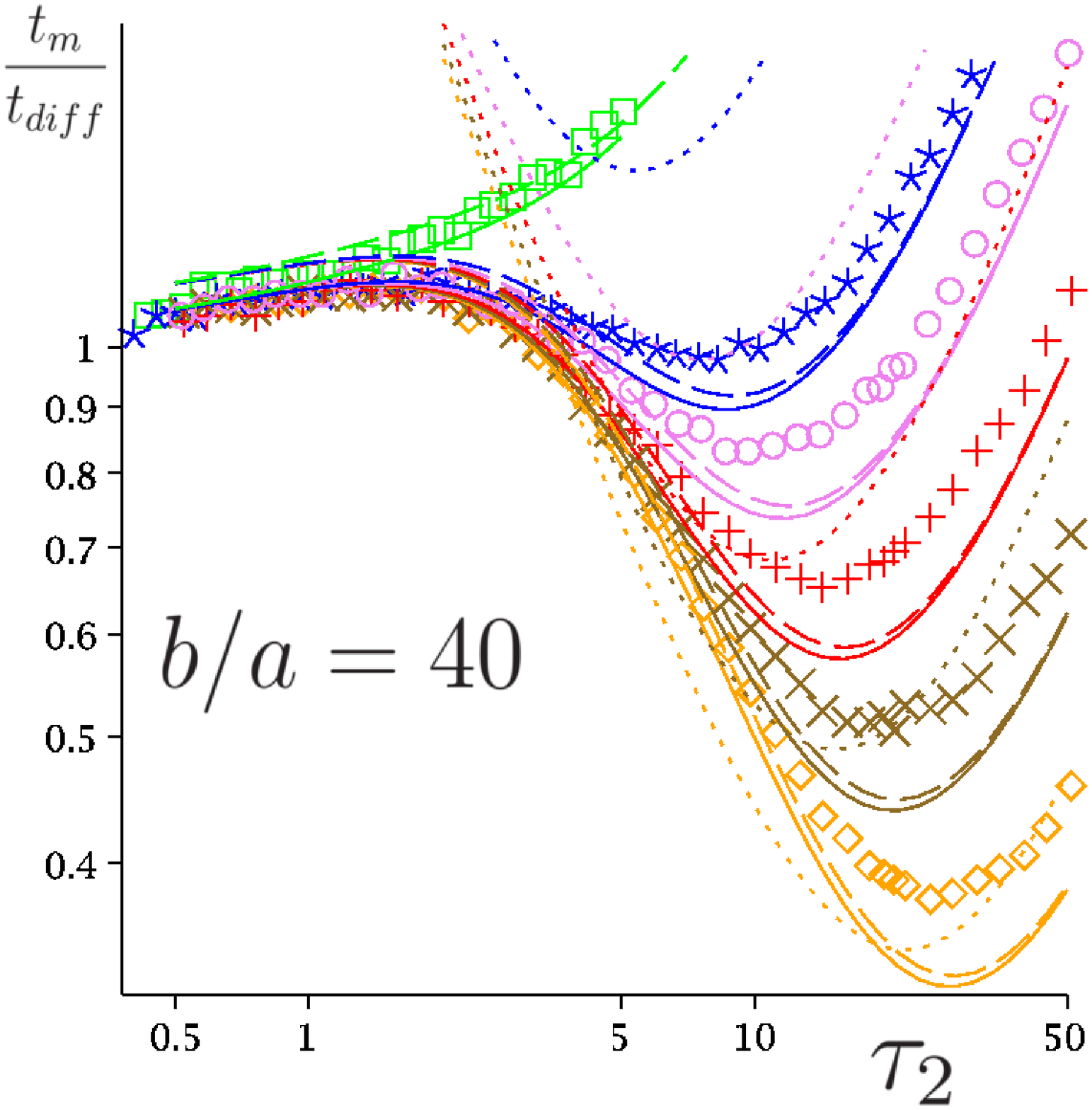}{}
{Diffusive mode in 3 dimensions. $t_m /t_{\rm diff}$ as a function of $\tau_2$ for different values of the ratio $b/a$ (logarithmic scale). 
The full analytical form given in \cite{LeGros}  (plain lines) is plotted against 
the simplified expression \refm{tmb3dvd} (dashed lines), the simplified expression with $\tau_1=0$   (dotted line),
 and  numerical simulations (symbols) for the following values of the parameters (arbitrary units): 
$a=1$ (green, $\square$), $a=5$ (blue, $\star$), $a=7$ (purple, $\circ$), $a=10$ (red, $+$), $a=14$ (brown, $\times$), 
$a=20$ (orange, $\diamond$). $\tau_1=6$ everywhere except for the small dots, $V=1$, $D=1$. $t_m /t_{\rm diff}$ presents a minimum only for $a>a_c\simeq 4$. 
}{3dvd_courbes}

There is a range of parameters for which intermittence is favorable, as indicated by  \refi{3dvd_courbes}.
Both the analytical expression for $t_m^{opt}$ in the regime without intermittence 
and with intermittence \refm{tmopt3dvd} scale as $b^3$. 
However, the dependence on $a$ is different. 
In the diffusive regime, $t_m \propto a^{-1}$,
whereas in the intermittent regime $t_m \propto a^{-2}$.
This enables us to define   a critical $a_c$, such that  
when $a>a_c$, intermittence is favorable: 
 $a_c \simeq 6.5 \frac{D}{V}$ is the value for which the gain \refm{3dvdgain} is 1.

\subsubsection{Ballistic mode}

We now discuss the last case, where the detection phase 1 is modeled by a ballistic mode \refi{modes_generic}. 
Since an explicit analytical determination of the search time seems out of reach, 
a numerical exploration of  the parameter space is needed to identify 
the regimes where the search time can be minimized. 
Approximation schemes are then developed  in each  
regime to obtain analytical expressions (more details are given in \cite{LeGros}).

\po{Numerical study}

The numerical analysis performed in \cite{LeGros} puts forward  two strategies minimizing the search time, 
depending on a critical value $v_l^{c}$ to be determined later on~:
\begin{itemize}
 \item when $v_l > v_l^{c}$, $\tau_1^{opt} \to \infty$ and $\tau_2^{opt} \to 0$. 
In this regime   intermittence is not favorable.
\item when $v_l < v_l^{c}$, $\tau_1^{opt} \to 0$,  and $\tau_2^{opt}$ is finite. 
In this regime the optimal strategy is  intermittent.
\end{itemize}

\po{Regime without  intermittence (single state ballistic searcher)~: $\tau_2 \to 0$}\label{r3dvv1}

Following the same argument as in two dimensions, without intermittence the best strategy is obtained in
the limit $\tau_1 \to \infty$  in order to minimize oversampling of the search space. 
Following the derivation of (\ref{topt2}) (see \cite{LeGros} for details), 
it is found that the search time reads~: 
\begin{equation}\label{3dvvtmsansinter}
 t_{bal}=\frac{4 b^3}{3 a^2 v_l}.
\end{equation}

\po{Regime with  intermittence }

In the regime of favorable intermittence, the numerical study suggests 
that the best strategy is realized for  $\tau_1 \to 0$.
In this  regime  $\tau_1 \to 0$, the phase 1 can be well approximated by a diffusion 
with effective diffusion coefficient $D_{\rm eff}=v_l^2\tau_1/3$ . 
The analytical expression  $t_m$ derived in \refm{tmb3dvd} can then be used, and  yields  for $\tau_1 = 0$and  $b \gg a$~: 
\begin{equation}
 t_m=\frac{b^3\sqrt{3}}{V^2\tau_2^2}\left(\frac{\sqrt{3}a}{V\tau_2}-\tanh\left(\frac{\sqrt{3}a}{V\tau_2}\right)\right)^{-1}.
\end{equation}
then one finds straightforwardly that $\tau_2^{opt} = \frac{\sqrt{3}a}{Vx}$, 
where  $x$ is solution of $x\tanh(x)^2+2\tanh(x)-2x=0$, that is $x \simeq 1.606$.
Using this optimal  value of $\tau_2$ in the expression of $t_m$, 
one finally obtains~:
\begin{equation}\label{tminter}
 t_m^{opt} = \frac{2}{\sqrt{3}}\frac{x}{\tanh(x)^2}\frac{b^3 }{a^2 V} \simeq 2.18 \frac{b^3 }{a^2 V}.
\end{equation}
These expressions show  a good agreement with numerical  simulations (see \cite{LeGros}).

\po{Discussion of the critical value $ v_l^{c}$}\label{r3dvv2}

The gain is given by~: 
\begin{equation}
gain=\frac{t_{bal}}{t_m^{opt}} \simeq 0.61 \frac{V}{v_l}.
\end{equation}
As in two dimensions, it is trivial that $v_l^{c} < V$, and the critical value
$v_l^c$ can be defined as the value of $v_l$ such that  $gain=1$.  This yields~:
\begin{equation}
v_l^{c} \simeq 0.6 V.
\end{equation}
Importantly, $v_l^c$ neither depends on  $b$ nor $a$.
Simulations are in good agreement with this result, 
except for a small numerical shift.

\subsubsection{Conclusion in dimension three}

For the three possible modelings of 
the detection mode (static, diffusive and ballistic) in three dimensions, 
there is a regime where the optimal strategy is  intermittent.  
Remarkably, and as was the case in one and two dimensions,  
the optimal time to spend in the fast non-reactive phase 2 is independent 
of the modeling of the detection mode and reads $\tau_2^{opt} \simeq 1.1 \frac{a}{V}$. 
Additionally, while the mean first passage time to the target scales as $b^3$, 
 the optimal values of the durations of the two phases do
not depend on the target density $a/b$.

\bibliographystyle{apsrmp}

\end{document}